\documentclass[rmp,aps,amsfonts,amssymb,nofootinbib,floats,twocolumn]{revtex4}

\usepackage{graphicx}
\usepackage{epsfig}
\usepackage{color}

\usepackage{bm}
\usepackage{amsmath}
\usepackage{amssymb}

\begin{document}
\title{Magnetic pyrochlore oxides}

\author{Jason S. Gardner}
\email{jsg@nist.gov}
\affiliation{NIST Center for Neutron Research, National Institute of Standards and Technology
Gaithersburg, MD 20899-6102 and Indiana University, 2401 Milo B. Sampson Lane, Bloomington, 
IN 47408-1398, USA }
\author{Michel J. P. Gingras}
\email{gingras@gandalf.uwaterloo.ca}
\affiliation{Department of Physics and Astronomy,
University of Waterloo, Waterloo, ON, N2L 3G1 
and Canadian Institute for Advanced Research, 180 Dundas St. W., Toronto, ON, M5G 1Z8, Canada}
\author{John E.  Greedan}
\email{greedan@univmail.cis.mcmaster.ca} \affiliation{Department of Chemistry
 and Brockhouse Institute for Materials Research, McMaster University, Hamilton, ON, L8S 4M1,
Canada}

\begin{abstract}

Within the past 20 years or so, there has occurred an explosion of interest in the magnetic behavior
of pyrochlore oxides of the type  $A_{2}^{3+}$$B_{2}^{4+}$O$_{7}$
where $A$ is a rare-earth ion and $B$ is usually a transition metal.
Both the $A$ and $B$ sites form a network of corner-sharing tetrahedra which is the quintessential framework
for a geometrically frustrated magnet.
In these systems the natural tendency to form long range ordered
ground states in accord with the Third Law
is frustrated, resulting in some novel short range ordered alternatives
such as spin glasses, spin ices and spin liquids and much new physics.
This article attempts to review the myriad of properties found in pyrochlore
oxides, mainly from a materials perspective,
but with an appropriate theoretical context.

\end{abstract}

\date{\today}

\maketitle

\tableofcontents

\part{Introduction}

\label{sec:intro}

\section{Overview}

Competing interactions, or frustration, are common in systems of interacting degrees
of freedom. The frustration arises from the fact that each of the interactions in
competition tends to favor its own characteristic spatial correlations. A more operational
definition classifies a system as frustrated whenever it cannot minimize its total
classical energy by minimizing the interaction energy between each pair of interacting
degrees of freedom, pair by pair.  Frustration is ubiquitous in condensed matter physics.
It arises in liquid crystals, in superconducting Josephson junction arrays
in a magnetic field, in molecular crystals such as solid N$_2$, and in magnetic thin
films. Frustration is important outside the realm of condensed matter physics in, 
for example, the competition between nuclear forces and long-range electrostatic Coulomb
interactions between protons which  is believed to lead to the formation
of a so-called ``nuclear
pasta" state of spatially modulated nuclear density in stellar interiors.
This enhances the scattering cross-section for neutrinos with nuclear
matter and may be a factor in the mechanism of supernovae explosions.
Yet, perhaps, the most popular context
for frustration is in magnetic systems. This review is concerned with the many interesting
and often exotic magnetic and thermodynamic
phenomena that have been observed over the past twenty years in a broad family of
geometrically frustrated magnetic materials, the pyrochlore oxides. 
 Before launching into the review per se,
 we first briefly outline what is meant by geometric magnetic frustration
and comment on the scientific motivation for investigating geometrically
frustrated magnetic systems.

Although the details will be provided later, here we give the reader a brief
introduction to the key issues in frustrated magnetic systems. One can
distinguish two classes of frustrated magnetic systems: those in which
the frustration is {\it geometric} and those in which it is {\it random}.  
The simplest example of geometric
frustration is that of Ising spins which can only
 point in two possible (up or down) directions,
interacting via nearest-neighbor antiferromagnetic exchange and which lie at the
vertices of an equilateral triangle as shown in Fig.~\ref{GFMinanutschell}.
 Under those conditions, it is
impossible for the three spins to align mutually antiparallel to each other.
 When many triangles
are condensed to form an edge-sharing triangular lattice, a massive level of
configurational spin disorder results, giving rise to an extensive residual zero temperature
entropy and no phase transition down to absolute
zero temperature~\cite{Wannier:1950,Houtappel:1950}.   This example is one of geometric frustration since it is the regular periodic structure of
the (here triangular) space lattice that imposes the constraint inhibiting the development
of a long range ordered state given antiferromagnetic interactions between the spins.  The same
system with Ising spins on a triangular lattice, but with ferromagnetic interactions, is not
frustrated and displays a second order phase transition to long range order at nonzero
temperature.

One can distinguish two types of random frustration. The first one arises due to competing interactions
among degrees of freedom in full thermodynamic equilibrium, as in the case of liquid crystals
and the nuclear pasta mentioned above.  Although it is not commonly referred to as such, one could
speak of ``dynamical frustration" or ``annealed frustration", by analogy to annealed disorder.
Since there is rarely only one type of interaction setting a sole characteristic length scale in
any realistic system, annealed frustration is really the norm. In this context, the most dramatic
signature of dynamical frustration occurs when multiple length scales develop, as in
modulated phases of condensed matter, for example the case of stripes in the copper
oxide superconducting materials, or in the case of reentrant phase transitions, as in some
liquid crystal materials. In other words, a system with competing interactions will often attempt
to resolve the underlying frustration by developing non-trivial spatial correlations, such as
modulated structures, whenever full dynamical equilibrium is maintained.  We will return
below to the topic of multiple dynamical degrees of freedom and annealed frustration
when we discuss the role that lattice degrees of freedom may play in geometrically
frustrated magnetic systems.

The other case of random frustration arises when the randomness is ``quenched", i.e.
frozen in.  This occurs when a subset of the degrees of freedom evolve on a time scale
that is, for all practical purposes, infinitely long compared to the predominant degrees of
freedom under focus. Consider the example above of the triangular lattice Ising
antiferromagnet.  There, the position of the atoms is quenched and only the magnetic
(spin) degrees of freedom are dynamical. This is an example of quenched 
frustration. However, the frustration in that case is said to be {\it geometric}, not random.
Quenched random frustration arises in systems where the frozen degrees of freedom (e.g.
positions of the magnetic atoms) are not related by periodic translational invariance.
One example of random frustration
is that of magnetic iron (Fe) atoms in a gold (Au) matrix. In AuFe, the
interaction between the Fe moments is mediated by conduction electrons via the
RKKY (Rudermann-Kittel-Kasuya-Yoshida) $J_{\rm RKKY}(r)$ interaction, which can be
either ferromagnetic or antiferromagnetic depending on the distance $r$ between two magnetic Fe
atoms. Depending on their relative separation distance, some trios of Fe spins will have
either one or three of their $J_{\rm RKKY}$ interactions antiferromagnetic, and will
therefore be frustrated.  Because the atomic positions are frozen-in, this is an example of
{\it quenched random frustration}. Quenched random frustration is a crucial ingredient in 
the physics of spin glass materials~\cite{Binder:1986}. These systems exhibit a transition between a
paramagnetic state of thermally fluctuating spins to a glass-like state of spins frozen in
time, but random in direction. Random frustration can also arise when the magnetic
moments in an otherwise disorder-free geometrically frustrated system are diluted, or
substituted, with non-magnetic atoms~\cite{Binder:1986,Villain:1979}.

While it is common to invoke exchange interactions between spins as the source of their
magnetic coupling, there are cases, discussed in some detail below,
where magnetostatic dipole-dipole interactions play a crucial role. Because of their 
long-range nature and angular dependence, dipolar interactions are intrinsically frustrated,
 irrespective of
the lattice dimensionality or topology. Indeed, two dipole moments
${\bm \mu}_1$ and ${\bm {\mu}}_2$ at positions ${\bm r}_2$ and ${\bm r}_1$
(${\bm r}_{12} = {\bm r}_2 - {\bm r}_1$)  reach their minimum energy for
${\bm \mu}_1$  $\parallel {\bm \mu}_2$ $\parallel {\bm r}_{12}$.
Obviously, this condition cannot be met for any system of more than two dipole
moments that are not all located on a perfectly straight line. As a result, even systems
where long-range collinear ferromagnetism is due to dipolar interactions, are in fact
frustrated, and display quantum mechanical zero point
fluctuations~\cite{Corruccini:1993,White:1993}. This example
illustrates that even systems which achieve a seemingly simple, conventional long range
order can be subject to frustration of their underlying microscopic interactions.
Being intrinsically frustrated, the introduction of randomness
via the substitution of dipole-carrying
atoms by non-magnetic (diamagnetic) species, can lead for sufficiently large dilution to a
spin glass phase~\cite{Binder:1986,Villain:1979}.

\section{Motivation for the study of frustration}

Magnetism and magnetic materials are pervasive in everyday life, from
electric motors to hard disk data storage. From a fundamental perspective, magnetic
materials and theoretical models of magnetic systems have, since the original works of
Ising and Potts, offered physicists perhaps the best test bench to investigate the
broad fundamental concepts, even at times universal ones, underlying collective
phenomena in nature. The reasons for this are threefold. Firstly, from an experimental
perspective, magnetic materials present themselves in various aspects. They can be
metallic, insulating or semiconducting. As well, the magnetic species may reside on crystalline
lattices which are spatially anisotropic, providing examples of quasi-one or
 quasi-two dimensional systems and permitting an exploration 
of the role that spatial dimensionality plays in phase
transitions. Secondly, and perhaps most importantly, magnetic materials can be investigated
via a multitude of experimental techniques that can probe many aspects of
magnetic and thermodynamic phenomena by exploring spatial and temporal correlations
over a range of several decades of length or time scales. 
Finally, from a theoretical view point, magnetic
materials can often be described by well-defined microscopic Hamiltonian models which,
notwithstanding the mathematical complexity commonly associated in solving them,
allow in principle to develop a theoretical framework that can be used to interpret experimental
phenomena. This close relationship between theory and experiment has been a key
characteristic of the systematic investigation of magnetism since its incipiency. Such
a symbiotic relationship between experiment and theory has been particularly strong in the
context of investigations of frustrated magnetism systems. Indeed, a number of
experimental and theoretical studies in frustrated magnetic systems were originally
prompted by theoretical proposals. We briefly mention two. The first pertains to the
proposal that frustrated antiferromagnets with a non-collinear ordered state may display a
transition to a paramagnetic state belonging to a different (``chiral") universality class
from conventional O(N) universality for collinear magnets~\cite{Kawamura:1988}.
The second, perhaps having stimulated the largest effort, is that some frustrated
quantum spin systems may lack conventional semi-classical long-range order altogether
and possess instead a quantum disordered ``spin liquid" state breaking no global spin or
lattice symmetries~\cite{Anderson:1973,Anderson:1987}.

\begin{figure}[t]
\begin{center}
\includegraphics[width=6cm,angle=0,clip=390]{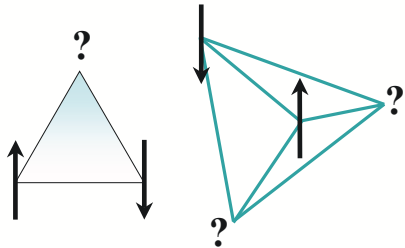}
\end{center}
\caption{Antiferromagnetically coupled spins arranged on a triangle or tetrahedron
are geometrically frustrated.}
\label{GFMinanutschell}
\end{figure}

Systems where magnetic moments reside on
lattices of corner-sharing triangles or tetrahedra are subject to geometric magnetic frustration,
as illustrated in Fig.~\ref{GFMinanutschell}, and are expected to be ideal
candidates for exhibiting large quantum mechanical spin fluctuations and the emergence of novel,
exotic, magnetic ground states.  

The cubic pyrochlore oxides, $A_2$$B_2$O$_7$, have attracted
significant attention over the past twenty years because the $A$ and $B$ ions
reside on two distinct
interpenetrating lattices of corner-sharing tetrahedra as shown in Fig.~\ref{Fig:Unit cell}.
Henceforth, we shall refer to the lattice of corner-sharing tetrahedra simply as
the pyrochlore lattice, as has become customary since the mid 1980's.
 If either $A$, $B$ or both are magnetic, and the nearest-neighbor exchange
interaction is antiferromagnetic, the system is highly geometrically frustrated. As a result,
antiferromagnetically coupled classical Heisenberg spins on the pyrochlore lattice do
not develop long range order at any nonzero temperature, opening up new avenues for novel,
intrinsically quantum mechanical, effects to emerge at low temperatures.

In all real systems, interactions beyond isotropic nearest-neighbor exchange are at
play. Exchange beyond nearest-neighbors, symmetric and antisymmetric 
(Dzyaloshinskii-Moriya) exchange, dipolar interactions and magnetoelastic couplings are all concrete
examples of corrections to the isotropic nearest-neighbor exchange part of the
Hamiltonian, $\cal H_{\rm H}$. These additional interactions, $\cal H'$, are typically weaker than
$\cal H_{\rm H}$ and, often, two or more  terms in $\cal H'$ compete to dictate the development of the
spin-spin correlations as the system is cooled from the paramagnetic phase.
The specific details
pertaining to the material under consideration determine whether
 the perturbations $\cal H'$ are able,
ultimately, to  drive the system into a semi-classical
long-range ordered state, or whether quantum fluctuations
end up prevailing.  It is for this reason
that many magnetic pyrochlore oxides have been found to
display a gamut of interesting and unconventional magnetic and thermodynamic behaviors.
Examples of phenomena exhibited by $A_2B_2$O$_7$ materials include
spin glass freezing in Y$_2$Mo$_2$O$_7$, spin liquid behavior in Tb$_2$Ti$_2$O$_7$,
disordered spin ice (not to be confused with spin glass) phenomenology in Ho$_2$Ti$_2$O$_7$
and Dy$_2$Ti$_2$O$_7$, ordered spin ice in Tb$_2$Sn$_2$O$_7$, order-by-disorder in Er$_2$Ti$_2$O$_7$, unconventional anomalous
Hall effect in metallic Nd$_2$Mo$_2$O$_7$, superconductivity in Cd$_2$Re$_2$O$_7$
and the Kondo-like effect in Pr$_2$Ir$_2$O$_7$, to name a few.

The above examples illustrate the broad diversity of phenomena exhibited by pyrochore
oxides.  There exist a number of general reviews concerned with pyrochlore structure materials
such as those by \textcite{Subramanian:1983}, \textcite{Greedan:1992a}, and \textcite{Subramanian:1993} and more focused reviews for example on the rare earth titanates, \cite{Greedan:2006}. As well, pyrochlore
structure materials are often featured in more general reviews of geometrically frustrated magnetic materials, such as those by \textcite{Ramirez:1994}, \textcite{Schiffer:1996}, \textcite{Greedan:2001} and others~\cite{Diep:1994, Diep:2004, Gaulin:1994}.  The purpose of this article is to provide, given the space limitations, a comprehensive review of both experimental and theoretical investigations of the $A_2B_2$O$_7$ materials
over the past twenty years or so. Whenever possible, we outline what we perceive as promising avenues for future work.
Thus, space limitations preclude detailed discussion of other very interesting highly frustrated magnetic materials based on
lattices of corner-shared triangles such as kagome systems and garnets, or the spinels and
Laves phases, which are also based on the pyrochlore lattice. However, we will refer to those materials in specific contexts
whenever appropriate in order to either contrast or relate to the behavior observed in the
pyrochlore oxides.

This review is organized as follows: after a brief historical overview and some generalities on
geometrically frustrated magnets we review and discuss
experimental data from the most static (i.e. long range ordered phases)
to the most dynamic (i.e. spin liquid phases). Then follows a brief
section on the superconducting compounds found among pyrochlore oxides,
although frustration has not been demonstrated to play a role
here.  Finally, we report on salient results from experiments performed under
the influence of applied magnetic field or pressure.
We conclude the review by a brief discussion of perceived avenues for future research.

\section{Historical Perspective}

The term frustration, meaning the inability of a plaquette of spins with an
odd number of antiferromagnetic bonds to adopt a minimum energy collinear spin
configuration, first appeared in a paper on spin glasses by \textcite{Toulouse:1977}  and soon after,
 in one by  \textcite{Villain:1977}.  However, it is said that
the concept of frustration as a key ingredient in the physics of spin glasses was first
introduced by Anderson who apparently wrote on a blackboard at a summer school in
Aspen in 1976 {\it ``Frustration is the name of the game''}
~\cite{Chandra:1995,Harrison:2004}.  It is interesting that the notion of frustration, random frustration to be more precise, had to await the discovery of spin glasses before it was introduced in magnetism.

Indeed, geometrically frustrated antiferromagnets, without being labeled as
such, had been studied much earlier. In 1950, \textcite{Wannier:1950} and \textcite{Houtappel:1950}
investigated the exact solution of the Ising antiferromagnet on the triangular lattice, finding
that no transition to long range order occurs at positive temperature and that the system possesses an
extensive ground state entropy. Soon after, following on a previous work by \textcite{Neel:1948},
\textcite{Yafet:1952} investigated the problem of antiferromagnetic spin arrangements in spinel ferrite
materials and the role of magnetic interactions between the octahedral and the tetrahedral sites in
determining the ground state spin structure, where the octahedral
sites form a pyrochlore lattice. Motivated by the problem of magnetic ordering for spins
on the octahedral sites of normal spinels and the related one of ionic ordering in inverse spinels,
\textcite{Anderson:1956} investigated the physics of antiferromagnetically coupled Ising spins on what we now recognize as the pyrochlore lattice. This work appears to have been the first to identify clearly
the peculiar magnetic properties of this system and, in particular, noted
the existence of a macroscopic number of
ground states, and the connection with the problem of the zero point proton entropy in
hexagonal water ice previously investigated by \textcite{Pauling:1935}.  We will return to the
connection between geometrically frustrated Ising spins on the pyrochlore lattice and the
water ice problem later.

Quantum mechanics made its first noticeable entry in the realm of geometrically
frustrated magnetism in 1973 when \textcite{Anderson:1973} proposed that a regular
two-dimensional triangular lattice of spins $S=1/2$ coupled via nearest-neighbor
antiferromagnetic exchange could, instead of displaying a three sublattice 120 degree N\'eel
ordered state, possess an exotic {\it resonating valence bond} liquid of spin singlets. 
This proposal was further examined and its conclusion supported by more detailed
calculations by~\textcite{Fazekas:1974}.
While \textcite{Anderson:1973} had written that there were no experimental realizations of an $S=1/2$ antiferromagnet with isotropic exchange, the situation would change dramatically almost twenty-five years later with the 1987 discovery of the copper-oxide based high-temperature
superconductors, some of which possess a parent insulating state with
antiferromagnetically coupled $S=1/2$ spins~\cite{Anderson:1987}.

The discovery of spin glasses by \textcite{Cannella:1972}
prompted a flurry of experimental and theoretical activity aimed at understanding the
nature of the observed spin freezing and determining whether a true thermodynamic
{\it spin glass} phase
transition occurs in these systems. Particularly relevant to this review,
\textcite{Villain:1979} investigated the general problem of insulating
spin glasses and the role of weak dilution of the $B$-site pyrochlore lattice in spinels.
The extensive ground state degeneracy characterizing this system for
classical Heisenberg spins interacting via nearest-neighbor antiferromagnetic exchange
$J$ was noted. He speculated that such a system would remain in a paramagnetic-like state down to
zero temperature, giving rise to a state at temperatures $T\ll J$ that he called a
{\it collective paramagnet} $-$ a classical spin liquid of sorts.
\textcite{Villain:1979} also considered the case of tetragonal spinels where there are two exchange
couplings, $J$ and $J'$, perpendicular and parallel to the tetragonal axis.
More recently, theoretical~\cite{Yamashita:2000,Tchernyshyov:2002}
 and experimental~\cite{Lee:2000}  studies have investigated the cubic to tetragonal distortion
in spinels, which leads to a relief of the magnetic frustration and allows the system
to develop long range magnetic order.
 Villain discussed how the effect of a {\it weak}
dilution of the magnetic moments may be dramatically different in the case of cubic
spinels compared to those with a slight tetragonal distortion. In particular, he argued that
in the case of the cubic spinel, weak dilution would leave the system a collective
paramagnet, while for the tetragonal spinel the ground state would be glassy.

While there had been some experimental studies on non-collinear magnets
between the work of Villain and the mid-to-late 1980's~\cite{Coey:1987},
the high level of research activity in the field of frustrated magnetism really only
started following the discovery of the unconventional magnetic
properties of SrCr$_{9-x}$Ga$_{3+x}$O$_{19}$
(SCGO) by \textcite{Obradors:1988}.
In the context of oxide pyrochlores, the topic of this review, experimental and theoretical interests developed in the mid-1980's with reports of spin glass behavior in the
apparently well-ordered material Y$_2$Mo$_2$O$_7$ and accelerated rapidly
after 1990.

\section{Theoretical Background}

\label{Theory}

The amount of theoretical work dedicated to the topic of highly frustrated systems, particularly quantum spin systems, has exploded over the past ten years. It is beyond our scope
here to discuss more than a fragment and in fact the field is so active that
a dedicated review would at this time be warranted.
Short of such being available, we refer the reader to a somewhat recent book edited by \textcite{Diep:2004} and a forthcoming one by edited by \textcite{Mila:2008}.
In this section, we merely present the simplest and
 most generic theoretical arguments  of relevance to the pyrochlore systems.
The focus of the discussion is on
the effects of perturbations on a
nearest-neighbor Heisenberg pyrochlore antiferromagnet.

One simple and highly convenient measure of the level of frustration in
magnetic system is the  so-called {\it frustration index} $f$ ~\cite{Ramirez:1994}, defined as
\begin{equation}
f \equiv \frac{ \vert \theta_{\rm CW} \vert}{T^*} \; .
\end{equation}
\noindent Here $\theta_{\rm CW}$ is the Curie or Curie-Weiss temperature,
 defined from the high temperature paramagnetic
response of the system, i.e. the high temperature, linear part of the inverse DC susceptibility.
The temperature $T^*$ is
the critical temperature $T_c$ (or N\'eel temperature, $T_{\rm N}$)
at which the system ultimately develops long range spin order.
In the case of freezing into a glassy state, which may correspond
to a genuine thermodynamic spin glass transition, $T^*$ would be
the freezing temperature $T_f$.
The more frustrated a system is, the lower $T^*$ is  compared to $\theta_{\rm CW}$,
 hence an ideal spin liquid system would have $f=\infty$.
In most real materials there are, however, perturbative interactions, $\cal H'$
that ultimately intervene at low temperature and lead to the development of
order out of the spin liquid state. The rest of this section discusses some
of the most common perturbations.

\textcite{Anderson:1956} and \textcite{Villain:1979} were the first
to anticipate the absence of long range order at nonzero temperature
 in  the Ising ~\cite{Anderson:1956} and Heisenberg~\cite{Villain:1979}
pyrochlore antiferromagnet. The Hamiltonian for the Heisenberg
antiferromagnet model is
\begin{equation}
{\cal H}_{\rm H}= -J \sum_{\langle i,j\rangle} {\bm S}_i \cdot {\bm S}_j	\; ,
\label{HHeis}
\end{equation}
where we take for convention that $J<0$ is antiferromagnetic while $J>0$ is ferromagnetic.
The spins ${\bm S}_i$ can be, at least within the model, taken either
as classical and of fixed length $\vert \bm S_i \vert=1$ or quantum.
A model with $n=1$, $n=2$ or $n=3$ components of ${\bm S}$ 
corresponds to the Ising, XY and Heisenberg model, respectively.
Since the pyrochlore lattice has cubic symmetry, the Ising and XY models with a uniform
global Ising easy axis direction $\hat z$ or XY global plane are not physical.
One can however, define local $\hat z_i$ Ising directions or local XY planes with
normal $\hat z_i$ that are parallel to the four cubic $\langle 111\rangle$
directions.

A mean field theory calculation for the Heisenberg Hamiltonian  by
\textcite{Reimers:1991a}  provided indirect confirmation of Anderson
and Villain's result by finding that a nearest-neighbor model with
classical Heisenberg spins on the pyrochlore lattice
has two (out of four) branches of exactly degenerate
soft (critical) modes throughout the Brillouin zone.
In other words, there is an extensive number of soft modes in this
system and no unique state develops long range order at the
mean field critical temperature. In early Monte Carlo simulations,
\textcite{Reimers:1992} found no sign of a transition
in this system, either to long range order or to a more complex type, such as nematic order,
down to a temperature of $J/20$, where $J$ is the nearest-neighbor
antiferromagnetic exchange constant.  This conclusion was confirmed in numerical studies by
\textcite{Moessner:1998a,Moessner:1998}.
This behavior of the nearest-neighbor pyrochlore antiferromagnet
differs markedly from that of the  nearest-neighbor
classical Heisenberg antiferromagnet on the two-dimensional kagome lattice where co-planar order-by-disorder develops as the temperature drops below $\sim J/100$ \cite{Chalker:1992,Reimers:kagome}.
However, the Mermin-Wagner-Hohenberg theorem forbids a
phase transition at nonzero temperature in the kagome Heisenberg antiferromagnet.
While no transition is seen in the pyrochlore Heisenberg
system ~\cite{Reimers:1992,Moessner:1998a,Moessner:1998},
the pyrochlore lattice with easy-plane (XY) spins, where
the easy planes are perpendicular to the local
$\langle 111\rangle$ directions, displays a phase transition to long range
order ~\cite{Bramwell:1994,Champion:2003}.
Finally, in agreement with Anderson's prediction ~\cite{Anderson:1956},
Monte Carlo simulations find that the Ising antiferromagnet, or equivalently the
Ising ferromagnet with local $\langle 111\rangle$ directions,
which is the relevant model to describe spin ice materials,
does not order down to the lowest temperature ~\cite{Harris:1998}.

\textcite{Moessner:1998a,Moessner:1998} investigated
the conditions under which an $n$-component classical spin ${\vec S}_i$
with $\vert {\vec S}_i\vert=1$ arranged on a lattice of corner-sharing
units of $q$ sites and interacting with $q-1$ neighbors in each unit
can exhibit thermally-induced order-by-disorder.
Their analysis allows one to rationalize why the Heisenberg $n=3$
kagome  antiferromagnet develops co-planar order and why
the XY pyrochlore antiferromagnet (with spins in a global XY
plane) develops collinear order ~\cite{Moessner:1998}.
From their analysis, one would predict a phase transition to
long range nematic order for $q=3,n=3$ in three spatial dimensions.
Interestingly, such a situation arises in the $S=1/2$
Na$_4$Ir$_3$O$_8$ hyperkagome ~\cite{Okamoto:2007} and Gd$_3$Ga$_5$O$_{12}$
garnet~\cite{Schiffer:1994,Schiffer:1995a,Petrenko:1998,Dunsiger:2000,Yavorskii:2006}. 
%As may have been anticipated from
%\textcite{Moessner:1998a,Moessner:1998},
%nematic order at finite temperature for a classical version
%of the hyperkagome Heisenberg antiferromagnet seems to occur~\cite{Hopkinson:2007}.
Moving away from the strictly nearest-neighbor model, one would
generally expect interactions beyond nearest-neighbors
to drive a transition to an ordered state, at least for classical spins~\cite{Reimers:1992}.
While this has been studied via Monte Carlo simulations~\cite{Reimers:crit,Mailhot:1993},
it has not been the subject of very many investigations.
In that context, we  note that a paper by \textcite{Pinettes:2002} studies a
classical Heisenberg pyrochlore
antiferromagnet with an additional exchange interaction
$J'$ that interpolates between the pyrochlore lattice $(J'=0)$
and the face-centered cubic lattice $(J'=J)$.
It is found that for $J'/J$ as low as $J'/J \sim 0.01$,
the system orders into a collinear state whose energy is greater
than the classical incommensurate ground state predicted by mean field theory.  
A Monte Carlo study of another classical model of a pyrochlore lattice
with antiferromagnetic nearest-neighbor exchange and
ferromagnetic next nearest-neighbor exchange finds an ordered
state with mixed ordered and dynamical character~\cite{Tsuneishi:2007}.
Because the pyrochlore lattice lacks bond inversion symmetry,
Dzyaloshinsky-Moriya (DM) spin-spin
interactions arising from spin-orbit coupling
are allowed by symmetry and can give rise to various ordered
states depending on the orientation of the DM vector~\cite{Elhajal:2005}.

As we will discuss below,
a number of pyrochlore materials exhibit glassy behavior where the
spins freeze at low temperature into a state where their orientation
is random in space, but frozen in time ~\cite{Binder:1986}. It is not yet fully understood whether this glassy behavior is due to the intrinsic nature of the systems considered, as is the case for the spin ices, or if it originates from random disorder (e.g.~antisite disorder, random bonds or other) as
in conventional random disorder spin glasses~\cite{Binder:1986}.
Only a very few theoretical studies have explored the effect of random
disorder in antiferromagnetic pyrochlores with classical Heisenberg
spins and with small and homogoneously random deviations of exchange $J_{ij}$
from the average antiferromagnetic exchange $J$~\cite{Bellier:2001,Saunders:2007}.
Interestingly, random bonds on a single tetrahedron lift the degeneracy
and drive an order-by-(random)-disorder to a collinear spin arrangement.
However, on the lattice, the system remains frustrated and there is no
global order-by-disorder~\cite{Bellier:2001}.  Considering this model in more detail,
\textcite{Saunders:2007} find compelling evidence of
a conventional spin glass transition. \textcite{Sagi:2005} have investigated
how coupling with the spins to the lattice may lead to a freezing behavior
and trapping of the spins into metastable states in an otherwise disorder-free
nearest-neighbor classical pyrochlore antiferromagnet.
\textcite{Villain:1979} discussed the role of non-magnetic impurities
on the pyrochlore lattice antiferromagnet.
He argued that the collective paramagnetic behavior of the otherwise
disorder-free system should survive up to a finite  concentration
of impurities. A similar conclusion was reached on the basis of
heuristic arguments and simulations for the site-diluted two-dimensional
kagome  Heisenberg antiferromagnet~\cite{Shender:1993}.
It would be valuable
if the problems of site-disorder and dilute random bonds
in the antiferromagnetic pyrochlore lattice were explored
in more detail.  We note in passing that the problem of disorder in  quantum variants
of the kagome Heisenberg antiferromagnet has attracted some attention~\cite{Lauchli:2007}.

In insulating $A_2$$B_2$O$_7$, where $A$ is a 4f rare-earth trivalent
 ion, the inter-rare-earth exchange interactions
are often small, contributing only $10^0-10^1$~K to
the Curie-Weiss $\theta_{\rm CW}$ temperature.
This is because the unfilled magnetically active 4f orbitals are strongly shielded by the 5s, 5p
and 5d orbitals and their direct exchange overlap or superexchange overlap is small.
At the same time, trivalent rare earths often possess large magnetic moments and hence the magnetostatic dipole-dipole energy scale is significant compared to $\theta_{\rm CW}$ and
needs to be incorporated into Eq.~(\ref{HHeis}),

\begin{eqnarray}
{\cal H}_{\rm int} & = &
-\frac{1}{2}\sum_{(i,j)}
 {\cal J}_{ij}\, {\bm J}_i \cdot {\bm J}_j + \\ \nonumber
        &  & \left ( \frac{\mu_0}{4\pi} \right)
        \frac{{(g_{\rm L}\mu_{\rm B}})^2}   {2 {r_{\rm nn}}^3 }
        \sum_{(i,j)}
        \frac{( {\bm J}_i\cdot {\bm J}_j - 3 {\bm J}_i\cdot \hat r_{ij}\hat r_{ij}\cdot {\bm J}_j)}
                {(r_{ij}/r_{\rm nn})^3} \; .
\label{Hdip}
\end{eqnarray}

Here we have taken the opportunity to generalize Eq.~(\ref{HHeis}) to the case where
the total angular momentum ${\bm J}_i={\bm L}_i+{\bm S}_i$ is included.
$g_{\rm L}$ is the Land\'e factor and
${\bm r}_j - {\bm r}_i = \vert {\bm r}_{ij} \vert \hat r_{ij}$,
where ${\bm r}_i$ is the position of magnetic ion $i$.
In general, one needs to include the role of the single-ion crystal field, ${\cal H}_{\rm cf}$,
to the full Hamiltonian, ${\cal H}={\cal H}_{\rm cf}+{\cal H}_{\rm int}$.

A mean field study of ${\cal H}_{\rm int}$, with ${\cal H}_{\rm cf}=0$ and
with only nearest-neighbor antiferromagnetic ${\cal J}_{\rm nn}$ and
dipolar interactions of approximately 20\% the strength of
${\cal J}_{\rm nn}$, was done by \textcite{Raju:1999} to investigate the type of order
expected in Gd$_2$Ti$_2$O$_7$. The assumption
${\cal H}_{\rm cf}=0$ is a reasonable first approximation
for Gd$_2$Ti$_2$O$_7$ since the electronic
ground state of Gd$^{3+}$ is $^8$S$_{7/2}$ with $L=0$.
\textcite{Raju:1999} argued that this model
has an infinite number of soft modes with
arbitrary ordering wave vector along $[111]$ and that the transition
at $\sim 1$~K \  could possibly proceed via order-by-disorder.
However, \textcite{Palmer:2000} argued that quartic terms in the
mean field Ginzburg-Landau theory would select
a ${\bm k}_{\rm ord}=000$ ordering wavevector
at the critical temperature $T_c$ that
also corresponds to the zero temperature classical ground state.
Subsequent mean field calculations~\cite{Cepas:2004,Enjalran:2003} showed that there are in fact
two very closely spaced transitions and that the transition
at the highest temperature is at a unique ordering wave vector
$( {\bm k}_{\rm ord}=\frac{1}{2}\frac{1}{2}\frac{1}{2} )$ that had been missed
in previous calculations ~\cite{Raju:1999,Palmer:2000}.
A Monte Carlo simulation~\cite{Cepas:2005} found a single first order
transition.
It appears likely that the small temperature separation between
the two transitions that is
predicted by mean field theory is not
 resolved in the Monte Carlo simulation.
Irrespective of the subtleties associated with the transition,
it is generally accepted that the classical ground state of
the nearest-neighbor pyrochlore Heisenberg antiferromagnet with
weak dipolar interactions is the so-called Palmer-Chalker
ground state~\cite{Palmer:2000,Cepas:2004,Enjalran:2003,DelMaestro:2004, Wills:2006,DelMaestro:2007}.
Starting from the Palmer-Chalker ground state, a $1/S$ expansion calculation
shows that all conventional magnetic spin-wave-like
excitations of this system are gapped and that the $1/S$ quantum
fluctuations are negligible~\cite{DelMaestro:2004,DelMaestro:2007}.
For Ho and Dy-based pyrochlore oxides, the crystal field Hamiltonian
${\cal H}_{\rm cf}$ produces a single-ion ground state doublet and the
system maps onto a classical dipolar Ising model. We shall return to
this model when reviewing the problem of spin ice in
Section~\ref{sec:spin-ice}.

\begin{figure}[t]
\begin{center}
\includegraphics[width=8cm,angle=0]{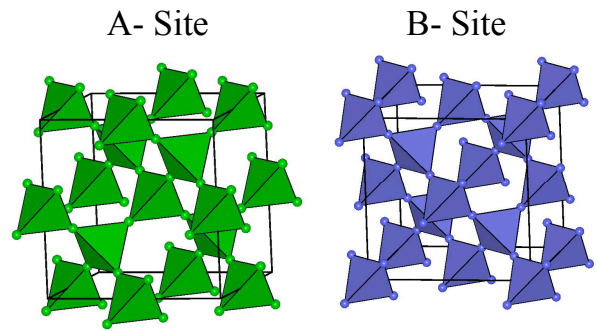}
\end{center}
\caption{The $A$ and $B$ sublattices of the cubic
 pyrochlore materials. Either or both sublattices can be magnetic.}
\label{Fig:Unit cell}
\end{figure}

Another important perturbation of relevance to real materials
is the coupling between spin and lattice degrees of freedom. Spin-lattice coupling
can lead to a redistribution of the frustration among the
magnetic bonds such that  the energy cost for distorting the lattice is more than compensated by the energy difference between that of the more strongly coupled
antiparallel spins on the compressed bonds compared to the
higher energy (frustrated, parallel) spins on the dilated
bonds~\cite{Yamashita:2000,Tchernyshyov:2002}.  It is anticipated that such coupling can lead to a global lifting
of magnetic degeneracy and drive a transition to long range magnetic
order that is accompanied by a cubic to tetragonal lattice deformation.
\textcite{Villain:1979} had also discussed a number of significant differences in the thermodynamic properties and the response to random disorder of a tetragonal spinel compared to
that of a cubic spinel.  The problems of magnetic order, lattice distortion, coupling to an external
magnetic field and magnetization plateaus in oxide spinels are
currently very topical ones~\cite{Lee:2000,Tchernyshyov:2002,Penc:2004,Bergman:2006}.
However, in pyrochlore oxides, interest in the role of the lattice
degrees of freedom is just starting to attract attention~\cite{Sagi:2005,Ruff:2007}.

We close with a brief discussion of the quantum $S=1/2$ pyrochlore Heisenberg antiferromagnet.
This is obviously an extremely difficult problem and there is no concensus
yet about the nature of the ground state.
It appears to have been first studied by \textcite{Harris:1991} who found
indicators that the system may form a dimerized state at low temperature.
By applying perturbation theory to the density matrix operator,
\textcite{Canals:1998,Canals:2000} found evidence for a quantum spin liquid
in the $S=1/2$ pyrochlore antiferromagnet.
From  exact diagonalizations on small clusters, they also found that the
singlet-triplet gap is filled with a large number of singlets, similar
to what is observed in the $S=1/2$ kagome lattice
 antiferromagnet~\cite{Lecheminant:1997}.
Calculations using the contractor-renormalization (CORE)
 method~\cite{Berg:2003} find a ground state that breaks lattice symmetry,
analogous to that suggested by \textcite{Harris:1991},
and with a singlet-triplet gap filled with singlets, similar
to \textcite{Canals:1998}.
Moving away from strictly numerical approaches,
analytical calculations based on large-$N$
~\cite{Moessner:2006,Tchernyshyov:2006}
and large-$S$ expansions~\cite{Hizi:2007}
as well as variational calculations away from an exact dimer
covering ground state solution~\cite{Nussinov:2007} have
all been very recently carried out.

\part{Materials}

\label{Section:Mat}

While this review is focused on the three dimensional
pyrochlore lattice, which is shown in Fig. \ref{Fig:Unit cell}, there are of course other
lattice types which provide the conditions for geometric frustration.
Among these, in two spatial dimensions are the edge-sharing and
corner-sharing triangular lattices - the latter best known as the
Kagome lattice - and in three spatial dimensions one has
garnets, langasites and face centered cubic lattices among others.
There exists a considerable and important literature devoted to these
materials which lies, unfortunately, outside the scope of this work.

A network of corner-sharing tetrahedra is found in several common mineral types in nature.
The octahedrally coordinated $B$-site of the spinel
class of minerals ($AB_2$O$_4$) forms a network of
corner-sharing tetrahedra.  The $A$- and the $B$-site can be
 filled by many ions, but when the $B$-site is occupied by
a trivalent magnetic transition metal ion, magnetic frustration
 can play a role in its bulk properties.  The trivalent chromates
have recently sparked considerable
interest~\cite{Ueda:2005,Lee:2000,Lee:2002,Matsuda:2007}
in this area of magnetic frustration.  Another common family of materials
 that contain a network of corner-sharing tetrahedra are the cubic
Laves phases, such as YMn$_2$~\cite{Ballou:2001,Shiga:1993}.

As already emphasized, in this review, we will restrict ourselves
to the cubic pyrochlore oxides with the general formula $A_2$$B_2$O$_7$,
where $A$ is a trivalent rare earth which includes the lanthanides,
Y and sometimes Sc, and $B$ is either a transition metal or
a p-block metal ion.  It should be mentioned here that it is possible to form 
$A^{2+}_2$$B^{5+}_2$O$_7$ (2+,5+)
pyrochlores, however these are not  common and have not been studied in great detail.
Most of this manuscript will deal with the (3+,4+) variety of pyrochlores,
 but when appropriate the (2+,5+) will be discussed.

\section{Crystal Structure}

\label{Section:Str}

\subsection{Space Group and Atomic Positions}

Pyrochlore materials take their name from the mineral NaCaNb$_2$O$_6$F pyrochlore,
the structure of which was first reported by~\textcite{Gaertner:1930}. 
The name, literally
``green fire'', alludes to the fact that the mineral shows a green
color upon ignition.  Most synthetic pyrochlores of interest here are oxides that
crystallize in the space group $Fd \bar{3}m$ (No. 227).
 As pointed out by \textcite{Subramanian:1993},
confusion can exist due to the fact that there exist four
possible choices for the origin. Currently, standard practice
is to formulate oxide pyrochlores as $A_2$$B_2$O$_6$O$'$
and to place the $B$ ion at $16c$ (the origin of the
second setting for $Fd \bar{3}m$ in the
International Tables of Crystallography), $A$ at $16d$,
 O at $48f$ and O$'$ at $8b$ as shown in Table \ref{Table:spacegroup}.
Note that there is only one adjustable positional parameter  $x$
 for the O atom in $48f$. There are at least two important
 implications of these structural details.  One concerns
the topology of the
$16c$ and $16d$ sites and the
other the coordination geometry of the O ligands about
the two metal sites. Both the $16c$ and $16d$
sites form a three dimensional array of corner-sharing
tetrahedra as shown in Fig. \ref{Fig:Unit cell}, thus
giving rise to one of the canonical geometrically frustrated lattices.

\begin{table}
  \begin{tabular}{||c|c|c|c||}
    \hline
  Atom & Wyckoff & Point  &  Minimal \\
  & Position & Symmetry & Coordinates\\ \hline
   $A$  & $16d$ & $\bar{3}$m (D$_{3d}$) & ${1}\over{2}$, ${1}\over{2}$, ${1}\over{2}$ \\ \hline
   $B$  & $16c$ & $\bar{3}$m (D$_{3d}$)  & 0, 0, 0\\ \hline
   $O$  & $48f$ & mm (C$_{2v}$) & $x$, ${1}\over{8}$, ${1}\over{8}$\\ \hline
   $O'$  & $8b$ &  $\bar{4}$3m (T$_{d}$)  & ${3}\over{8}$, ${3}\over{8}$, ${3}\over{8}$\\
    \hline
  \end{tabular}
  \caption{The crystallographic positions for the space group,
$Fd \bar{3}m$ (No. 227) suitable for the cubic
pyrochlore $A_2B_2$O$_6$O$'$ with origin at $16c$.}
  \label{Table:spacegroup}
\end{table}

\subsection{Local Environment of the $\bm A$ and $\bm B$-Sites}

\label{local-environment}

\begin{figure}[t]
\begin {center}
\includegraphics[width=4cm]{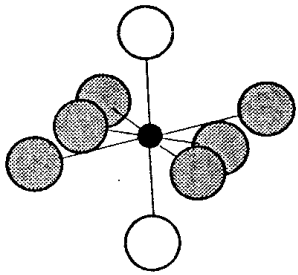}
\end{center}
\caption{The coordination geometry of the  $A$-site by O
(shaded spheres) and ${\rm O}'$(open spheres) atoms.
The ${\rm  O}' - A - {\rm O}'$ unit is oriented normal to the average plane of a  puckered six-membered ring.  This bond
is one of the shortest seen between a rare earth and oxygen ions.}
\label{Fig:A-site}
\end{figure}

The coordination geometry about the two metal sites is controlled by the value of {\it x} for the O atom in the {\it 48f} site.  For {\it x} = 0.3125 one has a perfect octahedron about {\it 16c}
and {\it x} = 0.375 gives a perfect cube about {\it 16d}.
In fact {\it x} is usually found in the range  0.320 to 0.345
and the two geometries are distorted from the ideal polyhedra.
For a typical pyrochlore the distortion about the $B$ ion or {\it 16c}
site is relatively minor. The  $\bar{3}m$(D$_{3d}$) point
symmetry requires that all six $B$ - O bonds must be of equal
length. The O - $B$ - O angles are  distorted only slightly
from the ideal octahedral values of 90$^{\circ}$ ranging
between 81$^{\circ}$ and 100$^{\circ}$. The distortion of the
A site geometry from an ideal cube is on the other hand very large.
This site is depicted in Fig.~\ref{Fig:A-site} and is best described
as consisting of a puckered six-membered ring of O atoms with two
O$'$ atoms forming a linear O$'$ - $A$ - O$'$ stick oriented normal to
the average plane of the six-membered ring.
The $A$ - O and $A$ - O$'$ bond distances are very different.
While typical $A$ - O values are 2.4 -2.5 {\AA}, in accord with
the sum of the ionic radii, the $A$ - O$'$ bonds are amongst the
shortest known for any rare earth oxide, $\approx$ 2.2 {\AA}.
Thus, the $A$-site has very pronounced axial symmetry,
the unique axis of which is along a local
$\langle 111\rangle$ direction.
This in turn has profound implications for the crystal field
 at the $A$-site which determines much of the physics found
in the pyrochlore materials.

Recall that the spin-orbit interaction is large in rare earth
ions and the total angular momentum,
${\bm J} = {\bm L} + {\bm S}$
is a good quantum number.
For a given ion one can apply Hund's rules to determine
the isolated (usually) electronic ground state.
 Electrostatic and covalent bonding effects originating
from the local crystalline environment,
the so-called crystal field (CF), lift the $2J + 1$
degeneracy of the ground state. A discussion of modern methods for calculation of the CF
for f-element ions is beyond the scope of this review but is described in several monographs
for example, one by~\textcite{Hufner:1978}.
For our purposes we will assume that the single
ion energy levels and wave functions,
the eigenvectors and eigenenergies of the CF
Hamiltonian ${\cal H}_{\rm cf}$ have been suitably determined
either through ab initio calculations or
from optical  or neutron spectroscopy (see for example \textcite{Rosenkranz:2000}
and more recently~\textcite{Mirebeau:2007}).

${\cal H}_{\rm cf}$  can be expressed either in terms of the so-called tensor operators
or the ``operator equivalents'' due to \textcite{Stevens:1952}.  
The two approaches are contrasted by ~\textcite{Hufner:1978}.  While the tensor operators are more convenient
 for ab-initio calculations, the latter are better suited to our purposes here.
In this formalism, ${\cal H}_{\rm cf}$ is expressed in terms of
polynomial functions of the $J_{iz}$ and $J_{i\pm}$, with $J_{i\pm}=J_{ix}\pm J_{iy}$,
which are components of the $\bm J_i$ angular momentum operator.
 The most general expression for ${\cal H}_{\rm cf}$ is:
\begin{equation}
{\cal H}_{\rm cf}= \sum_i \sum_{l,m} B_l^m O_l^m( {\bm J}_i)      \; ,
\end{equation}
where, for example, the operator equivalents are $O_2^0=3{J_z}^2-J(J+1)$ and
$O_6^6 = {J_+}^6 + {J_-}^6$.  The full CF Hamiltonian for -3m (D$_{3d}$) point symmetry
involves a total of six terms for $l=2,4$ and 6~\cite{Greedan:1992a}.
In fact, due to the strong axial symmetry of the $A$-site, described previously,
it can be argued that the single $l=2$ term, $B_2^0$, plays a major role
 in the determination of the magnetic anisotropy of the ground state.
In the Stevens formalism, $B_2^0$ = $A_2^0 \langle {\rm r}^2\rangle\alpha_J$(1-$\sigma_2)$,
where $A_2^0$ is a point charge lattice sum representing the CF strength, $\langle {\rm r}^2\rangle$ is the expectation value of r$^2$ for the 4f electrons, $\sigma_2$
is an electron shielding factor and $\alpha$$_J$ is the Stevens factor~\cite{Stevens:1952}.
This factor changes sign in a systematic pattern throughout the lanthanide series,
being positive for $A$= Sm, Er, Tm and Yb and negative for all others.
So, the sign of $B_2^0$ depends on the product $\alpha$$_J$$A_2^0$ and $A_2^0$
is known from measurements of the electric field gradient from
for example $^{155}$Gd M\"{o}ssbauer studies to be positive
for pyrochlore oxides \cite{Barton:1979}.  Thus, $B_2^0$ should
be positive for $A$ = Sm, Er, Tm and Yb and negative for all others.
From the form of $B_2^0$ above, it is clear that states of different
$|M_J\rangle$  do not mix and that the energy spectrum will consist
of a ladder of states with either  
 $|M_{J_{(\rm min)}}\rangle$ ($B_2^0$ $>$ 0)
or $|M_{J_{(\rm min)}}\rangle$ ($B_2^0$ $<$ 0)  as the ground state.
Note that the former constitutes an easy plane and the latter
an easy axis with respect to the quantization axis which is
$<$111$>$ for pyrochlores.  A comparison of the known
anisotropy for the $A_2$Ti$_2$O$_7$ and $A_2$Sn$_2$O$_7$ materials,
that is, easy axis for $A$ = Pr, Nd, Tb, Dy and Ho and easy plane for
$A$ = Er and Yb with the sign of  $B_2^0$ shows a remarkable agreement
with this very simple argument. 
~[Note that only the Stevens formalism works here. 
The tensor operator definition of $B_2^0$ 
is  $B_2^0=A_2^0$$\langle {\rm r}^2 \rangle$,
so this quantity is always positive for pyrochlore oxides,
 independent of the rare earth $A$. We do not maintain that
the actual ground state wave function can be obtained within such a
simple model (although the agreement for $A$ = Dy and Ho is remarkable),
rather, that the overall anisotropy can be predicted without a detailed calculation.]

\subsection{Alternative Views of the Pyrochlore Structure}

\begin{figure}
\begin {center}
\includegraphics[width=6cm]{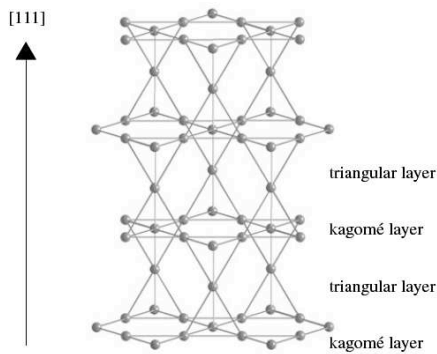}
\end{center}
\caption{Alternating kagome 
 and triangular planar layers stacked along the [111] direction of the pyrochlore lattice.}
\label{Fig:Alt_view}
\end{figure}

An important feature of the pyrochlore structure relevant to the central
theme of this review  is the fact that 
%insight provided by viewing the pyrochlore 
the {\it 16c} or {\it 16d} sites form layers
stacked along the [111] direction as shown Fig.~\ref{Fig:Alt_view}.
From this perspective, the pyrochlore lattice is seen to consist of
alternating kagome and triangular planar layers.
As well, from the viewpoint of chemical bonding, the pyrochlore structure can be
described as either an ordered defect fluorite (CaF$_2$)
or as two interpenetrating networks, one of composition $B_2$O$_6$,
which is a network of corner-sharing metal-oxygen octahedra
(topologically equivalent to that found in the perovskite structure)
and the other of composition $A_2$O$'$, which forms a
zig-zag chain through the large channels formed by the $B_2$O$_6$ network.
These models have been described in detail in several of the
previous reviews and the discussion here
focuses on a few important issues.
While there is a topological relationship with
the perovskites, the pyrochlore lattice is considerably more rigid.
For example the $B$ - O - $B$ angle in which the O atom is shared between two octahedra
is restricted to a very narrow range in pyrochlores, usually from
about 127$^{\circ}$ to 134$^{\circ}$, with only little influence from
the size of the $A$ ion. On the other hand this angle can range from
180$^{\circ}$ to 140$^{\circ}$ in perovskites and there is a strong
variation in the radius of the $A$ ion.  The interpenetrating network description allows an understanding of the formation of so-called defect or ``rattling'' pyrochlores, such
as KOs$_2$O$_6$, in which the entire $A_2$O$'$ network is missing
and where the large K$^+$ ion occupies the  $8b$ site which is
normally occupied by O$'$ in the standard pyrochlore structure.
The disordered fluorite model is useful in cases where a disordered,
defect fluorite structure phase competes with the ordered pyrochlore,
such as for the $A_2$Zr$_2$O$_7$ series where $A/B$ site mixing is observed.

\subsection{Phase Stability}

There are over 15 tetra-valent ions that can reside on the $B$-site of a cubic pyrochlore  at room temperature. Pyrochlore oxides of most interest here are formed
with a rare earth ion in the $A$-site and transition or
 p-block elements in the $B$-site. The rare earth
elements form a series in which the trivalent state
predominates and through which the ionic radius decreases
systematically with increasing atomic number (the so-called lanthanide contraction).
Nonetheless, it is uncommon to find that the pyrochlore
 phase is stable for all $A$ for a given $B$.

\begin{figure}
\begin {center}
\includegraphics[width=9cm]{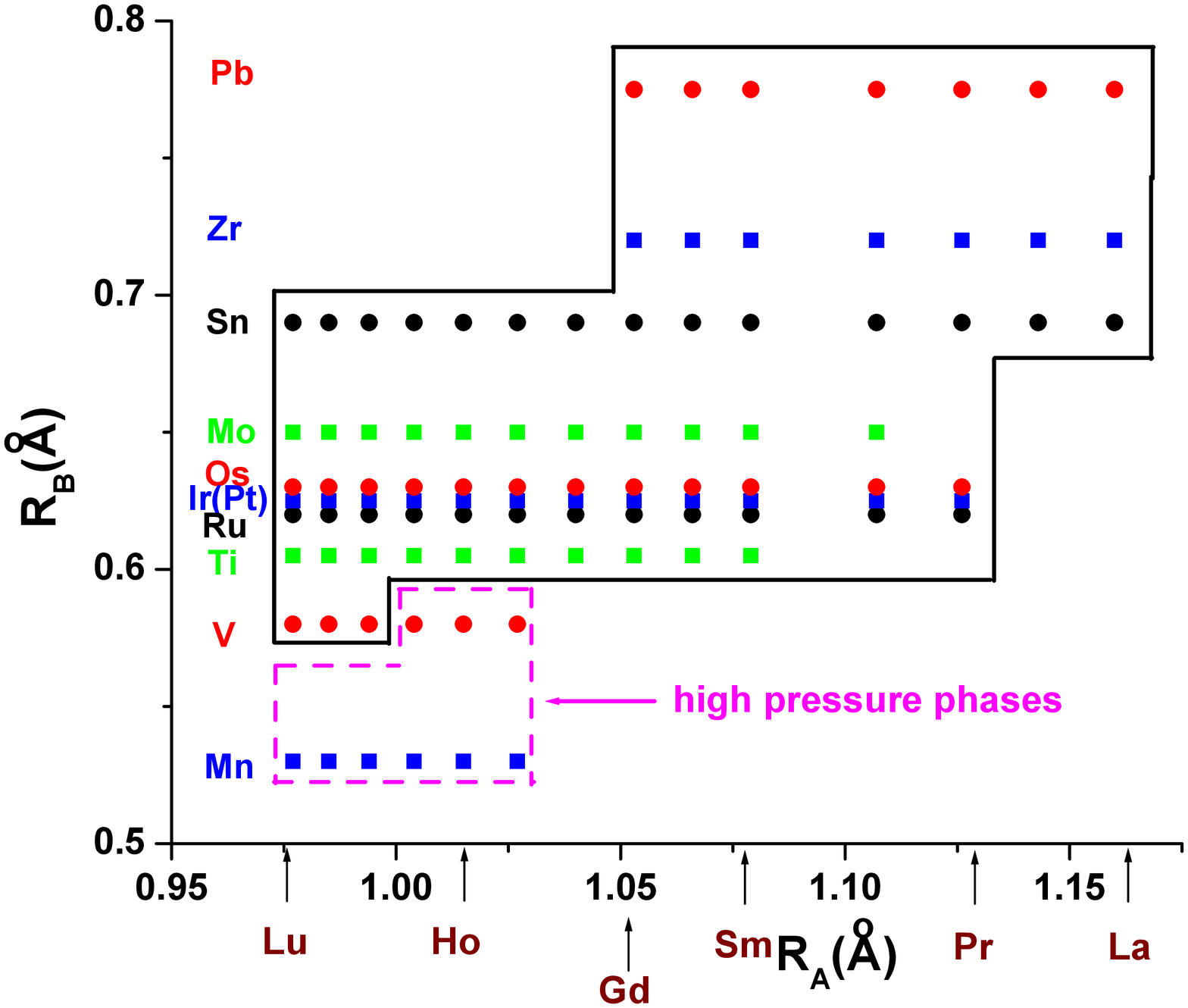}
\end{center}
\caption{A structure-field or stability-field map for $A_2$$B_2$O$_7$ materials. 
The $B$ = Pt series are also high pressure phases. Adapted from \textcite{Subramanian:1993}.}
\label{Fig:StruFie}
\end{figure}

The most efficient way to display the range of stability of pyrochlores in the space of the ionic radii is through a structure-field map or structure-stability map.  Figure \ref{Fig:StruFie} is an abridged version of a comprehensive map published by \textcite{Subramanian:1993}. First it is worth noting that the $B$ = Sn series is the only  one to form for all rare earths.  There would also appear to
 be minimal and maximal ionic radius ratios, $A^{3+}$/$B^{4+}$,
 which define the stability limits for pyrochlore phases of this type.
 For $B$ ions with very small radii, such as Mn$^{4+}$, the pyrochlore
phase can be prepared only using high pressure methods.
This is in strong contrast to perovskites of the type $A^{2+}$Mn$^{4+}$O$_3$
 which can be prepared easily under ambient oxygen pressures.
The first series stable under ambient pressure is that for $B$ = V,
but only for the three smallest members, Lu, Yb and Tm.
The series can be extended using high pressure synthesis~\cite{Troyanchuk:1990}.
For the largest $B$ ion, Pb$^{4+}$, the smallest rare earth for which a stable
pyrochlore is found appears to be Gd. Thus, an ambient pressure stability
range defined in terms of radius ratios, $A^{3+}$/$B^{4+}$, (RR) would
extend between 1.36 and 1.71 with marginal stability expected as those
limits are approached. There are exceptions, for example Pr$_2$Ru$_2$O$_7$ 
is clearly stable with a RR of 1.82,
yet Pr$_2$Mo$_2$O$_7$ does not exist with a RR = 1.73.

Also of relevance to this review are issues of defect formation in pyrochlores.
This is an active area of research in materials science where defect pyrochlores 
which show high ionic conductivities are of interest.
 While the defect levels are often small and  difficult to quantify via experiment,
 some computational guidelines have appeared recently~\cite{Minervini:2002}.
The general trend is that defect concentrations are predicted to increase as the RR
 approaches its lower limit, for example as the $B$ radius increases within any given
series, and that the predicted levels are generally of the order of 1\% or less.

\section{Sample Preparation and Characterisation}

\label{Section:Prep}

\begin{figure}
\begin {center}
\includegraphics[width=8cm]{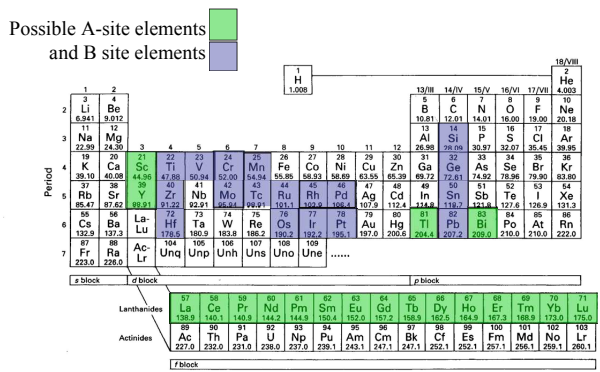}
\end{center}
\caption{Elements known to produce the (3+,4+) cubic pyrochlore oxide phase.}
\label{Fig:Ptable}
\end{figure}

Polycrystalline samples of pyrochlore oxides readily form
by a standard solid state process, where
 stoichiometric proportions of high purity ($>$ 99.99\%)
 rare earth and transition metal oxides are calcined
 at 1350$^{\circ}$C over several days in air with intermittent grindings.
Elements known to exist
in the cubic (3+,4+) pyrochlore phase are highlighted in the
periodic table shown in Fig. \ref{Fig:Ptable}.
Several series of compounds with particularly volatile elements,
 like tin, require excess material to account for
losses during the heating processes and others require
 heating under special atmospheres to form the desired oxidation state.
 For example the molybdates require a vacuum furnace or a
reducing atmosphere to form the Mo$^{4+}$ state and the
rare earth manganese (IV) oxide pyrochlores are not stable
at ambient pressure and require a high pressure synthesis
technique~\cite{Subramanian:1988}.
Specialized growth conditions will be discussed below,
as relevant.  When high chemical homogeneity is required,
the sol-gel method \cite{Kido:1991} has been successful.

Crystals of a several oxide pyrochlores were grown in the 60's and 70's
by flux methods~\cite{Wanklyn:1968}.
The size and quality of the crystals grown by these methods vary,
and it is often difficult to obtain
voluminous crystals, especially for neutron scattering experiments.
The `floating zone'  method of crystal growth was first used by~\textcite{Gardner:1998}
to produce large single crystals of Tb$_2$Ti$_2$O$_7$.  Later the entire titanate series was produced by others~\cite{Balakrishnan:1998,Tang:2006,Zhou:2007} and
within the last few years,  groups have started to grow many other pyrochlores in image furnaces including molybdates, ruthenates and stannates~\cite{Miyoshi:2001,Taguchi:2002,Wiebe:2004}.
Single crystals have also been produced by several other methods
 including vapour transport~\cite{He:2006}, hydrothermal techniques~\cite{Chen:1998} and KF flux~\cite{Millican:2007}.

Powder samples of the $\beta$-pyrochlores (``defect'' pyrochlores)
 are formed at lower temperature ($\approx$ 500$^{\circ}$C)
in an oxidizing atmosphere~\cite{Yonezawa:2004}.
 Small single crystals are also available~\cite{Schuck:2006}.

Most oxide pyrochlores are cubic at room temperature
with a lattice parameter between 9.8~\AA~and 10.96~\AA~\cite{Subramanian:1983}.
Of these, the titanates are probably the most studied, followed by the zirconates, but many are still poorly represented in the literature.  Both the titanate and  zirconate pyrochlores were first reported by \textcite{Roth:1956}.  Since then, over 1000 papers have been published on the two families.

\section{Metal Insulator Transitions in the Oxide Pyrochlores }

\label{Section:M-I}

\begin{figure}[t]
  \begin {center}
 \scalebox{0.8}
{
  \includegraphics[width=8cm]{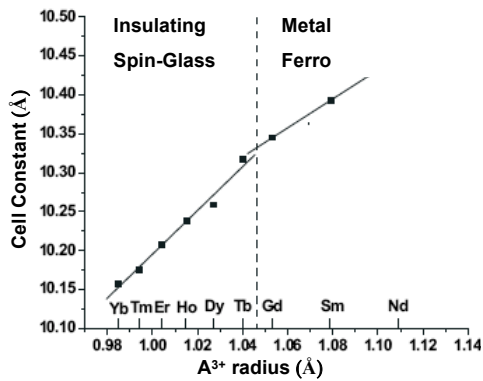}}
   \caption{Variation of unit cell constant
and physical properties of the series $A_2$Mo$_2$O$_7$
with the $A^{3+}$ radius~\cite{Ali:1989}}
  \label{Fig:Irr_MI}
  \end{center}
\end{figure}

\begin{figure}[t]
  \begin {center}
  \scalebox{0.7}{
 \includegraphics[width=8.7cm]{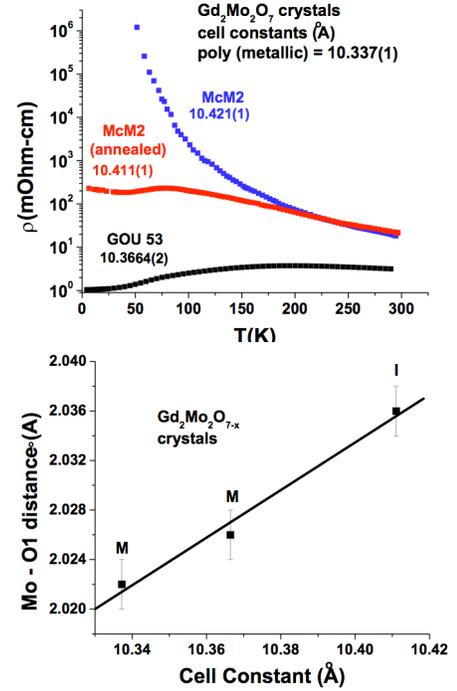}}
   \caption{Top: Comparison of electrical resistivity data for
 three single crystals of Gd$_2$Mo$_{2-2x}^{4+}$Mo$_{2x}^{3+}$O$_{7-x}$
 as a function of unit cell constant, $a_0$.
Bottom: Variation of the Mo - O distance with increasing cell 
constant for Gd$_{2}$Mo$_{2}$O$_{7-x}$ crystals.}
 \label{Fig:GdMo_MI}
  \end{center}
\end{figure}

Two types of metal-insulator (MI) transitions occur in the oxide pyrochlore family.
First, are those that, as a function of a thermodynamic variable (e.g. temperature,
magnetic field, pressure) change their transport properties,
for example Tl$_2$Mn$_2$O$_7$~\cite{Fujinaka:1979}.
 Secondly, there are the {\it series} of compounds where the
room temperature character changes from metal to
 insulator as the rare earth ion changes, for example
the molybdate ~\cite{Greedan:1986} or iridate series~\cite{Yanagishima:2001}.
There appears to be very little controversy over
the first class of MI transition, however for the second type, the exact
position of this transition is a topic of debate.
 For example, in the molybdenum pyrochlore series,
 studies of their bulk properties have indicated a
strong correlation with the magnetism and electrical
 transport properties, i.e., the ferromagnets were
 metallic while the paramagnets were insulating~\cite{Greedan:1987,Ali:1989}.
Indeed, the dependence on the lattice constant, $a_0$,
on the $A^{3+}$ radius showed a distinct break at the MI boundary;
see Fig. \ref{Fig:Irr_MI}.    In some subsequent studies however,
the Gd phase is clearly insulating~\cite{Kezsmarki:2004,Cao:1995}.
The initial studies were carried out on polycrystalline samples,
prepared between 1300 - 1400$^{\circ}$C and in at least one case,
in a CO/CO$_2$ ``buffer gas" mixture which fixes the oxygen partial
pressure during synthesis~\cite{Greedan:1986}
Several subsequent studies have used single crystals grown
by various methods above 1800$^{\circ}$C, including melt
and floating zone growths~\cite{Raju:1995,Kezsmarki:2004,Moritomo:2001}.   While the polycrystalline samples have been fairly well characterized, including elemental analysis, thermal gravimetric weight gain and measurement of the cubic lattice constant, $a_0$,
 this is less true of the single crystals. The differences
between poly and single crystalline samples can be monitored
most simply using the unit cell constant as illustrated in Fig.~\ref{Fig:GdMo_MI}
  in which unpublished data~\cite{Raju:1995} for a selection
of single crystals of Gd$_{2}$Mo$_{2}$O$_{7}$ are plotted.
Note that as $a_0$ increases the samples become more insulating.
 From accurate structural data for the powders and single crystals,
it has been determined that the increase in $a_0$ correlates
with an increase in the Mo - O distance as shown Fig.~\ref{Fig:GdMo_MI}.
The most likely origin of this systematic increase is the substitution
of the larger Mo$^{3+}$ (0.69~{\AA}) for the smaller Mo$^{4+}$(0.64~{\AA})
which can arise from oxygen deficiency in the crystals resulting in the
 formula Gd$_2$Mo$_{2-2x}^{4+}$Mo$_{2x}^{3+}$O$_{7-x}$.
Note that other defect mechanisms, such as vacancies on either
 the $A$ or Mo sites would result in the oxidation of Mo$^{4+}$ 
to the smaller Mo$^{5+}$ (0.61~{\AA})
 which would lead to a reduction in $a_0$.
 Further evidence for this mechanism comes from the fact
that the crystal with the intermediate cell constant in Fig.~\ref{Fig:GdMo_MI} 
 was obtained from the most insulating crystal by annealing in a CO/CO$_2$
 buffer gas at 1400$^{\circ}$C for 1-2 days.
Thus, insulating single crystals of Gd$_{2}$Mo$_{2}$O$_{7}$ are oxygen deficient.
 This analysis is fully consistent with the report of \textcite{Kezsmarki:2004}
 who find that ``Gd$_{2}$Mo$_{2}$O$_{7}$" is insulating but can
 be doped into the metallic state by 10 percent Ca substitution on the Gd site.
 Ca-doping will of course add holes to the Mo states which compensate for
 the electron doping due to the oxygen deficiency in the ``as grown''
 crystals.  After considering all the studies, we conclude that the
MI boundary for stoichiometric $A_{2}$Mo$_{2}$O$_{7}$
is indeed between $A$ = Gd and Tb.

\begin{figure}[t]
  \begin {center}
 \scalebox{0.7}{
  \includegraphics[width=9cm]{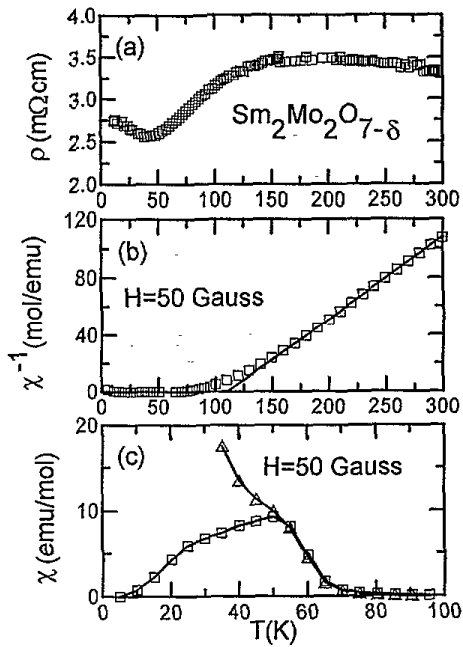}}
   \caption{Properties of an oxygen deficient single
 crystal of Sm$_{2}$Mo$_{2}$O$_{7-x}$ (a) Metallic behavior.
(b) Inverse susceptibility showing a positive (FM) Curie-Weiss temperature.
 (c) Susceptibility showing $T_c$ = 65~K and a zero field cooled (squares) field cooled
  (triangles) divergence at 50~K~\cite{Cao:1995}.}
  \label{Fig:SmMo_prop}
  \end{center}
\end{figure}

Single crystals for the other ferromagnetic metals, $A$ = Nd and Sm, are also reported and these are subject to the same oxygen deficiency and electron doping but, as they are farther from
the MI boundary, the effects on the physical properties are less drastic.
 For example the transport, optical and magnetic properties,
 reported by \textcite{Cao:1995} for Sm$_{2}$Mo$_{2}$O$_{7}$
were for a crystal with $a_0$ = 10.425(1)~{\AA}
compared with the polycrystal with $a_0$ = 10.391(1)~{\AA}.
While the material is still metallic (albeit a poor metal),
the ferromagnetic Curie temperature is reduced by $\approx$ 30~K
relative to the powder,  to 65~K.
Note also the ZFC/FC divergence below 50~K which suggests
 a spin glassy component.
Very similar results were reported for other $A$ = Sm single crystals,
 $a_0$ = 10.4196(1) {\AA}, 
for example by 
\textcite{Park:2003} and by \textcite{Moritomo:2001}, 
which also show $T_c$ $\sim$ 65~K and
a ZFC/FC divergence below 20~K.

The  oxide pyrochlores can display a wide range of bulk properties including oxygen ion conductivity,
superconductivity, ferroelectricity and unusual magnetic
 behavior (e.g. spin liquid and spin ice).
However these are intimately related to their stoichiometry,
electrical and crystallographic properties,
as shown in the previous paragraphs.  Whenever possible, we encourage one to measure, and report, these basic properties
at room temperature or as a function of temperature.

\part{Experimental Results}

\section{Long-Range Ordered Phases}

Many magnetic oxide pyrochlores enter a long range ordered state.
 These ordered phases are often promoted by single-ion anisotropy, anisotropic spin-spin
interactions and/or further neighbor
 exchange interactions.
Despite the fact that they develop long range order, these systems are still
of significant interest,  providing magnetic systems with axial
and planar symmetry, quantum order by disorder transition,
 polarised moments and partially ordered systems, for example.

\subsection{Gd$_2$Ti$_2$O$_7$ and Gd$_2$Sn$_2$O$_7$}
\label{Sec:GdTiO}

\begin{figure}[t]
\begin{center}
\includegraphics[width=9cm,angle=0,clip=20]{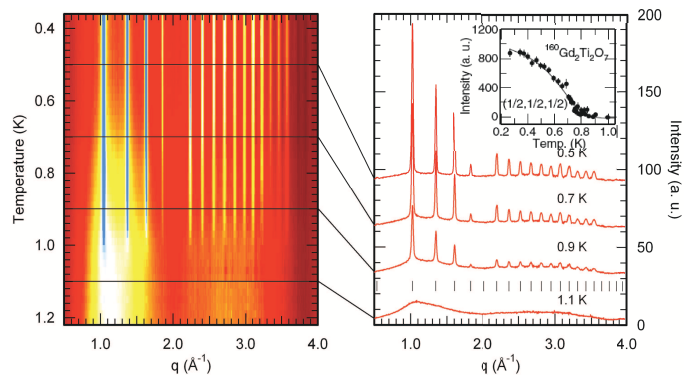}
\end{center}
\caption{Left: False colour map of the temperature dependence
of the magnetic diffraction from Gd$_2$Ti$_2$O$_7$.
As the temperature is lowered the broad diffuse scattering
 centered at 1.2 \AA$^{-1}$ sharpens into peaks.
Right: Diffraction patterns at specific temperatures
 are shown along with tic marks indicating where Bragg
reflections should be for the proposed ground state.
At the lowest temperatures the
$\frac{1}{2}\frac{1}{2}\frac{1}{2}$ reflection
has some intensity and this is highlighted in the inset~\cite{Stewart:2004a}.}
\label{GdTiO_diff}
\end{figure}

Antiferromagnetically coupled Heisenberg spins on a pyrochlore lattice are expected to be highly
frustrated because the exchange energy can be minimised
in an infinite number of  ways.  Classical~\cite{Moessner:1998,Canals:1998} 
and quantum calculations~\cite{Canals:1998,Canals:2000}
 show that the system should remain in a collective paramagnetic state
 with short-range spin-spin correlations at any non-zero temperature.
 As discussed in Section \ref{Theory}, \textcite{Raju:1999} 
reported an infinite number of degenerate spin
 configurations near the mean field transition temperature when
dipolar interactions are considered.
However,~\textcite{Palmer:2000}
showed that a four-sublattice state with the ordering vector 
$\bm k_{\rm ord}$=000 could be stabilized.
More recent mean field theory calculations~\cite{Cepas:2004,Enjalran:2003}
found a higher temperature second order
transition to a phase with three ordered sublattices and 1 disordered one.
At a slightly lower temperature another transition was found to the Palmer-Chalker state.
In contrast, Monte Carlo simulations by \textcite{Cepas:2005} find a single
strongly first order transition to the Palmer-Chalker state. As
we shall see  below, the experimental situation is even more
complex.

%
%\begin{figure}[t]
%\begin{center}
%\includegraphics[width=8.3cm,angle=0,clip=20]{GdTiO_Structure.eps}
%\end{center}
%\caption{The proposed magnetic structures for Gd$_2$Ti$_2$O$_7$ at base temperature.
%Left: A $1- {\bm k}$ structure where the diffuse scattering originates from one spin on every
%tetrahedron being dynamic.  This structure is then made up of alternating
%ordered kagome  planes separated by 
%a disordered triangular lattice. Right: A $4-{\bm k}$ structure where the
%disordered spins all belong to the same tetrahedron in a unit cell.
% This structure was chosen over the one on the left  as a result of
%polarized neutron diffraction results that indicated that the disordered
%spins were nearest-neighbors~\cite{Stewart:2004}.}
%\label{GdT_Str}
%\end{figure}

The S = 7/2 Gd$^{3+}$ ion, with its half-filled 4f-shell and zero orbital
momentum is the best example of a Heisenberg spin amongst the rare earths.
The bulk susceptibility for both Gd$_2$Ti$_2$O$_7$ and
Gd$_2$Sn$_2$O$_7$ follows the Curie-Weiss law with an effective
moment close to the theoretical limit of 7.94~$\mu_B$ for the free
ion~\cite{Raju:1999,Bondah-Jagalu:2001} and a Curie-Weiss
temperature of $\theta_{\rm CW} \sim -10$~K.
  Both compounds however enter a magnetically
 ordered state at $\approx$ 1~K, resulting in a frustration index of about 10.

Early specific heat and AC susceptibility
measurements showed that Gd$_2$Ti$_2$O$_7$
entered an ordered phase at 0.97~K~\cite{Raju:1999}.
A subsequent specific heat study
showed two transitions in zero applied field,
at 0.97 and 0.7~K~\cite{Ramirez:2002} and additional phase transitions, induced in applied fields.
These phases were later confirmed by single crystal magnetisation work by \textcite{Petrenko:2004}.

The nature of these ordered states is  hard to determine by neutron
diffraction due to the high absorption cross section of gadolinium.
 However a small, enriched, sample of $^{160}$Gd$_2$Ti$_2$O$_7$ has been studied.
 In the initial work at 50~mK, a partially ordered noncollinear
antiferromagnetic structure,
%with a single propagation vector ${\bm k}_{\rm ord} =\frac{1}{2}\frac{1}{2}\frac{1}{2}$, 
was
proposed~\cite{Champion:2001}. This unusual spin configuration
had  3/4 of the Gd$^{3+}$ spins ordered within kagome planes,
separated by the triangular plane of Gd ions (see Fig. \ref{Fig:Alt_view}),
which remain disordered down to the lowest temperature.
This model was later modified when the diffuse
scattering and a new magnetic reflection at 
$\frac{1}{2}\frac{1}{2}\frac{1}{2}$,
was observed and studied~\cite{Stewart:2004}.  The magnetic scattering
from $^{160}$Gd$_2$Ti$_2$O$_7$ is shown in
Fig.~\ref{GdTiO_diff}.
 Very broad diffuse scattering is seen above 1~K.
Below this temperature, Bragg peaks occur at the reciprocal
lattice positions indicated by the tic marks (vertical bars) in the right panel.
 The lowest angle magnetic reflection, the very weak
$\frac{1}{2}\frac{1}{2}\frac{1}{2}$ peak 
at  $\vert{\mathbf Q}\vert$ = 0.6 {\AA}$^{-1}$ only gains intensity below the second transition.
 At all temperatures, diffuse magnetic scattering is observed which
diminishes as the temperature decreases and the Bragg peaks grow.
The diffuse magnetic scattering, centred at
 $\vert{\mathbf Q}\vert \approx$~1.2~\AA$^{-1}$
indicates that the correlation length of the disordered
spins is $\approx$~3.5~\AA, the nearest-neighbor distance and
not 7~\AA~appropriate for the model proposed earlier by~\textcite{Champion:2001}.  
This picture of two different Gd moments is consistent
with $^{155}$Gd M\"{o}ssbauer experiments~\cite{Bonville:2003}.  Most interestingly,
this is not the state proposed by \textcite{Palmer:2000}.

\begin{figure}
\begin{center}
\includegraphics[width=7.5cm,angle=0,clip=20]{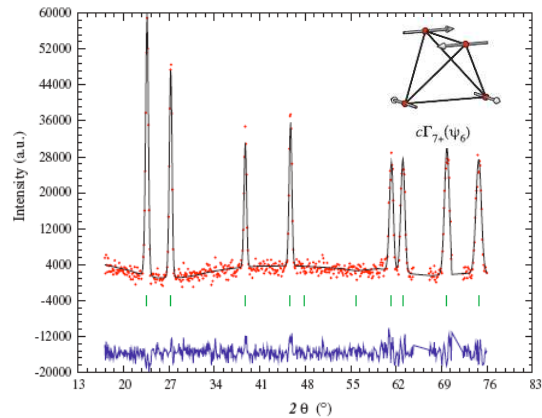}
\end{center}
\caption{Neutron powder diffraction pattern from Gd$_2$Sn$_2$O$_7$
at 0.1~K, after a nuclear background (1.4~K) data set was subtracted~\cite{Wills:2006}.
 Inset: One of the
3 $\Gamma_{7+}$ basis vectors that describe the data
(see fit), consistent with the Palmer-Chalker model~\cite{Palmer:2000}.}
\label{Fig_GdSn_NPD}
\end{figure}

The isostructural stannate, Gd$_2$Sn$_2$O$_7$, has received
comparatively less attention. As mentioned above, the paramagnetic
properties are similar to Gd$_2$Ti$_2$O$_7$~\cite{Matsuhira:2002}.
However, only one transition, at 1.0~K, and only one Gd moment has
been observed in Gd$_2$Sn$_2$O$_7$~\cite{Bonville:2003}.
Recent neutron diffraction measurements by \textcite{Wills:2006},
shown in Fig. \ref{Fig_GdSn_NPD}, have confirmed that Gd$_2$Sn$_2$O$_7$
undergoes only one transition down to the lowest temperatures and
that the ground state is the one predicted theoretically for a Heisenberg
pyrochlore antiferromagnet with dipolar interactions by \textcite{Palmer:2000}.

The difference in these two materials highlights the extreme sensitivity of
 the ground state to small changes in composition
and of the consequential details in the spin-spin interactions
in these highly frustrated materials.
\textcite{Wills:2006} proposed
that a different relative magnitude in the two types of
third-neighbour superexchange interaction may be responsible for
the different ground states in these otherwise very similar Gd compounds~\cite{DelMaestro:2007}.

\begin{figure}
\begin{center}
\includegraphics[width=6cm,angle=0,clip=20]{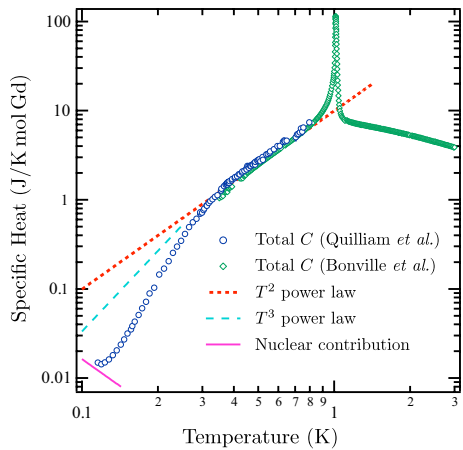}
\end{center}
\caption{Specific heat of Gd$_2$Sn$_2$O$_7$ from \textcite{Quilliam:2007} (circles) and \textcite{Bonville:2003} (diamonds).  The T$^2$ power law (dotted line) and a conventional T$^3$ power law(dashed line) for a three dimensional antiferromagnet are plotted.  The upturn seen below 150~mK results from the nuclear electric quadrupole interaction~\cite{Quilliam:2007}.}
\label{Fig_GdSn_SpHt}
\end{figure}

To understand the magnetic ground states further and the true cause of the differences between Gd$_2$Ti$_2$O$_7$ and Gd$_2$Sn$_2$O$_7$ one has to investigate the excitation spectrum.
Specific heat is a bulk probe that often provides a reliable insight
into the low lying excitations.  The specific heat, C$_v$, of
these two compounds just below the transition temperature
is described well by an anomalous power law, C$_{v}$ $\propto$ $T^2$, that had been
presumed to be associated with an unusual energy-dependence of the density of
states~\cite{Raju:1999,Ramirez:2002,Bonville:2003,Bonville:2003a}.
Electron spin resonance (ESR)~\cite{Hassan:2003,Sosin:2006,Sosin:2006a},
M\"{o}ssbauer~\cite{Bertin:2002},
muon spin relaxation~\cite{Bonville:2003a,Yaouanc:2005,Dunsiger:2006}
and neutron spin echo~\cite{Ehlers:2006} have provided
many interesting results.  In all these measurements,
spin fluctuations were observed much below the ordering temperature
of the compound, seemingly consistent with the unusual energy-dependence of
 the density of states invoked to interpret the specific heat measurements.
\textcite{Sosin:2006,Sosin:2007} failed to confirm the unusually
large planar anisotropy measured by~\textcite{Hassan:2003},
however they found a small gap $\approx$ 1~K that coexists
with a paramagnetic signal in Gd$_2$Ti$_2$O$_7$, however no paramagnetic spins were observed at low temperatures in Gd$_2$Sn$_2$O$_7$.  Recently ~\textcite{Quilliam:2007},
remeasured the specific heat of Gd$_2$Sn$_2$O$_7$ and found that
the $C_v\propto T^2$ behavior does not hold below 400~mK
(see Fig.~\ref{Fig_GdSn_SpHt}) and that C$_{v}$ decreases
 exponentially below 350~mK, providing evidence for a gapped
spin-wave spectrum due to anisotropy resulting from
single-ion effects and long-range dipolar interactions~\cite{DelMaestro:2007}.

The current experimental understanding of these Heisenberg antiferromagnets is at 
at this time incomplete.
 Gd$_2$Sn$_2$O$_7$ appears to be a good example of the model considered by
\textcite{Palmer:2000}, and Gd$_2$Ti$_2$O$_7$ is
perhaps a small perturbation from this.
 However, an explanation for the low lying excitations as probed by
muon spin relaxation in both compounds still eludes theorists~\cite{DelMaestro:2007}.
Understanding these ubiquitous, low temperature spin dynamics in a
relatively simple model magnet, like the Gd-based Heisenberg-like
antiferromagnet on the pyrochlore lattice, 
would likely help elucidate the nature of the low
temperature spin dynamics in other frustrated magnets.
We discuss in Section \ref{Sec:Mag_perp}
some recent work on Gd$_2$Ti$_2$O$_7$ in a magnetic field.

\subsection{Er$_2$Ti$_2$O$_7$}

\label{Sec:ErTiO}

The Er$^{3+}$ single ion crystal field ground state in Er$_2$Ti$_2$O$_7$ 
is a Kramers doublet with a magnetic moment of 3.8 $\mu_{\rm B}$
and 0.12 $\mu_{\rm B}$ perpendicular and parallel to the local
$\langle 111 \rangle$ axis, respectively.
Er$_2$Ti$_2$O$_7$,  like Yb$_2$Ti$_2$O$_7$,
is a realization of an XY-like system on the pyrochlore
lattice. However, while Yb$_2$Ti$_2$O$_7$
 has a small ferromagnetic Curie-Weiss temperature
($\theta_{\rm CW} \sim$ +0.75 K)~\cite{Hodges:2002},
Er$_2$Ti$_2$O$_7$
has a large negative  Curie-Weiss temperature
($\theta_{\rm CW} \sim$ - 22~K)~\cite{Champion:2003},
 and is therefore a pyrochlore XY antiferromagnet.
 The lowest-lying crystal field states of Er$_2$Ti$_2$O$_7$  are at
6.38 meV (74.1 K) and 7.39 meV (85.8 K) above the ground states~\cite{Champion:2003}.

Specific heat measurements show a sharp feature at approximately $T_{\rm N}\sim1.2$~K~\cite{Blote:1969,Champion:2003,Siddharthan:1999}.  Neutron scattering below $T_{\rm N}$ confirms the existence of a N\'eel long-range ordered state with propagation vector ${\bm k}_{\rm ord} =000$.
% which means that
%each tetrahedron has the same static spin configuration
%below $T_{\rm N}$.
The neutron scattering data
strongly suggest a second order phase transition
at $T_{\rm N}$, with the scattered intensity $I$ vanishing at $T_{\rm N}$
as $I(T) \propto (T_{\rm N}-T)^{2\beta}$ with critical exponent $\beta\approx 0.33$,
characteristic of the three-dimensional XY model.

\begin{figure}[t]
  \begin {center}
\includegraphics[width=4cm]{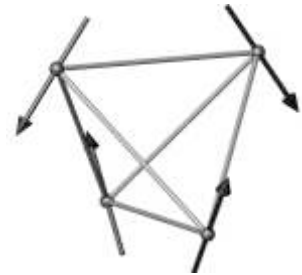}
    \caption{Ground state spin configuration determined for Er$_2$Ti$_2$O$_7$~\cite{Poole:2007}.
 In a macroscopic sample six symmetry equivalent 
spin configurations will co-exist.}
  \label{GS-Er2Ti2O7}
  \end{center}
\end{figure}

As the transition is continuous, the system is expected to order in only one
of the irreducible representations of the Er$^{3+}$ site representation.
A refinement of the magnetic structure finds that only the two basis
vectors $\psi_2$ or $\psi_3$ of $\Gamma_5$, or a superposition of both, is
consistent with the observed magnetic intensitites of the Bragg peaks.
Results from a 
spherical neutron polarimetry study find that the zero magnetic field
ordered state is described almost entirely by $\psi_2$ with very little
admixing from $\psi_3$.
The spin configuration corresponding to $\psi_2$ is
shown in Fig.~\ref{GS-Er2Ti2O7}~\cite{Poole:2007}.
The experimental observations of (i) an ordering into the noncoplanar
$\psi_2$ structure via (ii) a second order phase transition are both surprising.

As discussed in Sections~\ref{Theory} and \ref{Sec:GdTiO},
 isotropic exchange plus dipolar interactions select by themselves
a state with spins perpendicular to the local $\langle 111\rangle$ axes.
Hence, easy plane (XY) anisotropy should further enhance the stability of this state.
This was explicitly shown in calculations~\cite{DelMaestro:2007}
pertaining to Gd$_2$Sn$_2$O$_7$ which does display a Palmer-Chalker
ground state~\cite{Wills:2006}.
However a Palmer-Chalker ground state for Er$_2$Ti$_2$O$_7$ is seemingly robustly rejected by recent neutron
polarimetry experiments~\cite{Poole:2007}.
Since nearest-neighbor antiferromagnetic exchange,
dipolar interactions and the XY nature of the magnetic moments
should conspire together to produce a Palmer-Chalker ground state
in Er$_2$Ti$_2$O$_7$, other interactions or effects
must be at play to stabilize
the experimentally observed $\psi_2$ ground state~\cite{Poole:2007}.

An early numerical study of the classical XY antiferromagnet on the pyrochlore lattice
had found a strongly first order transiton into a long-range ordered state with
propagation vector  ${\bm k}_{\rm ord}$=000~\cite{Bramwell:1994}.
This finding was subsequently confirmed by
another study~\cite{Champion:2003,Champion:2004}.
 In an earlier study,
\textcite{Bramwell:1994} had only identified a macroscopic degeneracy
of discrete ground states, while there is in fact a continuous degree of
freedom for antiferromagnetically coupled XY spins on a single tetrahedron, resulting
in a continuous ground state degeneracy for a macroscopic
sample~\cite{Champion:2003,Champion:2004}.  Monte Carlo simulations of a classical $\langle 111\rangle$
pyrochlore XY antiferromagnet find that the degeneracy is lifted by
an order-by-disorder transition driven by thermal fluctuations with, most
interestingly, the system developing long range order in the same
$\psi_2$ state as found experimentally in Er$_2$Ti$_2$O$_7$.
The Monte Carlo results can be rationalized on the basis of a classical spin wave
calculation~\cite{Champion:2004}.
There, it is found that there are zero energy spin wave modes over planes
in the Brillouin zone for which the spin wave  wavevector ${\bm q}$
obeys ${\bm q}\cdot {\bf a}=0$
and  ${\bm q}\cdot  ( {\bf b} - {\bf c})=0$,
where ${\bf a}$, ${\bf b}$ and ${\bf c}$ are
are the basis vectors of the primitive rhombohedral unit cell of the
pyrochlore lattice.

On the basis of the results from Monte Carlo and classical spin wave calculations,
\textcite{Champion:2004}
proposed that the $\psi_2$ in
Er$_2$Ti$_2$O$_7$ is perhaps stabilized by zero-point quantum fluctuations,
arguing that their effects are
captured by the classical spin wave argument.
However, problems remain with this interpretation. As mentioned above,
the transition at $T_{\rm N}\sim 1.2$ K is second order in the experiment
while simulations show a strong first order transition~\cite{Champion:2004}.
Also, the density of zero energy classical spin wave states is a constant, a
result that is incompatible with the experimental specific heat data
which shows a $T^3$ temperature dependence below 1 K.
 Inelastic neutron scattering reveals no evidence for such
low lying excitations required to generate the $T^3$ temperature
 dependence of the specific heat.
The experimental results and models presented above suggest
 a well established long-range ordered ground state for Er$_2$Ti$_2$O$_7$.
 Yet, muon spin relaxation data reveal
that below $T_{\rm N}$, the muon polarization relaxation rate remains large
($2\times 10^6$~${\mu{\rm s}}^{-1}$)
and essentially temperature independent down to 20 mK,
indicative of a dynamic ground state~\cite{Lago:2005}.
This behavior reminds one of low-temperature muon data in several
pyrochlores including the  ``ordered''
Gd-pyrochlores and the ``dynamic''  Tb$_2$Ti$_2$O$_7$.

Unlike Er$_2$Ti$_2$O$_7$,   Er$_2$(GaSb)O$_7$ does not appear
to order down to 50 mK~\cite{Blote:1969}.
It is not clear at this stage whether
the randomness on the $B$-sites leads to a disordered
 and glassy phase, i.e. an XY-like spin glass. Results on Er$_2$Sn$_2$O$_7$
will  be discussed in Section \ref{Sec:ErSn}.

\subsection{Tb$_2$Sn$_2$O$_7$}

\label{TbSnO}

Until recently the magnetic and electrical
properties of the stannates had received little attention.
 This is probably due to the slightly more problematic
synthesis and the lack of large, high quality, single crystals.

\begin{figure}[t]
\begin{center}
\includegraphics[width=8.5cm,angle=0,clip=20]{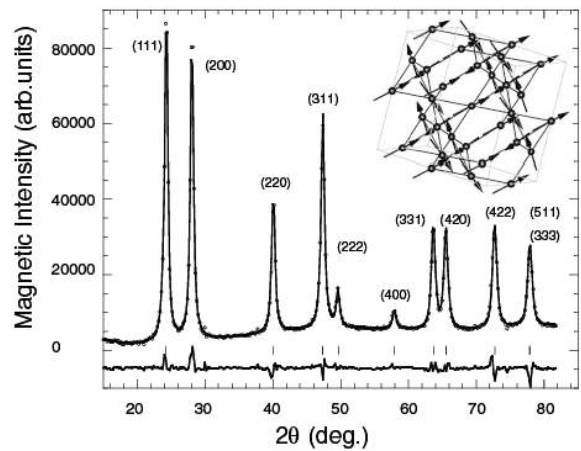}
\end{center}
\caption{Magnetic scattering from Tb$_2$Sn$_2$O$_7$ at 100~mK.
Inset: the proposed ``ordered'' spin ice structure
with spins canted off the $\langle 111\rangle$ by 13$^{\circ}$~\cite{Mirebeau:2005}.  The broad scattering at the
 base of the Bragg reflections is  indicative of spin clusters
which reach $\approx$ 200~\AA~at base temperature.}
\label{Fig_TbSn_NPD}
\end{figure}

Like many frustrated magnets, the high temperature (T $>$ 40~K)
inverse susceptibility is very linear.  In Tb$_2$Sn$_2$O$_7$, 
this yields a Curie-Weiss temperature
of approximately $\theta_{\rm CW} \sim -12$~K and an effective moment of 9.8 $\mu_B$,
which agrees well with the value of 9.72~$\mu_B$ for the $^7$F$_6$
ground state of Tb$^{3+}$~\cite{Bondah-Jagalu:2001}.
Reminiscent of the cooperative paramagnet Tb$_2$Ti$_2$O$_7$
(see Section \ref{sec:Spin_Liquid}), the inverse susceptibility
deviates downwards at $\approx$ 30~K upon cooling.
However, on cooling further, the system enters a frozen
ferromagnetic state at 0.87~K.  This was
first reported by \textcite{Matsuhira:2002} from low
temperature susceptibility measurements and then
by \textcite{Mirebeau:2005} who determined the magnetic
structure from neutron diffraction measurements.
In this work by \textcite{Mirebeau:2005},
a crossover from antiferromagnetic correlations to
ferromagnetic correlations was observed at 2~K.
This was followed, at a lower temperature,
by a transition into an ordered state and the appearance of
 Bragg peaks, see Fig. \ref{Fig_TbSn_NPD}.
Rietveld analysis determined that the spins are canted
 by 13.3$^{\circ}$ off the local $\langle 111\rangle$
directions with an ordered moment of 5.9 $\mu_B$,
significantly reduced from the free ion moment of 9 $\mu_B$, 
presumably due to crystalline electric field effects.

\begin{figure}[t]
\begin{center}
\includegraphics[width=6.5cm,angle=0,clip=20]{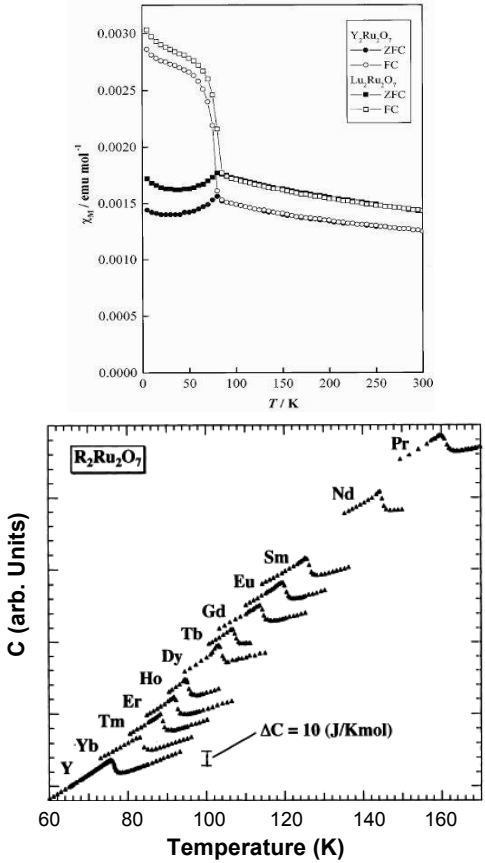}
\end{center}
\caption{Top: The temperature dependence of the susceptibility for Y and Lu ruthenate.
Both data sets show the Ru$^{4+}$ moments ordering at $\approx$ 75~K.
Bottom: Heat capacity measurements over the entire rare earth ruthenate
 series showing how the Ru$^{4+}$ ordering temperature increases as
the size of the rare earth ion decreases~\cite{Ito:2001}.}
\label{YRu_Sus}
\end{figure}

Ferromagnetically coupled spins
 on the pyrochlore lattice fixed along the local $\langle 111\rangle$
 directions gives rise to a highly degenerate state that 
has become  known as the spin ice state (see Section \ref{sec:spin-ice}).
By comparing the Tb$^{3+}$ moment value  deduced from
 neutron diffraction (5.9 $\mu_B$) to the smaller value 
 deduced from the nuclear specific heat (4.5 $\mu_B$) 
\textcite{Mirebeau:2006}
 suggested that slow collective fluctuations, on the time
scale of 10$^{-4}$-10$^{-8}$s, coexist with the ordered state
at the lowest temperatures~\cite{Mirebeau:2006}.

Subsequently, two groups led by
\textcite{DalmasdeReotier:2006} and \textcite{Bert:2006}
have suggested that the ground state is not static.
 They claim that the absence of any oscillations,
even at the shortest times, in the muon spin relaxation
spectrum precludes the existence of a static moment.
\textcite{DalmasdeReotier:2006}
argue that both neutron and muon data are consistent
 with a characteristic spin relaxation time of $\approx{10}^{-10}$s
whilst \textcite{Bert:2006} suggest that the dynamics result
from fluctuations of clusters of correlated spins with the
ordered spin ice structure. On the other hand,
the results from the neutron spin echo study of
 \textcite{Chapuis:2007} suggest no spin dynamics below 0.8~K.

\textcite{Mirebeau:2006} have suggested that the $\approx$3\%
expansion in the lattice constant between Tb$_2$Sn$_2$O$_7$ and Tb$_2$Ti$_2$O$_7$
results in the dipolar interaction overcoming the weakened
exchange interaction allowing Tb$_2$Sn$_2$O$_7$ to freeze.
Clearly more experimental work is needed on this compound to
determine the exact nature of the ground state.
Recently, from inelastic neutron scattering studies,
\textcite{Mirebeau:2007} found that the composition of the first
two crystal field levels of Tb$_2$Sn$_2$O$_7$ are
inverted when compared to those of Tb$_2$Ti$_2$O$_7$.  
This cannot be explained by simple calculations 
of the local crystal field environment and more work is needed.

\subsection{${\it A}_2$Ru$_2$O$_7$ (${\it A}$~=~Y, Gd, Dy, Ho, Er and Tl)}

All but the largest, $A$ = La and Ce, rare-earth ruthenates as
well as $A=$ Bi and $A=$ Tl 
form as cubic pyrochlores~\cite{Bertaut:1959,Yoshii:1999, Subramanian:1983}.
 The lattice parameters vary from 10.07 to 10.355~\AA~across the
rare earth series with the $A$ = Bi and Tl compounds at the high end of this range.
 Crystals of several ruthenates have been made by hydrothermal, vapour transport
and floating zone methods~\cite{Sleight:1972,Zhou:2007}.
Historically, the ruthenium pyrochlore oxides  have been extensively
studied for their novel conductivity~\cite{Pike:1977,Carcia:1982}
and catalytic activity~\cite{Horowitz:1983}, but until recently, very
little was known about their magnetic properties.
 Most of the ruthenates are semiconductors at room temperature
with hopping energies of $\sim$ 0.1  eV, but the Bi based sample
is metallic with Pauli paramagnetism.  Ru$^{4+}$ (low spin 4d$^4$) has 
an $S=1$ moment and the  
Ru sublattice appears to order between 70 and 160~K, depending on the rare earth
(see Fig. \ref{YRu_Sus}).
However, $\theta_{CW}$ for  Y$_2$Ru$_2$O$_7$
is reported to be -1250~K although it is not clear that this
was obtained in the paramagnetic regime~\cite{Gurgul:2007}.
Studies of diamagnetic $A$ = Y and Lu provide evidence
of the magnetic nature of the Ru$^{4+}$ ion.  These indicate a magnetic transition
at $\approx$~75~K, leading to a high frustration index, {\it f} $\approx$ 17.
 The magnetic susceptibilities measured under zero field cooled (ZFC)
and under field cooled (FC) conditions  show a different temperature dependence
(see Fig.~\ref{YRu_Sus}).  This is consistent with the observation of a
canted antiferromagnetic structure from neutron diffraction~\cite{Ito:2001,Kmiec:2006}.
 Inelastic neutron scattering measurements by \textcite{vanDuijn:2007}
reveals the development of a large gap in the excitation spectrum below
the ordering temperature.
%and this has led them to
%suggest that the transition is driven  by quantum fluctuations.

\begin{figure}[t]
\begin{center}
\includegraphics[width=8.5cm,angle=0,clip=20]{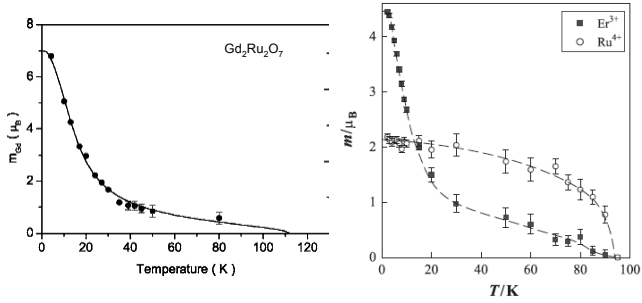}
\end{center}
\caption{Left: The temperature dependence of the Gd moment from Gd-M\"{o}ssbauer~\cite{Gurgul:2007}
 on Gd$_2$Ru$_2$O$_7$. Right:  The temperature dependence of the Ru$^{4+}$ and Er$^{3+}$
moments in Er$_2$Ru$_2$O$_7$ as measured by neutron diffraction~\cite{Taira:2003}.}
\label{Mom_Ru}
\end{figure}

Er$_2$Ru$_2$O$_7$ and Gd$_2$Ru$_2$O$_7$ have been studied by powder
neutron diffraction~\cite{Taira:2003} and M\"{o}ssbauer~\cite{Gurgul:2007}. 
 In both cases, the Ru sublattice orders at the temperature indicated by the peak
in the specific heat plotted in Fig.~\ref{YRu_Sus}.
  This long range ordered state is a ${\bm k}_{\rm ord}$=000 type.
 This is also true for Y$_2$Ru$_2$O$_7$~\cite{Ito:2001}.
In the Er and Gd ruthenates, the rare earth moment is polarised by
the ordered Ru-sublattice, via
 the rare earth - Ru$^{4+}$ exchange interactions.
 This reveals itself in a small but finite moment on the $A$-site as soon
as the ruthenium orders.  This moment grows with decreasing temperature
before the rare earth ion orders in its own right at 10~K (Er) and 40~K (Gd).
A refinement of the low temperature (3~K) neutron diffraction data from Er$_2$Ru$_2$O$_7$
 suggests that the magnetic moments, both Er and Ru, are in a plane oriented
 towards the same $[100]$ direction, but antiparallel to each other.
The refined moments for Ru$^{4+}$ saturate at the theoretical limit
for the $S$=1 ion (2 $\mu_B$) at 90~K, but at 3~K the moment on the Er site
is only 4.5 $\mu_B$, 50\% of the predicted value, and is probably reduced by
crystalline electric field effects, although the signal has not yet
saturated at the lowest temperature studied (see Fig.~\ref{Mom_Ru}).

\begin{figure}[t]
\begin{center}
\includegraphics[width=7cm,angle=0,clip=20]{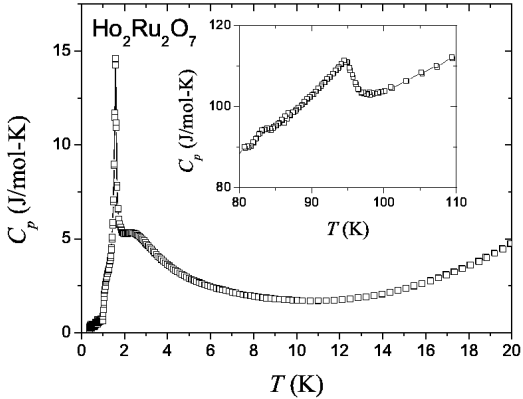}
\end{center}
\caption{Specific heat of polycrystalline Ho$_2$Ru$_2$O$_7$
as a function of temperature around the Ru$^{4+}$ (inset)
 and Ho$^{3+}$ (main) ordering temperatures~\cite{Gardner:2005}.}
\label{Fig:HoRu_Cp}
\end{figure}

Since \textcite{Bansal:2002} reported a possible spin-ice ground
state (see Section \ref{sec:spin-ice}) in Ho$_2$Ru$_2$O$_7$
and Dy$_2$Ru$_2$O$_7$, there have been several studies
 of the magnetic behaviour of Ho$_2$Ru$_2$O$_7$~\cite{Bansal:2003,Wiebe:2004,Gardner:2005}.
 In this compound, unlike the other heavy rare earth ruthenates, a small difference is
 seen between FC and ZFC when the rare earth ion orders.
 This happens at  1.4~K when the Ho$^{3+}$ sublattice enters a long-range 
ordered state, unlike a true spin ice material.
Lower temperature (T $<$ 20 K) specific heat (see Fig. \ref{Fig:HoRu_Cp})
 also highlights this transition.
As seen in a number of geometrically frustrated systems,
a broad feature (centered here at $\approx$3~K) precedes
the sharp lambda-like transition at a slightly lower temperature.
These features are related to the build-up of short- and long-range
correlations between Ho spins, respectively.
The `{\it ordered}' spin ice has the holmium moments canted off the local
$\langle 111\rangle$
axes presumably due to the local fields generated by the ordered Ru$^{4+}$ sublattice.
The ordered structure of Gd$_2$Ru$_2$O$_7$ is similar to Ho$_2$Ru$_2$O$_7$.
In both cases the full moment of the rare earth ion was not measured at the
lowest temperatures investigated~\cite{Gurgul:2007,Wiebe:2004}.

\begin{figure}[t]
\begin{center}
\includegraphics[width=8.5cm,angle=0,clip=20]{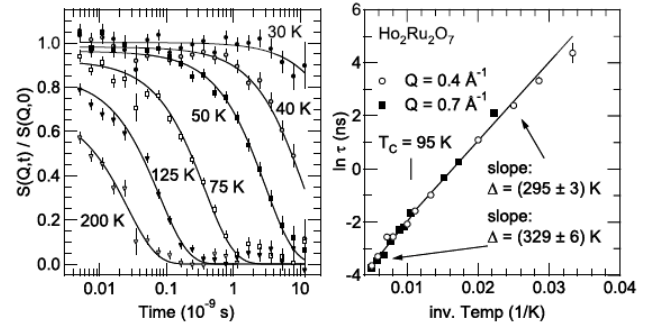}
\end{center}
\caption{Neutron spin echo spectra from Ho$_2$Ru$_2$O$_7$ at 
$\vert{\mathbf Q}\vert$~=~0.4 \AA$^{-1}$
 as measured on NG5NSE (left) and the dependence of the relaxation
 time on temperature (right) from two different positions in reciprocal space. 
The lines through the spectra on the left are a simple exponential, 
where the characteristic relaxation time is used in the Arrhenius plot
 on the right~\cite{Gardner:2005}.  }
\label{Fig:HoRu_NSE}
\end{figure}

The magnetic entropy associated with Ho$_2$Ru$_2$O$_7$ at these low temperatures reaches neither $R\ln 2$ (expected for a completely ordered doublet system) nor 
$R(\ln 2 - {1\over{2}} \ln {{3}\over{2}} )$ 
(expected for spin ice).
 In fact, the magnetic entropy of Ho$_2$Ru$_2$O$_7$~\cite{Gardner:2005}
resembles that of the dipolar spin ices in an applied field of 1~T.
 Several neutron scattering~\cite{Wiebe:2004,Gardner:2005} and AC
susceptibility~\cite{Gardner:2005} experiments have investigated the
spin dynamics in Ho$_2$Ru$_2$O$_7$.  The nature of the Ho-moments is
governed by the crystalline electric field levels  which are slightly
perturbed from those measured in Ho$_2$Ti$_2$O$_7$~\cite{Rosenkranz:2000}
especially after the ordering of the Ru$^{4+}$ sublattice~\cite{Gardner:2005}.
 This is most apparent in neutron spin echo measurements where a 35~K change
in the gap to the first excited state is observed
once the Ru$^{4+}$-ions order~\cite{Gardner:2005} (see Fig. \ref{Fig:HoRu_NSE}).

While Tl$_2$Ru$_2$O$_7$ is cubic with a room temperature lattice parameter
 of 10.18 \AA, when it is cooled below 120~K, the structure transforms to
an orthorhombic structure.
 Concomitantly, a metal-insulator transition occurs and the resistivity
changes by 5 orders of magnitude.  A sudden drop in the susceptibility
also occurs at this temperature~\cite{Lee:2006}.
These data have been interpreted as the three-dimensional 
%3-D% 
analogue to the more commonly known 
one-dimensional $S=1$ Haldane gap system~\cite{Haldane:1983}.
 Specifically, it is believed that the structural
 phase transition creates effective $S=1$, chains of Ru-ions which do not dimerize.
 These chains are arguably created due to the orbital ordering of the Ru 4d electrons.

Three (2+,5+) pyrochlores of interest are Cd$_2$Ru$_2$O$_7$~\cite{Wang:1998},
 Ca$_2$Ru$_2$O$_7$~\cite{Munenaka:2006} and Hg$_2$Ru$_2$O$_7$~\cite{Yamamoto:2007},
with an $S=3/2$ Ru$^{5+}$ ion and non-metallic transport properties.
Hg$_2$Ru$_2$O$_7$, like Tl$_2$Ru$_2$O$_7$,  has a metal-insulator transition
that is concomitant with a structural phase transition.
Single crystals of Ca$_2$Ru$_2$O$_7$ have been grown by the hydrothermal
method~\cite{Munenaka:2006} and the effective paramagnetic moment is
 only 0.36 $\mu_B$, about one order of magnitude smaller that the
 theoretical value for the $S=3/2$ ion.  At 23~K, there is clear irreversibility
between FC and ZFC susceptibilities indicative of a spin glass transition and
reminiscent of Y$_2$Mo$_2$O$_7$~\cite{Greedan:1986}.

\begin{table}[t]
\begin{center}
\begin{tabular}{||c|c|c|c|c|c||}
\hline
{{\it A}} &
{a}$_{0}$ ({\AA}) &
{T}$_{c}${(K)} &
{$\theta_{\rm CW} $(K)} &
{$\theta_{\rm CW} $/T}$_{c}$ &

Reference \\
\hline
Sc&
9.586(3)&
15&
-&
-&
\cite{Troyanchuk:1988}\\
\hline
Sc&
9.5965(4)&
20(1)&
77(3)&
3.9&
\cite{Greedan:1996a} \\
\hline
Y&
9.912(3)&
-&
-&
-&
\cite{Fujinaka:1979} \\
\hline
Y&
9.912(3)&
-&
-&
-&
\cite{Troyanchuk:1988}\\
\hline
Y&
9.901&
20(5)&
50(10)&
2.5&
\cite{Subramanian:1988} \\
\hline
Y&
9.919(2)&
15(1)&
42(2)&
2.8&
\cite{Reimers:1991} \\
\hline
Y&
9.91268(3)&
15&
50&
3.3&
\cite{Shimakawa:1999} \\
\hline
Lu&
9.814(3)&
9&
-&
-&
\cite{Troyanchuk:1988}\\
\hline
Lu&
9.815&
23(5)&
70(10)&
3.0&
\cite{Subramanian:1988} \\
\hline
Lu&
9.82684(3)&
15&
60&
4.0&
\cite{Shimakawa:1999} \\
\hline
In&
9.717(3)&
132&
-&
-&
\cite{Troyanchuk:1988}\\
\hline
In&
9.727(1)&
120&
150&
1.25&
\cite{Raju:1994} \\
\hline
In&
9.70786(9)&
120&
155&
1.29&
\cite{Shimakawa:1999} \\
\hline
Tl&
9.890(3)&
120&
-&
-&
\cite{Fujinaka:1979} \\
\hline
Tl&
9.892(1)&
121(1)&
155&
1.28&
\cite{Raju:1994} \\
\hline
Tl&
9.89093(7)&
122&
175&
1.43&
\cite{Shimakawa:1999}  \\
\hline
\end{tabular}
\caption{Collected values for unit cell constants, $a_0$, measured Curie temperatures,
$T_c$, and Curie-Weiss temperatures,
$\theta_{\rm CW}$, for $A_2$Mn$_2$O$_7$ phases with $A$ = Sc, Y, Lu, In and Tl.}
\label{Table:ScMnO_a}
\end{center}
\end{table}

\subsection{${\it A}_2$Mn$_2$O$_7$ (${\it A}$ ~=~Sc, Y, Tb - Lu and Tl) }

\label{Sec:AMnO}

\begin{table}[b]
  \begin{tabular}{||c|c|c|c||}
    \hline
   & $\theta_{\rm CW}$ (K) & $T_c$ (K)  & $\theta_{\rm CW}$ / $T_c$  \\  \hline
   EuO  & 80 & 69 & 1.2 \\ \hline
  EuS  & 19 & 16.6  & 1.2 \\ \hline
   Lu$_2$V$_2$O$_7$  & 83 & 74 & 1.1\\ \hline
   YTiO$_3$ & 33 & 29 & 1.1 \\
    \hline
  \end{tabular}
  \caption{Comparison of the Curie-Weiss and ordering
temperatures for selected ferromagnetic insulators.}
  \label{Table:Typ_FM}
\end{table}

The rare earth manganese (IV) pyrochlores are not stable
at ambient pressure at any temperature and
must be prepared using high pressure methods.
The earliest published report is by ~\textcite{Fujinaka:1979}
who synthesized the pyrochlores $A_2$$B_2$O$_7$
with $A$ = Y, Tl and $B$ = Cr, Mn using temperatures
 in the range 1000 - 1100$^{\circ}$C and pressures from 3 - 6 GPa.
 Reports of the high pressure synthesis of several $A_2$Mn$_2$O$_7$
phases appeared from
the groups  of
\textcite{Troyanchuk:1988} and \textcite{Subramanian:1988}
about ten years later. \textcite{Troyanchuk:1988}
prepared the phases
$A$ = Sc, Y, In, Tl and Tb-Lu with pressures between 5 - 8 GPa
and temperatures 1000~-~1500$^{\circ}$C. \textcite{Subramanian:1988}
 were able to prepare $A$ = Y and Dy-Lu using much lower pressures
and temperatures with a hydrothermal method in sealed gold tubes
including NaClO$_3$, NaOH and H$_2$O at 3 kbar (0.3~GPa) and 500$^{\circ}$C.
\textcite{Greedan:1996a} synthesized $A$ = Sc using a tetrahedral anvil
 press at 850$^{\circ}$C and 60 kbar (6~GPa).
Subsequently, \textcite{Shimakawa:1999} used a
hot isostatic press apparatus at 1000$^{\circ}$C - 1300$^{\circ}$C
and only 0.4~kbar (0.04~GPa) for the series $A$ = In, Y and Lu
while the $A$ = Tl material required 2.5~GPa and 1000$^{\circ}$C
in a piston-cylinder apparatus~\cite{Shimakawa:1999}.

\begin{figure}[t]
\begin{center}
  \scalebox{0.99}{
\includegraphics[width=7cm,angle=0,clip=20]{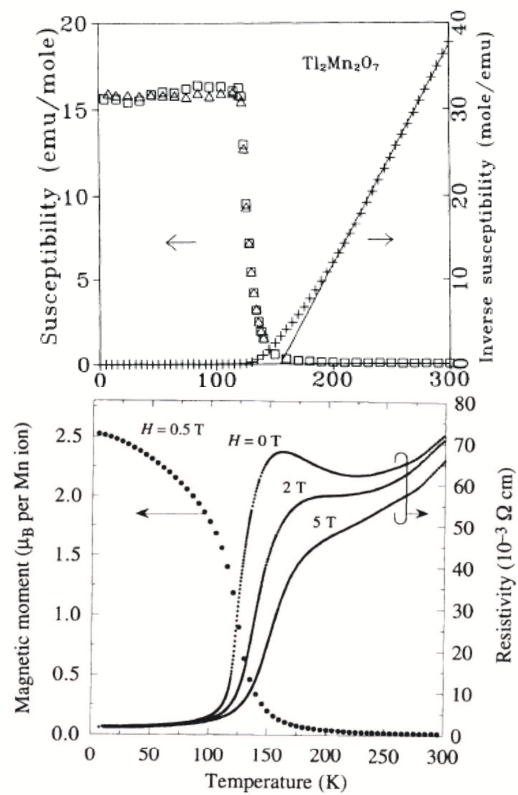}}
\end{center}
\caption{Top: Magnetic properties 
 of Tl$_2$Mn$_2$O$_7$ showing evidence for ferromagnetic order in
 0.005 T~\cite{Raju:1994}.
Bottom: Evidence for a giant magneto resistance
 effect in Tl$_2$Mn$_2$O$_7$~\cite{Shimakawa:1996}.}
\label{Fig_TlMn_Bulk}
\end{figure}

The unit cell constants from all known preparations of $A_2$Mn$_2$O$_7$
phases along with preliminary magnetic characterization data are
collected in Table \ref{Table:ScMnO_a} for $A$ = Sc, Y, Lu, In and Tl, i.e.,
 where only the Mn-sublattice is magnetic.  Given the variety of
 preparative conditions, the agreement among the various groups
for these key parameters is excellent, suggesting that sample to
sample compositional variation is small.  As seen in the case of
the molybdate pyrochlores (Section \ref{Section:M-I}), both the
unit cell constant, $a_0$, and $T_c$ are very sensitive to composition.
All of these materials appear to be ferromagnets, consistent
with the observation of a positive
Curie-Weiss temperature and an
apparent Curie temperature as shown for example in
Fig.~\ref{Fig_TlMn_Bulk}
for  Tl$_2$Mn$_2$O$_7$~\cite{Raju:1994}.
 Note, however, that there appear to be two separate classes
of ferromagnetic materials, if sorting is done by the ratio
$\theta_{\rm CW}/T_c$. The compounds $A$ = Tl and In have $T_c$
 values of 120~K and $\theta_{\rm CW}/T_c$ ratios of about 1.25 - 1.3.
Of course mean field theory gives $T_c=\theta_{\rm CW}$
for a ferromagnet.
Empirically, for most ferromagnets this ratio is larger than 1, 
with typical values between 1.1 and 1.2 as shown in
Table \ref{Table:Typ_FM} for some selected ferromagnetic insulators.
As well, one can estimate a theoretical  ratio calculated
using the Rushbrooke-Wood relationship for $T_c$ and the
mean field theory
for $\theta_{\rm CW}$, assuming the same exchange constant
$J$ (nearest-neighbors only)
 and $S = 3/2$, appropriate for Mn$^{4+}$, which gives
$T_c$/$\theta_{\rm CW}$ = 1.4~\cite{Rushbrooke:1958}.
On the other hand the $A$ = Sc, Y and Lu compounds show
 corresponding ratios within the range 3 - 4, values well
outside those seen for typical ferromagnets and well outside
 the expected theoretical limits. This suggests that these
 phases present a much more complex situation than that of a
simple ferromagnet. Interestingly, when the $A$ element has a
magnetic moment ($A$ = Tb - Yb), the properties, especially the
$\theta_{\rm CW}/T_c$ ratio, return to typical simple ferromagnetic
levels just above 1 as shown in Table \ref{Table:TbMnO_a}.

\begin{table}[t]
\begin{center}
%\begin{tabular}{|p{13pt}|l|l|l|l|l|}
\begin{tabular}{||c|c|c|c|c|c||}
\hline
{$A$}&
{a}$_0$ ({\AA})&
{T}$_{c}${(K)}&
{$\theta_{\rm CW}$}{(K)}&
{$\theta_{\rm CW}$/T}$_{c}$ &
Reference \\
\hline
Tb&
9.972(3)&
38&
-&
-&
\cite{Troyanchuk:1988} \\
\hline
Dy&
9.929&
40(5)&
43(2)&
1.08&
\cite{Subramanian:1988} \\
\hline
Ho&
9.906(3)&
24&
-&
-&
\cite{Troyanchuk:1988} \\
\hline
Ho&
9.905&
37(5)&
33(5)&
1&
\cite{Subramanian:1988} \\
\hline
Ho&
9.907(1)&
37(1)&
37(1)&
1.0&
\cite{Greedan:1996} \\
\hline
Er&
9.888(3)&
24&
-&
-&
\cite{Troyanchuk:1988} \\
\hline
Er&
9.869&
35(5)&
40(5)&
1.15&
\cite{Subramanian:1988}\\
\hline
Tm&
9.852(3)&
14&
-&
-&
\cite{Troyanchuk:1988} \\
\hline
Tm&
9.847&
30(5)&
56(8)&
1.8&
\cite{Subramanian:1988} \\
\hline
Yb&
9.830(3)&
22&
-&
-&
\cite{Troyanchuk:1988} \\
\hline
Yb&
9.830&
35(5)&
41(3)&
1.17&
\cite{Subramanian:1988} \\
\hline
Yb&
9.838(1)&
37(1)&
44(1)&
1.19&
\cite{Greedan:1996}\\
\hline
\end{tabular}
\caption{Collected values for unit cell constants, $a_0$
Curie temperatures, $T_c$, and
Curie-Weiss temperatures, $\theta_{\rm CW}$,
for $A_2$Mn$_2$O$_7$ phases with $A$ = Tb - Yb.}
\label{Table:TbMnO_a}
\end{center}
\end{table}

\subsubsection{Tl$_2$Mn$_2$O$_7$}

Tl$_2$Mn$_2$O$_7$ has attracted considerable attention
following the discovery of giant
 magneto-resistance (GMR)~\cite{Shimakawa:1996,Cheong:1996,Subramanian:1996}.
While this topic is  outside the main area of this review,
having little to do with geometric frustration,
some comments are in order.
Figure \ref{Fig_TlMn_Bulk}
 shows the evidence for this effect~\cite{Shimakawa:1996}.
It was soon realized that the mechanism for this
GMR must be different from that for the well-studied
perovskite manganates~\cite{Schiffer:1995}.
In the latter, double exchange between Mn$^{3+}$ and Mn$^{4+}$
is in part involved in both the metallic behavior and the ferromagnetism.
From a combination of accurate measurements of Mn - O1 distances
and core level spectroscopies, it was determined that only Mn$^{4+}$
 is present in Tl$_2$Mn$_2$O$_7$~\cite{Subramanian:1996,Rosenfeld:1996,Kwei:1997}.
 The ferromagnetism arises from superexchange, not double exchange,
and the metallic properties are due to accidental overlap
 between the Tl 6s band and the Mn 3d band.
The coupling between magnetic and transport properties results
 from abnormally strong incoherent scattering of the conduction
electrons due to spin fluctuations which accompany the FM orderin
GMR thus arises from the large field dependence of $T_c$.
Interestingly, doping of the Tl site with In or Sc, the parent
 compounds of which are both insulating, greatly enhances the GMR
effect as does doping with Tl-vacancies or even
Cd~\cite{Cheong:1996,Ramirez:1997,Alonso:2000,Velasco:2002}.

Neutron diffraction confirmed the ferromagnetic long range
order for this material, finding an ordered moment of
2.91(4)  $\mu_B$ per Mn$^{4+}$ ion, in excellent
 agreement with the 3.0 $\mu_B$ expected for an $S=3/2$ state.
 Small angle neutron scattering is fully consistent with the
formation of domain walls and spin waves below
$T_c$ = 123.2(3)~K~\cite{Lynn:1998,Raju:1994}.
 These results are of interest when compared to data from
 the insulating materials $A$ = Y, Ho and Yb, below.

\subsubsection{Y$_2$Mn$_2$O$_7$}

\begin{figure}[t]
\begin{center}
\includegraphics[width=9.5cm,angle=0,clip=20]{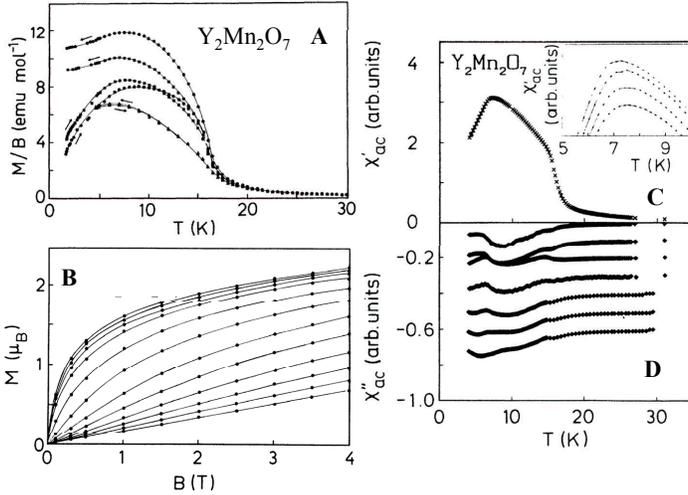}
\end{center}
\caption{Bulk properties of Y$_2$Mn$_2$O$_7$.
Top left: DC susceptibility showing $T_c$ = 16~K
in  0.15 mT (circle), 0.56 mT (square) and 10 mT (triangle).  Bottom left:
Magnetization versus applied field at various
temperatures (K) from top to bottom: 1.8, 5, 7.5, 10, 15, 20,
25, 30, 35, 40, 45 and 50~K.  Right:  AC susceptibility.
 (C) $\chi'$ = 20~Hz. The inset shows the $\chi'$ maximum for
various frequencies, top to bottom: 20, 100, 200 and 1000~Hz.
 (D) $\chi''$ at various frequencies, from top to bottom:
20, 40, 80, 100, 200, 500 and 1000~Hz. The curves are
each shifted by -0.1 from the preceding one~\cite{Reimers:1991}.}
\label{Fig_YMn_Bulk}
\end{figure}

As shown by \textcite{Reimers:1991}, this material exhibits many of the features of a ferromagnet,
including an apparent $T_c \approx 16(1)$~K from DC
magnetization studies and saturation with applied magnetic
 field at low temperatures (see Fig. \ref{Fig_YMn_Bulk}).
 Note that saturation is not reached even at $\mu_{0}H$ = 4~T and
that the moment is close to 2 $\mu_B$ rather than the
3 $\mu_B$ expected for an $S=3/2$ ion.

\begin{figure}[b]
\begin{center}
\includegraphics[width=8.8cm,angle=0,clip=20]{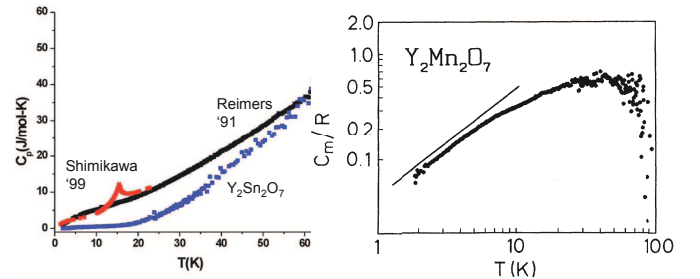}
\end{center}
\caption{Left: Comparison of the total heat capacity
for Y$_2$Mn$_2$O$_7$ reported
by \textcite{Reimers:1991} (black) and \textcite{Shimakawa:1996}
 (red) and the diamagnetic material, Y$_2$Sn$_2$O$_7$ (blue).
  Right: The magnetic component of the heat capacity.
Note the absence of a sharp peak and the linear
dependence at low temperatures~\cite{Reimers:1991}.}
\label{Fig_YMn_SpHt}
\end{figure}

Behavior quite atypical of ferromagnetism is found in the
AC  susceptibility and the heat capacity is shown in
Figs.~\ref{Fig_YMn_Bulk} and \ref{Fig_YMn_SpHt}.
First, while $\chi'$ does indeed show a sharp increase below 16~K, the dominant feature is a broad maximum centered at about 7~K which shows a strong frequency dependence, typical of spin glassiness. Controversy exists concerning measurements of the specific
heat.  \textcite{Reimers:1991} report no lambda-type anomaly
 near 16~K and a linear temperature dependence below 7~K,
behavior again typical of spin glasses.
They also reported that about 60 percent of the total
 entropy of $R \ln 4$ (attained by 100~K) is removed above 16~K.
 On the other hand \textcite{Shimakawa:1999} find a lambda
anomaly at 15~K which they have interpreted as evidence for
 long range ferromagnetic order.  Figure \ref{Fig_YMn_SpHt}
compares both results and those for a non-magnetic, lattice
matched material, Y$_2$Sn$_2$O$_7$.   
Although the agreement in the data
between the two groups is good, apart from the
weak lambda anomaly, Fig.~\ref{Fig_YMn_SpHt} shows
considerable magnetic heat capacity both above and below
 the lambda feature when compared to Y$_2$Sn$_2$O$_7$.
Clearly, this is not consistent with a simple ferromagnetic transition.
As well, even the observation of a lambda anomaly in the specific
heat is no guarantee of a phase transition to a conventional long-range ordered
state as has been demonstrated for other pyrochlores like
Yb$_2$Ti$_2$O$_7$ and the garnet,
Yb$_3$Ga$_5$O$_{12}$~\cite{Blote:1969,Hodges:2002,DalmasdeReotier:2003}.
While the results of Fig.~\ref{Fig_YMn_SpHt}
cannot be easily reconciled, the claim by \textcite{Shimakawa:1999}
that the heat capacity demonstrates, unequivocally,
an ordered state is not substantiated and should be investigated further.

\begin{figure}[t]
\begin{center}
\includegraphics[width=7cm,angle=0,clip=20]{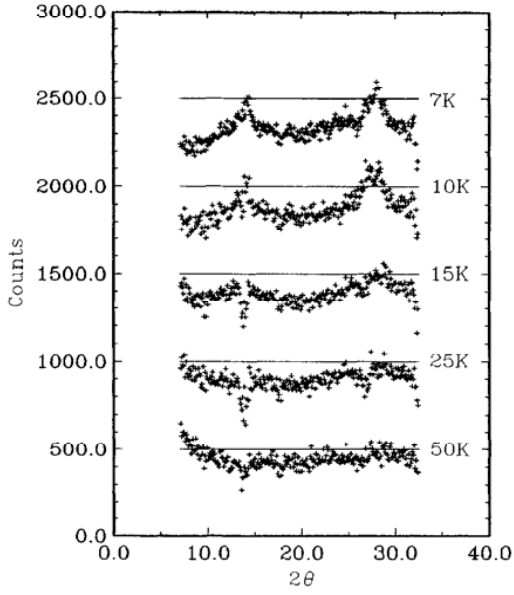}
\end{center}
\caption{Neutron diffraction data for Y$_2$Mn$_2$O$_7$
showing difference plots (data at 200K subtracted) at various temperatures.
Note the buildup of broad peaks at the $111$ and $222$ Bragg positions~\cite{Reimers:1991}.}
\label{Fig_YMn_NPD}
\end{figure}

\begin{figure}[t]
\begin{center}
  \scalebox{0.8}{
\includegraphics[width=7cm,angle=0,clip=20]{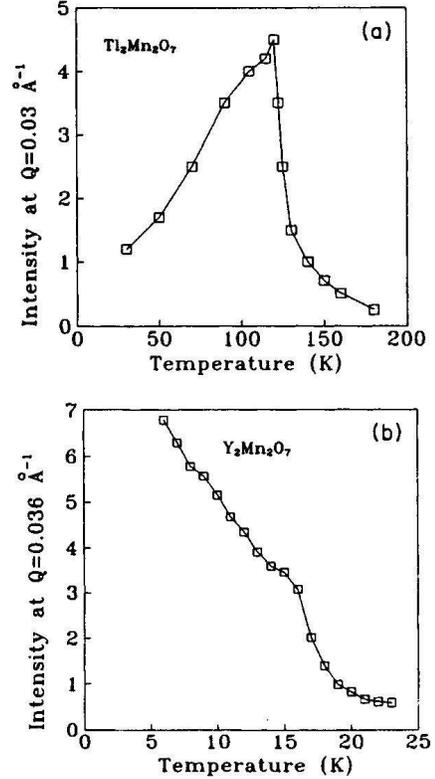}}
\end{center}
\caption{Comparison of the temperature dependence of the
 total small angle neutron scattering (SANS) intensity
 data for the ferromagnet Tl$_2$Mn$_2$O$_7$ and
Y$_2$Mn$_2$O$_7$  at $\vert{\mathbf Q}\vert$ = 0.03 {\AA}$^{-1}$~\cite{Raju:1994,Greedan:1996}.
Similar results for Tl$_2$Mn$_2$O$_7$ were obtained by \textcite{Lynn:1998}.}
\label{Fig_YMn_SANS}
\end{figure}

Neutron scattering data from \textcite{Reimers:1991}
and \textcite{Greedan:1991} also provide evidence for a
quite complex ground state.
 Elastic scattering data show mainly broad features
below 50~K (see Fig.~\ref{Fig_YMn_NPD}).
Enhanced magnetic scattering forms below 15~K at
 nuclear Bragg positions corresponding to the 111
and 222 at 13.9$^{\circ}$ and 28.0$^{\circ}$ respectively.
These data were further analyzed via Fourier transformation
 to give the real space spin-spin correlation function
which suggested that the nearest-neighbor exchange was
weakly antiferromagnetic while further neighbor exchange
 is strongly ferromagnetic. This is consistent with the
 observed positive Curie-Weiss temperatures and the apparent
failure to order in a simple ferromagnetic ground state.

\begin{figure}
\begin{center}
  \scalebox{0.8}{
\includegraphics[width=8cm,angle=0,clip=20]{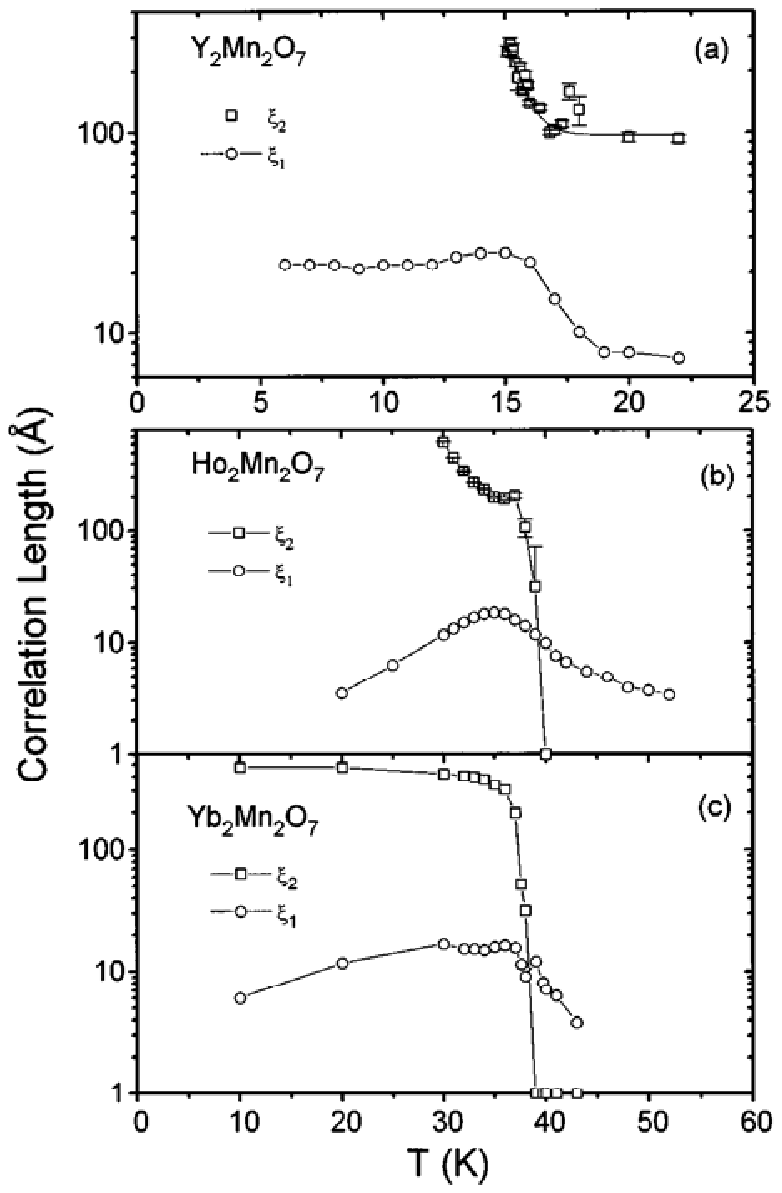}}
\end{center}
\caption{Temperature dependence of the correlation
lengths $\xi_1$ and $\xi_2$ for Y$_2$Mn$_2$O$_7$,
Ho$_2$Mn$_2$O$_7$ and Yb$_2$Mn$_2$O$_7$~\cite{Greedan:1996}.
}
\label{Fig_YMn_CorrL}
\end{figure}

Small angle neutron scattering (SANS) data from \textcite{Greedan:1991,Greedan:1996}  are also unusual and in fact unprecedented.  These are compared with corresponding data for
ferromagnetic Tl$_2$Mn$_2$O$_7$ in
Fig.~\ref{Fig_YMn_SANS}~\cite{Raju:1994,Lynn:1998}.
At $\vert{\mathbf Q}\vert$ = 0.030 {\AA}$^{-1}$ the ferromagnet (Tl$_2$Mn$_2$O$_7$)
shows a decrease in scattering intensity with decreasing temperature
below $T_c$ which is ascribed to scattering from spin waves
which diminish in intensity as the long range order is
established and the cluster size moves out of the SANS window.
 On the other hand, the behavior of Y$_2$Mn$_2$O$_7$
is the opposite, showing an increase below the apparent
 ordering temperature of 15~K, indicating that the
population of subcritical clusters does not diminish
 significantly to low temperatures. In addition the full
range of data could not be fitted to a simple Lorentzian
as is normally the rule for ferromagnets, as for example in
Tl$_2$Mn$_2$O$_7$~\cite{Lynn:1998}, but only
to a Lorentzian plus Lorentzian-squared (L+L$^2$)
law which  involves two correlation lengths, $\xi_1$ and $\xi_2$,

\begin{equation}
S(Q) = {A\over(Q + 1/\xi_1)} + {B\over(Q + 1/\xi_2)^2}.
\label{Equ:correlation}
\end{equation}

The L + L$^2$ law is often found for systems
in which there is competition between ferromagnetic
order and random-field disorder and where $\xi_1$ is
associated with the ferromagnetic correlations
and $\xi_2$ with the random fields~\cite{Rhyne:1985,Arai:1985,Aharony:1983}.
As shown in Fig. \ref{Fig_YMn_CorrL}, for Y$_2$Mn$_2$O$_7$
the two correlation lengths differ by nearly an order of
 magnitude over the temperature range investigated.
$\xi_1$ shows a buildup through a rounded maximum
to a value near 20~\AA.  Similar broad maxima are
found in nearly ferromagnetic,
 metallic glasses~\cite{Rhyne:1985,Rhyne:1985a}.
 $\xi_2$ quickly diverges to a resolution limited value
below 15~K.  The most straightforward interpretation is
that $\xi_1$ measures the temperature evolution of
ferromagnetic correlations that never realize a 
long-range ordered state due to the intervention of
 random fields as monitored by $\xi_2$.
 In this case, both tendencies have onsets near the same temperature.
Note that there is no anomaly near 7~K, the maximum in the
 real part of the AC susceptibility. While a detailed
 understanding of the SANS results for this material
is not yet available, there is little in these data
to indicate that Y$_2$Mn$_2$O$_7$ behaves as a conventional ferromagnet.

\subsubsection{Ho$_2$Mn$_2$O$_7$ and Yb$_2$Mn$_2$O$_7$}

\begin{figure}[b]
\begin{center}
  \scalebox{1}{
\includegraphics[width=8.9cm,angle=0,clip=20]{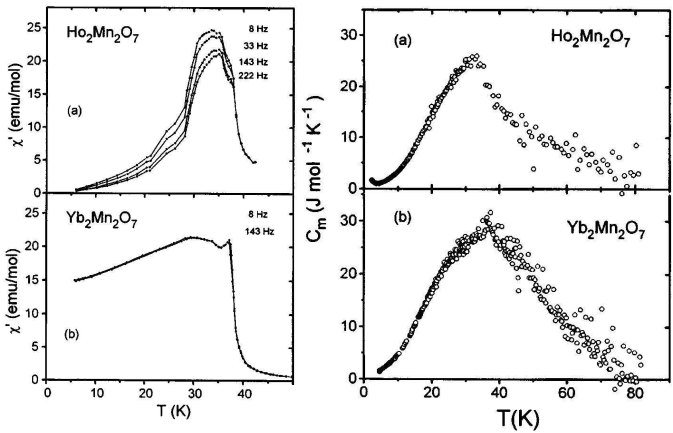}}
\end{center}
\caption{Bulk properties of Yb$_2$Mn$_2$O$_7$ and Ho$_2$Mn$_2$O$_7$.
 Left column is the real part of the
AC susceptibility for and right column is the magnetic component
 of the heat capacity~\cite{Greedan:1996}.
}
\label{Fig_HoMn_Bulk}
\end{figure}

Among the remaining $A_2$Mn$_2$O$_7$ pyrochlores only those with
 $A$ = Ho and Yb have been studied in detail~\cite{Greedan:1996}.
As seen from Table \ref{Table:Typ_FM}, the $\theta_{\rm CW}/T_c$ ratio
is within the range typically seen for ferromagnets and both show
positive Curie-Weiss temperatures which indicates ferromagnetic rather
than ferrimagnetic correlations between the two sublattices.
The apparent $T_c$ values are significantly higher than those
for $A$ = Y, Lu or Sc indicating that the $A$ - Mn exchange has
a strong influence on the ordering temperature.
Unlike Y$_2$Mn$_2$O$_7$, both materials show magnetic
 saturation at 5~K for modest applied fields $\mu_0$H $>$ 2~T
 with values of 12.4 $\mu_B$ and 9.2 $\mu_B$ per formula
unit for $A=$ Ho and Yb, respectively. Assuming ferromagnetic
 $A$ - Mn coupling and that the Mn sublattice moment saturates
 with the full spin-only value of 3.0 $\mu_B$ per Mn$^{4+}$,
the $A$ sublattice saturation moments are 3.2 $\mu_B$ and 1.6 $\mu_B$
per Ho$^{3+}$ and Yb$^{3+}$ ion, respectively.
These are considerably smaller than the full saturation
 moments, $g_{\rm L} J$, for free Ho$^{3+}$ (10.0 $\mu_B$) and
Yb$^{3+}$ (4.0 $\mu_B$), and indicate the
strong influence of crystal fields on the ground state.
A similar Yb moment is found in the isostructural compound, 
Yb$_2$V$_2$O$_7$~\cite{Soderholm:1980}.

\begin{figure}[t]
\begin{center}
  \scalebox{1}{
\includegraphics[width=7cm,angle=0,clip=20]{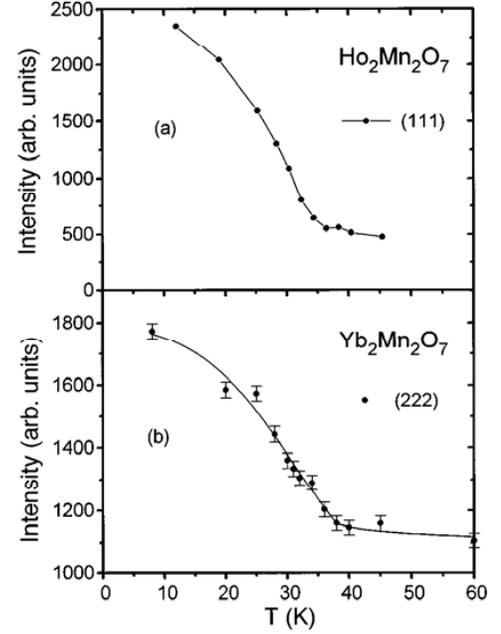}}
\end{center}
\caption{Temperature dependence of magnetic Bragg peaks
 for Yb$_2$Mn$_2$O$_7$ and Ho$_2$Mn$_2$O$_7$ showing order
 parameter like behavior and an apparent $T_c$ of 38(1)~K
 for both~\cite{Greedan:1996}.}
\label{Fig_HoMn_Bragg}
\end{figure}

In addition to DC susceptibility and magnetization,
AC  susceptibility, heat capacity and neutron scattering
data are available but these do not lead straightforwardly
 to a consistent interpretation. For example, the real part
of the AC susceptibility data shown in Fig.~\ref{Fig_HoMn_Bulk}
indicates divergent behavior for the two materials.
While there is an apparent $T_c$  at 38(1)~K with
a broad maximum just above at 30~K for both materials, 
the data for Ho$_2$Mn$_2$O$_7$ show a strong frequency dependence
while those for Yb$_2$Mn$_2$O$_7$ do not.  Thus, the Ho$_2$Mn$_2$O$_7$
phase appears to show a so-called re-entrant spin glass state
where glassy behavior develops at a temperature below $T_c$.

Heat capacity (see Fig.~\ref{Fig_HoMn_Bulk}),
and neutron scattering (see Fig.~\ref{Fig_HoMn_Bragg}),
data seem to suggest a very complex magnetic ground state for both materials.
  No lambda anomaly consistent with true long range order has been
observed but resolution limited magnetic Bragg peaks are seen for
both materials, the temperature dependence of which is order parameter
like and consistent with $T_c$ = 38(1)~K for both compounds.
However, the magnetic structure for Yb$_2$Mn$_2$O$_7$ and Ho$_2$Mn$_2$O$_7$
 is still unknown. SANS data are similar to those for Y$_2$Mn$_2$O$_7$
 in that the L + L$^2$ law holds and the two correlation lengths
have different temperature dependences (see Fig.~\ref{Fig_YMn_CorrL}).
Interestingly, the $B$ coefficient in Equ. \ref{Equ:correlation}, which measures the contribution
from the L$^2$ term associated with the $\xi_2$ correlation
length shows a temperature dependence  akin to that of an order parameter
for all three materials.

More work on the ordered manganese pyrochlores is needed to understand fully their magnetic ground states.  The role of random fields arising from the geometrically
 frustrated Mn sublattice, but whose microscopic origin is not understood, play an important role in the
determination of the magnetic ground state which appears
to be rather inhomogeneous. The $A$ = Y, Ho and Yb compounds do not
behave like simple ferromagnets and studies of their
spin dynamics are warranted.  In addition, diffraction studies
using modern instruments  should be performed on $A$ = Ho and Yb to determine
 the ordered component of the ground state.

\subsection{${\it A}_2$Ir$_2$O$_7$ (${\it A}$ ~=~Nd~-~Yb)}

\begin{figure}[t]
\begin{center}
 \scalebox{0.7}{
\includegraphics[width=9cm,angle=0,clip=20]{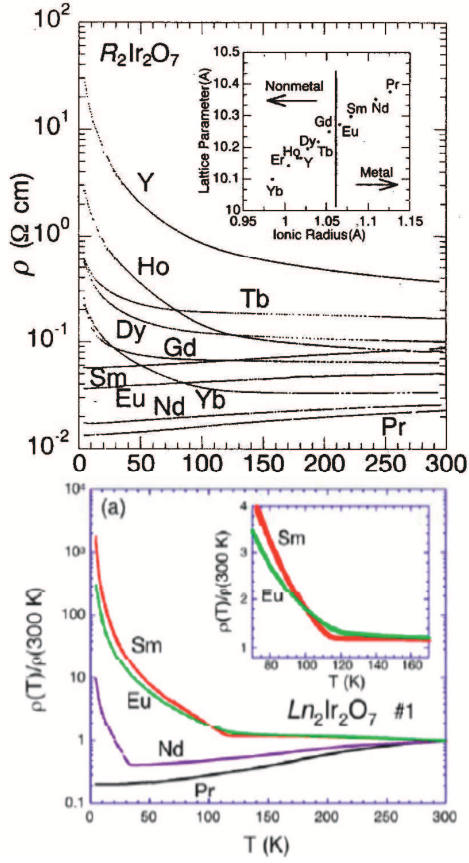}}
\end{center}\caption{Top: Resistivity data for the 
$A_2$Ir$_2$O$_7$ series showing the metal-insulator crossover between 
$A$ = Eu and A = Gd~\cite{Yanagishima:2001}. 
 Bottom:  Evidence for metal/insulator transitions with decreasing temperature for 
$A$ = Eu, Sm and Nd but not for Pr~\cite{Matsuhira:2007}.}
\label{Fig:A2Ir2_MI}
\end{figure}

In the rare earth iridium(IV) pyrochlores, iridium has an electronic configuration
5d$^5$ which is expected to be low spin,
resulting in an $S=1/2$ system and
this series is thus of considerable interest.
The pyrochlore structure is reported to be
stable for $A$ = Pr - Lu, one of the widest
stability ranges for any pyrochlore system~\cite{Subramanian:1993}.
  This series of compounds not only shows a
metal-insulator (MI) transition across the rare
 earth series at room temperature,
like that discussed in Section~\ref{Section:M-I}
for the molybdates, but several individual compounds
 do show a MI transition as a function of temperature.
 These materials were first studied in the early 70's
by \textcite{Sleight:1972}, but have not been studied
systematically until fairly recently.
Early reports on the electrical transport behavior
were contradictory.  One group, \textcite{Lazarev:1978}, reported room temperature resistivities in the range of poor metals while \textcite{Subramanian:1993} argued that
 the entire series of compounds were low activation
energy semiconductors.  Also, among the earliest
 measurements were specific heat studies for
$A$ = Er and Lu which showed no magnetic anomalies
up to 20 K but a rather large gamma coefficient which is not expected for materials reported to
be insulating~\cite{Blacklock:1980}.
Interest in these materials was re-kindled
in 2001 with the report of magnetic anomalies
above 100~K for the $A$ = Y, Lu, Sm and Eu series
members by \textcite{Taira:2001} and the study
of \textcite{Yanagishima:2001} who found a 
crossover from metallic to insulating behavior with
decreasing $A$ radius and with the metal-insulator 
boundary between $A$ = Eu and Gd.  Representative results are shown in Fig. \ref{Fig:A2Ir2_MI}.

\begin{figure}
\begin{center}
  \scalebox{1}{
\includegraphics[width=5cm,angle=0,clip=20]{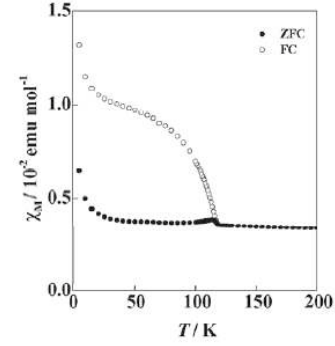}}
\end{center}
\caption{Magnetic susceptibility for Sm$_2$Ir$_2$O$_7$ showing evidence for a magnetic
 anomaly near 120~K~\cite{Taira:2001}.}
\label{Fig_Sm2Ir2_mag}
\end{figure}

There exists a controversy concerning the connection between
the magnetic and metallic properties of these materials.
Initial reports by \textcite{Yanagishima:2001} indicated that the metallic series
 members, $A$ = Pr, Nd, Sm and Eu were not magnetic.
This is contradicted by \textcite{Taira:2001},
 as can be seen  in Fig.~\ref{Fig_Sm2Ir2_mag}, where the metallic Sm$_2$Ir$_2$O$_7$
also clearly shows a magnetic transition.

In subsequent work, it was found that
hole doping of Y$_2$Ir$_2$O$_7$ by substitution of Ca$^{2+}$ for Y$^{3+}$ induces
 an insulator to metal transition at doping levels
of about 15 atomic percent~\cite{Fukazawa:2002}.
The magnetic transition disappears with the onset
 of metallic behavior. One other important issue
 addressed here is the apparent large gamma coefficient
of the specific heat which persists at low temperatures,
first noted by \textcite{Blacklock:1980}. \textcite{Fukazawa:2002}
found gamma to tend toward zero, 0.0(5) mJ/K$^2$~mol~Ir, at
0.4~K and thus concluded that Y$_2$Ir$_2$O$_7$ is indeed a Mott insulator.

Nonetheless, the issue of the persistence of magnetic order
into the metallic samples remains an important question.
 Another attempt to  address this problem has appeared
 very recently by \textcite{Matsuhira:2007}.
 These authors have investigated the correlation
between sample quality and their transport and magnetic properties.
 These studies have revealed MI transitions as the temperature
 is lowered for $A=$ ~ Eu, Sm and Nd.
However, Pr$_2$Ir$_2$O$_7$ appears
 to remain metallic down to the lowest temperatures
investigated (see Fig. \ref{Fig:A2Ir2_MI}).  
We return to Pr$_2$Ir$_2$O$_7$ in Section \ref{PIO}.
 For one sample, Sm$_2$Ir$_2$O$_7$, the onset temperatures
for both the MI and magnetic transition were shown to be identical,
 suggesting that the two phenomena are
intimately connected.
 The samples showing these effects were prepared by a different
 route than those from \textcite{Yanagishima:2001}
 and \textcite{Fukazawa:2002}  which involved firing
 of the starting materials in air.  \textcite{Matsuhira:2007}
used Pt tubes in evacuated silica with a 10 percent
excess of IrO$_2$ and many regrinding sequences. X-ray powder
 diffraction showed single phase samples with well resolved
K$\alpha_1$/K$\alpha_2$ splittings at high angles,
indicative of good crystallinity. No evidence for changes
in the diffraction pattern were observed below the apparent
MI transitions from which the authors suggest that these are
continuous rather than first order. As well, the nature of
the magnetic transition
is still unclear. The observation
of a lambda-type anomaly in the specific heat at $T_m$
indicated by the susceptibility is evidence for a long
range ordered antiferromagnetic ground state but more work is needed.
For $S = 1/2$ systems, neutron diffraction can be a
 challenge but studies on a single crystal should be definitive.
 Clearly, studies of this very interesting pyrochlore series are
 at an early stage and  more work is needed to resolve the
discrepancies among the various groups.

\noindent \underline {Related 5d Pyrochlores.}

Very little has been reported about other pyrochlore oxides
 based on 5d transition elements.
 The synthesis of $B$ = Os (5d$^4$) and Pt (5d$^6$) pyrochlores
with $A$ = Pr - Lu (including Sc and In for 
$B$ = Pt) have been
reported~\cite{Hoekstra:1968,Lazarev:1978}.
The $B$ = Pt series can only be prepared using high pressures,
 4 GPa, and high temperatures, 1200$^{\circ}$C~\cite{Hoekstra:1968}.
Room temperature electrical resistivity values are reported for
the $B$ = Os series and all are in the range for poor metals,
very similar to the initial reports for the 
$B$ = Ir compounds described above.  Cd$_2$Os$_2$O$_7$ 
was first studied by \textcite{Sleight:1974} and later by others for its unusual MI transition at 226~K~\cite{Mandrus:2001}.  This suggests that a closer study of this series could be very interesting.

\subsection{${\it A}_2$Mo$_2$O$_7$ (${\it A}$ ~=~ Gd,~Nd and Sm)}

The earliest report of the existence of rare earth molybdenum
 pyrochlores appears to be that of  \textcite{McCarthy:1971}
from a study of the Eu - Mo - O and Sm - Mo - O phase diagrams
in which the pyrochlore compounds  were observed.
\textcite{Hubert:1974} was the first to report magnetic
susceptibility for Y$_2$Mo$_2$O$_7$ in the range 300 to 1000~K.
 \textcite{Ranganathan:1983} synthesized the series
 $A$ = Gd - Lu along with solid solutions in which the $A$-site
contained various ratios of Nd/Yb and Nd/Er  and reported magnetic
susceptibility and limited electrical transport data for 77~K - 300~K.
The observation of positive Curie-Weiss
($\theta_{\rm CW}$) temperatures led the authors
 to suggest that some of these materials might be ferromagnets,
 for example Sm$_{2}$Mo$_{2}$O$_{7}$~\cite{Mandiram:1980}.
 As well, samples rich in Nd appeared to be metallic~\cite{Ranganathan:1983}.
This situation was clarified by \textcite{Greedan:1987}
 who showed that Mo(IV) pyrochlores for $A$ = Nd, Sm and Gd were
 indeed ferromagnets with $T_c$ = 97, 93 and 83~K, respectively,
which was attributed to the ordering of the Mo(IV) moments,
 an unprecedented observation.  As mentioned in
Section~\ref{Section:M-I},
 there is a link between the ionic radius of the A-ion and
 the electrical transport properties of the molybdenum
pyrochlores which in turn are strongly correlated to the magnetic properties.

The existing evidence appears to support the view that the
properties of ferromagnetism and metallic behavior are fundamentally
linked in these materials. A proposal for the origin of the
ferromagnetism (and antiferromagnetism for Y$_2$Mo$_2$O$_7$)
 has been advanced by \textcite{Solovyev:2003}.
 While the argument is complex and has a number of elements,
 the key idea is that the Mo t$_{2g}$ states are split by
the axial crystal field component into a$_{1g}$ and e$_{g}$ states.
These states are effected differently by systematic changes in
 crystal structure in proceeding from $A$ = Nd to Y.
For $A$ = Nd and Gd, the  e$_{g}$ band is found to be relatively
broad and can support intinerant spin up electrons, rather
in analogy to the double exchange mechanism in the manganate
perovskites which selects a ferromagnetic ground state.
 However, for Y$_2$Mo$_2$O$_7$, the  e$_{g}$ states become
 more localized, while the a$_{1g}$ band broadens.
This is found to favor an antiferromagnetic ground state.
 The marked differences in magnetism and transport properties
are very surprising given that the structural changes in moving
 from $A$ = Nd $\rightarrow$ Gd$\rightarrow$ Y are very subtle.
 For example the Mo - O - Mo bond angle changes
from 131.4$^\circ$[Nd] to 130.4$^\circ$[Gd]
to 127.0$^\circ$[Y]~\cite{Moritomo:2001,Reimers:1988}.

\subsubsection{Gd$_2$Mo$_2$O$_7$}

\begin{figure}[t]
\begin{center}
  \scalebox{0.7}{
\includegraphics[width=8.5cm,angle=0,clip=20]{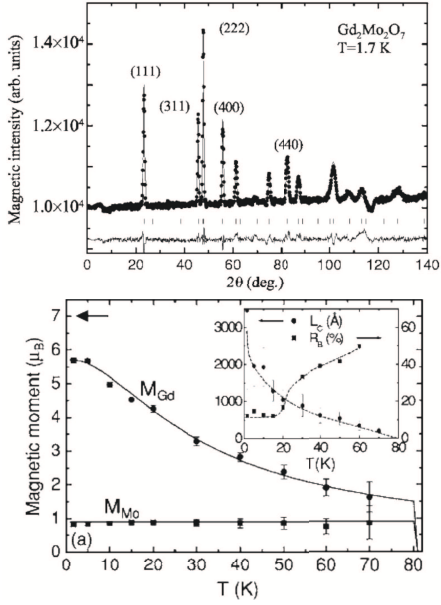}}
\end{center}
\caption{Neutron powder diffraction results for
 Gd$_{2}$Mo$_{2}$O$_{7}$ by \textcite{Mirebeau:2006a}.
Top:  The magnetic scattering at 1.7~K after a
90~K data set has been subtracted.  The solid line is a fit
to a model with collinear ferromagnetic coupling between
Gd and Mo moments.  Bottom: The temperature dependence
of the ordered moments for the Gd and Mo sublattices.
The inset shows the temperature development of the
correlation lengths (L) obtained from the widths of
 the Bragg peaks and the Bragg R for the magnetic structure.}
\label{GdMo_NPD}
\end{figure}

Gd$_2$Mo$_2$O$_7$ has attracted much interest due to its position
near the metal/insulator (MI) boundary (see Fig.~\ref{Fig:Irr_MI})
in this molybdate series.  As already mentioned, the electrical transport properties
are dependent on the stoichiometry of the sample (see Fig.~\ref{Fig:GdMo_MI}).
 Stoichiometric samples are ferromagnetic and metallic with $T_c$ near 80~K.
These show a giant negative magneto-resistance exceeding
 40 percent below 15~K~\cite{Troyanchuk:1998}.
Insulating samples are oxygen deficient and electron doped.
It has been shown by \textcite{Hanasaki:2006}
that insulating samples of Gd$_{2}$Mo$_{2}$O$_{7}$
can be driven metallic by application of high pressure
with the MI boundary occurring between 3~and 4~GPa 
(see Section~\ref{Sec:pressure}).
Extensive specific heat studies have been carried out on
 polycrystalline samples for both $A$ = Sm and Gd by \textcite{Schnelle:2004}.
 Unlike previous studies by \textcite{Raju:1992},
 clear maxima were observed for both materials near 75~K,
indicative of long range magnetic order on the Mo sublattice.
For Gd$_2$Mo$_2$O$_7$, a step-like anomaly at 11.3~K superimposed
on a dominant Schottky peak suggests partial ordering of the Gd spins,
 which was supported by measurement of the entropy.

Until quite recently, there has been little information regarding
the magnetic structure of either the $A$ = Sm or Gd molybdate pyrochlores,
due to the high neutron absorption cross sections for both elements.
\textcite{Mirebeau:2006a} have solved this problem with the use
of $^{160}$Gd substituted materials.  The room temperature unit cell constant for this sample
is 10.3481(2) \AA, which indicates only a very slight oxygen doping.
The magnetic diffraction pattern at 1.7~K is shown in
 Fig.~\ref{GdMo_NPD} along with the temperature
dependence of the sublattice moments and correlation lengths.
 The absence of a (200) reflection is consistent with a collinear magnetic structure and the best fit occurs for
a ferromagnetic coupling of the Gd and Mo sublattices.
This agrees with the earliest magnetization studies
of ~\textcite{Sato:1986} who found that the bulk saturation
moment could only be understood in terms of a ferromagnetic Gd - Mo coupling.

From the right panel of Fig.~\ref{GdMo_NPD} it is clear that the
Mo sublattice orders above 80~K and that the Gd moments
 are polarized due to the Gd-Mo coupling.
 The ordered moments on both sublattices are significantly
 smaller than the spin only values of 2.0 $\mu_{B}$ and 7.0 $\mu_{B}$
 for Mo$^{4+}$ and Gd$^{3+}$, respectively, although the Mo
moments are similar to those found for the $A$ = Nd phase.
Muon spin relaxation data~\cite{Apetrei:2007}, 
which are most sensitive to the Gd spins,
indicate that strong spin fluctuations persist below $T_c$
 and as low as 6.6 K. This is consistent with
the $^{155}$Gd M\"{o}ssbauer data of \textcite{Hodges:2003}
and also, perhaps, with the specific heat data of \textcite{Schnelle:2004}.
 The conclusion is that Gd$_{2}$Mo$_{2}$O$_{7}$ is perhaps an
 unconventional ferromagnet with strong spin fluctuations.
This maybe analogous to the situation encountered for manganate pyrochlores
in Section~\ref{Sec:AMnO}.
 The remarkable properties of this material under applied
pressure will be described in Section \ref{Sec:pressureA2Mo2O7}

\subsubsection{Nd$_2$Mo$_2$O$_7$}

\begin{figure}[t]
\begin{center}
  \scalebox{0.9}{
\includegraphics[width=9cm,angle=0,clip=20]{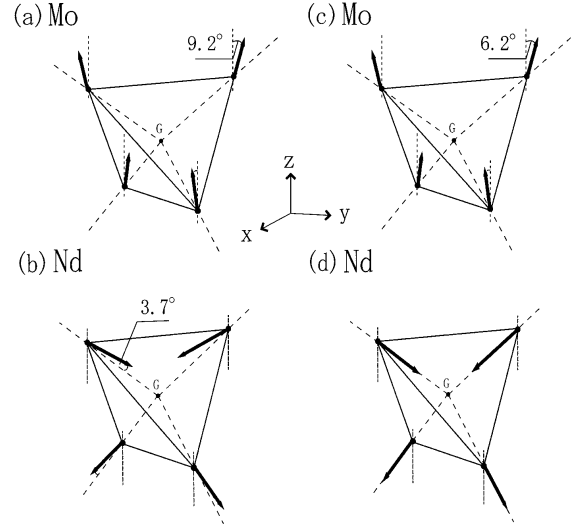}}
\end{center}
\caption{Proposed spin configurations [(a),(b) and (c),(d)]
for the Mo and Nd moments from neutron diffraction on a
 single crystal of Nd$_{2}$Mo$_{2}$O$_{7}$ at 4~K~\cite{Yasui:2001}.
}
\label{Fig_NdMo_Str}
\end{figure}

\noindent\underline{Magnetism.}
It is well established that the Nd$_2$Mo$_2$O$_7$ phase is
 a metallic ferromagnet with $T_c$ = 95~K for stoichiometric,
polycrystalline samples. That the ferromagnetism is due to the
Mo sublattice was first demonstrated by~\textcite{Greedan:1991}
from powder neutron diffraction studies.
A canted ferromagnetic structure on the Mo sites with an ordered moment of 1.1 $\mu_B$
 was proposed from analysis of data at 53~K which are dominated
by magnetic scattering from the Mo.  Subsequent studies by
\textcite{Yasui:2001} using a single crystal with
$T_c$ = 93~K have provided a better defined magnetic structure for both the Nd and Mo sublattices.
 The magnetic structure at 4~K is shown in Fig.~\ref{Fig_NdMo_Str}.
The Mo spins are non-collinear making an angle of 9$^{\circ}$ with the
local $z$ axis which is parallel to a four fold axis of the crystal.
 The Mo moment is 1.2 $\mu_B$, very close to that found earlier
 and slightly greater than half the value expected for a
spin only $S=1$ ion. The Nd moments are aligned nearly along
 the principal 3-fold rotation axes of the tetrahedra,
in a 2-in/2-out pattern. Nd moments at 4~K are 1.5 $\mu_B$ and
 the relative orientations of the Nd and Mo sublattice moment
directions are antiferromagnetic.

\begin{figure}
\begin{center}
  \scalebox{0.9}{
\includegraphics[width=7cm,angle=0,clip=20]{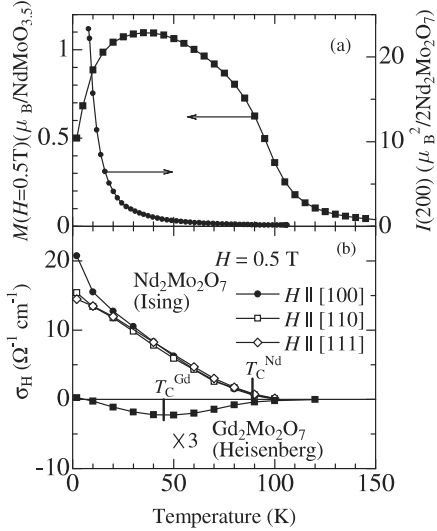}}
\end{center}
\caption{ (a) Temperature dependence of the bulk magnetic
moment and the  (200)  magnetic reflection which tracks
 the development of the chiral order on the Nd sites in
 Nd$_{2}$Mo$_{2}$O$_{7}$.
(b) Temperature dependence of Hall conductivities for
Nd$_{2}$Mo$_{2}$O$_{7}$ and Gd$_{2}$Mo$_{2}$O$_{7}$.
Note the conventional behavior for $A$ = Gd while the
 conductivity remains large and finite at low
temperature for $A$ = Nd~\cite{Taguchi:2004}.
}
\label{Fig_NdMo_Bulk}
\end{figure}

\noindent\underline{Anomalous Hall Effect.}
Much attention has been focussed on Nd$_{2}$Mo$_{2}$O$_{7}$
 since the observation of the unprecedented behavior of
the so-called anomalous Hall effect (AHE) independently
by \textcite{Katsufuji:2000} and \textcite{Taguchi:2001}.
In ferromagnets, the transverse or Hall resistivity has
two contributions, the ``ordinary" Hall coefficient, $R_o$,
which is proportional to the applied magnetic field $B$
 and the ``anomalous" coefficient,
$R_s$ which is proportional to the sample magnetization,
$M$, as in Eq.~(\ref{Equ:AHE}).

\begin{equation}
                    \rho_{H} = R_{o}B + 4\pi R_{s}M	.
 \label{Equ:AHE}
\end{equation}

The usual behavior is for the AHE contribution to vanish at
low temperatures in ferromagnetic metals with collinear
 spin configurations.  However,
in the case of Nd$_2$Mo$_2$O$_7$, the AHE
actually increases and remains large and finite at
$T=0$ (see Fig.~\ref{Fig_NdMo_Bulk}).
In one interpretation, this behavior has been attributed to
the spin chirality (the two-in, two-out state is 6-fold degenerate)
of the Nd$^{3+}$ moment configuration which acts as an effective magnetic
field and effects the carrier dynamics in the same way as a real
 magnetic field~\cite{Taguchi:2001,Taguchi:2004}.
A strong point of this argument is the contrast between this AHE and that in the Heisenberg, $A$ = Gd phase where spin chirality is not present.
For the $A$ = Gd material the AHE vanishes at
 low temperatures as with a normal ferromagnetic metal.
However, this interpretation has been questioned
by \textcite{Yasui:2003a} and \textcite{Kageyama:2001} who argue,
based on extensive neutron scattering and specific heat studies,
that the origins of the AHE for Nd$_2$Mo$_2$O$_7$ are much more
complex than thought originally and that the spin chirality
 mechanism alone cannot provide a quantitative explanation
and may at best play only a minor role ~\cite{Kageyama:2001,Yasui:2003,Yasui:2006}.
  On the other hand, \textcite{Kezsmarki:2005} interpret magneto-optical
 data on both Nd$_{2}$Mo$_{2}$O$_{7}$ and Gd$_{2}$Mo$_{2}$O$_{7}$
in favor of an important contribution from spin chirality.
 At present this controversy appears to be unresolved.
Further comment on this problem is presented in
Section~\ref{Sec:AMo} below.

\subsubsection{Sm$_2$Mo$_2$O$_7$}

Stoichiometric samples of this material generally show $T_c$ = 93~K, although as already mentioned in Section~\ref{Section:M-I},
for oxygen deficient samples this value can be significantly
 reduced (see Fig.~\ref{Fig:SmMo_prop}).
A crystal of Sm$_2$Mo$_2$O$_7$ with a $T_c$ = 73~K
 was reported by \textcite{Taguchi:1999} to show also a
 finite anomalous Hall Effect  coefficient at T~=~0~K.
A giant negative magnetoresistance of 13 percent near $T_c$
and 18 percent at 4~K has been reported~\cite{Taguchi:2000},
but little more is known about Sm$_2$Mo$_2$O$_7$.

\subsubsection{${\it A}_2$(Mo${\it B})_2$O$_7$}

\label{Sec:AMo}

Studies of $B$-site substituted compounds are very sparse.
\textcite{Troyanchuk:1998} have reported the preparation
of V - doped Gd$_2$Mo$_2$O$_7$ with $B$-site composition Mo$_{1.2}$V$_{0.8}$.
Gd$_2$Mo$_2$O$_7$ is a ferromagnetic metal while the corresponding
Gd - V pyrochlore does not exist even under high pressures.
 Nonetheless, the $A_2$V$_2$O$_7$ series members, which include only
 $A$ = Lu, Yb and Tm at ambient pressure are all ferromagnetic
 semiconductors with $T_c$ in the range 70~K - 75~K~\cite{Bazuev:1976,Shinike:1977,Soderholm:1980}. At this particular composition, the solid solution is
 insulating with resistivities in the range of $\Omega$-cm but
 remains ferromagnetic with a slightly enhanced $T_c$ relative
to the undoped material and a GMR effect of about 40 percent
seen for the pure material has been destroyed~\cite{Troyanchuk:1998}.

In another case, the AHE has been studied for Nd$_2$Mo$_2$O$_7$
with Ti substitutions for Mo up to the $B$-site composition
Mo$_{1.7}$Ti$_{0.3}$~\cite{Kageyama:2001}.
At this doping level the compound is still metallic.
This group had shown earlier that the AHE has two contributions,
 proportional to the magnetizations of Mo and Nd separately,
that is,~\cite{Yoshii:2000,Iikubo:2001}

\begin{equation}
 \rho_{H} = R_{o}B + 4\pi R_{s}M({\rm Mo}) + 4\pi R'_{s}M({\rm Nd})
\label{AHE2}
\end{equation}

\noindent Of these two components, the doping experiments showed that
 $R_s$ changes sign from positive to negative with increasing
Ti content while $R'_s$ does not.
This observation was taken as evidence that the
mechanism for AHE in these materials is more complex
than the pure chirality model advanced initially~\cite{Taguchi:2001}.

\section{Spin Glass Phases}

The spin glass state is one where the combination of (i) randomness and (ii) frustration
prevents the development of conventional long range magnetic order characterized
by delta-function magnetic Bragg peaks~\cite{Binder:1986}.
There are a number of experimental pathways to the spin glass state.
The classic case involves dilute concentrations of magnetic atoms in a metallic,
diamagnetic host such as Au$_{1-x}$Fe$_x$ where $x\sim 0.05$, already mentioned in
Section I. Here the effective coupling between Fe moments
is mediated by the Au conduction electrons, giving rise to an RKKY interaction
whose sign depends on the distance between two Fe moments.
This leads to competing random ferro- and antiferromagnetic constraints at each magnetic site.
Another approach is to introduce disorder into a magnetically ordered
but geometrically frustrated material by dilution of the magnetic sites with diamagnetic
ions to levels near the percolation threshold for the particular lattice.
An example is Eu$_{1-x}$Sr$_x$S where Sr$^{2+}$ dilution of the face centered cubic Eu$^{2+}$ sites
destroys the ferromagnetic ground state and a spin glass state emerges near $x = 0.5$.
Similar to the latter example is LiHo$_x$Y$_{1-x}$F$_4$, where
Ho$^{3+}$ is an effective Ising spin and where the predominant
interactions are dipolar.  Dipolar interactions are intrinsically frustrated since they can be
either ferromagnetic or antiferromagnetic depending on the orientation of the
vector $\bm r_j-\bm r_i$ between spins at positions $\bm r_i$ and $\bm r_j$.
As a result, substitution of Ho by Y generates random frustration.
Indeed, LiHo$_x$Y$_{1-x}$F$_4$ is an Ising ferromagnet for $x$ down to about $x\sim 0.2$.
For $x\lesssim 0.2$, a dipolar spin glass state develops~\cite{Reich:1990,Wu:1993},
although this conclusion has recently been questioned~\cite{Jonsson:2007}.

One question that has attracted much attention in the field of spin glasses
is whether or not a thermodynamic spin freezing transition occurs at nonzero temperature
for physical dimensions. While it was determined some time ago that there
is no spin glass transition in two dimensions, a seemingly definite conclusion
for three-dimensional (3D) systems has only recently become
available, thanks to the work of
\textcite{Ballesteros:2000} who
pioneered the use of the scaled correlation length method
in finite-size scaling analysis of spin glass models.  A large majority of real magnetic systems are better described by Heisenberg spins
rather than Ising spins and it was believed that the 3D XY and Heisenberg spin glass models
do not have a transition. The explanation for the experimentally observed transition
in real systems therefore had to invoke random anisotropy as the mechanism
responsible for driving the system into the
Ising spin glass universality class~\cite{Binder:1986}.
However, recent extensive Monte Carlo simulations on the 3D XY and Heisenberg
models employing the \textcite{Ballesteros:2000}
method find some compelling evidence for a phase
transition at nonzero temperature in these two systems~\cite{Lee:SG2003}, although
the lower-critical dimension for these models for a nonzero transition temperature
seems to be very close to three~\cite{Lee:SG2007}.

From a theoretical perspective, models with couplings of random signs, such
as the so-called Edwards-Anderson model, have attracted by far the most
attention. One generally believes that as long as there is a transition
at nonzero temperature, the universality class should be the same irrespective
of the details of the model, i.e., either continuous or discrete distributions of
random bonds or a randomly diluted frustration.

\begin{figure}[t]
\makebox[3.7in]
{
\makebox[1.8in][l]{
\parbox{0.1in}{\includegraphics[width=1.9in]{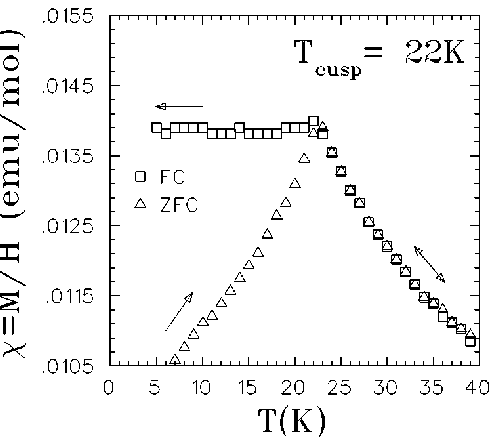}}}
\makebox[0.01in][l]{ }
\makebox[2.5in][l]{
\parbox{0.1in}{\includegraphics[width=1.89in]{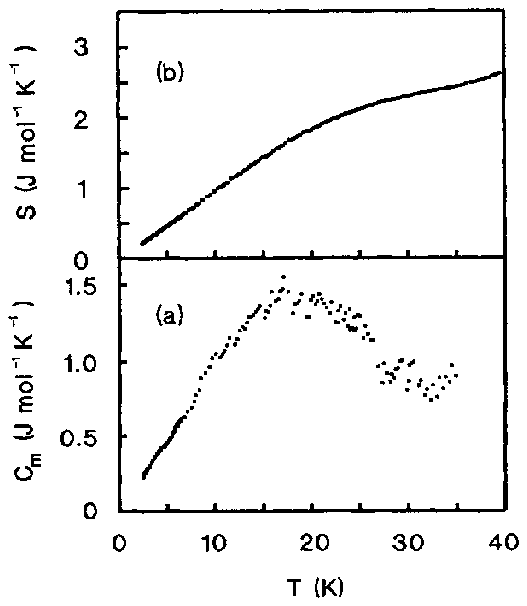}}}
}
\caption{Left:  DC magnetic susceptibility for Y$_2$Mo$_2$O$_7$
at an applied field of 100 Oe~\cite{Gingras:1997}.
Right:  Heat capacity (a) and entropy (b) for Y$_2$Mo$_2$O$_7$.
 Note the broad maximum below 20~K and the linear dependence
on temperature below 7~K~\cite{Raju:1992}.}
\label{YMO_FC-ZFC}
\end{figure}

Experimentally, a number of signatures are taken as indicative of the spin glass
state, such as the frequency dependence of $\chi'$ in the AC
susceptibility, the linear temperature dependence of the low temperature
heat capacity or the $\mu$SR line shape, but the definitive
 approach is the measurement of the temperature dependence
 of the nonlinear magnetic susceptibility, $\chi_3$~\cite{Binder:1986}.
The DC magnetization, $M_z(T,B_z)$, can be expanded as a Taylor series of the applied magnetic field $B_z$ as

\begin{equation}
M_z(T,B_z) \approx \chi_1(T)B_z - \chi_3(T){B_z}^3 \;,
\end{equation}

where $T$ is the temperature.  At a second order spin glass transition, one expects $\chi_3(T)$ to show a power-law
divergence as $\chi_3(T)\sim (T-T_f)^{-\gamma}$,
with $\gamma$ a critical exponent
characterizing the spin glass transition at the freezing temperature
$T_f$~\cite{Binder:1986,Gingras:1997}.
Other critical exponents can also be determined by measuring the full nonlinear
magnetic field dependence of $M_z(B_z,T)$ ~\cite{Binder:1986,Gingras:1997}.

With this very minimal background material in hand, we can now discuss
the spin glass behaviors observed in some of the pyrochlore oxides.

\begin{figure}[b]
\makebox[3.5in]
{
\makebox[1.6in][l]{
\parbox{0.01in}{\includegraphics[width=1.7in]{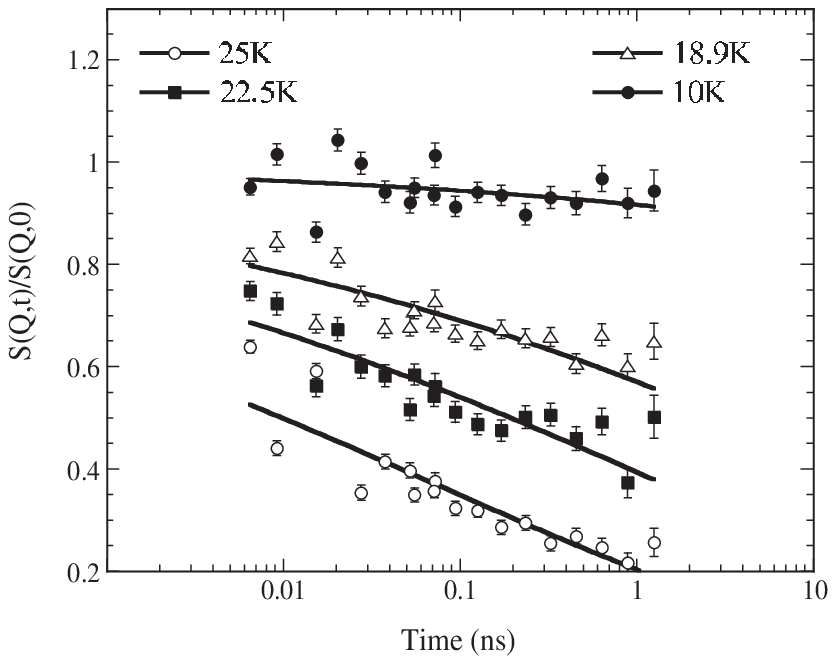}}}
\makebox[0.005in][l]{ }
\makebox[1.98in][l]{
\parbox{0.01in}{\includegraphics[width=1.9in]{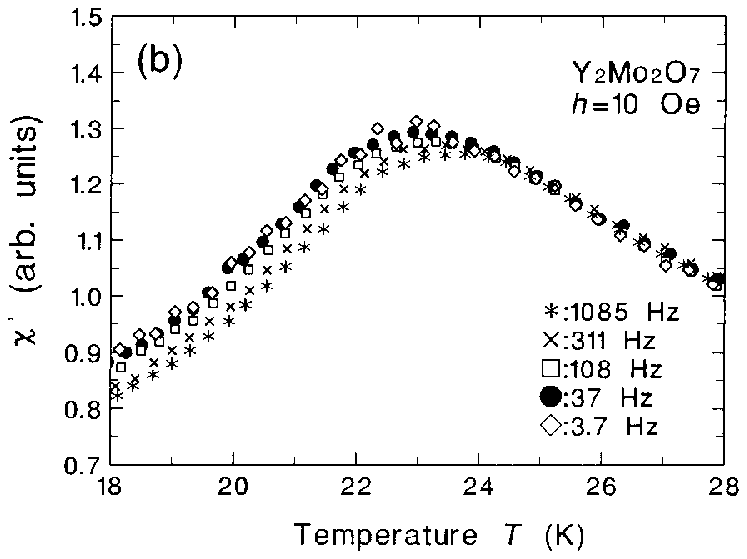}}}
}
\caption{Left:  Neutron spin echo results for Y$_2$Mo$_2$O$_7$ at
temperatures spanning $T_f$ = 22.5~K determined from static magnetization data.
 Note that within this time window, spin freezing
is not fully established until 10~K~\cite{Gardner:2001,Gardner:2004}.
 Right:  Frequency dependent AC susceptibility, $\chi'$,
for Y$_2$Mo$_2$O$_7$ showing classical spin glass behavior~\cite{Miyoshi:2000}.}
\label{YMO_Freq_dep}
\end{figure}

\subsection{Y$_2$Mo$_2$O$_7$ and Tb$_2$Mo$_2$O$_7$ }

\subsubsection{Y$_2$Mo$_2$O$_7$}

\label{Sec:YMO}

The initial report by \textcite{Greedan:1986},  showing a canonical spin glass behavior,
(see Fig.~\ref{YMO_FC-ZFC}a) for Y$_2$Mo$_2$O$_7$  has sparked considerable interest which continues to the present time. Neutron powder diffraction data,
albeit of moderate resolution, could be analyzed in terms of a fully ordered pyrochlore model~\cite{Reimers:1988}.  Specific heat studies by \textcite{Raju:1992} showed only a broad maximum near the apparent $T_f=22$~K and a linear dependence on temperature at low $T$, another feature typical of spin glasses (see Fig. \ref{YMO_FC-ZFC}b).  AC susceptibility~\cite{Miyoshi:2000}, thermo-remanent magnetization by~\textcite{Dupuis:2002}
and ~\textcite{Ladieu:2004} provide more evidence for the canonical
 spin glass behavior of this material. The AC susceptibility in nearly zero field shows the classic frequency dependence (see Fig.~\ref{YMO_Freq_dep}).

\begin{figure}[t]
\begin{center}
  \scalebox{0.9}{
\includegraphics[width=7cm,angle=0,clip=20]{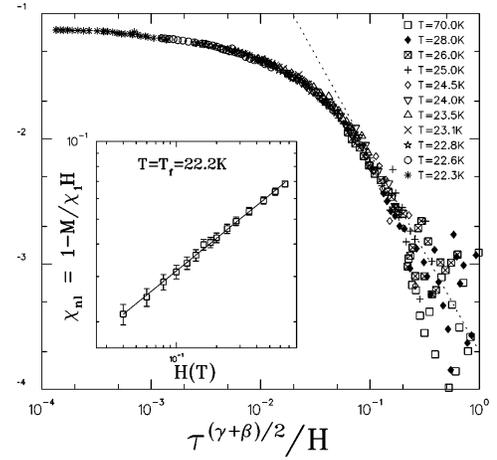}}
\end{center}
\caption{Nonlinear magnetization analyzed according to a scaling model
for a $T_f=22$~K and $\gamma=2.8$ and $\beta=0.75$.
Inset shows a log-log plot which allows determination
 of the critical parameters with $\delta = 4.73$ which
follows from the scaling law with the above
$\gamma$ and $\beta$ values~\cite{Gingras:1997}.}
\label{YMO_Sus}
\end{figure}

The best evidence for the nearly canonical spin glass character came from measurements of the non-linear susceptibility, $\chi_3$, by \textcite{Gingras:1996,Gingras:1997}.
 The non-linear magnetization could be analyzed, (see Fig.~\ref{YMO_Sus}) according to a scaling model for phase transitions, yielding critical exponents $\gamma$, $\beta$ and $\delta$ which satisfied the scaling relationship

\begin{equation}
               \delta = 1 + \gamma/\beta
\label{Equ:SG}
\end{equation}

\noindent with values $T_f$ = 22~K, $\gamma$ = 2.9(5),  $\beta$ = 0.8(2) and $\delta \approx 4.7$.
These agree well with those found for conventional spin glasses with dilute magnetic centers and
positional disorder~\cite{Binder:1986}, making Y$_2$Mo$_2$O$_7$ indistinguishable from such systems.

\begin{figure}[t]
\begin{center}
  \scalebox{0.9}{
\includegraphics[width=7cm,angle=0,clip=20]{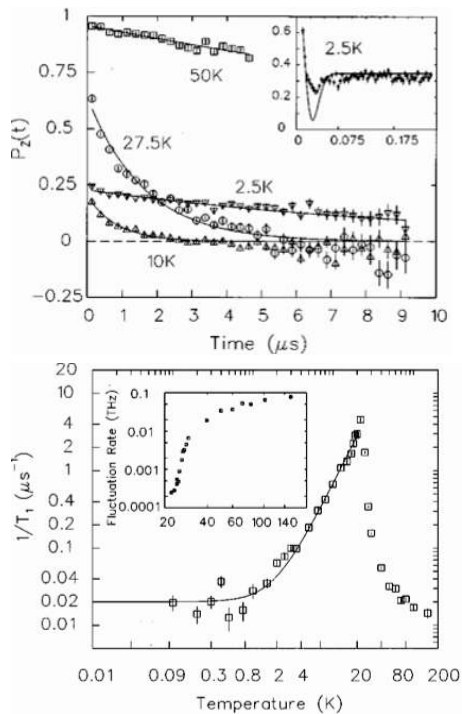}}
\end{center}
\caption{The muon spin relaxation rate,
$1/T_1$ versus temperature. Note the critical slowing down near
$T_f$ and the presence of a small but finite relaxation rate
 persisting to very low temperatures~\cite{Dunsiger:1996}.}
\label{YMO_muon}
\end{figure}

Strong evidence for spin freezing comes also from studies of spin dynamics by muon spin relaxation~\cite{Dunsiger:1996}. Fig.~\ref{YMO_muon} shows $1/T_1$,  the muon spin relaxation rate, as a function of temperature.  These data show features typical of disordered spin-frozen systems such as a critical slowing down as $T_f$ is approached from above followed by a sharp decrease. A finite  $1/T_1$ persists down to the lowest temperatures studied, but this is at the resolution limit of this technique and more needs 
to be done to conclusively determine if the spins are still dynamic in the mK temperature range.

\begin{figure}[b]
\begin{center}
  \scalebox{0.95}{
\includegraphics[width=7cm,angle=0,clip=20]{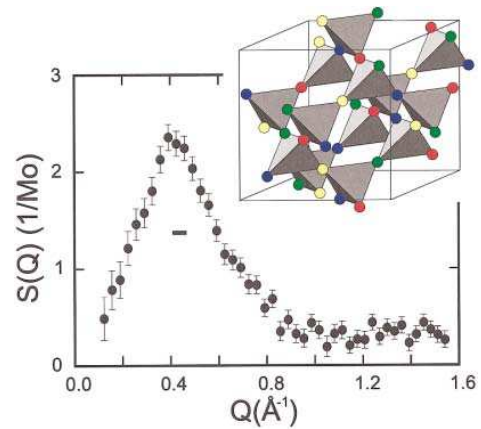}}
\end{center}
\caption{Elastic neutron scattering for Y$_2$Mo$_2$O$_7$ shown
as the difference between data at 1.8~K and 50~K.
The broad peak suggests the short range order model
 shown in the inset~\cite{Gardner:1999a}.
}
\label{Fig_YMo_NPD}
\end{figure}

Neutron scattering experiments,~\textcite{Gardner:1999a}, provide more insight into the possible local ordering and the spin dynamics over a wide temperature range. Figure \ref{Fig_YMo_NPD} shows the $\vert{\mathbf Q}\vert$ dependence of the magnetic elastic scattering at 1.8~K. The obvious feature is a broad peak centered at $\vert{\mathbf Q}\vert$ = 0.44 {\AA}$^{-1}$.  In terms of unit cell dimensions, this value of $\vert{\mathbf Q}\vert$ corresponds to 2$\pi$/d(110), which involves the cubic face diagonal.  From the half-width at half maximum (HWHM) a correlation
length can be estimated as $\xi$ = 1/HWHM = 5 {\AA}.  These two observations are consistent with a  four sublattice structure as depicted in the inset of Fig. \ref{Fig_YMo_NPD}.

Inelastic scattering data give a detailed picture of the  evolution of the spin dynamics in terms of four distinct regimes:
(1) For $T>200$~K there are no discernable correlations
in either space or time. (2) Between 200~K and $T=2T_f$
spatial correlations build up, peaking at $\vert{\mathbf Q}\vert = 0.44$~{\AA}$^{-1}$,
 but the energy distribution is broad. (3) Within the interval
$T_f < T < 2T_f$,
the spatial correlations are no longer evolving but the spin fluctuation rate decreases proportionally to
$(T - T_f)$ in a manner appropriate to the approach to a phase transition. (4) Below $T_f$ the spin fluctuation rate is small and changes very little, consistent with a highly spin frozen state.

The neutron spin echo technique provides information on spin dynamics over a shorter time scale than $\mu$SR, from about 1 to 10$^{-2}$ ns. Within this time window, a different view of the spin dynamics emerges~\cite{Gardner:2001,Gardner:2004}.  The left panel of Fig. \ref{YMO_Freq_dep} shows the time dependence of the normalized intermediate scattering function, $S(Q,t)/S(Q,0)$ for temperatures
 spanning $T_f$ (22.5~K) determined from static magnetization studies.
While the data for 10~K show spin-frozen behavior, there are still
significant spin dynamics below T$_f$. Fits to these data indicate
that spin relaxation times increase by a factor of 10$^7$ between
25~K and 10~K, from 0.07 ns to 8 x 10$^5$ ns.

\begin{figure}[t]
\begin{center}
  \scalebox{0.85}{
\includegraphics[width=8cm,angle=0,clip=20]{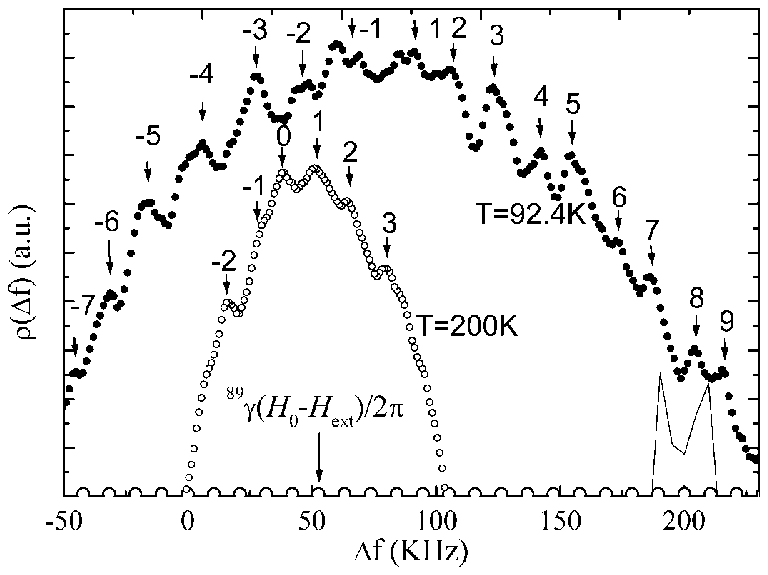}}
\end{center}
\caption{Y-NMR data for Y$_2$Mo$_2$O$_7$ showing multiple peaks arising
below 200K, indicating the presence of multiple Y sites ~\cite{Keren:2001}
}
\label{Fig_YMo_NMR}
\end{figure}

The evidence presented above raises a fundamental question: why is Y$_2$Mo$_2$O$_7$ such a typical spin glass?  In general, insulating antiferromagnetic spin glasses involve dilution
 of the magnetic centers by diamagnetic ions at a concentration below the percolation limit which introduces both positional disorder and frustration simultaneously, two conditions considered necessary for the establishment of the spin glass ground state.  In Y$_2$Mo$_2$O$_7$ the magnetic frustration is provided by the magnetic lattice topology but is there also disorder of some subtle type?  While powder neutron diffraction data show that the average structure is well described by the fully ordered pyrochlore model~\cite{Reimers:1988,Lozano:2007}, local structure probes such as EXAFS, and NMR  have suggested the presence of disorder at some level in this material. Mo K edge EXAFS data were interpreted by \textcite{Booth:2000} to show that the  variance in the Mo - Mo nearest-neighbor distance
was about ten times larger that that for the Mo - O and Mo - Y distances.  From this and a number of assumptions it was estimated that the level of disorder introduced in a pair-wise exchange constant would be of the order of 5 percent. This was judged to be about a factor of five too small when compared with predictions from the \textcite{Sherrington:1975} mean field theory for spin glasses. \textcite{Keren:2001} carried out $^{89}$Y NMR experiments over a limited temperature range.  Data obtained above 200~K (the Curie-Weiss temperature) showed a smooth and broad resonance but at 200~K and 92.4~K, a large number of peaks of small amplitude become superimposed on the broad feature (see Fig.~\ref{Fig_YMo_NMR}).  This was interpreted in terms of a distribution of $^{89}$Y environments due to localized ``lattice"  distortions and led to speculation that these distortions are frustration driven.  A similar experiment was performed using muon spin relaxation by \textcite{Sagi:2005} down to 20~K. The width of the internal field distribution as detected by the muon grows upon cooling at a rate which cannot be explained by the increasing susceptibility alone. Therefore, they conclude that the width of the distribution of coupling constants also grows upon cooling, or in other words, the lattice and spin degrees of freedom are involved in the freezing process in Y$_2$Mo$_2$O$_7$.

 A model of Heisenberg spins  with magnetoelastic coupling was invoked to account for the high freezing temperature.  The derivative of the exchange strength relative to bond length was found to be 0.01~eV/\AA~and the elastic constant to be 0.1~eV/\AA$^2$. In ZnCr$_2$O$_4$, 
where magnetoelastic coupling appears to be important,  these values are 0.04~eV/\AA~and 6.5~eV/\AA$^2$~respectively~\cite{Sushkov:2005}.  Other theoretical studies that have considered classical  pyrochlore Heisenberg antiferromagnets with random variations of the exchange couplings include~\textcite{Bellier:2001}  and \textcite{Saunders:2007}.

At the time of writing, the microscopic mechanism behind the spin glass transition in Y$_2$Mo$_2$O$_7$ is unresolved.  Ideally, one would like to find a means to reconcile the diffraction studies with the local probe results.  One possibility is the application of the neutron pair distribution function method~\cite{Billinge:2004,Proffen:2003}. Studies using this approach are currently on-going~\cite{Lozano:2007}.

\textcite{Dunsiger:1996a} showed that a 20\% dilution of the magnetic Mo$^{4+}$ site with nonmagnetic Ti$^{4+}$ reduces the freezing temperature but increases the residual muon relaxation rate, indicating an increased density of states for magnetic excitations near zero energy.  While there are no other published reports on $B$-site substitution in Y$_2$Mo$_2$O$_7$,  there may be much to learn from such efforts.  As already noted, the microscopic origin of the spin glass behavior  in this material is not understood. What is known suggests that Y$_2$Mo$_2$O$_7$ falls into the category of a bond disordered spin glass rather than a site disordered glass such as Eu$_{1-x}$Sr$_x$S, for example.
In the context of a picture in which Y$_2$Mo$_2$O$_7$ is an ``intrinsic'' random bond spin glass, it would be of interest to track $T_f$ as a function of the doping of the Mo site with a non-magnetic ion such as Ti$^{4+}$.  Interesting physics may emerge at low temperature close to the percolation limit
 which is 39 percent of magnetic site occupancy for the pyrochlore lattice.

\subsubsection{Tb$_2$Mo$_2$O$_7$}

Although this material has received much less attention than
the Y-based pyrochlore, there are strong similarities.  Tb$_2$Mo$_2$O$_7$ is an inherently more complex system with two geometrically frustrated magnetic sublattices. Also, the $A$ = Tb phase is just on the insulating side of the MI boundary for this series and is not a bulk ferromagnet although the Curie-Weiss temperature is reported to be $\theta_{\rm CW}=+17$K~\cite{Sato:1986}.
Among the few detailed studies of transport properties, the material is shown to exhibit an unusual magnetoresistance with both positive (10 percent) and negative (30 percent) MR at low and high fields,
respectively~\cite{Troyanchuk:1988}.  This behavior lacks an explanation to date.

\begin{figure}[t]
\begin{center}
  \scalebox{0.6}{
\includegraphics[width=9cm,angle=0,clip=20]{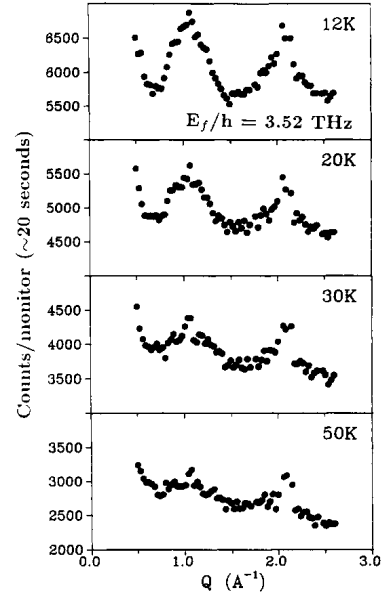}}
\end{center}
\caption{Diffuse magnetic elastic scattering for Tb$_2$Mo$_2$O$_7$ 
at various temperatures ~\cite{Gaulin:1992}.
}
\label{Fig_TMo_NPD}
\end{figure}

Magnetization and neutron scattering studies indicated spin glass like behavior with $T_f$ = 25K and intense diffuse magnetic scattering~\cite{Greedan:1990,Gaulin:1992}.  The diffuse magnetic scattering sets in below 50~K (see Fig.~\ref{Fig_TMo_NPD}),  showing two very broad peaks near $\vert{\mathbf Q}\vert$= 1.0 {\AA}$^{-1}$ and 2.0 {\AA}$^{-1}$.  This pattern is quite different from that for Y$_2$Mo$_2$O$_7$ (see Section~\ref{Sec:YMO}) and indicates that the Tb sublattice scattering is dominant here.  Note also the upturn at low Q.

\begin{figure}[t]
\begin{center}
  \scalebox{0.8}{
\includegraphics[width=9cm,angle=0,clip=20]{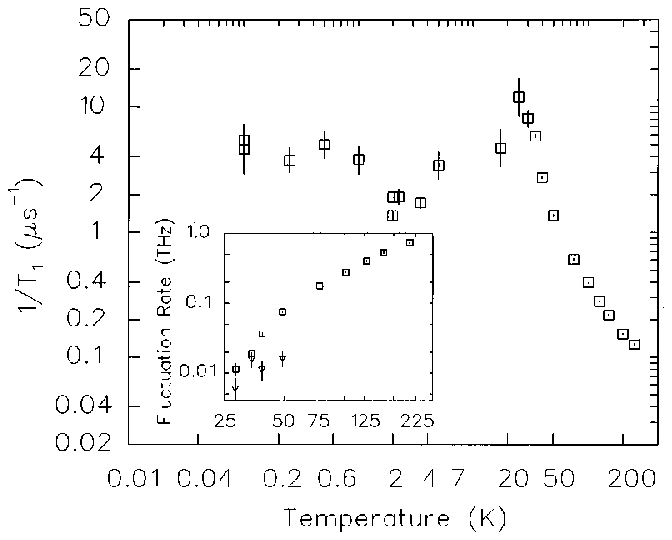}}
\end{center}
\caption{The muon spin relaxation rate vs temperature for
Tb$_2$Mo$_2$O$_7$ in an applied field of 5 mT.
The inset shows the Tb moment fluctuation rate
above the spin freezing point compared with neutron
data indicated by the inverted triangles~\cite{Dunsiger:1996}.
}
\label{Fig_TMo_muon}
\end{figure}

SANS data were also reported over the Q-range 0.019 {\AA}$^{-1}$ to 0.140 {\AA}$^{-1}$. Subtraction of the 300~K data from low temperature data shows that non-zero SANS appeared only above $\vert{\mathbf Q}\vert$ = 0.14 {\AA}$^{-1}$, so it is unlikely that this SANS tail is of ferromagnetic origin~\cite{Greedan:1990}. However, these results might be in conflict with a recent report from \textcite{Apetrei:2007} who observe SANS above $\vert{\mathbf Q}\vert$ = 0.25~{\AA}$^{-1}$.  This issue should be clarified.  From inelastic neutron scattering, the Tb spins have been shown to be fluctuating at about 0.02 THz above $T_f$, but spin freezing within the experimental time window was seen below 25~K~\cite{Gaulin:1992}.  Spin relaxation studies were extended into the $\mu$SR time window (see Fig.~\ref{Fig_TMo_muon}). Comparing with Fig.~\ref{YMO_muon},  which shows corresponding data for Y$_2$Mo$_2$O$_7$, one notes some similarities and differences.  First, the data above $T_f$ are very similar,
indicating a critical slowing down behavior, followed by a clear maximum and a
 subsequent decrease. Note, however, that the relaxation time below about 1~K
remains very large (5 $\mu$s), relative to that for $A$ = Y (0.02 $\mu$s).
This difference is attributed to the larger moment of Tb$^{3+}$ relative
to Mo$^{4+}$ being roughly proportional to the ratio of the squares of the moments.
These results show that the Tb spins are involved in
the freezing and that there exists an appreciable density of states for low energy
 magnetic excitations for these materials which are accessible even at very low temperatures.

\subsubsection{Other ${\it A}_2$Mo$_2$O$_7$}

\begin{figure}[t]
\begin{center}
  \scalebox{0.65}{
\includegraphics[width=9cm,angle=0,clip=20]{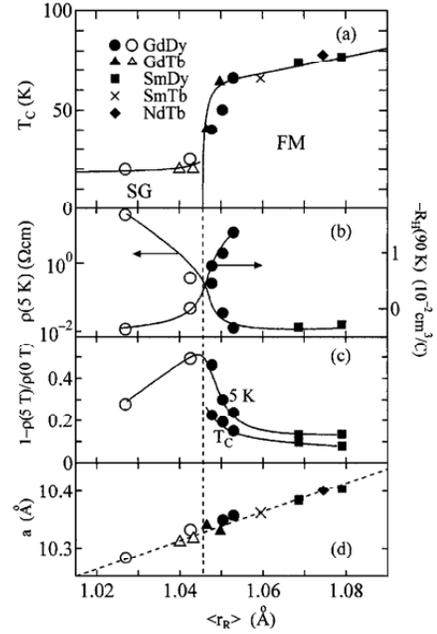}}
\end{center}
\caption{A phase diagram for ($A_{1-x}A'_x$)Mo$_2$O$_7$ pyrochlores.
 (a) $T_c$ or $T_f$. (b) Resistivity at 5K. (c) Magneto-resistance at
5K and $T_c$ (d) Unit cell constant~\cite{Katsufuji:2000}.
}
\label{Fig_AAMo_bulk}
\end{figure}

In fact there is relatively little published information on other members of this series for $A$ = Dy, Ho, Er, Tm, Yb or Lu.  Most show DC and AC susceptibility anomalies near $T=22$~K, as in the $A$ = Y material~\cite{Raju:1995}.  \textcite{Miyoshi:2000} show AC susceptibility data for $A$ = Ho and Tb.
The $A$ = Yb material has been studied using $^{170}$Yb M\"{o}ssbauer measurements.
The major results are that evidence was found for a lower than axial symmetry at the Yb site, suggesting some local disorder and that the local magnetic fields acting on the Yb nucleus show a random distribution, consistent with the bulk spin glass properties~\cite{Hodges:2003}.

There have been a few studies involving mixed occupation of the $A$ site.  \textcite{Katsufuji:2000}  investigated materials of the type $A_{1-x}A'_x$  with $A$ = Nd, Sm and Gd and $A'$ = Tb and Dy.
That is, solid solutions between the ferromagnetic phases, $A$ = Nd, Sm and Gd and the spin glass materials $A'$ = Tb and Dy. The results are summarized in Fig.~\ref{Fig_AAMo_bulk} where the FM metal to spin glass insulator transition occurs at an average $A$-site radius between
 that for Gd$^{3+}$ [1.053 {\AA}] and Tb$^{3+}$ [1.04 {\AA}].  This is fully consistent with the results from the pure samples.  Two studies have been reported in which the $A$-site is substituted
 with non magnetic ions and the results are markedly different.  In some very early work by \textcite{Sato:1987}, solid solutions of the type $A$ = Y$_{1-x}$La$_x$ were investigated.  Compositions were chosen to span the average $A$ site radius from Nd to  Y, intending to cover the ferromagnet to spin glass transition range.  Surprisingly, perhaps, all samples showed only spin glass behavior
 with $T_f$ values not far from that for $A$ = Y, $T_f \sim 22$~K.
 Even the compositions $A$ = Y$_{0.5}$La$_{0.5}$ and Y$_{0.6}$La$_{0.4}$,
which have an average $A$ site radius equivalent to Nd and Sm, respectively,
showed no ferromagnetic transition, although the Curie temperatures
were positive, +41~K and +31~K respectively. Unfortunately, no electrical
transport data have been reported.

\textcite{Hanasaki:2007} chose solid solutions with $A$ = Eu (Eu$^{3+}$is, technically, a non-magnetic ion as $J=0$) and $A'$ = Y or La.  The pure $A$ = Eu is reported to be metallic and ferromagnetic with $T=50$~K or so~\cite{Kezsmarki:2006}.  The data are from an image furnace grown crystal and 
unfortunately no unit cell constant is reported, so some electron doping is likely.  \textcite{Hanasaki:2007} report ferromagnetic behavior for the La-doped materials but with a re-entrant spin glass transition at 22~K.  $T_c$ for the Eu$_{0.85}$La$_{0.15}$ material appears to be about 65~K. For this $A$-site composition the average $A$-site radius is equivalent to that of Sm$^{3+}$ and one might expect a higher value. Perhaps, electron doping is playing a role again.  The phases with $A$ = (Eu, Y) are spin glasses.  These results are quite new and interesting but the system should be investigated further.
Finally, the discrepancy between the results of \textcite{Sato:1987} where the $A$-site involves La and Y and those of \textcite{Hanasaki:2007}, just described, is difficult to understand. One obvious difference
is that the variance of the average radius is of course greater for the $A$ = La, Y combination than for Eu, La or Eu, Y. Thus, the influence of the $A$-site composition on the magnetic
and transport properties is apparently subtle and merits closer investigation.

\section{Spin Ice Phases}

\label{sec:spin-ice}

The spin ice phenomenology in the pyrochlore oxides was discovered in 1997 in Ho$_2$Ti$_2$O$_7$~\cite{Harris:1997,Harris:1998a}.  This system possesses a ferromagnetic Curie-Weiss temperature $\theta_{\rm CW} \sim +2$ K. It was thus quite surprising that this material does not develop long
range order down to 50 mK, with neutron scattering revealing only broad
diffuse scattering features for temperatures much lower than
$\theta_{\rm CW}$.

In Ho$_2$Ti$_2$O$_7$ the strong axial crystal electric field acting on
Ho$^{3+}$ gives rise to an almost ideal, classical Ising
spin. Because of symmetry, the local Ising (quantization) axis is
along the local cubic $\langle 111 \rangle$ directions, such that on a
tetrahedron, a spin can only point ``in'', towards the middle of the
opposing triangular face, or oppositely, hence ``out'' of the
tetrahedron.  While geometric frustration is generally associated with
antiferromagnetic interactions, here frustration arises for
ferromagnetic ones. Indeeed, there are six ``two in/two out'' spin configurations that minimize
 the ferromagnetic exchange energy on an individual tetrahedron, and thus six ground states.
There is an infinity of such ground states for  a macroscopically large sample, and there is
therefore an extensive ground state entropy. This entropy, $S_0$, can
be estimated by borrowing the argument used by Pauling for estimating
the residual proton configuration disorder in common hexagonal water ice~\cite{Pauling:1935}.
The main point is to consider the difference between the number of
constraints necessary to determine a ground state and the number of degrees of freedom that the system possesses. Consider Anderson's Ising pyrochlore antiferromagnet, onto which the local $\langle 111\rangle$ pyrochlore Ising model maps, as discussed above~\cite{Anderson:1956}.  The ground state condition is ``underconstrained'', demanding only that the total magnetization of the four Ising spins on
each tetrahedron be zero. Six of the $2^4=16$ possible spin
configurations satisfy this condition. Counting $2^4$ configurations for
each tetrahedron gives, for a system of $N$ spins and $N/2$
tetrahedra, a total number of microstates, $\tilde{\Omega}=(2^4)^{N/2}=4^{N}$.
This number drastically overestimates the exact total, $\Omega=2^N$.
The reason is that each spin is shared between two tetrahedra, hence the above 16 configurations on each tetrahedron are not independent. Following Pauling's argument, we allocate $2^2=4$ states
per tetrahedron and, assuming that $6/16$ of them satisfy the constraint,
this leads to a ground state degeneracy $\Omega_0=\{2^2(6/16)\}^{N/2}=(3/2)^{N/2}$. The corresponding entropy $S_0=k_{\rm B}\log(\Omega_0)=(Nk_{\rm B})/2\log(3/2)$ is of course just Pauling's original result.

\begin{figure}[b]
  \begin {center}
\includegraphics[width=8cm]{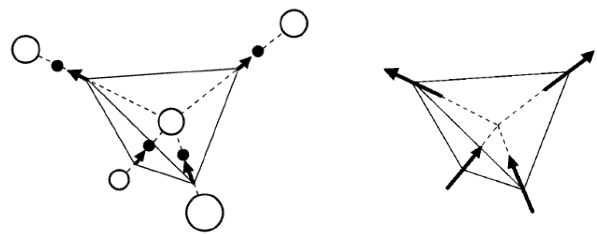}
    \caption{
Illustration of the equivalence of the water ice rule
``two protons near, two protons far'' and the spin ice rule ``two
spins in, two spins out''. The diagram (left) illustrates a water
molecule in the tetrahedral coordination of the ice structure with
proton positions located by displacement vectors that occupy a
lattice of linked tetrahedra. In spin ice (right) the displacement
vectors are replaced by rare earth moments (``spins'') occupying
the pyrochlore lattice, which is the dual lattice (i.e.
the lattice formed by the mid-points of the bonds) of the oxide
lattice in cubic ice.}
\label{spinice}
  \end{center}
\end{figure}

%%%%%%%%%%%%%%%%%%%%%%%%%%%%%%%%%%%%%%%%%%%%%%%%%%%%%%%%%%%%%%%%%%%%%%%

Not only is the residual entropy of ferromagnetic $\langle 111 \rangle$ spins on the pyrochlore lattice the same as Pauling's entropy for water ice but, as shown in Fig.~\ref{spinice}, there is also a rather direct connection between the spin configurations in the pyrochlore problem and that of the proton positions in water ice. For this reason, the term {\it spin ice} was coined.  In anticipation of the forthcoming discussion of the physics at play in Tb$_2$Ti$_2$O$_7$, it is worthwhile to comment on a case where the nearest-neighbor exchange interactions are antiferromagnetic for a situation with local $\langle 111\rangle$ spins. In that case, the ground state consists of all spins pointing in or out of a reference tetrahedron.  Hence, there are in that case only two ground states related by a global spin inversion symmetry, and a second order transition in the universality class of the (unfrustrated) three-dimensional Ising model is expected. Earlier, Anderson had noticed the connection between the statistical mechanics of antiferromagnetically coupled Ising spins on the pyrochlore lattice and Pauling's model of proton disorder in water ice.  However, in Anderson's model, the Ising spins share a common (global) $z-$axis direction, and frustration arises as usual for antiferromagnetic interactions with spins on triangular or
tetrahedral units. However, since the pyrochlore lattice has cubic symmetry, the $x$, $y$ and $z$ directions are equivalent, and this renders Anderson's global antiferromagnetic Ising model unrealistic. It is the
local nature of the quantization direction that is crucial, and which is the origin of the frustration for ferromagnetic interactions and for the ``elimination'' of the frustration for antiferromagnetic exchange. To see this, consider the following toy-model Hamiltonian

 \begin{eqnarray} H & = &  - J \sum_{\langle i,j
\rangle } {\bm S}_i \cdot  {\bm S}_j
    \;   - \Delta \sum_i (\hat z_i \cdot {\bm S}_i)^2 ,
 \end{eqnarray}

with classical Heisenberg spins ${\bm S}_i$ on the sites $i$ of the pyrochlore lattice, interacting via nearest-neighbor exchange coupling $J$.  The second term is a single ion anisotropy interaction with $\Delta$ the anisotropy parameter and $\hat z_i$  a unit vector in the local $\langle111\rangle$ direction at site $i$. For $J=0$ and $\Delta>0$, the energy is lower if ${\bf S}_i$ points along $\hat z_i$, and one refers to this as an Ising anisotropy.  The case $\Delta <0$ would be referred to as an XY model~\cite{Bramwell:1994,Champion:2004}.  However, as discussed in Section~\ref{local-environment}, the real microscopic crystal field Hamiltonian is more complicated than that considered by \textcite{Bramwell:1994} and \textcite{Champion:2004} where the limit $\Delta \rightarrow \infty$ was taken.

Hence, we assume an extreme Ising limit, $\Delta/J \rightarrow \infty$.
${\bm S}_i$ is then confined to be either parallel or antiparallel to $\hat z_i$. To implement this
energetic single-ion constraint, we write ${\bm S}_i = \sigma_i^{z_i} \hat z_i$.  Injecting this back into $H$ above, we obtain

\begin{eqnarray} H & = &  - J    \sum_{\langle i,j \rangle  }
(\hat z_i \cdot \hat z_j) \sigma_i^{z_i}\sigma_j^{z_j}  \nonumber \\
  & = &   + J/3  \sum_{\langle i,j \rangle } \sigma_i^{z_i}\sigma_j^{z_j}
  , \end{eqnarray} 
where we have taken $\hat z_i \cdot \hat z_j=-1/3$  for two distinct cubic $[111]$ directions.  
One sees that for ferromagnetic $J$ ($J>0$), the now ``global''
 $\sigma_i^{z_i}$ Ising variables map onto an equivalent Ising
 antiferromagnet (Anderson's model, \textcite{Anderson:1956})with coupling $J/3$, and is therefore frustrated.
Conversely, for antiferromagnetic $J$ ($J<0$), the minimum energy state
consists of all spins pointing in (say $\sigma_i^{z_i}=+1$ $\forall$ $i$) or
all out ($\sigma_i^{z_i}=-1$ $\forall$ $i$).

\begin{figure}[t]
  \begin {center}
  \includegraphics[height=6.cm,width=7cm]{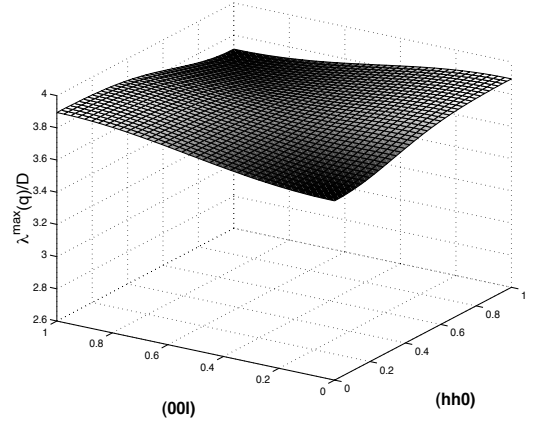}
\caption{Dipolar spin ice model:
the scaled maximum eigenvalues, $\lambda^{max}({\bm q})/D$, in the $(hhl)$ plane.
The dipole-dipole interactions are treated with the Ewald approach.
Here, the exchange coupling ${\cal J}_{\rm nn}$ was set to zero.  While the spectrum is very flat, suggesting a weak propensity towards long-range order, a maximum at a unique ordering wavevector
${\bm q}_{\rm  max}  = 001$.}
  \label{Ewald-spectrum}
  \end{center}
\end{figure}

As will be discussed in Section \ref{DyandHo}, Dy$_2$Ti$_2$O$_7$ is also identified as a
spin ice material. It turns out that both Ho$^{3+}$ and Dy$^{3+}$ carry a sizeable magnetic
moment, $\mu$, of approximately 10$\mu_{\rm B}$ in the crystal field ground state
 of Ho$_2$Ti$_2$O$_7$ and Dy$_2$Ti$_2$O$_7$. Thus, these systems
have a magnetostatic dipole-dipole interaction $D = \mu_0 \mu^2/(4\pi {r_{\rm nn}}^3)$
of approximately $D\sim 1.4$ K at the nearest-neighbor distance $r_{\rm nn}$.
Since $D$ is approximately the same as $\theta_{\rm CW}$, it is surprising
that the long-range and complex nature of the dipolar interactions do not lift
the spin ice degeneracy and drive the system
to long range order at a critical temperature $T_c\sim D$.
In fact, numerical and theoretical studies have compellingly demonstrated that
it is {\it precisely} the mathematical form of the dipole-dipole interactions
that is at the origin of the spin ice phenomenology in rare-earth pyrochlores.
In particular, it is the ferromagnetic character of the dipolar interactions at the
nearest-neighbor distance, and not a ferromagnetic nearest-neighbor exchange 
(which turns out to be antiferromagnetic, as analysis of experimental
data on Ho$_2$Ti$_2$O$_7$ and Dy$_2$Ti$_2$O$_7$ have revealed) that is
primarily the source of frustration.  Indeed, as discussed above, if it were not for the dipolar interactions,
the nearest-neighbor antiferromagnetic interactions alone in Ho$_2$Ti$_2$O$_7$ and  Dy$_2$Ti$_2$O$_7$ would drive the system into a long-range ordered phase.

The minimal model, called the dipolar spin ice model (DSM), that is needed to investigate these
questions includes nearest-neighbor exchange (first term) and long range magnetic dipole interactions (second term) in.

\begin{eqnarray}
\label{Hdsm}
H&=&-J\sum_{\langle
ij\rangle}{\bm S}_{i}^{{\hat z}_{i}}\cdot{\bm S}_{j}^{ {\hat z}_{j}}
\nonumber \\ &+& Dr_{{\rm nn}}^{3}\sum_{i>j}\frac{{\bm S}_{i}^{{\hat
z}_{i}}\cdot{\bm S}_{j}^{{\hat z}_{j}}}{|{\bm r}_{ij}|^{3}} -
\frac{3({\bm S}_{i}^{{\hat
 z}_{i}}\cdot{\bm r}_{i j}) ({\bm S}_{j}^{{\hat z}_{j}}\cdot{\bm
r}_{ij})}{|{\bm r}_{ij}|^{5}}  .
\end{eqnarray}

For the open pyrochlore lattice structure, we expect further neighbor exchange to be very small, so these can be neglected as a first approximation. Recent Monte Carlo simulations seem to confirm this expectation~\cite{Ruff:2005,Yavorskii:2007}.  Here the spin vector ${\bm S}_{i}^{{\hat z}_{i}}$ labels the Ising moment of magnitude $\vert {\bm S}_{i}^{{\hat z}_{i}} \vert=1$ at lattice site $i$, 
oriented along the local Ising $\langle 111 \rangle$ axis ${{\hat z}_{i}}$. The
distance $\vert {\bm r}_{ij}\vert$ is measured in units of the nearest
neighbor distance, $r_{\rm nn}$. $J$ represents the exchange
energy and $D=(\mu_{0}\mu^{2}/(4\pi r_{\rm nn}^{3})$. Because of the
local Ising axes, the effective nearest-neighbor energy scale
is $J_{\rm eff}=J_{\rm nn}+D_{\rm nn}$
where
$J_{\rm nn}\equiv J/3$ and  $D_{\rm nn}\equiv 5D/3$,
since ${\hat z}_i\cdot{\hat z}_j = -1/3$ and $({\hat z}_i\cdot{\bf r}_{ij})  ({\bf r}_{ij}\cdot{\hat z}_j) = -2/3$.
If $D_{\rm nn} = 0$ and $J > 0 $~one obtains the spin ice model originally proposed
by \textcite{Harris:1997,Harris:1998} henceforth referred to as
the ``nearest-neighbour spin ice model''.

The condition $J_{\rm eff}>0$ is a simple criterion to assess whether a system displays a spin ice state.
Mean field theory~\cite{Gingras:2001} and Monte Carlo simulations~\cite{Melko:2004}
 find a critical value for the transition between ``all-in/all-out'' N\'eel order and spin ice
at $J_{\rm nn}/D_{\rm nn} $$\approx$ -0.901, hence quite close to the naive
nearest-neighbor estimate $J_{\rm nn}/D_{\rm nn} =-1$.
The dipolar interactions beyond nearest-neighbors provide a weak,
extra stabilization of the N\'eel phase over the spin ice state.

The success of the nearest-neighbor criterion in assessing whether a system should display spin ice phenomenology or not indicates that, to a large extent, dipolar interactions beyond nearest-neighbors are {\it self-screened}, as originally suggested by a mean field calculation~\cite{Gingras:2001} and Monte Carlo simulations~\cite{Hertog:2000}. The reason why dipolar spin ice systems obey the ice rules is not immediately apparent.  \textcite{Isakov:2005} have proposed that the ice rules result from the fact
that the dipolar interactions are, up to a perturbatively small and rapidly decaying function,  a real-space projector onto the manifold of ice-rule obeying states.  A possibly simpler explanation has recently been proposed which involves separating each point dipole into its constituent magnetic (monopole) charges and requiring that each tetrahedron is, magnetic charge wise, neutral, which automatically leads to the conclusion that all ice-rule obeying states are, again up to a small correction,  ground states of the dipolar interactions~\cite{Castelnovo:2007}.  Such a real-space argument is possibly not unrelated to the Ewald energy calculations in which the dipolar lattice sums are regularized by effective Gaussian charges either in  reciprocal space~\cite{Enjalran:2003} or in direct space~\cite{Melko:2004}.

\subsection{Dy$_2$Ti$_2$O$_7$ and Ho$_2$Ti$_2$O$_7$}

\label{DyandHo}

\begin{figure}[t]
\begin{center}
\includegraphics[height=6.7cm,width=9.3cm]{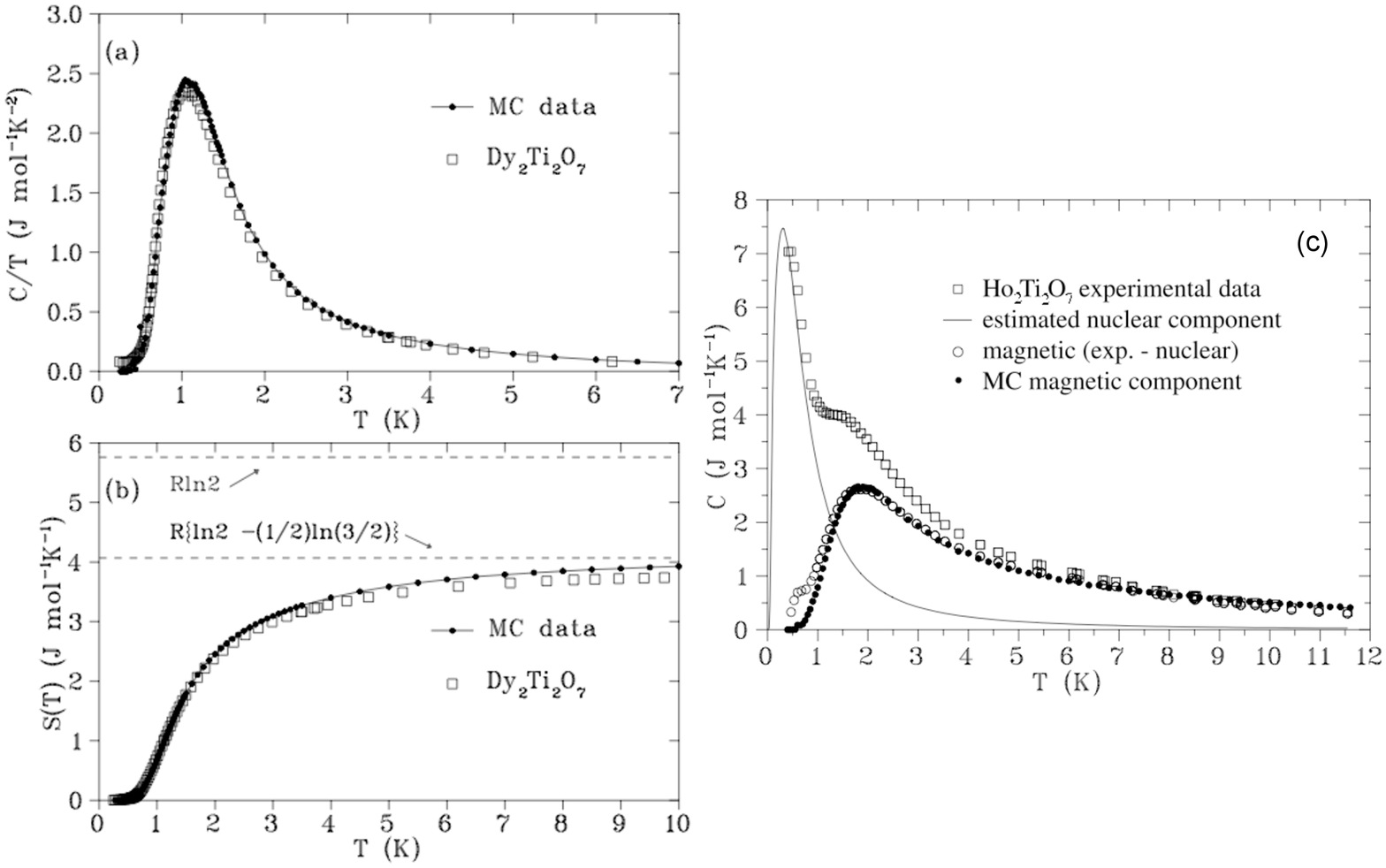}
 \caption{${\rm Dy_2Ti_2O_7}$: (a) Specific heat and (b) entropy versus
temperature,  measured by \textcite{Ramirez:1999}.
The recovered entropy at 10 K agrees reasonably well with Pauling's
entropy, $R/2\ln(3/3)$.  The data are compared to that calculated by
Monte Carlo simulations of \textcite{Hertog:2000} for
the dipolar spin ice model with exchange $J_{\rm nn} =-1.24$ K
 and dipolar coupling $D_{\rm nn}=2.35$ K. (c) ${\rm  Ho_2Ti_2O_7}$: The total specific heat
is shown by the empty squares and the expected
nuclear contribution by the solid  line. The
electronic contribution has been estimated by  subtracting these
two curves (open circles). Near to $0.7$ K,  this subtraction is  prone
to a large error (see text). Dipolar spin ice simulation results
are indicated by the filled circles
(from \textcite{Bramwell:2001a}).}
\label{Spinice-Cv}
\end{center}
\end{figure}

The Dy$^{3+}$ ion in Dy$_2$Ti$_2$O$_7$ was first identified as having strongly anisotropic properties from the maximum in the DC susceptibility~\cite{Blote:1969}  at $T\sim 0.9$ K followed by a precipitous
drop to essentially 0 at $T\sim 0.4$ K. This suggested a strongly anisotropic effective $g$ tensor with a $g_\perp \approx 0$ component perpendicular to the local $\langle 111\rangle$ axis, making Dy$^{3+}$ in effect a $\langle 111\rangle$ Ising model. This was confirmed by \textcite{Flood:1974} from direct measurement of the magnetic moment and from analysis of inelastic neutron scattering data by
Rosenkranz~\cite{Rosenkranz:2000}.  Calculations from crystal field theory, complemented by fitting crystal field parameters (CFPs) to susceptibility data also confirmed the strong Ising nature of Dy$^{3+}$ in  Dy$_2$Ti$_2$O$_7$~\cite{Jana:2002}.  Specifically, the single ion electronic ground state of Dy$^{3+}$ is a 
Kramers doublet of almost pure $\vert J=15/2, m_J=\pm 15/2\rangle$ separated from the first excited state by a gap $\Delta \approx 33$ meV ($\sim 380$ K) while the rescaled CFPs of \textcite{Rosenkranz:2000} and \textcite{Jana:2002}  finds a much smaller excitation gap $\Delta \approx 100$ cm$^{-1}$ ($\sim 140$ K).

\textcite{Blote:1969} measured the specific heat $C(T)$ of Dy$_2$Ti$_2$O$_7$ between 0.37 K and 1.3 K and found only  a broad maximum around 1.2 K and no sign of a sharp specific heat feature indicating a phase transition to long-range magnetic order. Most interestingly, a total magnetic entropy of only $\frac{3}{4}{\rm R}\ln(2)$ was found. Hence, a noticeable fraction of ${\rm R}\ln(2)$ expected for a Kramers doublet was missing.

Soon after the proposal of spin ice in Ho$_{2}$Ti$_2$O$_7$~\cite{Harris:1997,Harris:1998a}, \textcite{Ramirez:1999}  showed that the missing entropy in Dy$_2$Ti$_2$O$_7$ could be determined rather precisely by a measurement of the magnetic specific heat $C(T)$ between 0.4 K and and 12 K. Figure~\ref{Spinice-Cv} shows the temperature dependence of $C(T)/T$.  The magnetic entropy change, $\Delta S_{1,2}$, between temperatures $T_1$ and $T_2$ can be found by integrating $C(T)/T$ between these two temperatures:

\begin{eqnarray}
\Delta S_{1,2} \equiv  S(T_2)-S(T_1) = \int_{T_1}^{T_2} \frac{C(T)}{T} dT ~ .
\end{eqnarray}

Figure ~\ref{Spinice-Cv}b shows that the magnetic entropy recovered is about 3.9 J mol$^{-1}$ K$^{-1}$,
a value that falls considerably short of R$\ln(2) \approx 5.76$ J mol$^{-1}$ K$^{-1}$.  The difference, 1.86 J mol$^{-1}$ K$^{-1}$ is close to Pauling's estimate for the residual extensive entropy of water ice: (R/2)$\ln(3/2) \approx 1.68$ J mol$^{-1}$ K$^{-1}$, thus providing compelling thermodynamic evidence for the existence of an ice-rules obeying state in Dy$_2$Ti$_2$O$_7$.

As mentioned above, the large 10$\mu_{\rm B}$ moments of both Dy$^{3+}$ and Ho$^{3+}$ lead
to a critical role for magnetic dipole-dipole interactions in spin ices. This energy, $D_{\rm nn}$, at nearest-neighbor distances can be estimated from the effective Curie constant or from the single-ion crystal field doublet wavefunctions. The earlier theoretical studies~\cite{Siddharthan:1999,Hertog:2000}  of the DSM estimated

\begin{eqnarray}
D=\frac{5}{3}\left ( \frac{\mu_0}{4\pi} \right ) \frac{\mu^2}{{r_{\rm nn}}^3}
 \approx  +2.35\; {\rm K}
\end{eqnarray}

for both Ho$_2$Ti$_2$O$_7$ and Dy$_2$Ti$_2$O$_7$. Here $r_{\rm nn}=(a_0/4)\sqrt 2$ is the nearest-neighbor distance and $a$ is the size of the conventional cubic unit cell. As discussed above, the factor 5/3 originates from the orientation of the Ising quantization axes relative to the vector direction ${\bf r}_{\rm nn}$ that connects nearest-neighnor magnetic moments.  As we briefly discuss below, the current estimate on $D_{\rm nn}$ for Dy$_2$Ti$_2$O$_7$ is probably accurate to within 8\%$-$10\%.  Hence, the nearest-neighbor exchange, $J_{\rm nn}$, is the main unknown. It can be estimated from the high-temperature (paramagnetic) regime of the magnetic susceptibility, $\chi$,~\cite{Siddharthan:1999}
or of the specific heat, $C(T)$,~\cite{Jana:2002} data.  A different approach, followed by \textcite{Hertog:2000} and \textcite{Bramwell:2001a} has been to determine $J_{\rm nn}$ by fitting either the height of the specific heat, $C_{\rm p}$, peaks near 1 K or the temperature at which the peaks occurs, $T_{\rm p}$, against Monte Carlo simulations of the DSM.  Interestingly, fits of $T_{\rm p}$ or $C_{\rm p}$ allow for a consistent determination of $J_{\rm nn} \approx -1.24$ K for Dy$_2$Ti$_2$O$_7$. Figure~\ref{Spinice-Cv} shows the good agreement between Monte Carlo results and experimental data~\cite{Hertog:2000}.  Note here that the parameter $D_{\rm nn}$ sets the scale for the dipolar interactions, at
$\mu_0\mu^2/(4\pi {r_{\rm nn}}^3)$; the simulations themselves use true long-range dipole-dipole interactions implemented via the Ewald method~\cite{Melko:2004}.  These results show convincingly the spin ice phenomenology in Dy$_2$Ti$_2$O$_7$, and also in Ho$_2$Ti$_2$O$_7$, as we now discuss.

While Ho$_2$Ti$_2$O$_7$ was the first compound to be proposed as a spin ice,
specific heat  measurements proved initially less straightforward
to interpret than in Dy$_2$Ti$_2$O$_7$, and this led to some confusion.
Specifically, the rapid increase of the specific heat below 1 K was originally interpreted as an indication of a phase transition to a partially ordered state around a temperature of 0.6 K~\cite{Siddharthan:1999}.
Instead, it turns out that the anomalous low-temperature behavior of the specific heat  in Ho$_2$Ti$_2$O$_7$ is of nuclear origin. An anomalously large hyperfine interaction between the electronic
and nuclear spins for Ho commonly leads to a nuclear specific heat Schottky anomaly at 0.3 K.
A subtraction of the nuclear contribution from the total low-temperature
specific heat reveals the purely electronic specific heat~\cite{Bramwell:2001a}, $C(T)$.
The integration of $C(T)/T$ from  300 mK up to 10 K gave a magnetic entropy deficit of an amount close to Pauling's $R/2\ln(3/2)$ zero-point entropy, hence confirming, thermodynamically, that Ho$_2$Ti$_2$O$_7$ is indeed a spin ice~\cite{Cornelius:2001}.
Following the same procedure as the one used for Dy$_2$Ti$_2$O$_7$~\cite{Hertog:2000},
a comparison of $C(T)$ with Monte Carlo simulations allows to
estimate the exchange constant in Ho$_2$Ti$_2$O$_7$ as $J_{\rm nn} \sim -0.55$~K, 
an antiferromagnetic value~\cite{Bramwell:2001a}.

\begin{figure}[t]
\begin{center}
\includegraphics[height=12cm,width=6cm]{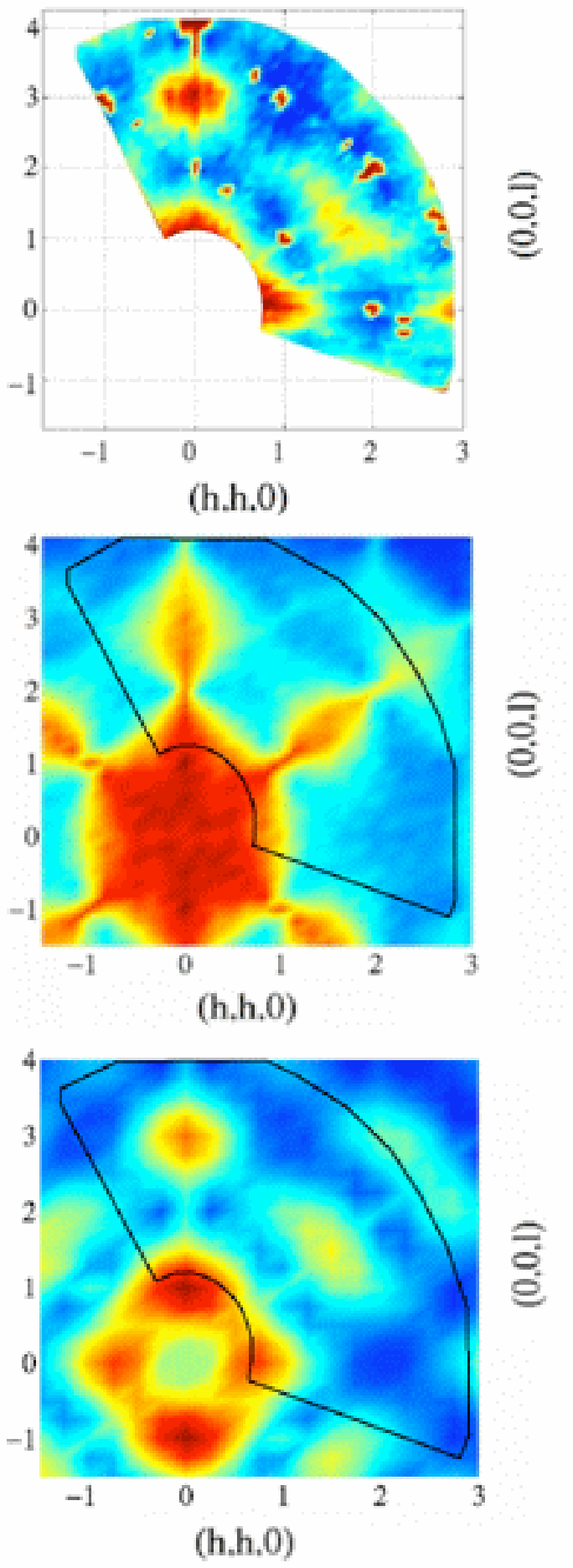}
\caption{ ${\rm Ho_2Ti_2O_7}$: Neutron scattering in the $hhl$
plane showing experimental data (upper panel; the sharp spots are
nuclear Bragg scattering with no magnetic component), compared
with Monte Carlo simulations of the near neighbour spin ice model
(middle panel) and the dipolar spin ice model (lower panel) \cite{Bramwell:2001a}.
Blue indicates the weakest and red-brown the strongest intensity.}
\label{Ho2Ti2O7-neutrons}
\end{center}
\end{figure}

While specific heat measurements on Ho$_2$Ti$_2$O$_7$ are problematic,
this is emphatically not so for neutron scattering experiments. Unlike dysprosium,
holmium has only one stable isotope whose neutron absorption
cross section is negligible. A comparison of the experimental neutron scattering intensity with
that calculated from Monte Carlo simulations of the dipolar spin ice model
with an exchange constant
$J_{\rm nn} \sim -0.55$~K determined as above shows excellent agreement.
Interestingly, both the experiment and Monte Carlo data differ substantially from
that calculated for the nearest-neighbor spin ice model
(see Fig.~\ref{Ho2Ti2O7-neutrons}).
This clearly shows that non-trivial spin correlations develop in
the material as it progressively freezes within the low-temperature spin ice
regime. Indeed, those correlations are the precursors of those that would
ultimately lead to long-range order, if not precluded by 
spin freezing~\cite{Melko:2001,Melko:2004}.

A puzzling question raised by the good agreement between Monte Carlo simulations of the DSM and the experimental results illustrated in Fig.~\ref{Spinice-Cv} is:  why don't the dipolar interactions drive a transition to long-range order at a critical temperature $T_c\sim D_{\rm nn} \sim 2$ K?
A partial answer can be found in the mean field theory calculations
of \textcite{Gingras:2001}.  It was found there that the ${\bf q}$ dependence of the softest branch
of critical modes in the dipolar spin ice model is very weakly dispersive, reaching a global
maximum eigenvalue $\lambda({\bf q}_{\rm max})$ at
$\bf q_{\rm max}$ = 001.
At the mean field level, this indicates that a transition to long range order should occur
at a critical temperature $T_c=\lambda({\bf q}_{\rm max})$,
 with the development of delta-function Bragg peaks below $T_c$.

First of all, this transition does not occur at $T_c\sim D_{\rm nn}$ because the soft (critical) mode at ${\bf q}_{\rm max}$ is in ``entropic'' competition with all the other quasi-soft modes at ${\bf q}\ne {\bf q}_{\rm max}$. The correlated spin ice regime, which one may dub a ``collective paramagnet''
in Villain's sense~\cite{Villain:1979}, albeit with utterly sluggish spin dynamics for $T\lesssim 1$ K
in Dy$_2$Ti$_2$O$_7$, has so much entropic disorder that it is not energetically
favorable to localize the system in phase space to a long-range ordered state. In the standard single spin-flip Monte Carlo simulations, this transition is not observed because the probability to flip a spin once the system enters a state where each tetrahedron obeys on average the ``two-in/two-out'' ice rule
is very small and decreases experimentally very fast with decreasing temperature~\cite{Melko:2004}.
The low-energy excitations deep in the spin ice state correspond to nonlocal closed loops
of spins flipping from ``in'' to ``out'' and vice versa so that the system, as
it experiences those excitations,
remains in a spin ice state~\cite{Barkema:1998}.
Using such loop excitations, Monte Carlo simulations~\cite{Melko:2001,Melko:2004}
found a transition to the long-range order predicted by mean field theory~\cite{Gingras:2001}.
In Monte Carlo simulations of the DSM, a strongly first order
transition occurs at a critical temperature $T_c\sim 0.07 D_{\rm nn}$ where
all the residual Pauling entropy is recovered through the pre-transitional
build-up of correlations and, mostly, via a large latent heat~\cite{Melko:2001,Melko:2004}.
For $D_{\rm nn} \sim 2.35$ K believed appropriate for Dy$_2$Ti$_2$O$_7$ (and Ho$_2$Ti$_2$O$_7$), this $T_c$ amounts to 160 mK.  However, to this date, no experimental work has observed a transition to long-range order in spin ice materials down to 60 mK (see for example, \textcite{Fukazawa:2002a}).  A common explanation for this absence of a transition in real spin ice materials is that equilibration is lost in the spin ice state (e.g. $T\lesssim 0.4$ K in Dy$_2$Ti$_2$O$_7$ and $T\lesssim 0.6 $K in Ho$_2$Ti$_2$O$_7$), with the real materials not ``benefitting'' from nonlocal dynamics as employed in the simulations~\cite{Melko:2001,Melko:2004}.  However, this explanation is certainly somewhat incomplete since, as we discuss further below, a number of experiments report spin dynamics
down to 20 mK. However, before we discuss experiments investigating the dynamics of spin ices,
we briefly comment on the spin-spin correlations in the spin ice regime of Dy$_2$Ti$_2$O$_7$.

\begin{figure}[b]
\begin{center}
\includegraphics[height=5.9cm,width=8.cm,angle=0]{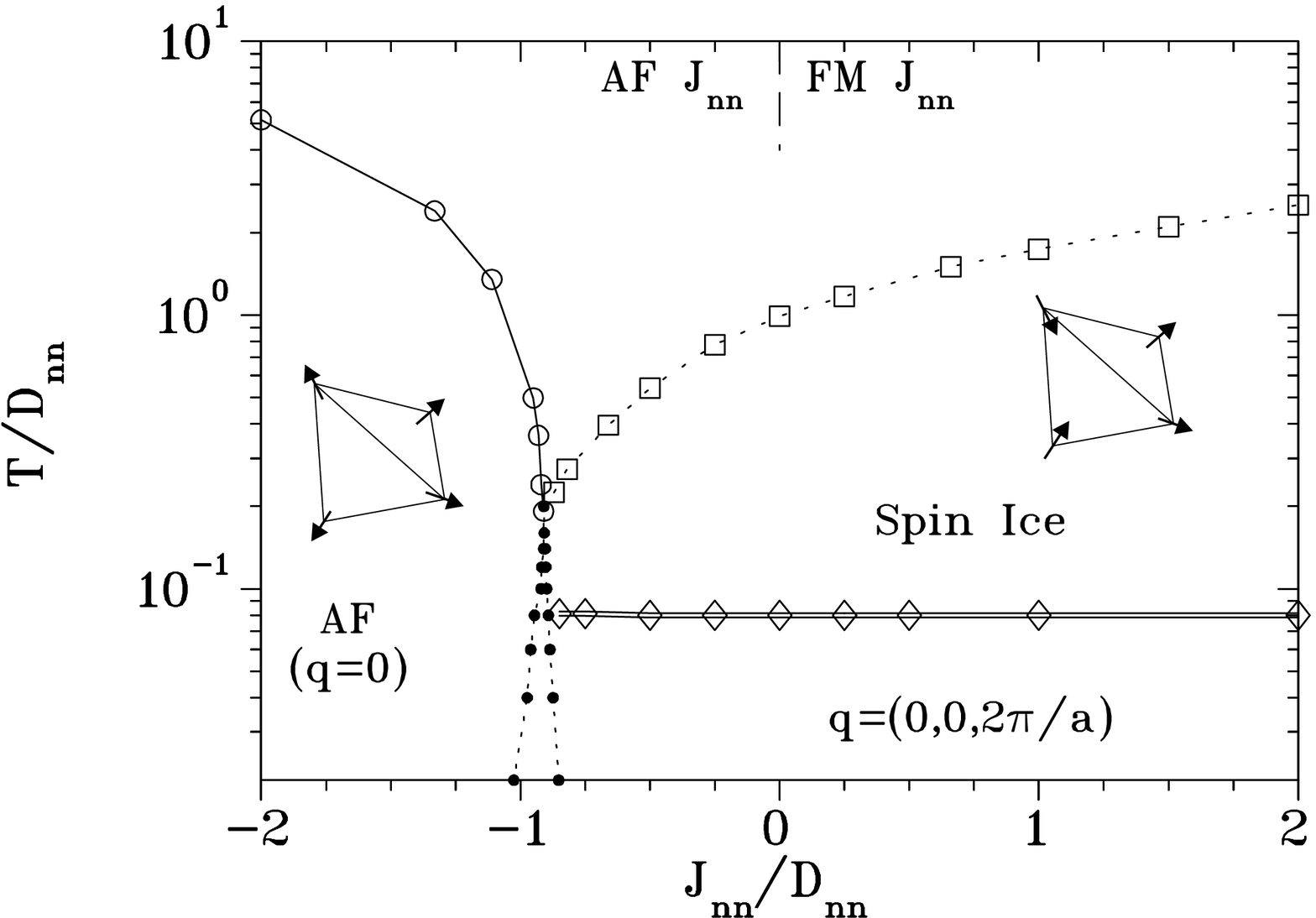}
\caption{Dipolar spin ice model: the phase diagram.  The
antiferromagnetic ground state is an all-spins-in or all-spins-out configuration
for each tetrahedron.  The spin ice configuration, which includes the
${\bf q}=001$ ground state,  is a two spins in-two spins out configuration
for each tetrahedron. The region encompassed between the quasi vertical dotted
lines displays hysteresis in the long-range ordered state selected
(${\bf q}=000$ vs.
${\bf q}=001$) as $J_{\rm nn}/D_{\rm nn}$ is varied at fixed temperature
$T$ (\textcite{Melko:2004}).
}
\label{SI-phasediag}
\end{center}
\end{figure}

Since the orignal work of \textcite{Ramirez:1999}, other measurements of the magnetic specific heat data of Dy$_2$Ti$_2$O$_7$ have been reported~\cite{Higashinaka:2002,Hiroi:2003,Higashinaka:2003,Ke:2007}.  Measurements of the magnetic specific heat of magnetic insulators are notoriously
difficult experiments owing to the  poor sample thermal conductivity.
It is perhaps for this reason that specific heat data of Dy$_2$Ti$_2$O$_7$ from different measurements show some disparity.  For example, the height of the magnetic specific heat at its maximum, $C_{\rm p}$,
for a powder sample~\cite{Ramirez:1999} differs by as much as 10\% compared to that
reported in \textcite{Higashinaka:2003} for a single crystal.  Ultimately, precise measurements of $C(T)$ are needed if one wants to make an accurate determination of the residual entropy in the system~\cite{Ke:2007}.

The Monte Carlo prediction of long range order developping at low temperature in the dipolar spin ice model and the experimentally observed collapse of the magnetic specific heat below 0.4 in Dy$_2$Ti$_2$O$_7$ raise the question of low temperature spin dynamics in spin ice materials.  In fact experimental studies of this question have led to the observation of a much richer
phenomenology than that which would have been naively expected.

\begin{figure}[t]
\begin{center}
\includegraphics[height=7.8cm,width=6cm,angle=0]{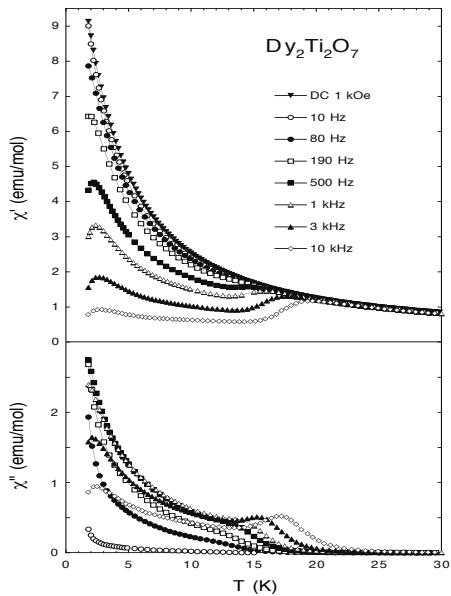}
\caption{${\rm Dy_2Ti_2O_7}$:  ac-susceptibility (upper panel:
$\chi'$, lower panel $\chi''$) as a function of temperature at
several frequencies, illustrating the ``15 K  peak''
from \textcite{Matsuhira:2001}.}
\label{AC-sus-Dy2Ti2O7}
\end{center}
\end{figure}

Measurements of the AC magnetic susceptibility $\chi(\omega)$ of Dy$_2$Ti$_2$O$_7$ down to 60 mK ~\cite{Fukazawa:2002a} and 100 mK~\cite{Matsuhira:2001} find that the real part of $\chi$, $\chi'$,
drops precipitously below a temperature of roughly 1 K for a frequency
of the order of 10 Hz (see Fig.~\ref{AC-sus-Dy2Ti2O7}).  At the same time, the imaginary part, $\chi''$,
shows a rounded maximum.  Both $\chi'$ and $\chi''$ remain essentially zero below 0.5 K down to the lowest temperature considered, hence signalling an essentially complete spin freezing of the system.
Thus, no signature of the transition to long range order predicted by numerical simulations~\cite{Melko:2001,Melko:2004} is observed.

The behavior of $\chi'$ and $\chi''$ near 1 K signals a spin freezing process analogous to what is observed in spin glasses. Indeed, measurements of the magnetization of Dy$_2$Ti$_2$O$_7$ show irreversibilities between zero field cooling (ZFC) and field cooling (FC) below 0.7 K ~\cite{Matsuhira:2001,Snyder:2004}.  Similar FC-ZFC irreversibilities occur in the Ho$_2$Sn$_2$O$_7$ spin
ice ~\cite{Matsuhira:2000}.  An analysis of the temperature dependence of the frequency $f_m$ at
which $\chi''$ displays a peak reveals a sort of thermally activated freezing behavior which was originally parametrized by an Arrhenius form with an activation energy of approximately 10 K.  However,
\textcite{Snyder:2004}  questioned the application of the Arrhenius law to these data.
On the other hand, Monte Carlo simulations that employ a standard single spin-flip Metropolis algorithm find that the fraction of accepted spin flips decreases with decreasing temperature according to a Vogel-Fulcher form $\exp[-A/(T-T^*)]$ with $T^* \sim 0.3$~K ~\cite{Melko:2004}. The dynamical freezing behavior seen in Dy$_2$Ti$_2$O$_7$ differs from the critical slowing down observed in conventional disordered spin glass materials~\cite{Binder:1986}. Interestingly, for the lowest temperature considered, $\chi''$ is cut-off on the low-frequency regime~\cite{Snyder:2004}, reminding one of what is observed in the LiHo$_x$Y$_{1-x}$F$_4$ Ising system $-$ a phenomenon that has been referred to as ``antiglass'' behavior~\cite{Reich:1987,Ghosh:2002}.  Another difference between the spin freezing in Dy$_2$Ti$_2$O$_7$ and that in conventional disordered spin glasses is the magnetic field dependence of $T_f$.  In spin glasses, $T_f$ decreases with increasing magnetic field strength, while in Dy$_2$Ti$_2$O$_7$, the opposite is seen for applied magnetic fields up to 5 kOe~\cite{Snyder:2004}.  Measurements on single crystals~\cite{Shi:2007} find that the spin freezing has a stronger frequency and magnetic field dependence for a field along the $[111]$ axis compared to  $[100]$, and the starting freezing frequency of the single crystal is higher than that of the powder sample~\cite{Snyder:2001}. While a quantitative understanding does not exists, the behavior at $T\lesssim 4$~K is qualitatively interpreted as a signature of the collective freezing as the system enters the low-temperature state where the ``two-in/two-out'' ice rules become fulfilled.

In addition to the low-temperature freezing in the ice-rule obeying state, two experimental studies coincidentally reported results from AC susceptibility $\chi(\omega)$ measurements in Dy$_2$Ti$_2$O$_7$ above 4K ~\cite{Snyder:2001,Matsuhira:2001}, finding another freezing process in a temperature range around 15 K, referred to as the `15 K feature'.  The signature of this freezing is only seen in the AC
susceptibility at finite frequency and not in the DC susceptibility data (see Fig.~\ref{AC-sus-Dy2Ti2O7}).
The maximum in the imaginary part, $\chi''$, is at 12 K for a frequency of 10 Hz
increasing to 17 K for a frequency of 10 kHz.  While the raw data of \textcite{Snyder:2001} and \textcite{Matsuhira:2001} are similar, they were analyzed somewhat differently.  \textcite{Snyder:2001} characterized the freezing via a single exponential relaxation while \textcite{Matsuhira:2001} characterized the AC susceptibility using  the so-called Davidson-Cole framework, based
on an underlying  distribution of time scales.   Notwithstanding these differences, both analyses agreed that there exists a typical time scale, $\tau(T)$, for this freezing phenomenon that is parametrized by an Arrhenius form,   $\tau \sim \exp(E_a/T)$,  with an activation barrier energy $E_a$ of the order of 200 K. A recent $\mu$SR study also finds that the relaxation rate of the muon spin polarization in a temperature range of 70 K to 280 K can be described by a typical relaxation rate $\lambda(T) \sim \exp(-E_a/T)$ with $E_a \sim 220$~K~\cite{Lago:2007}.  The $\exp(E_a/T)$ dependence of the typical time scale characterizing the dynamics around 15 K and the energy scale $E_a \sim 220$~K indicates that the relaxation involves transitions to and from the first excited doublet which constitute the main contribution to the spin dynamics in the temperature range above 20 K or so.  While this freezing phenomenon and energy scales characterizing it suggest an Orbach process~\cite{Finn:1961,Orbach:1961} that involves both the lattice degrees of freedom and the excited crystal field states, a concrete microscopic calculation has not yet been done.

One interesting aspect of the 15 K feature is its behavior when Dy$^{3+}$ is substituted by diamagnetic Y$^{3+}$ in Dy$_{2-x}$Y$_x$Ti$_2$O$_7$.  In particular, initial studies found that the 15 K peak in $\chi''(\omega)$  disappears  by $x=0.4$~\cite{Snyder:2001}.  This was originally interpreted as a sign that the 15 K feature is of collective origin.  However, in a subsequent study~\cite{Snyder:2004a}, it was found to re-emerge as $x$ is further increased and is almost as strong for $x=1.98$ as it is for $x=0$, but, interestingly, repositioned at a higher temperature of 22 K for a frequency of 1~kHz. This high temperature freezing feature is essentially a single-ion phenomenon, akin to superparamagnetic spin blocking. While this is not seen in AC measurements on Ho$_2$Ti$_2$O$_7$, it can be revealed by the application of a magnetic field~\cite{Ehlers:2003}.  Neutron spin echo experiments on the same material confirms the single-ion nature of the 15 K feature~\cite{Ehlers:2003}.

Another noteworthy aspect of the spin dynamics in Dy$_2$Ti$_2$O$_7$ is the temperature independence of the relaxation time $\tau$ between 5 K and 10 K ~\cite{Snyder:2003}. Below 5 K, the relaxation time becomes again sharply dependent on temperature upon approaching the spin ice freezing discussed above. This temperature independence of $\tau$ has been interpreted as a quantum tunneling effect between the up and down Ising spin states.  This was first observed in
Ho$_2$Ti$_2$O$_7$ using neutron spin echo~\cite{Ehlers:2003}.  Such temperature-independent relaxation in Dy$_2$Ti$_2$O$_7$ has also been seen in muon spin relaxation  ($\mu$SR)~\cite{Lago:2007}. However, there is a three-orders of magnitude difference between the relaxation rate measured in $\mu$SR and AC susceptibility. Perhaps this is because $\mu$SR relies on a local measurement that probes all wavevectors of the spin susceptibility.  At this time, this discrepancy is unresolved.

A further important topic pertaining to the dynamics of spin ice materials is that of the low-temperature
spin dynamics deep in the frozen spin ice state.  As discussed above, there is evidence that the electron spin flip dynamics are exponentially frozen out below 1 K in Dy$_2$Ti$_2$O$_7$, as well as in other spin ice materials.  There is, however, some indication of residual spin dynamics in these systems that survives down to the lowest temperatures. For example, $\mu$SR experiments find a relaxation rate of 0.2 $\mu{\rm s}^{-1}$ of the muon spin polarization at  a temperature of 20 mK~\cite{Lago:2007}.  This relaxation  has been ascribed to hyperfine coupling of the electronic and nuclear spins which induce 
a ``wobble''  around the local $\langle 111\rangle$ Ising directions.  Another work has suggested that the absence of a low-temperature nuclear specific heat anomaly in Dy$_2$Ti$_2$O$_7$ may indicate that the electron spin dynamics persist to the lowest temperature~\cite{Bertin:2002}. This argument would
suggest that Ho$_2$Ti$_2$O$_7$, which has a fully developed nuclear specific heat~\cite{Bramwell:2001a}, is completely static.  More experimental and theoretical work is required to fully understand the low-temperature dynamics of the spin ices.

Above, we touched on what role magnetic dilution plays on the spin dynamics and freezing phenomenon in Dy$_2$Ti$_2$O$_7$.  A recent magnetocaloric study of Dy$_2$Ti$_2$O$_7$ has found a crossover at a temperature of about 0.3 K to a low-temperature regime characterized by extremely slow relaxation~\cite{Orendac:2007}.  In addition, a dilution of Dy by 50\% of non-magnetic Y,
giving DyYTi$_2$O$_7$, leads to an increase of the relaxation time compared to pure Dy$_2$Ti$_2$O$_7$.  This is in contrast with the behavior seen above in the formation of the ice state, say above 2 K, where a nontrivial dependence of the relaxation time as a function of Dy concentration is observed. In particular, a level of magnetic dilution less than 12\% was found to accelerate the relaxation rate while a dilution level higher than 12\% was found to slow it down again, such that the relaxation rates are nearly the same for DyYTi$_2$O$_7$ and Dy$_2$Ti$_2$O$_7$~\cite{Snyder:2004a}.

\begin{figure}[t]
\begin{center}
\includegraphics[width=6cm,angle=0,clip=20]{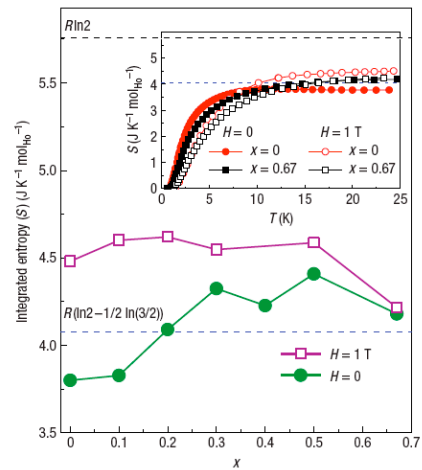}
\end{center}
\caption{The magnetic entropy for
Ho$_{2+x}$Ti$_{2-x}$O$_{7- \delta}$ (integrated from below $T\approx 1$ to 22~K)
as a function of stuffing at H=0 and 1 T. The  dashed
lines represent the predicted ice entropy and total spin entropy values.  Inset: The temperature dependence of the integrated
entropy for few compositions at H=0 and 1~T~\cite{Lau:2006}.}
\label{Fig:SSI_Ent}
\end{figure}

This discussion illustrates that the nature of the dynamics in spin ices and the role magnetic dilution is still poorly understood. In the context of magnetic dilution, we note that a recent paper reports a non-monotonic dependence of the residual entropy in Dy$_{2-x}$Y$_x$Ti$_2$O$_7$ and Ho$_{2-x}$Y$_x$Ti$_2$O$_7$.  The data are qualitatively explained by a generalization of Pauling's theory for the entropy of ice that incorporates site-dilution~\cite{Ke:2007}.  The topic of residual entropy in spin ice when the magnetic species are diluted is an interesting problem. One would naively expect that the extensive degeneracy would ultimately be lifted by disorder~\cite{Villain:1979}.
There are two obvious ways that one can envisage this happening.  Perhaps the most interesting is where dilution would lead to conventional long range order. For example, in water ice, KOH doping (which effectively removes protons) leads to a long-range ordered phase of ice called ice XI.  The proton vacancies created in the proton structure enhance the dynamics so that the system can develop long range order. Studies of diluted spin ice, such as Dy$_{2-x}$Y$_x$Ti$_2$O$_7$~\cite{Ke:2007}
and Ho$_{2-x}$Y$_x$Ti$_2$O$_7$~\cite{Ehlers:2006a}, have so far not reported any
signs that diamagnetic dilution leads to long range order. Most likely, the extensive degeneracy is lifted
at high $x$ with the system leaving the ice regime to become a dipolar Ising spin glass akin to
LiHo$_x$Y$_{1-x}$F$_4$~\cite{Reich:1987}.  Indeed, the systematic study of the development of dipolar spin glass in both Dy$_{2-x}$Y$_x$Ti$_2$O$_7$ and Ho$_{2-x}$Y$_x$Ti$_2$O$_7$ might help shed light on the paradoxical antiglass phenomenon in LiHo$_x$Y$_{1-x}$F$_4$~\cite{Reich:1987,Ghosh:2002}, or even  the possible absence of a spin glass phase altogether in that material~\cite{Jonsson:2007}.

On the theoretical front, the DSM of Eq.~(\ref{Hdsm}) with nearest-neighbor exchange and long range dipole-dipole interactions has been fairly successful in describing semi-quantitatively the thermodynamic properties of Dy$_2$Ti$_2$O$_7$ both in zero~\cite{Hertog:2000,Fennell:2004} and in nonzero magnetic field~\cite{Fukazawa:2002a,Ruff:2005,Tabata:2006}.
A detailed comparison between experimental results from measurements and Monte Carlo simulations~\cite{Ruff:2005}  for a field along the $[112]$ direction provides strong evidence that exchange interactions beyond nearest-neighbors are required to describe quantitatively the experimental
data (see also \textcite{Tabata:2006}).  This may come as a surprise given the already
good agreement that was first reported between Monte Carlo simulations of the DSM~\cite{Hertog:2000} and the specific heat measurements of \textcite{Ramirez:1999}.  However, since that work, as mentioned above,
several other zero field specific heat data sets have been reported~\cite{Higashinaka:2002,Hiroi:2003,Higashinaka:2003,Ke:2007} and these are no longer so-well described by the original DSM. 
A systematic study of more recent specific heat measurements in zero and nonzero field and
magnetization measurements for $[110]$ and $[111]$ fields have estimated first ($J_1$), second ($J_2$) and third ($J_3$) nearest-neighbor exchange parameters in Dy$_2$Ti$_2$O$_7$~\cite{Yavorskii-hexagons}.
Perhaps most interestingly, such a refinement of the spin Hamiltonian seemingly allows one to explain from a microscopic basis the diffuse scattering on the Brillouin zone boundary~\cite{Fennell:2004},
akin to the highly structured inelastic features found in the ZnCr$_2$O$_4$ spinel~\cite{Lee:2002}.

\subsection{${\it A}_2$Sn$_2$O$_7$ (${\it A}$ ~=~ Pr, Dy and Ho) }

Other pyrochlore oxides with similar properties include Ho$_2$Sn$_2$O$_7$, Dy$_2$Sn$_2$O$_7$ and Pr$_2$Sn$_2$O$_7$~\cite{Matsuhira:2000,Matsuhira:2002,Matsuhira:2004}.  These have not been studied in single crystal form, as far as we know, but bulk magnetic properties would suggest they are ferromagnetic with large Ising anisotropy and very slow dynamics.

Of these three compounds, only Ho$_2$Sn$_2$O$_7$ has been investigated using neutron scattering techniques on powder samples~\cite{Kadowaki:2002}.  In the inelastic spectrum they found very slow dynamics below 40~K and crystal field levels at 22 and 26~meV, suggesting the local Ho environment
 is very similar to that of Ho$_2$Ti$_2$O$_7$.  Analysis of the diffuse magnetic scattering also led the authors to conclude that Ho$_2$Sn$_2$O$_7$ was a dipolar-spin-ice material.

\subsection{Ho$_{2+x}$Ti$_{2-x}$O$_{7-\delta}$: stuffed spin ice}

Besides thermodynamic variables such as temperature and magnetic field, the chemical composition of the systems may be altered to study the relationship of structural and magnetic properties. To this end,  `stuffed' spin ices with general formula $A_{2+x}$B$_{2-x}$O$_{7-\delta}$ have recently been 
synthesized~\cite{Lau:2006,Schiffer:2007}, in which additional magnetic ions replace the non-magnetic Ti$^{4+}$ ions ($\delta>0$, as the oxygen content needs to be adjusted for charge balance).  The entire rare earth series of titanates have been formed~\cite{Lau:2006a} but to date, the only published data is on the `stuffed' spin ice (SSI) compounds.  \textcite{Lau:2007} have reported phase separation in these compounds, however the magnetic properties are very similar between samples.  Single crystals have been produced by~\textcite{Zhou:2007}.

\begin{figure}
\begin{center}
\includegraphics[width=8cm,angle=0,clip=20]{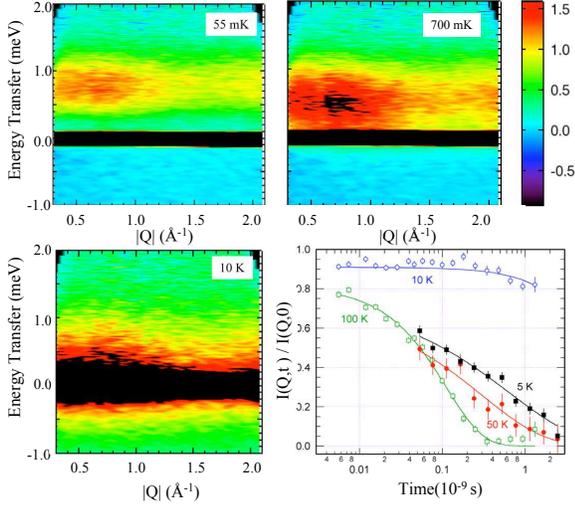}
\end{center}
\caption{Inelastic neutron scattering from
Ho$_{2.3}$Ti$_{1.7}$O$_{7-\delta}$~\cite{Zhou:2007}.
Data at 55 mK, 700 mK and 10 K, noting
the appearance of a gapped spin excitation at low temperatures.
Neutron spin echo results on stuffed (open symbols) and unstuffed (closed symbols) spin ice 
are depicted in the bottom right panel.
Within the neutron time window, the spin ice appears
static by 10~K (90\% of the spin are static),
 but the stuffed spin ice has persistent dynamics.}
\label{Fig:SSI_INS}
\end{figure}

The additional magnetic exchange pathways represent a major disturbance of the system, introducing positional disorder and, naively, an increased level of energetic constraints to the formation of the spin ice manifold.  Surprisingly, it has been found that the `stuffed' spin ice systems do not freeze, have short range correlations down to the lowest temperatures and, most interestingly, have the same entropy
per spin at low temperature as the `unstuffed' spin ice~\cite{Lau:2006,Zhou:2007}.  The origins of this residual entropy per spin in these systems is still debatable.

Another interesting feature of the residual entropy in the stuffed spin ice materials is its robustness to
an applied field.  As seen in Fig.~\ref{Fig:SSI_Ent} the entropy per spin with $x=0.67$ in 1~T and zero field are identical. This robust residual entropy in a very disordered sample like Ho$_{2.67}$Ti$_{1.33}$O$_{6.67}$ ($x=0.67$) needs investigating further.

Inelastic and quasielastic neutron scattering on polycrystalline and single crystalline Ho$_{2.3}$Ti$_{1.7}$O$_{7-\delta}$ have revealed subtle changes vis \`a vis the parent compound.  Diffuse scattering
is centered at 0.9 \AA$^{-1}$, but it is broader than the diffuse scattering seen in Ho$_{2}$Ti$_{2}$O$_{7}$.  Also, as seen in Fig.~\ref{Fig:SSI_INS}, the quasielastic scattering at temperatures above 10~K opens a gap resulting in a  low lying excitation, centered at 0.8~meV~\cite{Zhou:2007}.

More research is currently being done by several groups on `stuffed' spin ice.  It appears that the stuffed Dy-spin ices do not have residual entropy~\cite{Schiffer:2007}.

\section{Spin Liquid Phases}

\label{sec:Spin_Liquid}

An array of Heisenberg spins forming a three-dimensional pyrochlore lattice that interact among themselves via nearest-neighbor antiferromagnetic exchange interactions is theoretically predicted to remain disordered at  finite temperature for either classical~\cite{Moessner:2001,Moessner:1998,Moessner:1998a} or quantum spins~\cite{Canals:1998,Canals:2000}. Ising spins that are coupled antiferromagnetically on a pyrochlore lattice also possess a large ground state degeneracy characteristic of the spin liquid or cooperative paramagnetic state~\cite{Anderson:1956}.

\subsection{Tb$_2$Ti$_2$O$_7$}

\begin{figure}[b]
\makebox[3.2in]
{
\makebox[1.6in][l]{
\parbox{0.5in}{\includegraphics[width=1.7in]{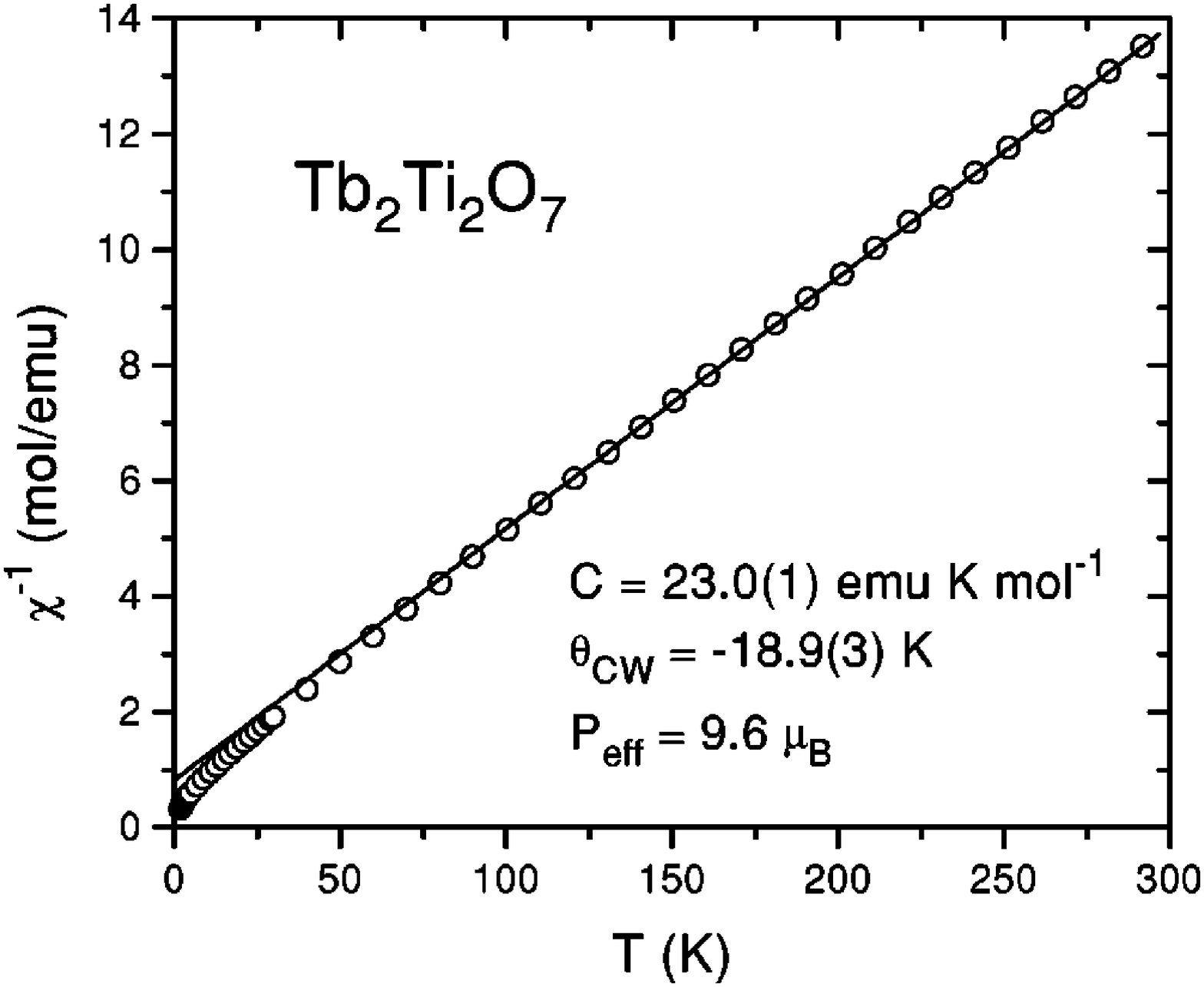}}}
\makebox[0.1in][l]{ }
\makebox[2in][l]{
\parbox{0.1in}{\includegraphics[width=1.85in]{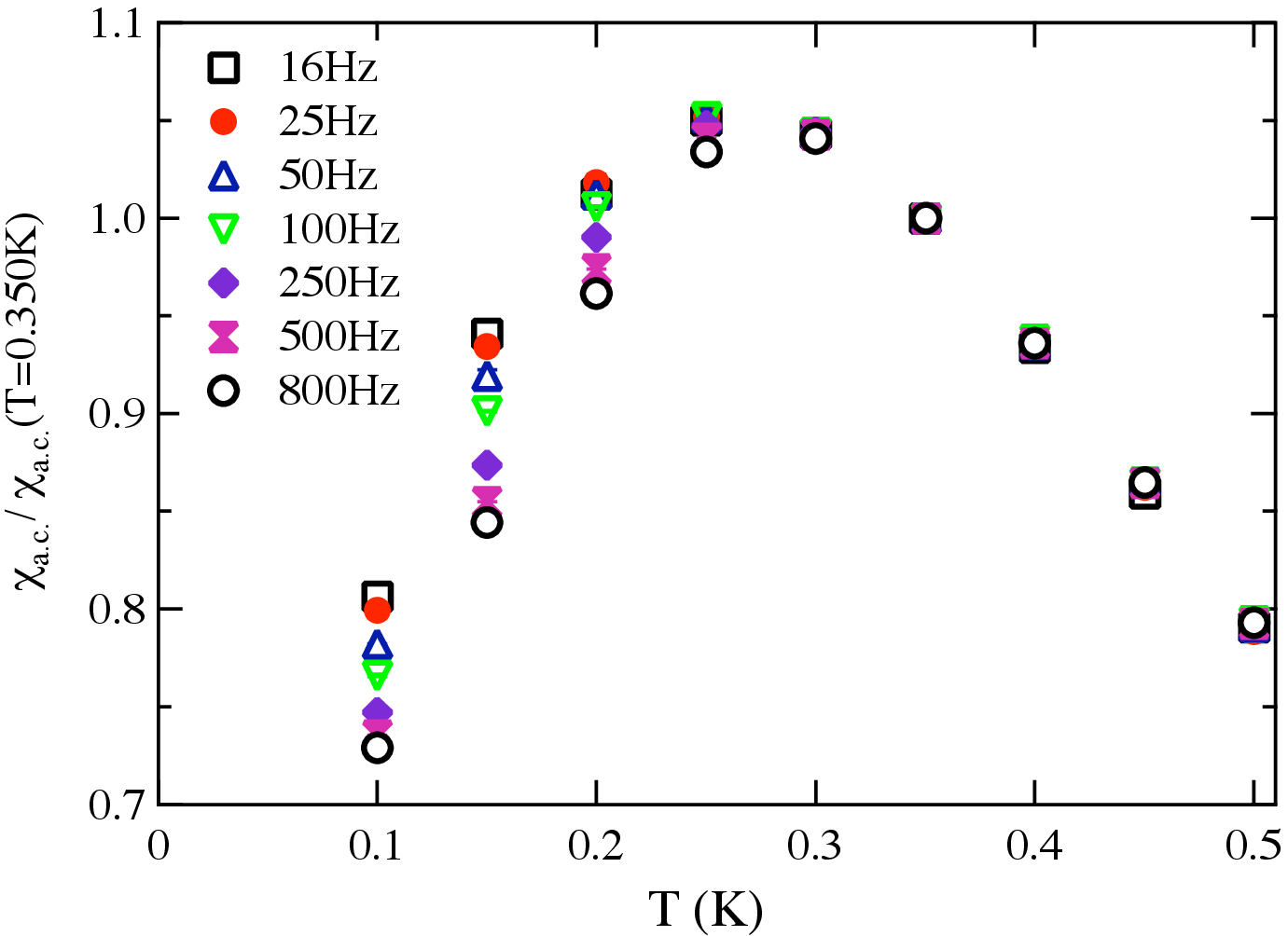}}}
}
\caption{Left:  The temperature dependance of the inverse molar
 susceptibility and the resulting fit to the Curie-Weiss law at
high temperature ($T> 250$~K).
Right:  The low temperature ($T<0.5$~K) dependence of the magnetic
 susceptibility as a function of frequency~\cite{Gardner:2003}.}
\label{TbT_Sus}
\end{figure}

Tb$_2$Ti$_2$O$_7$ is a well studied system where Tb$^{3+}$ ions are magnetic and the Ti$^{4+}$ ions are nonmagnetic. The temperature dependence of the magnetic susceptibility
of Tb$_2$Ti$_2$O$_7$ is described well by the Curie-Weiss law down to 50~K with $\theta _{\rm CW}$~=~-19~K and 9.6 $\mu _B$/Tb-ion. This effective moment is appropriate for the $^7$F$_6$ Tb$^{3+}$ ion. By studying a magnetically dilute sample, (Tb$_{0.02}$Y$_{0.98}$)$_2$Ti$_2$O$_7$, a value of -6~K was established as the crystal field contribution to  $\theta _{\rm CW}$.  Assuming a maximum of -2~K
for the dipolar interactions,  $\approx$ -11~K was proposed as a good estimate for the contribution of the exchange interactions to  $\theta _{\rm CW}$.  Figure \ref{TbT_Sus}a shows the temperature dependence of the bulk susceptibility down to 0.5~K.

\textcite{Han:2004} found, using neutron powder-diffraction  and x-ray absorption fine-structure, that the chemical structure of Tb$_2$Ti$_2$O$_7$ is well ordered with no structural transitions
between 4.5 and 600 K.  More recently however, \textcite{Ruff:2007} found a broadening of structural peaks in a high resolution,  single crystal, x-ray diffraction experiment below 20~K,
as expected just above a cubic-to-tetragonal transition, however this transition never fully develops even at 300~mK.

\begin{figure}[floatfix]
\begin{center}
\centerline{
\includegraphics[width=7.5cm,angle=0]{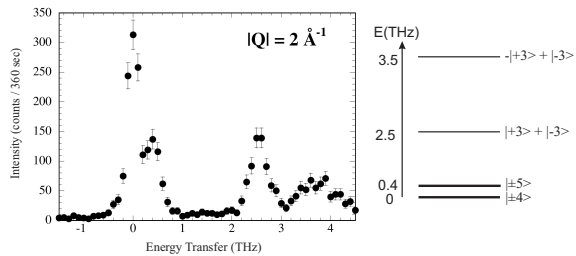}}
\end{center}
\caption{Left: A typical constant $\vert{\mathbf Q}\vert$ scan at 2~\AA$^{-1}$
 revealing low energy inelastic modes at 12 K on a powder sample of 
Tb$_2$Ti$_2$O$_7$.
Right: A schematic of the crystalline electric field levels~\cite{Gingras:2000}.}
\label{TbTCEF}
\end{figure}

Powder neutron diffraction measurements on Tb$_2$Ti$_2$O$_7$ clearly revealed two contributions to the scattering at 2.5~K~\cite{Gardner:1999}.  One of these consists of sharp nuclear Bragg peaks due to the crystalline order in the material and results in a cubic lattice parameter of 10.133~\AA.  A second diffuse, liquid-like, background is also present and was attributed to magnetic short range correlations. \textcite{Gardner:1999} used an  {\it isotropic} spin model, correlated over nearest spins only, to describe this scattering which captured the minimum in the forward direction,  $\vert{\mathbf Q}\vert \sim$~0~\AA$^{-1}$, as well as the approximate location of the peaks and valleys in the data.

Inelastic scattering experiments were also carried out on polycrystalline Tb$_2$Ti$_2$O$_7$~\cite{Gardner:2001} and a typical scan at $\vert{\mathbf Q}\vert$~=~2~\AA$^{-1}$ is shown in Fig.~\ref{TbTCEF}.  Three bands of excitations are clearly observed near 0.37, 2.53, and 3.50 THz and these 
were shown to be dispersionless above 30~K, a characteristic of a single ion effect associated with the rare earth site. A weak, but very interesting dispersion develops in the lowest energy band at temperatures below $\sim$ 20 K, which we will discuss later in this section.

A crystalline electric field (CEF) level scheme appropriate to the $^7$F$_6$ configuration of Tb$^{3+}$ in the $A$-site environment of Tb$_2$Ti$_2$O$_7$ was determined on the basis of these neutron and other bulk property data~\cite{Gingras:2000}. The ground state and first excited states are both doublets, with two singlets at much higher energies, as shown schematically in Fig.~\ref{TbTCEF}.  The $J^z$ eigenstates $\vert \pm 4 \rangle$  and  $\vert \pm 5 \rangle$ are believed to make up most of the weight of the ground state doublet and of the first excited state, respectively.   Such a CEF scheme means the moment can be considered extremely Ising in nature and pointing in the local $\langle 111 \rangle$  directions at temperatures $T\lesssim 20$~K.  More recently \textcite{Mirebeau:2007}  reported a series of neutron scattering measurements that  ``refined'' the CEF scheme of Tb$_2$Ti$_2$O$_7$ and 
compared it to Tb$_2$Sn$_2$O$_7$, which orders just below 1~K, and which was discussed in Section \ref{TbSnO}.

\begin{figure}[t]
\begin{center}
\includegraphics[width=8.8cm,angle=0]{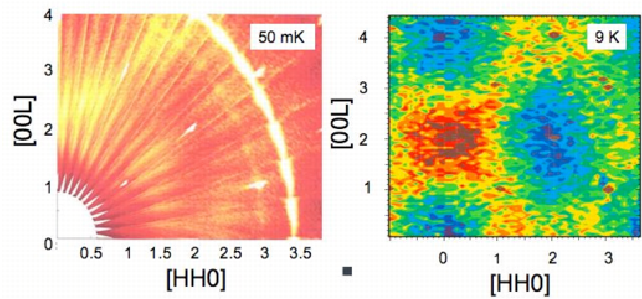}
\end{center}
\caption{Diffuse scattering from a single crystal of  Tb$_2$Ti$_2$O$_7$
 at 50~mK (left) and 9~K (right)~\cite{Gardner:2001,Gardner:2003}.
Sharp Bragg peaks can be seen at the appropriate reciprocal lattice
positions in the 50~mK data set, but at 9 K, these have been subtracted
out using a high temperature (100~K) data set.}
\label{TbT_Cry_diffuse}
\end{figure}

Large single crystals of Tb$_2$Ti$_2$O$_7$ were first grown using floating zone image furnace techniques by \textcite{Gardner:1998}.  The resulting single crystals enabled a series of experiments which could probe the four-dimensional dynamic structure factor $S({\bf Q}, \hbar \omega)$.  Figure \ref{TbT_Cry_diffuse} shows $S({\bf Q})=\int S({\bf Q}, \hbar \omega) d\omega$ measured at T=50~mK and at 9~K within the $(hhl)$ plane.  Bragg peaks are clearly seen at the allowed positions for the
$Fd \bar{3}m$ space group, that is $(hhl)$ being all even, or all odd, integers.  Also one clearly observes a ``checkerboard" pattern to the magnetic diffuse scattering.  This covers the entire Brillouin zone indicating the spins are correlated on a length scale much smaller than the unit cell $\sim$ 10.1~\AA.
However, the single crystal data are clearly {\it not} isotropic in reciprocal space, explaining the quantitative failure of the simple isotropic model used to analyze the earlier powder data~\cite{Gardner:2001}.  Also,the scattering does not conform to that expected for an Ising model with local
$\langle 111\rangle$ spin directions.  Specifically, an Ising model with  local
$\langle 111\rangle$ spins cannot simultaneously produce a large amount of scattering around 002 and null scattering at 000~\cite{Enjalran:2004}.

\begin{figure}[t]
\begin{center}
\centerline{
\includegraphics[width=7.0cm,angle=0]{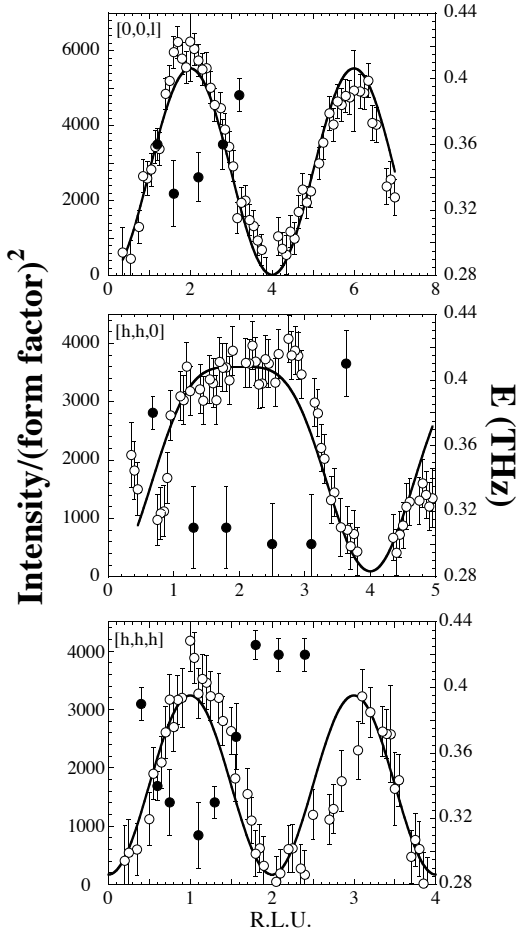}}
\end{center}
\caption{Cuts through the checkerboard board pattern of
diffuse magnetic scattering (open symbols) along high
 symmetry directions of the cubic lattice.  To complement these data,
the dispersion of the lowest lying magnetic excitation, at 4~K, is also
plotted (closed symbols).  Fits to the diffuse scattering are shown and discussed
in the text~\cite{Gardner:2001}.}
\label{TbT_Cry_INS}
\end{figure}

Single crystals also allow the study of the low-lying excitations as a function of ${\bf Q}$,~\cite{Gardner:2001,Gardner:2003} as opposed to the modulus  $\vert {\bf Q}\vert$  derived from powder samples.  As was observed in powders~\cite{Gardner:1999}, the dispersion develops below $\sim$ 25 K.  This low temperature dispersion is plotted along the three high symmetry directions within the (hhl) plane in Fig.~\ref{TbT_Cry_INS}, on which are overlaid cuts of $S({\bf Q})$ along [00l] (top), [hh0] (middle), and [hhh]  (bottom).  The diffuse scattering is well described by simple near-neighbor antiferromagnetic spin correlations on the pyrochlore lattice,
as shown by the solid lines in Fig.~\ref{TbT_Cry_INS}.  We also see that the minima in
the dispersion of this low lying magnetic mode (closed symbols) correspond exactly with peaks in $S({\bf Q})$, even though $S({\bf Q})$  is anisotropic in ${\bf Q}$.  \textcite{Gardner:2003} showed the gap between the first excited state and the ground state drops from 0.37~THz at 30~K to approximately 0.25~THz at 100~mK but does not soften further as the temperature is further decreased.

\begin{figure}[t]
\begin{center}
\centerline{
\includegraphics[width=7cm,angle=0]{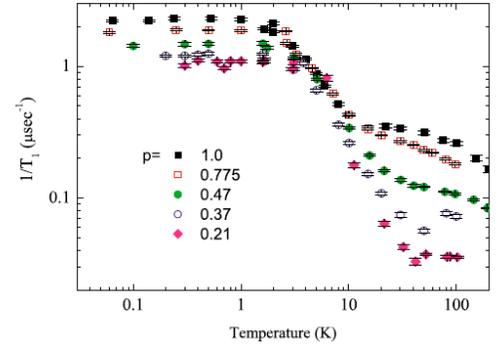}}
\end{center}
\caption{Temperature dependence of the muon
relaxation rate in a field of 50~G in various
samples of (Tb$_p$Y$_{1-p}$)$_2$Ti$_2$O$_7$~\cite{Keren:2004}.}
\label{Fig:TbT_muon}
\end{figure}

Neutron spin echo~\cite{Gardner:2003,Gardner:2004} and $\mu$SR~\cite{Gardner:1999,Keren:2004} measurements have also been carried out at mK temperatures to probe the dynamics of the magnetic moments.  Both show large, fluctuating moments down to at least 17~mK~\cite{Gardner:1999},
as expected for a cooperative paramagnet. However, below $T\approx 0.3$~K the $S({\bf Q},t)/S({\bf Q},0)$ baseline is no longer  zero indicating a proportion ($\approx$~20\%) of the magnetic moments are frozen in the neutron spin echo time window. This is consistent with the frequency dependence of the
 AC susceptibility shown in figure \ref{TbT_Sus}b~\cite{Gardner:2003,Luo:2001}.  However, we attribute this partial spin freezing to a small subset of magnetic moments distributed near a small number of defects. Evidence for spin freezing was also found independently in neutron scattering experiments performed on other single crystal samples~\cite{Yasui:2002}. In this case, hysteresis is observed in the scattering near 002 at temperatures less than $\sim$ 1.7 K, but again most of the scattering remains diffuse in nature even below 1.7 K.

Dilution studies by \textcite{Keren:2004}, perturbed the lattice through the percolation
 threshold but found that fluctuating moments prevail at all magnetic concentration as shown by the muon relaxation rate in Fig.~\ref{Fig:TbT_muon}.  They also noted that the magnetic coverage of the lattice  was important to the cooperative phenomenon but the percolation threshold played no critical role.

At first sight, it would be tempting to ascribe the failure of Tb$_2$Ti$_2$O$_7$ to develop long range order down to such a low temperature compared to $\theta_{\rm CW}$,  to the collective paramagnetic behavior of the classical pyrochlore Heisenberg antiferromagnet~\cite{Villain:1979,Moessner:1998,Moessner:1998a}. However, consideration of the single-ion crystal field ground state doublet of Tb$^{3+}$ in Tb$_2$Ti$_2$O$_7$ strongly suggests that, in absence of exchange
and dipolar interactions, bare Tb$^{3+}$ should be considered as an effective Ising spin~\cite{Rosenkranz:2000,Gingras:2000}.

Considering only nearest-neighbor exchange and long-range dipolar interactions,
the DSM of Eq.~(\ref{Hdsm}) would be the appropriate Hamiltonian to
describe Tb$_2$Ti$_2$O$_7$ as an Ising system.
On the basis of the estimated value of the nearest-neighbor exchange and dipolar
coupling, Tb$_2$Ti$_2$O$_7$ would be predicted to develop a
${\bm k}_{\rm ord}=000$ four
sublattice long-range N\'eel order at a critical temperature
$T_c\approx 1$~K~\cite{Hertog:2000}, in dramatic contradition with experimental
results.
Furthermore, the paramagnetic neutron scattering of such a classical Ising model is
inconsistent with that observed in Tb$_2$Ti$_2$O$_7$ at 9 K as shown in
Fig. \ref{TbT_Cry_diffuse}.
 Mean field theory calculations on a Heisenberg model with
dipolar and exchange interactions and with a simple single-ion
anistropy~\cite{Enjalran:2004}
as well as a random phase approximation
(RPA) calculation~\cite{Kao:2003} that considers
a semi-realistic description of the crystal-field levels of Tb$^{3+}$ show that
the strong scattering intensity around 002 should be extremely weak for
Ising spins. Rather, the intensity is caused by the finite
amount of fluctuations perpendicular
to the local $\langle 111\rangle$ directions~\cite{Enjalran:2004}.  In the RPA calculations~\cite{Kao:2003}, the scattering intensity near
002  originates from the contribution of the excited crystal field
levels at an energy $\sim 18$~K to the susceptibility, which is made
dispersive via the exchange and dipolar interactions.
The experimental observation that the diffuse scattering intensity is largely
unaltered down to a temperature of 50 mK, as shown in Fig.~\ref{TbT_Cry_diffuse}, 
suggests that the ``restored'' effective
isotropicity of the spins is an intrinsic part of the effective low-energy
theory that describes Tb$_2$Ti$_2$O$_7$.
In that context, a recent paper by~\textcite{Molavian:2007} argues, on the basis
of a calculation that considers non-interacting tetrahedra, that virtual
fluctuations between the ground state and excited crystal field levels
lead to an important renormalization of the coupling constants entering
the effective low-energy theory. In particular,~\textcite{Molavian:2007}
propose that, for the simple model considered, the system is pushed
to the spin ice side of the phase diagram by those quantum fluctuations.
In other words, within that description, Tb$_2$Ti$_2$O$_7$ is a sort
of quantum spin ice system where the spin ice like correlations
remain hidden down to $T\lesssim 500$~mK.

In this discussion of the lack of order in Tb$_2$Ti$_2$O$_7$,
it is not clear what is the role of the structural/lattice
fluctuations very recently reported by \textcite{Ruff:2007}.
 At this time, there is really no robust theoretical understanding of the physics at play in Tb$_2$Ti$_2$O$_7$ and more theoretical and experimental studies are required.

\subsection{Yb$_2$Ti$_2$O$_7$}

Ytterbium titanate is an insulator with lattice parameter $\it{a}_0$=10.026(1)~\AA~at room
temperature~\cite{Brixner:1964}.  Early work suggested an ordered magnetic state just below 0.2 K~\cite{Blote:1969}, where a sharp anomaly in the specific heat was observed.
 In a detailed magnetisation study, \textcite{Bramwell:2000}
found a Curie-Weiss temperature of 0.59(1)~K, indicative of weak ferromagnetic coupling and a free ion moment of 3.34(1) $\mu_B$.  \textcite{Hodges:2001} used M{\"o}ssbauer spectroscopy to investigate
the crystal field scheme and determined that the ground state Kramers
doublet was separated by 620~K from the first excited state producing
an easy plane anisotropy, like in Er$_2$Ti$_2$O$_7$.  They also found the
 effective paramagnetic moment to be 3.05(8) $\mu_B$.

\begin{figure}
\begin{center}
\includegraphics[width=6cm,angle=0,clip=390]{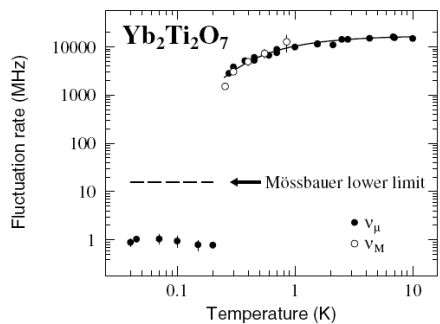}
\end{center}
\caption{Estimate fluctuation rate of the Yb$^{3+}$
moment derived from muon (filled circles) and M{\"o}ssbauer
 (open circles) spectroscopies~\cite{Hodges:2002}.
Note that the rate plateaus at 200~mK and does not go to zero.}
\label{YbTi_Moss}
\end{figure}

Hodges and co-workers~\cite{Hodges:2002}  extended their study of Yb$_2$Ti$_2$O$_7$ and found an abrupt change in the fluctuation rate of the Yb$^{3+}$ spin at 0.24~K but not a frozen ground state
as expected from the earlier studies~\cite{Blote:1969}.  Using muon spin relaxation and M{\"o}ssbauer spectroscopies, they concluded that the Yb$^{3+}$ spin fluctuations slow down by more than 3 orders of magnitude to several megahertz, without freezing completely.  This was confirmed in their neutron powder diffraction, where no extra Bragg intensity was reported below 0.24~K.  Single crystal neutron diffraction by \textcite{Yasui:2003} has revealed extra Bragg scattering below 0.24~K from a static ferromagnetic state, albeit with a reduced moment of 1.1 $\mu_{B}$.
The latter two studies motivated a polarised neutron scattering study by \textcite{Gardner:2004a}.
That work conclusively ruled out a frozen ferromagnetic state and confirmed
that the majority of the spin system continues to fluctuate below the 240~mK transition while 
a small amount of magnetic scattering was observed at the 111 Bragg position at 90~mK.

\subsection{Er$_2$Sn$_2$O$_7$}

\label{Sec:ErSn}

\begin{figure}[t]
\begin{center}
\includegraphics[width=7.cm,angle=0,clip=0]{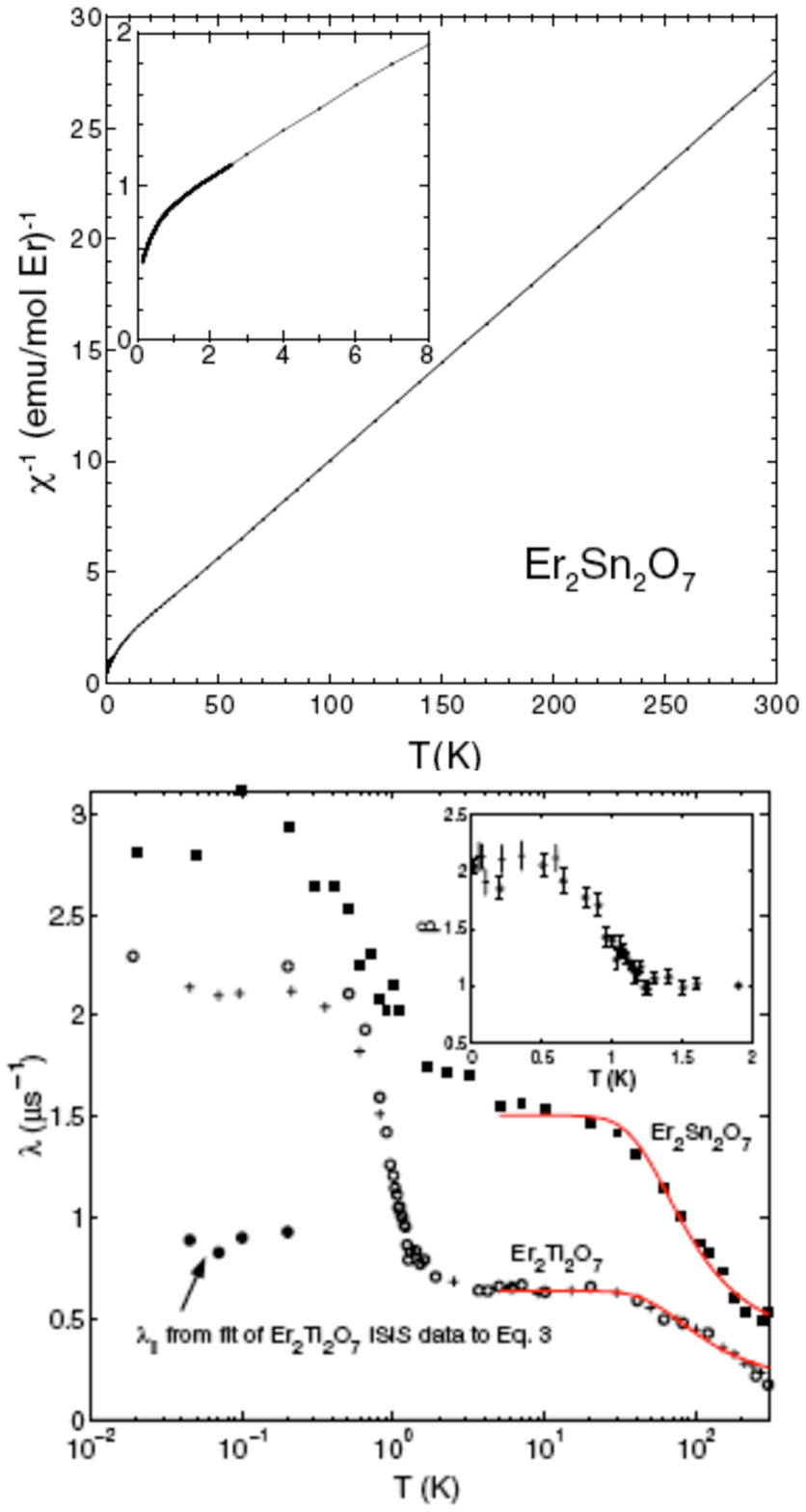}
\end{center}
\caption{Top: The temperature dependence of the inverse magnetic
 susceptibility from a powder sample of  Er$_2$Sn$_2$O$_7$~\cite{Matsuhira:2002}.
 The inset shows the data down to 0.13~K.
Bottom: Relaxation rates $\lambda$ derived from muon
spin relaxation measurements for Er$_2$Sn$_2$O$_7$
and Er$_2$Ti$_2$O$_7$~\cite{Lago:2005}.
Solid curves are the fit of the high temperature dat
 to an exponential activation function. $\lambda_{\parallel}$
is the longitudinal relaxation rate for  Er$_2$Ti$_2$O$_7$.
 Inset: thermal evolution of the order parameter critical 
exponent, $\beta$, for
 Er$_2$Ti$_2$O$_7$ showing the evolution from exponential
to Gaussian curve shape across $T_{\rm N}$. For  Er$_2$Sn$_2$O$_7$,
$\beta \approx 1$ for the entire temperature range.}
\label{ErSn_Chi_muon}
\end{figure}

Er$_2$Sn$_2$O$_7$ and Er$_2$Ti$_2$O$_7$ are thought to constitute experimental realizations of the highly frustrated XY antiferromagnet on the pyrochlore lattice.  Very little is known about Er$_2$Sn$_2$O$_7$ compared to the isostructural Er$_2$Ti$_2$O$_7$ compound which 
develops a non-collinear N\'eel state at 1.1~K, possibly via an order by disorder transition (see Section \ref{Sec:ErTiO}~\cite{Champion:2003}).  \textcite{Bondah-Jagalu:2001} reported a pronounced field
 cooled - zero field cooled divergence in the susceptibility at 3.4~K. The temperature and field dependence of the bulk magnetization of Er$_2$Sn$_2$O$_7$ was more recently
 published by~\textcite{Matsuhira:2002}.  At high temperatures $1/\chi(T)$ is linear and a fit
to the Curie-Weiss law results in a ground state moment
of 9.59~$\mu _B$ appropriate for the $^4$I$_{15\over2}$ ion.
It also gives a Curie-Weiss temperature, $\theta_{\rm CW} \approx -14$~K,
suggestive of antiferromagnetic correlations.  At 10~K $1/\chi(T)$  starts bending downwards,
presumably due to crystal field effects.  However, no anomaly indicative of a freezing into either a long
 or short range ordered system was seen down to 0.13~K, suggesting that geometrical frustration plays a large role in determining the magnetic ground state.  $\mu$SR studies
 by \textcite{Lago:2005} revealed persistent spin dynamics at 20~mK in both Er based samples.  While the temperature dependence of the muon relaxation rate is similar,
the depolarisation curve for Er$_2$Sn$_2$O$_7$ remains exponential down to the lowest temperatures suggesting the system fails to enter an ordered state, while the titanate takes on a gaussian relaxation at early times.

Neutron diffraction results by \textcite{Bramwell:2004} and \textcite{Shirai:2007}  would however suggest that Er$_2$Sn$_2$O$_7$ enters a  state with long-range order at $\approx$ 0.1~K.  This ordered ground state is different that that of Er$_2$Ti$_2$O$_7$, which appears to be stable in the Er$_2$(Sn,Ti)$_2$O$_7$ solid solution until 90\% substitution of Ti for Sn.

\subsection{Pr$_2$Ir$_2$O$_7$}

\label{PIO}

\begin{figure}[t]
\begin{center}
\includegraphics[width=7.4cm,angle=0,clip=390]{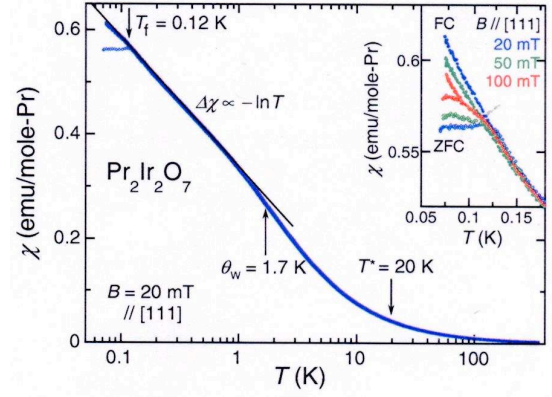}
\end{center}
\caption{DC susceptibility at a field of 20mT
applied along [111] for Pr$_2$Ir$_2$O$_7$.
 Note the $\ln(T)$ dependence expected for the Kondo effect and
(inset) evidence for spin freezing~\cite{Nakatsuji:2006}.}
\label{PrIr_sus}
\end{figure}

Recently, much attention has been devoted to the metallic material
Pr$_2$Ir$_2$O$_7$ which is reported to remain paramagnetic down to 0.3~K
by \textcite{Machida:2005}.  From magnetization, specific
heat and inelastic neutron scattering studies the Pr  ground state was determined to be a well isolated doublet with Ising-like moments oriented along the local $\langle 111\rangle$  axes.
The antiferromagnetically coupled spins have an energy scale of
$\approx$ 20~K and no freezing was observed.  Subsequent studies
carried out on crystals grown from a KF flux showed remarkably
complex behavior, including the Kondo effect and partial spin
freezing below 120~mK (see Fig.~\ref{PrIr_sus})~\cite{Nakatsuji:2006,Millican:2007}.
The authors have concluded that a spin liquid state exists
for the Pr moments but no studies of spin dynamics have yet been presented to support this claim.

Finally, a report of an unconventional anomalous Hall effect by \textcite{Machida:2007}
has appeared very recently and a chirality mechanism, similar to that proposed for Nd$_2$Mo$_2$O$_7$~\cite{Taguchi:2001,Taguchi:2004}, has been invoked.

\section{External Perturbations}

Many studies have been performed on geometrically frustrated magnets where an external pertubation
(apart from nonzero temperature) has been applied to the system.  These perturbations can take the form of common variables like magnetic field and the chemical  composition of the systems.  
Less common perturbations include applied pressure and irradiation damage and the effects
of these too have been investigated on pyrochlore oxides.  To date, no magnetic studies of irradiation damaged pyrochlores have been published, though several groups
have studied the structural properties of pyrochlores
after being bombarded with radiation~\cite{Sickafus:2000,Erwing:2004}.
Where appropriate, the results from changes in temperature or chemical composition have been mentioned in the previous sections and we will restrict ourselves here to those studies
where profound changes to the magnetic nature of the frustrated
 magnet have been observed.

\subsection{Magnetic Field}

\label{Sec:Mag_perp}

Most magnetic pyrochlore oxides have been subject to an applied field
and their behavior monitored. In some cases, drastic effects have been observed.
Indeed, one must remember that these systems are delicately
balanced between several competing energy scales, and a magnetic
field, even small fields, can alter the system significantly.
For example, in the spin ices, Ho$_2$Ti$_2$O$_7$ and Dy$_2$Ti$_2$O$_7$,
in a field as small as 100~Oe applied along $[001]$ or $[110]$,
 a few of the many ground states are favored creating metastable states,
 magnetization plateaus and slow spin dynamics but no elementary changes
are observed~\cite{Fennell:2005}.  In a few cases, the application of a
 field can drive the system into a new ground state.
These facts must be remembered when probing frustrated systems,
possibly even in the remanent field of a magnet.

\subsubsection{Gd$_2$Ti$_2$O$_7$}

\begin{figure}
\begin{center}
\includegraphics[width=8.5cm,angle=0,clip=390]{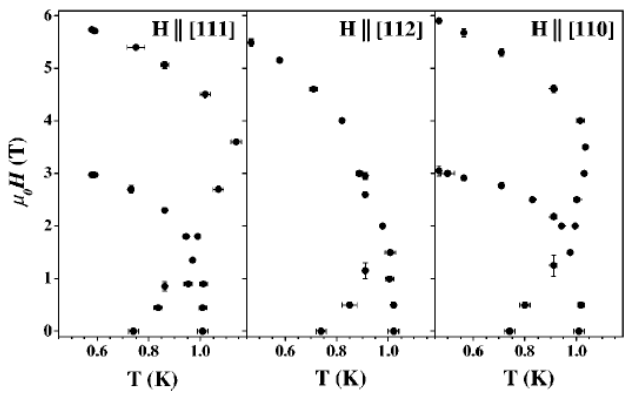}
\end{center}
\caption{Magnetic phase diagrams of the Gd$_2$Ti$_2$O$_7$ for three
different orientations of an applied magnetic field~\cite{Petrenko:2004}.
These phase boundaries roughly approximate those seen by \textcite{Ramirez:2002}
 on a powder sample.}
\label{Fig:Field_GTO}
\end{figure}

The magnetic field vs temperature phase diagram for the Heisenberg pyrochlore magnet
Gd$_2$Ti$_2$O$_7$ was first determined on polycrystalline samples using specific heat and AC susceptibility~\cite{Ramirez:2002}.  A mean field model was used to justify the five phases found
 below 1~K and in fields up to 8~T.  ~ \textcite{Petrenko:2004}
later reproduced these field induced transitions by measuring
the specific heat along 3 different crystallographic directions
on a single crystal (see Fig. \ref{Fig:Field_GTO}).

To determine the magnetic structure in these phases, neutron powder diffraction measurements have been performed.  Although significant differences can be seen in the diffraction
patterns in every phase,  only the zero field structures
have been modeled successfully~\cite{Stewart:2004,Shirai:2007}.

With the recent knowledge that Gd$_2$Sn$_2$O$_7$ is a better representation of a Heisenberg pyrochlore magnet with dipolar interactions and near neighbor exchange ~\cite{Wills:2006,DelMaestro:2007}  (see Section~\ref{Sec:GdTiO}), one would want to study it  in single crystalline form and in a magnetic field.

\subsubsection{Kagome ice}

By applying a magnetic field along the $[111]$ direction of single crystals of Dy$_2$Ti$_2$O$_7$
 or Ho$_2$Ti$_2$O$_7$ one can induce a new microscopically degenerate phase, known as kagome ice.  In this state the 2-in-2-out spin configuration, known as the ice-rules state, is still found,
however one spin is pinned by the field and the number of possible ground states is lowered~\cite{Moessner:2003}. As shown in Fig.~\ref{Fig:Alt_view}, the pyrochlore structure as observed along the [111] is comprised of stacked, alternating kagome and triangular planes. When the applied field exceeds 0.3~kOe the kagome ice phase is stable and exhibits a residual entropy of $\approx$~40\%
of that seen in zero field spin ice.

\textcite{Matsuhira:2002a} performed magnetisation and specific heat experiments on Dy$_2$Ti$_2$O$_7$, proving the existence of this phase. Since then,  several pieces of work have been done, mainly on Dy$_2$Ti$_2$O$_7$, with probes like AC susceptibility and specific heat~\cite{Higashinaka:2004,Sakakibara:2004,Tabata:2006}.  Very recently \textcite{Fennell:2007} have performed neutron diffraction measurements on Ho$_2$Ti$_2$O$_7$ in the kagome ice phase and speculate that a Kasteleyn transition takes place.  \textcite{Wills:2002} investigated  via analytical and Monte Carlo calculations an ice-like state in a kagome lattice with local Ising anisotropy before one was experimentally realized.  While the physics of kagome ice emerging from
a spin ice single crystal subject to a $[111]$ field is rather interesting, the behavior of spin ice subject to a magnetic field along one of the $[110]$, $[112]$ or $[100]$ directions is also worthy of a brief comment.

The pyrochlore lattice can be viewed as two sets of orthogonal chains of spins. One set is parallel to the $[110]$ direction and the other along $[1\bar 1 0]$. These two sets are referred to as $\alpha$ and $\beta$ chains, respectively.  The application of a field parallel to $[110]$ and of strength slightly larger than 1 T leads to a pinning of the spins on the $\alpha$ chain, effectively freezing those spins.  Both the net exchange and net dipolar field produced by the frozen spins on the $\alpha$ chains vanish by symmetry for the spins on the $\beta$ chains.  Since the $\beta$ chains are perpendicular to the field along $[110]$, the spins along those $\beta$ chains are free to interact only among themselves.  Just as a $[111]$ field leads to a decomposition of the pyrochlore lattice into weakly coupled kagome planes with their normal along $[111]$, a field along $[110]$ induces a ``magnetic break up''
of an otherwise cubic pyrochlore system into a set of
quasi one-dimensional $\beta$ chains predominantly weakly coupled by
dipolar interactions.  On the basis of Monte Carlo simulations~\cite{Ruff:2005,Yoshida:2004}  and chain mean field theory~\cite{Ruff:2005}, one expects that dipolar spin ice would have a transition to a long range ordered state, referred to as ${\bf q}={\rm X}$ order~\cite{Harris:1997}.
However, while neutron scattering experiments on Ho$_2$Ti$_2$O$_7$ and Dy$_2$Ti$_2$O$_7$ ~\cite{Fennell:2005} and specific heat measurements~\cite{Hiroi:2003a} on Dy$_2$Ti$_2$O$_7$ find some evidence for a transition with a field along $[110]$, the transition is not sharp. It is not  known whether the apparent failure of the system to develop long range ${\bf q}={\rm X}$ order is
due to either an imperfect alignment of the field along $[110]$ or 
the inability of the system to properly equilibrate upon approaching the putative transition.

The $[100]$ field problem was probably the first one studied theoretically~\cite{Harris:1998}.
It was predicted on the basis of Monte Carlo simulations on the nearest-neighbor spin ice model that the system should exhibit a field-driven transition between a gas-like weakly magnetized ice-rule obeying state and a strongly polarized state. It was originally argued that this transition is first order and terminates at a critical point, similarly to the gas-liquid critical point~\cite{Harris:1998}.
However, recent work that combines numerical and analytical calculations argues for a more exotic topological Kastelyn transition driven by the proliferation of defects in the ice-state as the temperature or the field strength is varied~\cite{Jaubert:2007}.
The same work speculates that there exists preliminary evidence for such a rounded Kastelyn transition in the neutron scattering data on Ho$_2$Ti$_2$O$_7$ subject to a $[100]$ field~\cite{Fennell:2005}.
On the other hand, while the Monte Carlo data of ~\textcite{Harris:1998}
find a sharp and a broad feature in the specific heat as a function of temperature for
sufficiently low (but nonzero) $[100]$ field, specific heat measurements~\cite{Higashinaka:2003a}
on Dy$_2$Ti$_2$O$_7$ only find one peak down to a temperature of 0.35 K. Perhaps long range dipolar interactions are of some relevance to the phenomenology at play for the case of a $[100]$ field.
Indeed, long range dipole-dipole interactions are definitely crucial to the qualitative physics
at play for fields along $[110]$ and near $[112]$~\cite{Ruff:2005}.
Clearly, more experimental and theoretical work on the problem of spin ice in a $[100]$ magnetic field is called for.

\subsubsection{Tb$_2$Ti$_2$O$_7$}

\begin{figure}[t]
\begin{center}
\includegraphics[width=8.5cm,angle=0,clip=390]{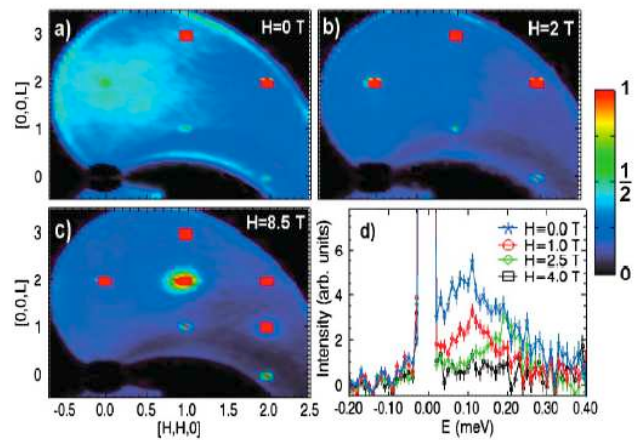}
\end{center}
\caption{Neutron scattering with $\Delta$E =
$\pm$0.5~meV from Tb$_2$Ti$_2$O$_7$ at 1~K~\cite{Rule:2006}.
 In zero field~(a), 2~T (b) and 8.5~T (c).
In panel d) constant $Q$ scans at 0.1~K are shown,
indicating that the low lying magnetic scattering (
slow spin dynamics) is quenched as the field increases.}
\label{Fig:TbT_elastic}
\end{figure}

Neutron diffraction and susceptibility data have recently been collected on the cooperative paramagnet 
 Tb$_2$Ti$_2$O$_7$ in high magnetic fields~\cite{Rule:2006,Ueland:2006}.
 In the AC  susceptibility work by \textcite{Ueland:2006} unusual cooperative effects and slow spin dynamics were induced by applying  fields as high as 9~T at temperatures above 1~K.

\textcite{Rule:2006} performed single crystal neutron scattering on Tb$_2$Ti$_2$O$_7$.  They were able to show that the short range spin-spin correlations, discussed in Section~\ref{sec:Spin_Liquid}~\cite{Gardner:2001,Gardner:2003}, are very dynamic in origin and that they become of longer range as a field is applied.  Applying a fairly small field of 1000 Oe, at 400~mK, causes the magnetic diffuse scattering 
to condense into peaks (see Fig.~\ref{TbT_Cry_diffuse}),  with characteristics similar to a polarised paramagnet. Figure \ref{Fig:TbT_elastic} shows similar data at 1~K.  In zero field,  Bragg peaks are seen at  113  and 222  from the underlying nuclear structure and broad diffuse scattering,
magnetic in origin is centered at 002.  In a field up to 2~T (not shown) the diffuse scattering sharpens
 into a 002  peak and in higher fields ($\mu_0$H $>2$ T) a magnetically ordered phase is induced, with
the development of true long range order accompanied by spin waves.

\subsection{High Pressure}

\label {Sec:pressure}

The study of pressure on samples in polycrystalline and single crystal form has become more common in recent years, with a paper on the pressure effect on the  magnetoresistive material Tl$_2$Mn$_2$O$_7$ being among the earliest ~\cite{Sushko:1996} in pyrochlore physics.  More recently, studies have been performed on the transport, structural and magnetic properties of other pyrochlores~\cite{Zhang:2007,Zhang:2006,Saha:2006,Mirebeau:2002}.

Sm$_2$Zr$_2$O$_7$ undergoes a structural distortion at low pressures.
However,  the distortion is not observed between 13.5 and 18~GPa.  Above 18 GPa, the pyrochlore structure is unstable and  a distorted fluorite structure, in which
the cations and anion vacancies are disordered, is observed~\cite{Zhang:2007}.   Similar things happen to Cd$_2$Nb$_2$O$_7$ at 12 GPa~\cite{Zhang:2006,Samara:2006}.  Gd$_2$Ti$_2$O$_7$ undergoes a subtle distortion of the lattice at 9 GPa at room temperature~\cite{Saha:2006}.
Neutron diffraction at 1.4~K on Ho$_2$Ti$_2$O$_7$ at pressures up to 6 GPa saw no change in the spatial correlations of this spin ice compound~\cite{Mirebeau:2004}, but a recent study of Dy$_2$Ti$_2$O$_7$ observed a small change in the magnetisation at 13 kbar~\cite{Mito:2007}.

\subsubsection{Tb$_2$Ti$_2$O$_7$}

\begin{figure}
\centerline{\psfig{file=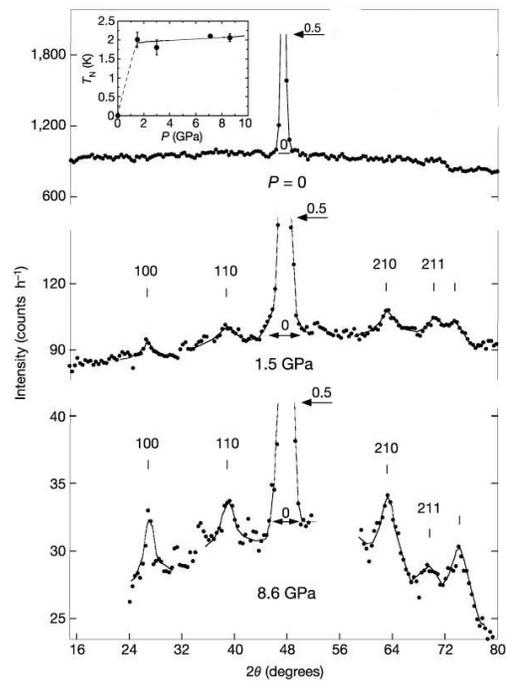,width=6.7cm,angle=0}}
\caption{Pressure induced magnetic Bragg scattering from
polycrystalline Tb$_2$Ti$_2$O$_7$ at $<$2~K.
 The data shows the pressure evolution of static correlations in the
 cooperative paramagnet.  The inset depicts the pressure dependence of the
 transition temperature~\protect\cite{Mirebeau:2002}.}.
\label{TbT_press}
\end{figure}

The  Tb$_2$Ti$_2$O$_7$ cooperative paramagnet has been studied by means of single-crystal
 and polycrystalline neutron diffraction for an unprecedented
range of thermodynamical parameters combining high pressures,
magnetic fields and low temperatures~\cite{Mirebeau:2002,Mirebeau:2004a}.
  \textcite{Mirebeau:2002} found a long range, magnetically ordered state
near 2~K for a hydrostatic pressure exceeding $\sim$ 1.5~GPa.
Powder neutron diffraction data as a function of pressure
 are shown in Fig.~\ref{TbT_press}.
 Weak magnetic Bragg peaks occur at positions not associated
with nuclear reflections although indexable to the $Fd \bar{3}m$ space group.
  They also found two small reflections that were attributed to a long range
modulation of the main structure.

To investigate the transition further, \textcite{Mirebeau:2004a} 
studied single crystals where the ordered magnetic moment may be tuned by means of the direction
of the anisotropic pressure component.  It was shown that an anisotropic pressure component 
is needed to suppress the spin liquid state,  but an anisotropic pressure alone with
no hydrostatice component produced no effect at 1.4 K.
 When an ordered phase was induced, the direction of the anisotropic
 pressure was important with a factor of 30 difference seen in the
 strength of the magnetic scattering between different directions~\cite{Mirebeau:2004a}.
One wonders if the rich behavior of
Tb$_2$Ti$_2$O$_7$ under pressure is related to the recent observations
of seemingly dynamical lattice effects in zero applied pressure~\cite{Ruff:2007}.

\subsubsection{${\it A}_2$Mo$_2$O$_7$ (${\it A}$ = Gd and Tb)}

\label {Sec:pressureA2Mo2O7}

\begin{figure}
\begin{center}
\includegraphics[width=6.4cm,angle=0,clip=390]{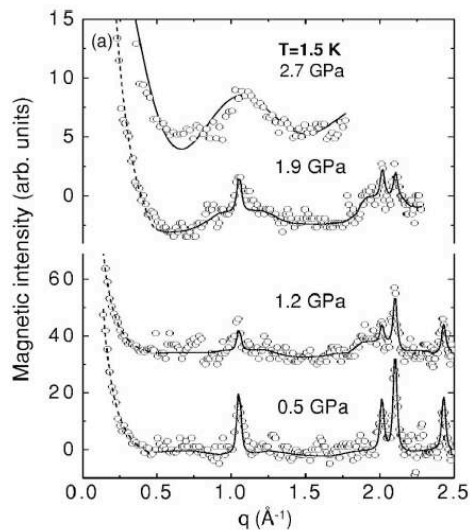}
\end{center}
\caption{Pressure dependence on the magnetic scattering from
 Gd$_{2}$Mo$_{2}$O$_{7}$. Note the broadening of the reflection near
 $q = 1$ {\AA}$^{-1}$ which is complete by 2.7 GPa~\cite{Mirebeau:2006a}}
\label{Fig:GdMo_press}
\end{figure}

The properties of Gd$_{2}$Mo$_{2}$O$_{7-x}$ change remarkably with pressure.
 \textcite{Kim:2005} applied moderate pressures of 1.6 GPa and found a
 significant decrease in the saturation moment at 4~K with increased hysteresis.
 \textcite{Mirebeau:2006a} and \textcite{Miyoshi:2006}
extended the pressure range and were able to drive the system into
 a glassy state at 2.7 GPa.  In the study by \textcite{Mirebeau:2006a}
the melting of the magnetic structure was monitored by neutron diffraction
(see Fig.~\ref{Fig:GdMo_press}). Thus, the application of pressure can change
the sign of the Mo$-$Mo exchange interaction from ferromagnetic to antiferromagnetic.
 In addition, a phase with mixed Tb/La substitution on the $A$-site,
 namely(La$_{0.2}$Tb$_{0.8}$)$_2$Mo$_2$O$_7$,
was studied by \textcite{Apetrei:2007a}.
This material, which has nearly the same unit cell constant
 as Gd$_{2}$Mo$_{2}$O$_{7-x}$ [10.3787(8)~\AA], shows ferromagnetism
 with $T_c$ = 58~K but with a different magnetic structure in which
 the Tb moments form the familiar two-in two-out Ising structure, but
there is still ferromagnetic coupling between the $A$ and $B$ magnetic sublattices.
 It is also remarked that the ordered moments are reduced significantly from
the free ion values by 50 percent for Tb, 30 percent for Gd and 70 percent for Mo.
 The case for Tb is not surprising and can be ascribed to crystal field effects,
 while the result for Mo is in line with previous
reports~\cite{Gaulin:1992,Greedan:1991,Yasui:2001}.

\part{Conclusions}

We have highlighted most of the advances made over the 
past 20 years in classifying the magnetic ground state 
in many geometrically frustrated magnets with the pyrochlore lattice.  
An extensive review on other frustrated magnetic systems like the spinels and kagome  
lattice magnets would be most welcomed.  Although significant advances have 
been made in the understanding and characterization of these pyrochlore magnets, 
much more work can be done.  As the systems get more complicated 
(i.e. through dilution of the magnetic sublattice or mixing of magnetic species)
 we hope and expect to see more new and exciting physics, but experimentalists will have to be more diligent in their  characterization of their samples, since it is now clear that the magnetic ground state is very sensitive to  small perturbations.  Understanding the role local disorder and spin-lattice interactions play in these materials, as well as the origin of the persistent spin dynamics seen well below 
apparent ordering temperatures, usually by a local probe such as $\mu$SR is crucial in advancing this field.  Another area that researchers should invest some time studying is the transport properties  at metal/insulator or superconducting  transitions.  While a connection to geometric frustration is often a feature of the introduction to papers involving superconducting pyrochlores  like Cd$_2$Re$_2$O$_7$~\cite{Sakai:2001,Hanawa:2001} and the series {\it M}Os$_2$O$_6$, where {\it M} = K, Rb and Cs~\cite{Yonezawa:2004,Yonezawa:2004a,Yonezawa:2004b}, there is no credible evidence yet that magnetism, frustrated or otherwise, plays any role in the superconducting phase.  Finally, while the models, created by theorists, for understanding geometrically frustrated systems
have become increasingly sophisticated, there persist many outstanding questions.
What is the origin of the 15~K anomaly in the titanate spin ices? 
Why does Tb$_2$Ti$_2$O$_7$ remain dynamic and what is the true nature
 of the Tb$^{3+}$ moments? What is the true ground state for Yb$_2$Ti$_2$O$_7$? Is the ground state of Er$_2$Ti$_2$O$_7$ truly selected by quantum order-by-disorder?  These must be addressed in the near future.

\section*{Acknowledgements}

We acknowledge here all our past and immediate collaborators, post-docs 
and students, too numerous to list individually, and with 
whom we have enjoyed working over the years. 
Their contributions to this field are clear from the large number of
citations of their work in the bibliography which follows.
We also thank our many colleagues for hundreds of useful discussions and e-mail correspondences.

%
%NIST requires the use of International System of units (SI) for dimensioned physical quantities, however in this paper, where we have relied extensively on other peoples data, we were not able to comply with this rule 100\% of the time.

We acknowledge the generous financial support of:
the NSERC of Canada, the Canada Research Chair program (M.G., Tier 1),
the Research Corporation, the Canadian Institute for Advanced Research,
Materials and Manufacturing Ontario,
the Ontario Innovation Trust and the Canada Foundation for Innovation.

%\bibliographystyle{apsrmp}

%\bibliography{MJPG_GGG_RMP2008}

\begin{thebibliography}{470}
\expandafter\ifx\csname natexlab\endcsname\relax\def\natexlab#1{#1}\fi
\expandafter\ifx\csname bibnamefont\endcsname\relax
  \def\bibnamefont#1{#1}\fi
\expandafter\ifx\csname bibfnamefont\endcsname\relax
  \def\bibfnamefont#1{#1}\fi
\expandafter\ifx\csname citenamefont\endcsname\relax
  \def\citenamefont#1{#1}\fi
\expandafter\ifx\csname url\endcsname\relax
  \def\url#1{\texttt{#1}}\fi
\expandafter\ifx\csname urlprefix\endcsname\relax\def\urlprefix{URL }\fi
\providecommand{\bibinfo}[2]{#2}
\providecommand{\eprint}[2][]{\url{#2}}





\bibitem[{\citenamefont{Aharony and Pytte}(1983)}]
{Aharony:1983}
\bibinfo{author}{\bibnamefont{Aharony},~\bibfnamefont{A.}} and
\bibinfo{author}{\bibfnamefont{E.}~\bibnamefont{Pytte}},
 \bibinfo{year}{1983},
 \bibinfo{journal}{Phys. Rev.} \textbf{\bibinfo{volume}{27}},
 \bibinfo{pages}{5872}.


   \bibitem[{\citenamefont{Ali \emph{et~al.}}(1989)}]
{Ali:1989}
\bibinfo{author}{\bibfnamefont{Ali}~\bibnamefont{N.}},
  \bibinfo{author}{\bibnamefont{M. P.}, \bibfnamefont{Hill}},
  \bibinfo{author}{\bibfnamefont{S.}~\bibnamefont{Labroo}}, and
  \bibinfo{author}{\bibfnamefont{J. E.}~\bibnamefont{Greedan}},
  \bibinfo{year}{1989},
  \bibinfo{journal}{J. Solid State Chem.} \textbf{\bibinfo{volume}{83}},
  \bibinfo{pages}{178}.


 \bibitem[{\citenamefont{Alonso \emph{et~al.}}(2000)}]
{Alonso:2000}
\bibinfo{author}{\bibnamefont{Alonso}, \bibfnamefont{J. A.}},
  \bibinfo{author}{\bibfnamefont{M. J.}~\bibnamefont{Martinez-Lope}},
  \bibinfo{author}{\bibfnamefont{M. T.}~\bibnamefont{Casais}}, and
  \bibinfo{author}{\bibfnamefont{J. L.}~\bibnamefont{Martinez}},
  \bibinfo{year}{2000},
  \bibinfo{journal}{Chem. Mater.} \textbf{\bibinfo{volume}{12}},
  \bibinfo{pages}{1127}.


\bibitem[{\citenamefont{Anderson}(1956)}]
{Anderson:1956}
\bibinfo{author}{\bibnamefont{Anderson},~\bibfnamefont{P. W.}},
\bibinfo{year}{1956},
\bibinfo{journal}{Phys. Rev.} \textbf{\bibinfo{volume}{102}},
\bibinfo{pages}{1008}.



\bibitem[{\citenamefont{Anderson}(1973)}]
{Anderson:1973}
  \bibinfo{author}{\bibfnamefont{Anderson}, \bibnamefont{P. W.}},
  \bibinfo{year}{1973},
  \bibinfo{journal}{Mater. Res. Bull.} \textbf{\bibinfo{volume}{8}},
  \bibinfo{pages}{153}.


 \bibitem[{\citenamefont{Anderson}(1987)}]
 {Anderson:1987}
   \bibinfo{author}{\bibfnamefont{Anderson}, \bibnamefont{P. W.}},
   \bibinfo{year}{1987},
   \bibinfo{journal}{Science} \textbf{\bibinfo{volume}{235}},
   \bibinfo{pages}{1196}.


  \bibitem[{\citenamefont{Apetrei \emph{et~al.}}(2007)}]
{Apetrei:2007}
\bibinfo{author}{\bibfnamefont{Apetrei}, ~\bibnamefont{A.}},
  \bibinfo{author}{\bibnamefont{I.}, \bibfnamefont{Mirebeau}},
  \bibinfo{author}{\bibfnamefont{I.}~\bibnamefont{Goncharenko}},
  \bibinfo{author}{\bibfnamefont{D.}~\bibnamefont{Andreica}}, and
  \bibinfo{author}{\bibfnamefont{P.}~\bibnamefont{Bonville}},
  \bibinfo{year}{2007},
  \bibinfo{journal}{J. Phys.: Condens. Matter} \textbf{\bibinfo{volume}{19}},
  \bibinfo{pages}{145214}.



  \bibitem[{\citenamefont{Apetrei \emph{et~al.}}(2007a)}]
{Apetrei:2007a}
\bibinfo{author}{\bibfnamefont{Apetrei},~\bibnamefont{A.}},
  \bibinfo{author}{\bibnamefont{I.}, \bibfnamefont{Mirebeau}},
  \bibinfo{author}{\bibfnamefont{I.}~\bibnamefont{Goncharenko}},
  \bibinfo{author}{\bibfnamefont{D.}~\bibnamefont{Andreica}}, and
  \bibinfo{author}{\bibfnamefont{P.}~\bibnamefont{Bonville}},
  \bibinfo{year}{2007a},
  \bibinfo{journal}{Phys. Rev. Lett.} \textbf{\bibinfo{volume}{97}},
  \bibinfo{pages}{206401}.



 \bibitem[{\citenamefont{Arai \emph{et~al.}}(1985)}]
{Arai:1985}
\bibinfo{author}{\bibnamefont{Arai}, ~\bibfnamefont{M.}},
  \bibinfo{author}{\bibfnamefont{Y.}~\bibnamefont{Ishikawa}},
  \bibinfo{author}{\bibfnamefont{N.}~\bibnamefont{Saito}}, and
  \bibinfo{author}{\bibfnamefont{H.}~\bibnamefont{Takei}},
  \bibinfo{year}{1985},
  \bibinfo{journal}{J. Phys. Soc. Japan} \textbf{\bibinfo{volume}{54}},
  \bibinfo{pages}{781}.


%
%  \bibitem[{\citenamefont{Babel \emph{et~al.}}(1967)}]
%{Babel:1967}
%\bibinfo{author}{\bibnamefont{Babel}, \bibfnamefont{D.}},
%  \bibinfo{author}{\bibfnamefont{G.}~\bibnamefont{Pausewang}}, and
%  \bibinfo{author}{\bibfnamefont{W.}~\bibnamefont{Viebahn}},
%  \bibinfo{year}{1967},
%  \bibinfo{journal}{Z. Naturforschung B} \textbf{\bibinfo{volume}{22}},
%  \bibinfo{pages}{1219}.

%
%  \bibitem[{\citenamefont{Babel}(1972)}]
%{Babel:1972}
%\bibinfo{author}{\bibnamefont{Babel},~\bibfnamefont{D.}},
% \bibinfo{year}{1972},
% \bibinfo{journal}{Z. Anorg. Allgem. Chemie} \textbf{\bibinfo{volume}{387}},
% \bibinfo{pages}{161}.



\bibitem[{\citenamefont{Balakrishnan \emph{et~al.}}(1998)}]
{Balakrishnan:1998}
\bibinfo{author}{\bibnamefont{Balakrishnan}, \bibfnamefont{G.}},
  \bibinfo{author}{\bibfnamefont{O. A.}~\bibnamefont{Petrenko}},
   \bibinfo{author}{\bibfnamefont{M. R.}~\bibnamefont{Lees}}, and
  \bibinfo{author}{\bibfnamefont{D. M$^c$K.}~\bibnamefont{Paul}},
  \bibinfo{year}{1998},
  \bibinfo{journal}{J. Phys.: Condens. Matter} \textbf{\bibinfo{volume}{10}},
  \bibinfo{pages}{L723}.



\bibitem[{\citenamefont{Ballesteros} \emph{et~al.}(2000)}]
{Ballesteros:2000}
  \bibinfo{author}{\bibfnamefont{Ballesteros}, \bibnamefont{H. G.}},
  \bibinfo{author}{\bibnamefont{A.}~\bibfnamefont{Cruz}},
  \bibinfo{author}{\bibnamefont{L. A.}~\bibfnamefont{Fern\'andez}},
  \bibinfo{author}{\bibnamefont{V.}~\bibfnamefont{Mart\'in-Mayor}},
  \bibinfo{author}{\bibnamefont{J.}~\bibfnamefont{Pech}},
  \bibinfo{author}{\bibnamefont{J. J}~\bibfnamefont{Ruiz-Lorenzo}},
  \bibinfo{author}{\bibnamefont{A.}~\bibfnamefont{Taran\'con}},
  \bibinfo{author}{\bibnamefont{P.}~\bibfnamefont{T\'ellez}},
  \bibinfo{author}{\bibnamefont{C. L.}~\bibfnamefont{Ullod}}, and
  \bibinfo{author}{\bibnamefont{C.}~\bibfnamefont{Ungil}},
 \bibinfo{year}{2000},
  \bibinfo{journal}{Phys. Rev. B} \textbf{\bibinfo{volume}{62}},
  \bibinfo{pages}{14237}.




\bibitem[{\citenamefont{Ballou}(2001)}]
{Ballou:2001}
\bibinfo{author}{\bibnamefont{Ballou},~\bibfnamefont{R.}},
 \bibinfo{year}{2001},
 \bibinfo{journal}{ Can. J. Phys.} \textbf{\bibinfo{volume}{79}},
 \bibinfo{pages}{1475}.




\bibitem[{\citenamefont{Bansal \emph{et~al.}}(2002)}]
{Bansal:2002}
\bibinfo{author}{\bibnamefont{Bansal}, \bibfnamefont{C.}},
  \bibinfo{author}{\bibfnamefont{H.}~\bibnamefont{Kawanaka}},
   \bibinfo{author}{\bibfnamefont{H.}~\bibnamefont{Bando}}, and
  \bibinfo{author}{\bibfnamefont{Y.}~\bibnamefont{Nishihara}},
  \bibinfo{year}{2002},
  \bibinfo{journal}{Phys. Rev. B} \textbf{\bibinfo{volume}{66}},
  \bibinfo{pages}{052406}.




\bibitem[{\citenamefont{Bansal \emph{et~al.}}(2003)}]
{Bansal:2003}
\bibinfo{author}{\bibnamefont{Bansal}, \bibfnamefont{C.}},
  \bibinfo{author}{\bibnamefont{K.}~\bibfnamefont{Hirofumi}},
   \bibinfo{author}{\bibnamefont{B.}~\bibfnamefont{Hiroshi}}, and
  \bibinfo{author}{~\bibnamefont{N.}~\bibfnamefont{Yoshikazu}},
  \bibinfo{year}{2003},
  \bibinfo{journal}{Physica B} \textbf{\bibinfo{volume}{329}},
  \bibinfo{pages}{1034}.




\bibitem[{\citenamefont{Barkema and Newman}(1998)}]
{Barkema:1998}
 \bibinfo{author}{\bibfnamefont{Barkema}, \bibnamefont{G. T.}}, and
  \bibinfo{author}{\bibnamefont{M. E. J.}~\bibfnamefont{Newman}},
 \bibinfo{year}{1998},
  \bibinfo{journal}{Phys. Rev. E} \textbf{\bibinfo{volume}{57}},
  \bibinfo{pages}{1155}.



\bibitem[{\citenamefont{Barton and Cashion}(1979)}]
{Barton:1979}
\bibinfo{author}{\bibnamefont{Barton}, \bibfnamefont{W. A.}} and
\bibinfo{author}{\bibfnamefont{J.}~\bibnamefont{Cashion}},
\bibinfo{year}{1979},
\bibinfo{journal}{J. Phys. C.} \textbf{\bibinfo{volume}{12}},
\bibinfo{pages}{2897}.


\bibitem[{\citenamefont{Bazuev \emph{et~al.}}(1976)}]
{Bazuev:1976}
\bibinfo{author}{\bibnamefont{Bazuev}, \bibfnamefont{G. V.}},
  \bibinfo{author}{\bibfnamefont{O. V.}~\bibnamefont{Makarova}},
   \bibinfo{author}{\bibfnamefont{V. Z.}~\bibnamefont{Oboldin}}, and
  \bibinfo{author}{\bibfnamefont{G. P.}~\bibnamefont{Shveikin}},
  \bibinfo{year}{1976},
  \bibinfo{journal}{Doklady Akademii Nauk SSSR} \textbf{\bibinfo{volume}{230}},
  \bibinfo{pages}{869}.


\bibitem[{\citenamefont{Bellier-Castella} \emph{et~al.}(2001)}]
{Bellier:2001}
  \bibinfo{author}{\bibfnamefont{Bellier-Castella}, \bibnamefont{L.}},
  \bibinfo{author}{\bibnamefont{M. J. P.}~\bibfnamefont{Gingras}},
  \bibinfo{author}{\bibnamefont{P. C. W.}~\bibfnamefont{Holdsworth}}, and
  \bibinfo{author}{\bibnamefont{R.}~\bibfnamefont{Moessner}},
 \bibinfo{year}{2001},
  \bibinfo{journal}{Can. J. Phys.} \textbf{\bibinfo{volume}{79}},
  \bibinfo{pages}{1365}.



\bibitem[{\citenamefont{Berg} \emph{et~al.}(2003)}]
{Berg:2003}
  \bibinfo{author}{\bibfnamefont{Berg}, \bibnamefont{E.}},
\bibinfo{author}{\bibnamefont{E.}~\bibfnamefont{Altman}}, and
 \bibinfo{author}{\bibnamefont{A.}~\bibfnamefont{Auerbach}},
 \bibinfo{year}{2003},
  \bibinfo{journal}{Phys. Rev. Lett.} \textbf{\bibinfo{volume}{90}},
  \bibinfo{pages}{147204}.


\bibitem[{\citenamefont{Bergman} \emph{et~al.}(2006)}]
{Bergman:2006}
  \bibinfo{author}{\bibfnamefont{Berman}, \bibnamefont{D. L.}},
  \bibinfo{author}{\bibnamefont{R.}~\bibfnamefont{Shindou}},
  \bibinfo{author}{\bibnamefont{G. A.}~\bibfnamefont{Fiete}}, and
  \bibinfo{author}{\bibnamefont{L.}~\bibfnamefont{Balents}},
 \bibinfo{year}{2006},
  \bibinfo{journal}{Phys. Rev. B} \textbf{\bibinfo{volume}{74}},
  \bibinfo{pages}{134409}.



%
%\bibitem[{\citenamefont{Bernu} \emph{et~al.}(1994)}]
%{Bernu:1994}
%  \bibinfo{author}{\bibfnamefont{Bernu}, \bibnamefont{B.}},
%  \bibinfo{author}{\bibnamefont{P.}~\bibfnamefont{Lecheminant}},
%  \bibinfo{author}{\bibnamefont{C.}~\bibfnamefont{Lhuillier}}, and
%  \bibinfo{author}{\bibnamefont{L.}~\bibfnamefont{Pierre}},
% \bibinfo{year}{1994},
%  \bibinfo{journal}{Phys. Rev. B} \textbf{\bibinfo{volume}{50}},
%  \bibinfo{pages}{10048}.



   \bibitem[{\citenamefont{Bert \emph{et~al.}}(2006)}]
{Bert:2006}
\bibinfo{author}{\bibnamefont{Bert}, \bibfnamefont{F.}},
  \bibinfo{author}{\bibfnamefont{P.}~\bibnamefont{Mendels}},
  \bibinfo{author}{\bibfnamefont{A.}~\bibnamefont{Olariu}},
  \bibinfo{author}{\bibfnamefont{N.}~\bibnamefont{Blanchard}},
  \bibinfo{author}{\bibfnamefont{G.}~\bibnamefont{Collin}},
  \bibinfo{author}{\bibfnamefont{A.}~\bibnamefont{Amato}},
  \bibinfo{author}{\bibfnamefont{C.}~\bibnamefont{Baines}}, and
  \bibinfo{author}{\bibfnamefont{A. D.}~\bibnamefont{Hillier}},
  \bibinfo{year}{2006},
  \bibinfo{journal}{Phys. Rev. Lett.} \textbf{\bibinfo{volume}{97}},
  \bibinfo{pages}{117203}.




  \bibitem[{\citenamefont{Bertaut \emph{et~al.}}(1959)}]
{Bertaut:1959}
\bibinfo{author}{\bibnamefont{Bertaut}, \bibfnamefont{E. F.}},
  \bibinfo{author}{\bibfnamefont{F.}~\bibnamefont{Forrat}}, and
  \bibinfo{author}{\bibfnamefont{M. C.}~\bibnamefont{Montmory}},
  \bibinfo{year}{1959},
  \bibinfo{journal}{Compt. Rend.} \textbf{\bibinfo{volume}{249c}},
  \bibinfo{pages}{276}.




   \bibitem[{\citenamefont{Bertin \emph{et~al.}}(2002)}]
{Bertin:2002}
\bibinfo{author}{\bibnamefont{Bertin}, \bibfnamefont{E.}},
  \bibinfo{author}{\bibfnamefont{P.}~\bibnamefont{Bonville}},
  \bibinfo{author}{\bibfnamefont{J.-P.}~\bibnamefont{Bouchaud}},
  \bibinfo{author}{\bibfnamefont{J. A.}~\bibnamefont{Hodges}},
  \bibinfo{author}{\bibfnamefont{J. P.}~\bibnamefont{Sanchez}}, and
  \bibinfo{author}{\bibfnamefont{P.}~\bibnamefont{Vulliet}},
  \bibinfo{year}{2002},
  \bibinfo{journal}{Eur. Phys. J. B} \textbf{\bibinfo{volume}{27}},
  \bibinfo{pages}{347}.




\bibitem[{\citenamefont{Billinge}(2004)}]
{Billinge:2004}
\bibinfo{author}{\bibnamefont{Billinge},~\bibfnamefont{S.J.L.}},
 \bibinfo{year}{2004},
 \bibinfo{journal}{Z. Kristallogr.} \textbf{\bibinfo{volume}{219}},
 \bibinfo{pages}{117}.


\bibitem[{\citenamefont{Binder and Young}(1986)}]
{Binder:1986}
\bibinfo{author}{\bibnamefont{Binder},~\bibfnamefont{K.}}, and
\bibinfo{author}{\bibfnamefont{A. P.}~\bibnamefont{Young}},
 \bibinfo{year}{1986},
 \bibinfo{journal}{Rev. Mod. Phys.} \textbf{\bibinfo{volume}{58}},
 \bibinfo{pages}{801}.



\bibitem[{\citenamefont{Blacklock and White}(1979)}]
{Blacklock:1979}
\bibinfo{author}{\bibnamefont{Blacklock},~\bibfnamefont{K.}} and
\bibinfo{author}{\bibfnamefont{H. W.}~\bibnamefont{White}},
 \bibinfo{year}{1979},
 \bibinfo{journal}{J. Chem. Phys.} \textbf{\bibinfo{volume}{71}},
 \bibinfo{pages}{5287}.




 \bibitem[{\citenamefont{Blacklock and White}(1980)}]
{Blacklock:1980}
\bibinfo{author}{\bibnamefont{Blacklock},~\bibfnamefont{K.}} and
\bibinfo{author}{\bibfnamefont{H. W.}~\bibnamefont{White}},
 \bibinfo{year}{1980},
 \bibinfo{journal}{J. Chem. Phys.} \textbf{\bibinfo{volume}{72}},
 \bibinfo{pages}{2191}.



  \bibitem[{\citenamefont{Bl{\"o}te \emph{et~al.}}(1969)}]
{Blote:1969}
\bibinfo{author}{\bibnamefont{Bl{\"o}te}, \bibfnamefont{H. W. J.}},
  \bibinfo{author}{\bibfnamefont{R. F.}~\bibnamefont{Wielinga}}, and
  \bibinfo{author}{\bibfnamefont{W. J.}~\bibnamefont{Huiskamp}},
  \bibinfo{year}{1969},
  \bibinfo{journal}{Physica} \textbf{\bibinfo{volume}{43}},
  \bibinfo{pages}{549}.




\bibitem[{\citenamefont{Bondah-Jagalu and Bramwell}(2001)}]
{Bondah-Jagalu:2001}
\bibinfo{author}{\bibnamefont{Bondah-Jagalu},~\bibfnamefont{V.}} and
\bibinfo{author}{\bibfnamefont{S. T.}~\bibnamefont{Bramwell}},
 \bibinfo{year}{2001},
 \bibinfo{journal}{Can. J. Phys.} \textbf{\bibinfo{volume}{79}},
 \bibinfo{pages}{1381}.



   \bibitem[{\citenamefont{Bonville \emph{et~al.}}(2003)}]
{Bonville:2003}
\bibinfo{author}{\bibnamefont{Bonville}, \bibfnamefont{P.}},
  \bibinfo{author}{\bibfnamefont{J. A.}~\bibnamefont{Hodges}},
  \bibinfo{author}{\bibfnamefont{M.}~\bibnamefont{Ocio}},
  \bibinfo{author}{\bibfnamefont{J. P.}~\bibnamefont{Sanchez}},
  \bibinfo{author}{\bibfnamefont{P.}~\bibnamefont{Vulliet}},
  \bibinfo{author}{\bibfnamefont{S.}~\bibnamefont{Sosin}}, and
  \bibinfo{author}{\bibfnamefont{D.}~\bibnamefont{Braithwaite}},
  \bibinfo{year}{2003},
  \bibinfo{journal}{J. Phys.: Condens: Matter} \textbf{\bibinfo{volume}{15}},
  \bibinfo{pages}{7777}.




  \bibitem[{\citenamefont{Bonville \emph{et~al.}}(2003a)}]
{Bonville:2003a}
\bibinfo{author}{\bibnamefont{Bonville}, \bibfnamefont{P.}},
  \bibinfo{author}{\bibfnamefont{J. A.}~\bibnamefont{Hodges}},
  \bibinfo{author}{\bibfnamefont{E.}~\bibnamefont{Bertin}},
  \bibinfo{author}{\bibfnamefont{J.-Ph.}~\bibnamefont{Bouchaud}},
  \bibinfo{author}{\bibfnamefont{M.}~\bibnamefont{Ocio}},
  \bibinfo{author}{\bibfnamefont{P.}~\bibnamefont{Dalmas de R\'eotier}},
  \bibinfo{author}{\bibfnamefont{L. -P..}~\bibnamefont{Regnault}},
  \bibinfo{author}{\bibfnamefont{H. M.}~\bibnamefont{R\o nnow}},
  \bibinfo{author}{\bibfnamefont{J. P.}~\bibnamefont{Sanchez}},
  \bibinfo{author}{\bibfnamefont{S.}~\bibnamefont{Sosin}},
  \bibinfo{author}{\bibnamefont{A.}~\bibfnamefont{Yaouanc}},
  \bibinfo{author}{\bibfnamefont{M.}~\bibnamefont{Rams}},  and
  \bibinfo{author}{\bibfnamefont{K.}~\bibnamefont{Kr\'{o}las}},
  \bibinfo{year}{2003a},
  \bibinfo{journal}{Cond-mat/0306470}.



 \bibitem[{\citenamefont{Booth \emph{et~al.}}(2000)}]
{Booth:2000}
\bibinfo{author}{\bibnamefont{Booth}, \bibfnamefont{C. H.}},
  \bibinfo{author}{\bibfnamefont{J. S.}~\bibnamefont{Gardner}},
  \bibinfo{author}{\bibfnamefont{G. H.}~\bibnamefont{Kwei}},
  \bibinfo{author}{\bibfnamefont{R. H.}~\bibnamefont{Heffner}},
  \bibinfo{author}{\bibfnamefont{F.}~\bibnamefont{Bridges}}, and
  \bibinfo{author}{\bibfnamefont{M. A.}~\bibnamefont{Subramanian}},
  \bibinfo{year}{2000},
  \bibinfo{journal}{Phys. Rev. B} \textbf{\bibinfo{volume}{62}},
  \bibinfo{pages}{R755}.



 \bibitem[{\citenamefont{Bramwell \emph{et~al.}}(1994)}]
{Bramwell:1994}
\bibinfo{author}{\bibnamefont{Bramwell}, \bibfnamefont{S. T.}},
\bibinfo{author}{\bibfnamefont{M. J. P.}~\bibnamefont{Gingras}}, and
\bibinfo{author}{\bibfnamefont{J. N.}~\bibnamefont{Reimers}},
\bibinfo{year}{1994},
\bibinfo{journal}{J. Appl. Phys.} \textbf{\bibinfo{volume}{75}},
\bibinfo{pages}{5523}.


 \bibitem[{\citenamefont{Bramwell \emph{et~al.}}(2000)}]
{Bramwell:2000}
\bibinfo{author}{\bibnamefont{Bramwell}, \bibfnamefont{S. T.}},
\bibinfo{author}{\bibfnamefont{M. N.}~\bibnamefont{Field}},
\bibinfo{author}{\bibfnamefont{M. J.}~\bibnamefont{Harris}}, and
\bibinfo{author}{\bibfnamefont{I. P.}~\bibnamefont{Perkin}},
\bibinfo{year}{2000},
\bibinfo{journal}{J. Phys.: Condens. Matter} \textbf{\bibinfo{volume}{12}},
\bibinfo{pages}{483}.



\bibitem[{\citenamefont{Bramwell and Gingras}(2001)}]
{Bramwell:2001}
\bibinfo{author}{\bibnamefont{Bramwell}, \bibfnamefont{S. T.}} and
\bibinfo{author}{\bibfnamefont{M. J. P.}~\bibnamefont{Gingras}},
\bibinfo{year}{2001},
\bibinfo{journal}{Science} \textbf{\bibinfo{volume}{294}},
\bibinfo{pages}{1495}.



\bibitem[{\citenamefont{Bramwell} \emph{et~al.}(2001a)}]
{Bramwell:2001a}
  \bibinfo{author}{\bibfnamefont{Bramwell}, \bibnamefont{S. T.}},
  \bibinfo{author}{\bibnamefont{M. J.}~\bibfnamefont{Harris}},
  \bibinfo{author}{\bibnamefont{B. C.}~\bibfnamefont{den Hertog}},
  \bibinfo{author}{\bibnamefont{M. J. P.}~\bibfnamefont{Gingras}},
  \bibinfo{author}{\bibnamefont{J. S.}~\bibfnamefont{Gardner}},
  \bibinfo{author}{\bibnamefont{D. F.}~\bibfnamefont{McMorrow}},
  \bibinfo{author}{\bibnamefont{A. R.}~\bibfnamefont{Wildes}},
  \bibinfo{author}{\bibnamefont{A. L.}~\bibfnamefont{Cornelius}},
  \bibinfo{author}{\bibnamefont{J. D. M.}~\bibfnamefont{Champion}},
  \bibinfo{author}{\bibnamefont{R. G.}~\bibfnamefont{Melko}}, and
  \bibinfo{author}{\bibnamefont{T.}~\bibfnamefont{Fennell}},
 \bibinfo{year}{2001a},
  \bibinfo{journal}{Phys. Rev. Lett.} \textbf{\bibinfo{volume}{87}},
  \bibinfo{pages}{047205}.





\bibitem[{\citenamefont{Bramwell \emph{et~al.}}(2004)}]
{Bramwell:2004}
\bibinfo{author}{\bibnamefont{Bramwell}, \bibfnamefont{S. T.}},
 \bibinfo{author}{\bibfnamefont{M.}~\bibnamefont{Shirai}}, and
\bibinfo{author}{\bibfnamefont{C.}~\bibnamefont{Ritter}},
\bibinfo{year}{2004},
\emph{\bibinfo{booktitle}{Institut Laue-Langevin Experimental Reports}},
\bibinfo{pages}{5-31-1496}.



\bibitem[{\citenamefont{Bombardi} \emph{et~al.}(2004)}]
{Bombardi:2004}
  \bibinfo{author}{\bibfnamefont{Bombardi}, \bibnamefont{A.}},
  \bibinfo{author}{\bibnamefont{J.}~\bibfnamefont{Rodriguez-Carvajal}},
  \bibinfo{author}{\bibnamefont{S.}~\bibfnamefont{Di Matteo}},
  \bibinfo{author}{\bibnamefont{F.}~\bibfnamefont{de Bergevin}},
  \bibinfo{author}{\bibnamefont{L.}~\bibfnamefont{Paolasini}},
  \bibinfo{author}{\bibnamefont{P.}~\bibfnamefont{Carretta}},
  \bibinfo{author}{\bibnamefont{P.}~\bibfnamefont{Millet}},  and
  \bibinfo{author}{\bibnamefont{R.}~\bibfnamefont{Caciuffo}},
  \bibinfo{year}{2004},
  \bibinfo{journal}{J. Phys. Soc. Jpn.} \textbf{\bibinfo{volume}{93}},
  \bibinfo{pages}{027202}.




\bibitem[{\citenamefont{Brixner}(1964)}]
{Brixner:1964}
\bibinfo{author}{\bibnamefont{Brixner}, \bibfnamefont{L. H.}},
\bibinfo{year}{1964},
\bibinfo{journal}{Inorg. Chem.} \textbf{\bibinfo{volume}{3}},
\bibinfo{pages}{1065}.


%
%\bibitem[{\citenamefont{Broholm} \emph{et~al.}(1990)}]
%{Broholm:1990}
%  \bibinfo{author}{\bibfnamefont{Broholm}, \bibnamefont{C.}},
%  \bibinfo{author}{\bibnamefont{G.}~\bibfnamefont{Aeppli}},
%  \bibinfo{author}{\bibnamefont{G. P.}~\bibfnamefont{Espinosa}},  and
%  \bibinfo{author}{\bibnamefont{A. S.}~\bibfnamefont{Cooper}},
%   \bibinfo{year}{1990},
%  \bibinfo{journal}{Phys. Rev. Lett.} \textbf{\bibinfo{volume}{65}},
%  \bibinfo{pages}{3173}.




\bibitem[{\citenamefont{Canals and Lacroix}(1998)}]
{Canals:1998}
\bibinfo{author}{\bibnamefont{Canals}, \bibfnamefont{B.}} and
\bibinfo{author}{\bibfnamefont{C.}~\bibnamefont{Lacroix}},
\bibinfo{year}{1998},
\bibinfo{journal}{Phys. Rev. Lett.} \textbf{\bibinfo{volume}{80}},
\bibinfo{pages}{2933}.


\bibitem[{\citenamefont{Canals and Lacroix}(2000)}]
{Canals:2000}
\bibinfo{author}{\bibnamefont{Canals}, \bibfnamefont{B.}} and
\bibinfo{author}{\bibfnamefont{C.}~\bibnamefont{Lacroix}},
\bibinfo{year}{2000},
\bibinfo{journal}{Phys. Rev. B.} \textbf{\bibinfo{volume}{61}},
\bibinfo{pages}{1149}.




\bibitem[{\citenamefont{Cannella  and  Mydosh}(1972)}]
{Cannella:1972}
\bibinfo{author}{\bibnamefont{Cannella}, \bibfnamefont{V.}} and
\bibinfo{author}{\bibfnamefont{J. A.}~\bibnamefont{Mydosh}},
\bibinfo{year}{1972},
\bibinfo{journal}{Phys. Rev. B} \textbf{\bibinfo{volume}{6}},
\bibinfo{pages}{4220}.




\bibitem[{\citenamefont{Cao \emph{et~al.}}(1995)}]
{Cao:1995}
\bibinfo{author}{\bibnamefont{Cao}, \bibfnamefont{N.}},
  \bibinfo{author}{\bibfnamefont{T.}~\bibnamefont{Timusk}},
  \bibinfo{author}{\bibfnamefont{N. P.}~\bibnamefont{Raju}},
  \bibinfo{author}{\bibfnamefont{J. E.}~\bibnamefont{Greedan}}, and
  \bibinfo{author}{\bibfnamefont{P.}~\bibnamefont{Gougeon}},
  \bibinfo{year}{1995},
  \bibinfo{journal}{J. Phys.: Condens.  Matter} \textbf{\bibinfo{volume}{7}},
  \bibinfo{pages}{2489}.



  \bibitem[{\citenamefont{Carcia \emph{et~al.}}(1982)}]
{Carcia:1982}
\bibinfo{author}{\bibnamefont{Carcia}, \bibfnamefont{P. F.}},
 \bibinfo{author}{\bibfnamefont{A.}~\bibnamefont{Ferretti}}, and
 \bibinfo{author}{\bibfnamefont{A.}~\bibnamefont{Suna}},
 \bibinfo{year}{1982},
 \bibinfo{journal}{J. Appl. Phys.} \textbf{\bibinfo{volume}{53}},
 \bibinfo{pages}{5282}.



 \bibitem[{\citenamefont{Cashion \emph{et~al.}}(1973)}]
 {Cashion:1973}
 \bibinfo{author}{\bibnamefont{Cashion}, \bibfnamefont{J. D.}},
 \bibinfo{author}{\bibfnamefont{D. B.}~\bibnamefont{Prowse}}, and
 \bibinfo{author}{\bibfnamefont{A.}~\bibnamefont{Vas}},
 \bibinfo{year}{1973},
 \bibinfo{journal}{J. Phys. C.} \textbf{\bibinfo{volume}{6}},
 \bibinfo{pages}{2611}.


%
% \bibitem[{\citenamefont{Castellan \emph{et~al.}}(2002)}]
%{Castellan:2002}
%\bibinfo{author}{\bibnamefont{Castellan}, \bibfnamefont{J. P.}},
%  \bibinfo{author}{\bibfnamefont{B. D. }~\bibnamefont{Gaulin}},
%  \bibinfo{author}{\bibfnamefont{J. }~\bibnamefont{van Duijn}},
%  \bibinfo{author}{\bibfnamefont{M. J.}~\bibnamefont{Lewis}},
%  \bibinfo{author}{\bibfnamefont{M. D.}~\bibnamefont{Lumsden}},
%  \bibinfo{author}{\bibfnamefont{R.}~\bibnamefont{Jin}},
%  \bibinfo{author}{\bibfnamefont{J.}~\bibnamefont{He}},
%  \bibinfo{author}{\bibfnamefont{S. E.}~\bibnamefont{Nagler}}, and
%  \bibinfo{author}{\bibfnamefont{D.}~\bibnamefont{Mandrus}},
%  \bibinfo{year}{2002},
%  \bibinfo{journal}{Phys. Rev. B} \textbf{\bibinfo{volume}{66}},
%  \bibinfo{pages}{134528}.



\bibitem[{\citenamefont{Castelnovo} \emph{et~al.}(2007)}]
{Castelnovo:2007}
  \bibinfo{author}{\bibfnamefont{Castelnovo}, \bibnamefont{C.}},
  \bibinfo{author}{\bibnamefont{R.}~\bibfnamefont{Moessner}},   and
  \bibinfo{author}{\bibnamefont{S. L.}~\bibfnamefont{Sondhi}},
 \bibinfo{year}{2008},
  \bibinfo{journal}{Nature} \textbf{\bibinfo{volume}{451}},
  \bibinfo{pages}{42}.



\bibitem[{\citenamefont{C\'epas and Shastry}(2004)}]
{Cepas:2004}
  \bibinfo{author}{\bibfnamefont{C\'epas}, \bibnamefont{O.}}, and
  \bibinfo{author}{\bibnamefont{B. S.}~\bibfnamefont{Shastry}},
 \bibinfo{year}{2004},
  \bibinfo{journal}{Phys. Rev. B} \textbf{\bibinfo{volume}{69}},
  \bibinfo{pages}{184402}.



 \bibitem[{\citenamefont{C\'epas} \emph{et~al.}(2005)}]
{Cepas:2005}
  \bibinfo{author}{\bibfnamefont{C\'epas}, \bibnamefont{O.}},
  \bibinfo{author}{\bibnamefont{A. P.}~\bibfnamefont{Young}}, and
  \bibinfo{author}{\bibnamefont{B. S.}~\bibfnamefont{Shastry}},
 \bibinfo{year}{2005},
  \bibinfo{journal}{Phys. Rev. B} \textbf{\bibinfo{volume}{72}},
  \bibinfo{pages}{184408}.



\bibitem[{\citenamefont{Chalker} \emph{et~al.}(1992)}]
{Chalker:1992}
  \bibinfo{author}{\bibfnamefont{Chalker}, \bibnamefont{J. T.}},
  \bibinfo{author}{\bibnamefont{P. C. W.}~\bibfnamefont{Holdsworth}},  and
  \bibinfo{author}{\bibnamefont{E. F.}~\bibfnamefont{Shender}},
 \bibinfo{year}{1992},
  \bibinfo{journal}{Phys. Rev. Lett.} \textbf{\bibinfo{volume}{68}},
   \bibinfo{pages}{855}.





\bibitem[{\citenamefont{Champion \emph{et~al.}}(2001)}]
{Champion:2001}
\bibinfo{author}{\bibnamefont{Champion}, \bibfnamefont{J. D. M.}},
  \bibinfo{author}{\bibfnamefont{A. S.}~\bibnamefont{Wills}},
  \bibinfo{author}{\bibfnamefont{T.}~\bibnamefont{Fennell}},
  \bibinfo{author}{\bibfnamefont{S. T.}~\bibnamefont{Bramwell}},
  \bibinfo{author}{\bibfnamefont{J. S.}~\bibnamefont{Gardner}}, and
  \bibinfo{author}{\bibfnamefont{M. A.}~\bibnamefont{Green}},
  \bibinfo{year}{2001},
  \bibinfo{journal}{Phys. Rev. B.} \textbf{\bibinfo{volume}{64}},
  \bibinfo{pages}{140407R}.



\bibitem[{\citenamefont{Champion \emph{et~al.}}(2003)}]
{Champion:2003}
\bibinfo{author}{\bibnamefont{Champion}, \bibfnamefont{J. D. M.}},
  \bibinfo{author}{\bibfnamefont{M. J.}~\bibnamefont{Harris}},
  \bibinfo{author}{\bibfnamefont{P. C. W.}~\bibnamefont{Holdsworth}},
  \bibinfo{author}{\bibfnamefont{A. S.}~\bibnamefont{Wills}},
  \bibinfo{author}{\bibfnamefont{G.}~\bibnamefont{Balakrishnan}},
  \bibinfo{author}{\bibfnamefont{S. T.}~\bibnamefont{Bramwell}},
  \bibinfo{author}{\bibfnamefont{E.}~\bibnamefont{{\v C}i$\check{\mbox{z}}$m{\'a}r}},
  \bibinfo{author}{\bibfnamefont{T.}~\bibnamefont{Fennell}},
  \bibinfo{author}{\bibfnamefont{J. S.}~\bibnamefont{Gardner}},
  \bibinfo{author}{\bibfnamefont{J.}~\bibnamefont{Lago}},
  \bibinfo{author}{\bibfnamefont{D. F.}~\bibnamefont{McMorrow}},
  \bibinfo{author}{\bibfnamefont{M.}~\bibnamefont{Orend{\'a}$\check{\mbox{c}}$}},
  \bibinfo{author}{\bibfnamefont{A.}~\bibnamefont{Orend{\'a}$\check{\mbox{c}}$ov{\'a}}},
  \bibinfo{author}{\bibfnamefont{D. M$^c$K.}~\bibnamefont{Paul}},
  \bibinfo{author}{\bibfnamefont{R. I.}~\bibnamefont{Smith}},
  \bibinfo{author}{\bibfnamefont{M. T. F.}~\bibnamefont{Telling}}, and
  \bibinfo{author}{\bibfnamefont{A.}~\bibnamefont{Wildes}},
  \bibinfo{year}{2003},
  \bibinfo{journal}{Phys. Rev. B.} \textbf{\bibinfo{volume}{68}},
  \bibinfo{pages}{020401}.



\bibitem[{\citenamefont{Champion and Holdsworth}(2004)}]
{Champion:2004}
\bibinfo{author}{\bibnamefont{Champion}, \bibfnamefont{J. D. M.}} and
\bibinfo{author}{\bibfnamefont{P. C. W.}~\bibnamefont{Holdsworth}},
\bibinfo{year}{2004},
\bibinfo{journal}{J. Phys.: Condens. Matter} \textbf{\bibinfo{volume}{16}},
\bibinfo{pages}{S665}.



%
%\bibitem[{\citenamefont{Chandra and Dou\c cot}(1988)}]
%{Chandra:1988}
%\bibinfo{author}{\bibnamefont{Chandra}, \bibfnamefont{P.}} and
%\bibinfo{author}{\bibfnamefont{B}~\bibnamefont{Dou\c cot}},
%\bibinfo{year}{1988},
%\bibinfo{journal}{Phys. Rev. B} \textbf{\bibinfo{volume}{38}},
%\bibinfo{pages}{9335}.




\bibitem[{\citenamefont{Chandra and Coleman}(1995)}]
{Chandra:1995}
\bibinfo{author}{\bibnamefont{Chandra}, \bibfnamefont{P.}} and
\bibinfo{author}{\bibnamefont{P.} \bibfnamefont{Coleman}},
  \bibinfo{year}{1995}, in
\emph{\bibinfo{booktitle}{
Strongly Interacting Fermions and High Temperature Superconductivity}},
  edited by \bibinfo{editor}{\bibfnamefont{B.}~\bibnamefont{Dou\c cot}} and
  \bibinfo{editor}{\bibfnamefont{J.}~\bibnamefont{Zinn-Justin}}
  (\bibinfo{publisher}{North-Holland}), p. \bibinfo{pages}{495}.




 \bibitem[{\citenamefont{Chapuis, \emph{et~al.}}(2007)}]
{Chapuis:2007}
\bibinfo{author}{\bibnamefont{Chapuis}, \bibfnamefont{Y.}},
  \bibinfo{author}{\bibfnamefont{A.}~\bibnamefont{Yaouanc}},
  \bibinfo{author}{\bibfnamefont{P.}~\bibnamefont{Dalmas de R\'{e}otier}},
  \bibinfo{author}{\bibfnamefont{S.}~\bibnamefont{Pouget}},
   \bibinfo{author}{\bibfnamefont{P.}~\bibnamefont{Fouquet}},
  \bibinfo{author}{\bibfnamefont{A.}~\bibnamefont{Cervellino}},  and
  \bibinfo{author}{\bibfnamefont{A.}~\bibnamefont{Forget}},
  \bibinfo{year}{2007},
  \bibinfo{journal}{J. Phys.: Condens. Matter} \textbf{\bibinfo{volume}{19}},
  \bibinfo{pages}{446206}.




\bibitem[{\citenamefont{Chen and Xu}(1998)}]
{Chen:1998}
\bibinfo{author}{\bibnamefont{Chen}, \bibfnamefont{D.}} and
\bibinfo{author}{\bibfnamefont{R.}~\bibnamefont{Xu}},
\bibinfo{year}{1998},
\bibinfo{journal}{Mat. Res. Bull.} \textbf{\bibinfo{volume}{33}},
\bibinfo{pages}{409}.




\bibitem[{\citenamefont{Cheong \emph{et~al.}}(1996)}]
{Cheong:1996}
\bibinfo{author}{\bibnamefont{Cheong}, \bibfnamefont{S.-W.}},
  \bibinfo{author}{\bibfnamefont{H. Y.}~\bibnamefont{Hwang}},
  \bibinfo{author}{\bibfnamefont{B.}~\bibnamefont{Batlogg}}, and
  \bibinfo{author}{\bibfnamefont{L. W.}~\bibnamefont{Rupp, Jr.}},
  \bibinfo{year}{1996},
  \bibinfo{journal}{Solid State Comm.} \textbf{\bibinfo{volume}{98}},
  \bibinfo{pages}{163}.



\bibitem[{\citenamefont{Coey}(1987)}]
{Coey:1987}
 \bibinfo{author}{\bibfnamefont{Coey}, \bibnamefont{J. M. D.}},
 \bibinfo{year}{1987},
  \bibinfo{journal}{Can. J. Phys. } \textbf{\bibinfo{volume}{65}},
  \bibinfo{pages}{1210}.




\bibitem[{\citenamefont{Cornelius and Gardner}(2001)}]
{Cornelius:2001}
  \bibinfo{author}{\bibfnamefont{Cornelius}, \bibnamefont{A. L.}}, and
  \bibinfo{author}{\bibnamefont{J. S.}~\bibfnamefont{Gardner}},
 \bibinfo{year}{2001},
  \bibinfo{journal}{Phys. Rev. B} \textbf{\bibinfo{volume}{64}},
  \bibinfo{pages}{060406}.



 \bibitem[{\citenamefont{Corruccini and White}(1993)}]
 {Corruccini:1993}
   \bibinfo{author}{\bibfnamefont{Corruccini}, \bibnamefont{L. R.}}, and
   \bibinfo{author}{\bibnamefont{S. J.}~\bibfnamefont{White}},
   \bibinfo{year}{1993},
   \bibinfo{journal}{Phys. Rev. B.} \textbf{\bibinfo{volume}{47}},
   \bibinfo{pages}{773}.


 \bibitem[{\citenamefont{Dalmas de R\'{e}otier, \emph{et~al.}}(2003)}]
{DalmasdeReotier:2003}
\bibinfo{author}{\bibnamefont{Dalmas de R\'{e}otier}, \bibfnamefont{P.}},
  \bibinfo{author}{\bibfnamefont{A.}~\bibnamefont{Yaouanc}},
    \bibinfo{author}{\bibfnamefont{P. C. M.}~\bibnamefont{Gubbens}},
   \bibinfo{author}{\bibfnamefont{C. T.}~\bibnamefont{Kaiser}},
  \bibinfo{author}{\bibfnamefont{C.}~\bibnamefont{Baines}}, and
  \bibinfo{author}{\bibfnamefont{P. J. C.}~\bibnamefont{King}},
  \bibinfo{year}{2003},
  \bibinfo{journal}{Phys. Rev. Lett.} \textbf{\bibinfo{volume}{91}},
  \bibinfo{pages}{167201}.


 \bibitem[{\citenamefont{Dalmas de R\'{e}otier, \emph{et~al.}}(2006)}]
{DalmasdeReotier:2006}
\bibinfo{author}{\bibnamefont{Dalmas de R\'{e}otier}, \bibfnamefont{P.}},
  \bibinfo{author}{\bibfnamefont{A.}~\bibnamefont{Yaouanc}},
  \bibinfo{author}{\bibfnamefont{L.}~\bibnamefont{Keller}},
  \bibinfo{author}{\bibfnamefont{A.}~\bibnamefont{Cervellino}},
  \bibinfo{author}{\bibfnamefont{B.}~\bibnamefont{Roessli}},
  \bibinfo{author}{\bibfnamefont{C.}~\bibnamefont{Baines}},
  \bibinfo{author}{\bibfnamefont{A.}~\bibnamefont{Forget}},
  \bibinfo{author}{\bibfnamefont{C.}~\bibnamefont{Vaju}},
  \bibinfo{author}{\bibfnamefont{P. C. M.}~\bibnamefont{Gubbens}},
  \bibinfo{author}{\bibfnamefont{A.}~\bibnamefont{Amato}}, and
  \bibinfo{author}{\bibfnamefont{P. J. C.}~\bibnamefont{King}},
  \bibinfo{year}{2006},
  \bibinfo{journal}{Phys. Rev. Lett.} \textbf{\bibinfo{volume}{96}},
  \bibinfo{pages}{127202}.




\bibitem[{\citenamefont{Del Maestro and Gingras}(2004)}]
{DelMaestro:2004}
  \bibinfo{author}{\bibfnamefont{Del Maestro}, \bibnamefont{A. G.}} and
  \bibinfo{author}{\bibnamefont{M. J. P.}~\bibfnamefont{Gingras}},
 \bibinfo{year}{2004},
  \bibinfo{journal}{J. Phys.: Condens. Matter} \textbf{\bibinfo{volume}{16}},
  \bibinfo{pages}{3339}.



\bibitem[{\citenamefont{Del Maestro and Gingras}(2007)}]
{DelMaestro:2007}
  \bibinfo{author}{\bibfnamefont{Del Maestro}, \bibnamefont{A. G.}} and
  \bibinfo{author}{\bibnamefont{M. J. P.}~\bibfnamefont{Gingras}},
 \bibinfo{year}{2007},
  \bibinfo{journal}{Phys. Rev. B} \textbf{\bibinfo{volume}{76}},
  \bibinfo{pages}{064418}.



\bibitem[{\citenamefont{den Hertog and Gingras}(2000)}]
{Hertog:2000}
  \bibinfo{author}{\bibnamefont{den Hertog}~\bibfnamefont{B. C.}} and
  \bibinfo{author}{\bibnamefont{M. J. P.}~\bibfnamefont{Gingras}},
 \bibinfo{year}{2000},
  \bibinfo{journal}{Phys. Rev. Lett. } \textbf{\bibinfo{volume}{84}},
  \bibinfo{pages}{3430}.


\bibitem[{\citenamefont{Diep}(1994)}]
{Diep:1994}
\bibinfo{author}{\bibnamefont{Diep}, \bibfnamefont{H. T., Ed.}},
  \bibinfo{year}{1994},
  \emph{\bibinfo{booktitle}{Magnetic Systems with Competing Interactions (Frustrated Spin Systems)}},
  (\bibinfo{publisher}{World Scientific, Singapore}).



\bibitem[{\citenamefont{Diep}(2004)}]
{Diep:2004}
\bibinfo{author}{\bibnamefont{Diep}, \bibfnamefont{H. T., Ed.}},
  \bibinfo{year}{2004},
  \emph{\bibinfo{booktitle}{Frustrated Spin Systems}},
  (\bibinfo{publisher}{World Scientific Publishing, Singapore}).




 \bibitem[{\citenamefont{Donohue \emph{et~al.}}(1965)}]
{Donohue:1965}
\bibinfo{author}{\bibnamefont{Donohue}, \bibfnamefont{P. C.}},
  \bibinfo{author}{\bibfnamefont{J. M.}~\bibnamefont{Longo}},
  \bibinfo{author}{\bibfnamefont{R. D.}~\bibnamefont{Rosenstein}}, and
  \bibinfo{author}{\bibfnamefont{L.}~\bibnamefont{Katz}},
  \bibinfo{year}{1965},
  \bibinfo{journal}{Inorg. Chem.} \textbf{\bibinfo{volume}{4}},
  \bibinfo{pages}{1152}.




\bibitem[{\citenamefont{Dunsiger \emph{et~al.}}(1996)}]
{Dunsiger:1996}
\bibinfo{author}{\bibnamefont{Dunsiger}, \bibfnamefont{S. R.}},
  \bibinfo{author}{\bibfnamefont{R. F.}~\bibnamefont{Kiefl}},
  \bibinfo{author}{\bibfnamefont{K. H.}~\bibnamefont{Chow}},
  \bibinfo{author}{\bibfnamefont{B. D.}~\bibnamefont{Gaulin}},
  \bibinfo{author}{\bibfnamefont{M. J. P.}~\bibnamefont{Gingras}},
  \bibinfo{author}{\bibfnamefont{J. E.}~\bibnamefont{Greedan}},
  \bibinfo{author}{\bibfnamefont{A.}~\bibnamefont{Keren}},
  \bibinfo{author}{\bibfnamefont{K.}~\bibnamefont{Kojima}},
  \bibinfo{author}{\bibfnamefont{G. M.}~\bibnamefont{Luke}},
  \bibinfo{author}{\bibfnamefont{W. A.}~\bibnamefont{MacFarlane}},
  \bibinfo{author}{\bibfnamefont{N. P.}~\bibnamefont{Raju}},
  \bibinfo{author}{\bibfnamefont{J. E.}~\bibnamefont{Sonier}},
  \bibinfo{author}{\bibfnamefont{Y.}~\bibnamefont{Uemura}}, and
  \bibinfo{author}{\bibfnamefont{W. D.}~\bibnamefont{Wu}},
  \bibinfo{year}{1996},
  \bibinfo{journal}{Phys. Rev. B} \textbf{\bibinfo{volume}{54}},
  \bibinfo{pages}{9019}.



  \bibitem[{\citenamefont{Dunsiger \emph{et~al.}}(1996a)}]
{Dunsiger:1996a}
\bibinfo{author}{\bibnamefont{Dunsiger}, \bibfnamefont{S. R.}},
  \bibinfo{author}{\bibfnamefont{R. F.}~\bibnamefont{Kiefl}},
  \bibinfo{author}{\bibfnamefont{K. H.}~\bibnamefont{Chow}},
  \bibinfo{author}{\bibfnamefont{B. D.}~\bibnamefont{Gaulin}},
  \bibinfo{author}{\bibfnamefont{M. J. P.}~\bibnamefont{Gingras}},
  \bibinfo{author}{\bibfnamefont{J. E.}~\bibnamefont{Greedan}},
  \bibinfo{author}{\bibfnamefont{A.}~\bibnamefont{Keren}},
  \bibinfo{author}{\bibfnamefont{K.}~\bibnamefont{Kojima}},
  \bibinfo{author}{\bibfnamefont{G. M.}~\bibnamefont{Luke}},
  \bibinfo{author}{\bibfnamefont{W. A.}~\bibnamefont{MacFarlane}},
  \bibinfo{author}{\bibfnamefont{N. P.}~\bibnamefont{Raju}},
  \bibinfo{author}{\bibfnamefont{J. E.}~\bibnamefont{Sonier}},
  \bibinfo{author}{\bibfnamefont{Y.}~\bibnamefont{Uemura}}, and
  \bibinfo{author}{\bibfnamefont{W. D.}~\bibnamefont{Wu}},
  \bibinfo{year}{1996a},
  \bibinfo{journal}{J. Appl. Phys.} \textbf{\bibinfo{volume}{79}},
  \bibinfo{pages}{6636}.



\bibitem[{\citenamefont{Dunsiger} \emph{et~al.}(2000)}]
{Dunsiger:2000}
  \bibinfo{author}{\bibnamefont{Dunsiger}~\bibfnamefont{S. R.}},
\bibinfo{author}{\bibnamefont{J. S. }~\bibfnamefont{Gardner}},
\bibinfo{author}{\bibnamefont{J. A. }~\bibfnamefont{Chakhalian}},
\bibinfo{author}{\bibnamefont{A. L. }~\bibfnamefont{Cornelius}},
\bibinfo{author}{\bibnamefont{J. }~\bibfnamefont{Jaime}},
\bibinfo{author}{\bibnamefont{R. F. }~\bibfnamefont{Kiefl}},
\bibinfo{author}{\bibnamefont{R. }~\bibfnamefont{Movshovich}},
\bibinfo{author}{\bibnamefont{W. A.}~\bibfnamefont{MacFarlane}},
\bibinfo{author}{\bibnamefont{R. I.}~\bibfnamefont{Miller}},
\bibinfo{author}{\bibnamefont{J. E.}~\bibfnamefont{Sonier}},  and
\bibinfo{author}{\bibnamefont{B. D.}~\bibfnamefont{Gaulin}},
   \bibinfo{year}{2000},
  \bibinfo{journal}{Phys. Rev. Lett.} \textbf{\bibinfo{volume}{85}},
  \bibinfo{pages}{3504}.




\bibitem[{\citenamefont{Dunsiger \emph{et~al.}}(2006)}]
{Dunsiger:2006}
\bibinfo{author}{\bibnamefont{Dunsiger}, \bibfnamefont{S. R.}},
  \bibinfo{author}{\bibfnamefont{R. F.}~\bibnamefont{Kiefl}},
  \bibinfo{author}{\bibfnamefont{J. A.}~\bibnamefont{Chakhalian}},
  \bibinfo{author}{\bibfnamefont{J. E.}~\bibnamefont{Greedan}},
  \bibinfo{author}{\bibfnamefont{W. A.}~\bibnamefont{MacFarlane}},
  \bibinfo{author}{\bibfnamefont{R. I.}~\bibnamefont{Miller}},
  \bibinfo{author}{\bibfnamefont{G. D.}~\bibnamefont{Morris}},
  \bibinfo{author}{\bibfnamefont{A. N.}~\bibnamefont{Price}},
  \bibinfo{author}{\bibfnamefont{N. P.}~\bibnamefont{Raju}}, and
  \bibinfo{author}{\bibfnamefont{J. E.}~\bibnamefont{Sonier}},
  \bibinfo{year}{2006},
  \bibinfo{journal}{Phys. Rev. B} \textbf{\bibinfo{volume}{73}},
  \bibinfo{pages}{172418}.




 \bibitem[{\citenamefont{Dupuis \emph{et~al.}}(2002)}]
{Dupuis:2002}
\bibinfo{author}{\bibnamefont{Dupuis}, \bibfnamefont{V.}},
  \bibinfo{author}{\bibfnamefont{E.}~\bibnamefont{Vincent}},
  \bibinfo{author}{\bibfnamefont{J.}~\bibnamefont{Hammann}},
  \bibinfo{author}{\bibfnamefont{J. E.}~\bibnamefont{Greedan}}, and
  \bibinfo{author}{\bibfnamefont{A. S.}~\bibnamefont{Wills}},
  \bibinfo{year}{2002},
  \bibinfo{journal}{J. Appl. Phys.} \textbf{\bibinfo{volume}{91}},
  \bibinfo{pages}{8384}.



\bibitem[{\citenamefont{Ehlers} \emph{et~al.}(2003)}]
{Ehlers:2003}
  \bibinfo{author}{\bibfnamefont{Ehlers}, \bibnamefont{G.}},
  \bibinfo{author}{\bibnamefont{A. L.}~\bibfnamefont{Cornelius}},
  \bibinfo{author}{\bibnamefont{M.}~\bibfnamefont{Orend\'ac}},
  \bibinfo{author}{\bibnamefont{M.}~\bibfnamefont{Kajnakov\'a}},
  \bibinfo{author}{\bibnamefont{T.}~\bibfnamefont{Fennell}},
  \bibinfo{author}{\bibnamefont{S. T.}~\bibfnamefont{Bramwell}}, and
  \bibinfo{author}{\bibnamefont{J. S.}~\bibfnamefont{Gardner}},
 \bibinfo{year}{2003},
  \bibinfo{journal}{J. Phys.: Condens. Matter} \textbf{\bibinfo{volume}{15}},
  \bibinfo{pages}{L9}.



\bibitem[{\citenamefont{Ehlers} \emph{et~al.}(2004)}]
{Ehlers:2004}
  \bibinfo{author}{\bibfnamefont{Ehlers}, \bibnamefont{G.}},
  \bibinfo{author}{\bibnamefont{A. L.}~\bibfnamefont{Cornelius}},
  \bibinfo{author}{\bibnamefont{T.}~\bibfnamefont{Fennell}},
  \bibinfo{author}{\bibnamefont{M.}~\bibfnamefont{Koza}},
  \bibinfo{author}{\bibnamefont{S. T.}~\bibfnamefont{Bramwell}}, and
  \bibinfo{author}{\bibnamefont{J. S.}~\bibfnamefont{Gardner}},
 \bibinfo{year}{2004},
  \bibinfo{journal}{J. Phys.: Condens. Matter} \textbf{\bibinfo{volume}{16}},
  \bibinfo{pages}{S635}.



\bibitem[{\citenamefont{Ehlers}(2006)}]
{Ehlers:2006}
\bibinfo{author}{\bibnamefont{Ehlers},~\bibfnamefont{G.}},
 \bibinfo{year}{2006},
 \bibinfo{journal}{J. Phys.: Condens. Matter} \textbf{\bibinfo{volume}{18}},
 \bibinfo{pages}{R231}.



\bibitem[{\citenamefont{Ehlers} \emph{et~al.}(2006a)}]
{Ehlers:2006a}
  \bibinfo{author}{\bibfnamefont{Ehlers}, \bibnamefont{G.}},
  \bibinfo{author}{\bibnamefont{J. S.}~\bibfnamefont{Gardner}},
  \bibinfo{author}{\bibnamefont{C. H.}~\bibfnamefont{Booth}},
  \bibinfo{author}{\bibnamefont{M.}~\bibfnamefont{Daniel}},
  \bibinfo{author}{\bibnamefont{K. C.}~\bibfnamefont{Kam}},
\bibinfo{author}{\bibnamefont{A. K.}~\bibfnamefont{Cheetham}},
  \bibinfo{author}{\bibnamefont{D.}~\bibfnamefont{Antonio}},
\bibinfo{author}{\bibnamefont{H. E.}~\bibfnamefont{Brooks}},
  \bibinfo{author}{\bibnamefont{A. L.}~\bibfnamefont{Cornelius}},
\bibinfo{author}{\bibnamefont{S. T.}~\bibfnamefont{Bramwell}},
  \bibinfo{author}{\bibnamefont{J.}~\bibfnamefont{Lago}},
  \bibinfo{author}{\bibnamefont{W.}~\bibfnamefont{Haussler}}, and
  \bibinfo{author}{\bibnamefont{N.}~\bibfnamefont{Rosov}},
 \bibinfo{year}{2006a},
  \bibinfo{journal}{Phys. Rev. B} \textbf{\bibinfo{volume}{73}},
  \bibinfo{pages}{174429}.



\bibitem[{\citenamefont{Elhajal} \emph{et~al.}(2005)}]
{Elhajal:2005}
  \bibinfo{author}{\bibfnamefont{Elhajal}, \bibnamefont{M.}},
\bibinfo{author}{\bibnamefont{B.}~\bibfnamefont{Canals}},
\bibinfo{author}{\bibnamefont{R.}~\bibfnamefont{Sunyer}}, and
 \bibinfo{author}{\bibnamefont{C.}~\bibfnamefont{Lacroix}},
 \bibinfo{year}{2005},
  \bibinfo{journal}{Phys. Rev. B} \textbf{\bibinfo{volume}{71}},
  \bibinfo{pages}{094420}.



\bibitem[{\citenamefont{Enjalran and Gingras}(2003)}]
{Enjalran:2003}
  \bibinfo{author}{\bibfnamefont{Enjalran}, \bibnamefont{M.}}  and
  \bibinfo{author}{\bibnamefont{M. J. P.}~\bibfnamefont{Gingras}},
  \bibinfo{year}{2003},
  \bibinfo{journal}{arXiv:cond-mat/0307152}.


\bibitem[{\citenamefont{Enjalran and Gingras}(2004)}]
{Enjalran:2004}
  \bibinfo{author}{\bibfnamefont{Enjalran}, \bibnamefont{M.}}  and
  \bibinfo{author}{\bibnamefont{M. J. P.}~\bibfnamefont{Gingras}},
 \bibinfo{year}{2004},
  \bibinfo{journal}{Phys. Rev. B} \textbf{\bibinfo{volume}{70}},
  \bibinfo{pages}{174426}.




  \bibitem[{\citenamefont{Ewing \emph{et~al.}}(2004)}]
{Erwing:2004}
\bibinfo{author}{\bibnamefont{Erwing}, \bibfnamefont{R. C.}},
  \bibinfo{author}{\bibfnamefont{W. J.}~\bibnamefont{Weber}}, and
  \bibinfo{author}{\bibfnamefont{J.}~\bibnamefont{Lian}},
  \bibinfo{year}{2004},
  \bibinfo{journal}{J. Appl. Phys.} \textbf{\bibinfo{volume}{95}},
  \bibinfo{pages}{5949}.



\bibitem[{\citenamefont{Fazekas  and Anderson}(1974)}]
{Fazekas:1974}
\bibinfo{author}{\bibnamefont{Fazekas}, \bibfnamefont{P.}} and
\bibinfo{author}{\bibfnamefont{P. W.}~\bibnamefont{Anderson}},
\bibinfo{year}{1974},
\bibinfo{journal}{Philos. Mag.} \textbf{\bibinfo{volume}{30}},
\bibinfo{pages}{432}.





\bibitem[{\citenamefont{Fennell} \emph{et~al.}(2004)}]
{Fennell:2004}
  \bibinfo{author}{\bibfnamefont{Fennell}, \bibnamefont{T.}},
  \bibinfo{author}{\bibfnamefont{O. A.} \bibnamefont{Petrenko}},
  \bibinfo{author}{\bibnamefont{B.}~\bibfnamefont{F\aa k}},
  \bibinfo{author}{\bibnamefont{S. T.}~\bibfnamefont{Bramwell}},
  \bibinfo{author}{\bibnamefont{M. }~\bibfnamefont{Enjalran}},
  \bibinfo{author}{\bibnamefont{T.}~\bibfnamefont{Yavors'kii}},
  \bibinfo{author}{\bibnamefont{M. J. P.}~\bibfnamefont{Gingras}},
  \bibinfo{author}{\bibnamefont{R. G.}~\bibfnamefont{Melko}},  and
  \bibinfo{author}{\bibnamefont{G.}~\bibfnamefont{Balakrishnan}},
  \bibinfo{year}{2004},
  \bibinfo{journal}{Phys. Rev. B} \textbf{\bibinfo{volume}{70}},
  \bibinfo{pages}{134408}.



\bibitem[{\citenamefont{Fennell \emph{et~al.}}(2005)}]
{Fennell:2005}
\bibinfo{author}{\bibnamefont{Fennell}, \bibfnamefont{T.}},
  \bibinfo{author}{\bibfnamefont{O. A.}~\bibnamefont{Petrenko}},
  \bibinfo{author}{\bibfnamefont{B.}~\bibnamefont{F\aa k}},
  \bibinfo{author}{\bibfnamefont{J. S.}~\bibnamefont{Gardner}},
  \bibinfo{author}{\bibfnamefont{S. T.}~\bibnamefont{Bramwell}}, and
  \bibinfo{author}{\bibfnamefont{B.}~\bibnamefont{Ouladdiaf}},
  \bibinfo{year}{2005},
  \bibinfo{journal}{Phys. Rev. B.} \textbf{\bibinfo{volume}{72}},
  \bibinfo{pages}{224411}.



  \bibitem[{\citenamefont{Fennell \emph{et~al.}}(2007)}]
{Fennell:2007}
\bibinfo{author}{\bibnamefont{Fennell}, \bibfnamefont{T.}},
  \bibinfo{author}{\bibfnamefont{S. T.}~\bibnamefont{Bramwell}},
  \bibinfo{author}{\bibfnamefont{D. F.}~\bibnamefont{McMorrow}},
  \bibinfo{author}{\bibfnamefont{P.}~\bibnamefont{Manuel}}, and
  \bibinfo{author}{\bibfnamefont{A. R.}~\bibnamefont{Wildes}},
  \bibinfo{year}{2007},
  \bibinfo{journal}{Nature Physics} \textbf{\bibinfo{volume}{3}},
  \bibinfo{pages}{566}.



\bibitem[{\citenamefont{Finn} \emph{et~al.}(1961)}]
{Finn:1961}
  \bibinfo{author}{\bibfnamefont{Finn}, \bibnamefont{C. P. B.}},
  \bibinfo{author}{\bibnamefont{R.}~\bibfnamefont{Orbach}}, and
  \bibinfo{author}{\bibnamefont{W. P.}~\bibfnamefont{Wolf}},
 \bibinfo{year}{1961},
  \bibinfo{journal}{Proc. Phys. Soc. } \textbf{\bibinfo{volume}{77}},
  \bibinfo{pages}{261}.



\bibitem[{\citenamefont{Flood}(1974)}]
{Flood:1974}
  \bibinfo{author}{\bibfnamefont{Flood}, \bibnamefont{D. J.}},
  \bibinfo{year}{1974},
  \bibinfo{journal}{J. Appl. Phys.} \textbf{\bibinfo{volume}{45}},
  \bibinfo{pages}{4041}.



 \bibitem[{\citenamefont{Fujinaka \emph{et~al.}}(1979)}]
{Fujinaka:1979}
\bibinfo{author}{\bibnamefont{Fujinaka}, \bibfnamefont{H.}},
  \bibinfo{author}{\bibfnamefont{N.}~\bibnamefont{Kinomura}},
  \bibinfo{author}{\bibfnamefont{M.}~\bibnamefont{Koizumi}},
  \bibinfo{author}{\bibfnamefont{Y.}~\bibnamefont{Miyamoto}}, and
  \bibinfo{author}{\bibfnamefont{S.}~\bibnamefont{Kume}},
  \bibinfo{year}{1979},
  \bibinfo{journal}{Mat. Res. Bull.} \textbf{\bibinfo{volume}{14}},
  \bibinfo{pages}{1133}.



\bibitem[{\citenamefont{Fukazawa and Maeno}(2002)}]
{Fukazawa:2002}
\bibinfo{author}{\bibnamefont{Fukazawa}, ~\bibfnamefont{H.}}, and
\bibinfo{author}{\bibfnamefont{Y.}~\bibnamefont{Maeno}},
 \bibinfo{year}{2002},
 \bibinfo{journal}{J.Phys. Soc. Jpn.} \textbf{\bibinfo{volume}{71}},
 \bibinfo{pages}{2578}.



\bibitem[{\citenamefont{Fukazawa} \emph{et~al.}(2002a)}]
{Fukazawa:2002a}
  \bibinfo{author}{\bibfnamefont{Fukazawa}, \bibnamefont{H.}},
  \bibinfo{author}{\bibnamefont{R. G.}~\bibfnamefont{Melko}},
  \bibinfo{author}{\bibnamefont{R.}~\bibfnamefont{Higashinaka}},
  \bibinfo{author}{\bibnamefont{Y.}~\bibfnamefont{Maeno}}, and
  \bibinfo{author}{\bibnamefont{M. J. P.}~\bibfnamefont{Gingras}},
 \bibinfo{year}{2002a},
  \bibinfo{journal}{Phys. Rev. B} \textbf{\bibinfo{volume}{65}},
  \bibinfo{pages}{054410}.



\bibitem[{\citenamefont{Gaertner}(1930)}]
{Gaertner:1930}
\bibinfo{author}{\bibnamefont{Gaertner},~\bibfnamefont{H. R.}},
 \bibinfo{year}{1930},
 \bibinfo{journal}{Neues Jb. Mineralog., Geol. Palaontol., Beilage-Bd. Abt.}
 \textbf{\bibinfo{volume}{A 61}},
 \bibinfo{pages}{1}.



\bibitem[{\citenamefont{Gardner \emph{et~al.}}(1998)}]
{Gardner:1998}
\bibinfo{author}{\bibnamefont{Gardner}, \bibfnamefont{J. S.}},
  \bibinfo{author}{\bibfnamefont{B. D.}~\bibnamefont{Gaulin}}, and
  \bibinfo{author}{\bibfnamefont{D. M.}~\bibnamefont{Paul}},
  \bibinfo{year}{1998},
  \bibinfo{journal}{J.  Crystal Growth} \textbf{\bibinfo{volume}{191}},
  \bibinfo{pages}{740}.



\bibitem[{\citenamefont{Gardner \emph{et~al.}}(1999)}]
{Gardner:1999}
\bibinfo{author}{\bibnamefont{Gardner}, \bibfnamefont{J. S.}},
  \bibinfo{author}{\bibfnamefont{S. R.}~\bibnamefont{Dunsiger}},
  \bibinfo{author}{\bibfnamefont{B. D.}~\bibnamefont{Gaulin}},
  \bibinfo{author}{\bibfnamefont{M. J. P.}~\bibnamefont{Gingras}},
  \bibinfo{author}{\bibfnamefont{J. E.}~\bibnamefont{Greedan}},
  \bibinfo{author}{\bibfnamefont{R. F.}~\bibnamefont{Kiefl}},
  \bibinfo{author}{\bibfnamefont{M. D.}~\bibnamefont{Lumsden}},
  \bibinfo{author}{\bibfnamefont{W. A.}~\bibnamefont{MacFarlane}},
  \bibinfo{author}{\bibfnamefont{N. P.}~\bibnamefont{Raju}},
  \bibinfo{author}{\bibfnamefont{J. E.}~\bibnamefont{Sonier}},
  \bibinfo{author}{\bibfnamefont{I.}~\bibnamefont{Swainson}}, and
  \bibinfo{author}{\bibfnamefont{Z.}~\bibnamefont{Tun}},
  \bibinfo{year}{1999},
  \bibinfo{journal}{Phys. Rev. Lett.} \textbf{\bibinfo{volume}{82}},
  \bibinfo{pages}{1012}.





\bibitem[{\citenamefont{Gardner \emph{et~al.}}(1999a)}]
{Gardner:1999a}
\bibinfo{author}{\bibnamefont{Gardner}, \bibfnamefont{J. S.}},
  \bibinfo{author}{\bibfnamefont{B. D.}~\bibnamefont{Gaulin}},
  \bibinfo{author}{\bibfnamefont{S.-H.}~\bibnamefont{Lee}},
  \bibinfo{author}{\bibfnamefont{C.}~\bibnamefont{Broholm}},
  \bibinfo{author}{\bibfnamefont{N. P.}~\bibnamefont{Raju}}, and
  \bibinfo{author}{\bibfnamefont{J. E.}~\bibnamefont{Greedan}},
  \bibinfo{year}{1999a},
  \bibinfo{journal}{Phys. Rev. Lett.} \textbf{\bibinfo{volume}{83}},
  \bibinfo{pages}{211}.



  \bibitem[{\citenamefont{Gardner \emph{et~al.}}(2001)}]
{Gardner:2001}
\bibinfo{author}{\bibnamefont{Gardner}, \bibfnamefont{J. S.}},
  \bibinfo{author}{\bibfnamefont{B. D.}~\bibnamefont{Gaulin}},
  \bibinfo{author}{\bibfnamefont{A. J.}~\bibnamefont{Berlinsky}},
  \bibinfo{author}{\bibfnamefont{P.}~\bibnamefont{Waldron}},
  \bibinfo{author}{\bibfnamefont{S. R.}~\bibnamefont{Dunsiger}},
  \bibinfo{author}{\bibfnamefont{N. P.}~\bibnamefont{Raju}}, and
  \bibinfo{author}{\bibfnamefont{J. E.}~\bibnamefont{Greedan}},
  \bibinfo{year}{2001},
  \bibinfo{journal}{Phys. Rev. B.} \textbf{\bibinfo{volume}{64}},
  \bibinfo{pages}{224416}.



  \bibitem[{\citenamefont{Gardner \emph{et~al.}}(2001a)}]
{Gardner:2001a}
  \bibinfo{author}{\bibnamefont{Gardner}, \bibfnamefont{J. S.}},
  \bibinfo{author}{\bibfnamefont{G.}~\bibnamefont{Ehlers}},
  \bibinfo{author}{\bibfnamefont{R. H.}~\bibnamefont{Heffner}}, and
  \bibinfo{author}{\bibfnamefont{F.} ~\bibnamefont{Mezei}},
  \bibinfo{year}{2001a},
  \bibinfo{journal}{J. Magn. Magn. Mater.} \textbf{\bibinfo{volume}{226-230}},
  \bibinfo{pages}{460}.



    \bibitem[{\citenamefont{Gardner \emph{et~al.}}(2003)}]
{Gardner:2003}
\bibinfo{author}{\bibnamefont{Gardner}, \bibfnamefont{J. S.}},
  \bibinfo{author}{\bibfnamefont{A.}~\bibnamefont{Keren}},
  \bibinfo{author}{\bibfnamefont{G.}~\bibnamefont{Ehlers}},
  \bibinfo{author}{\bibfnamefont{C.}~\bibnamefont{Stock}},
  \bibinfo{author}{\bibfnamefont{E. }~\bibnamefont{Segal}},
  \bibinfo{author}{\bibfnamefont{J. M.}~\bibnamefont{Roper}},
  \bibinfo{author}{\bibfnamefont{B.}~\bibnamefont{F\aa k}},
  \bibinfo{author}{\bibfnamefont{P. R.}~\bibnamefont{Hammar}},
  \bibinfo{author}{\bibfnamefont{M. B.}~\bibnamefont{Stone}},
  \bibinfo{author}{\bibfnamefont{D. H.}~\bibnamefont{Reich}}, and
  \bibinfo{author}{\bibfnamefont{B. D.}~\bibnamefont{Gaulin}},
  \bibinfo{year}{2003},
  \bibinfo{journal}{Phys. Rev. B.} \textbf{\bibinfo{volume}{68}},
  \bibinfo{pages}{180401}.



   \bibitem[{\citenamefont{Gardner \emph{et~al.}}(2004)}]
{Gardner:2004}
\bibinfo{author}{\bibnamefont{Gardner}, \bibfnamefont{J. S.}},
  \bibinfo{author}{\bibfnamefont{G.}~\bibnamefont{Ehlers}},
  \bibinfo{author}{\bibfnamefont{S. T.}~\bibnamefont{Bramwell}},  and
  \bibinfo{author}{\bibfnamefont{B. D.}~\bibnamefont{Gaulin}},
  \bibinfo{year}{2004},
  \bibinfo{journal}{J. Phys.: Condens: Matter} \textbf{\bibinfo{volume}{16}},
  \bibinfo{pages}{S643}.



   \bibitem[{\citenamefont{Gardner \emph{et~al.}}(2004a)}]
{Gardner:2004a}
\bibinfo{author}{\bibnamefont{Gardner}, \bibfnamefont{J. S.}},
  \bibinfo{author}{\bibfnamefont{G.}~\bibnamefont{Ehlers}},
  \bibinfo{author}{\bibfnamefont{N.}~\bibnamefont{Rosov}},
  \bibinfo{author}{\bibfnamefont{R. W.}~\bibnamefont{Erwin}},  and
  \bibinfo{author}{\bibfnamefont{C.}~\bibnamefont{Petrovic}},
  \bibinfo{year}{2004a},
  \bibinfo{journal}{Phys. Rev. B} \textbf{\bibinfo{volume}{70}},
  \bibinfo{pages}{180404(R)}.





     \bibitem[{\citenamefont{Gardner \emph{et~al.}}(2005)}]
{Gardner:2005}
\bibinfo{author}{\bibnamefont{Gardner}, \bibfnamefont{J. S.}},
  \bibinfo{author}{\bibfnamefont{A. L.}~\bibnamefont{Cornelius}},
    \bibinfo{author}{\bibfnamefont{L. J.}~\bibnamefont{Chang}},
  \bibinfo{author}{\bibfnamefont{M.}~\bibnamefont{Prager}},
  \bibinfo{author}{\bibfnamefont{Th.}~\bibnamefont{Br\"ukel}},  and
  \bibinfo{author}{\bibfnamefont{G.}~\bibnamefont{Ehlers}},
  \bibinfo{year}{2005},
  \bibinfo{journal}{J. Phys.: Condens: Matter} \textbf{\bibinfo{volume}{17}},
  \bibinfo{pages}{7089}.




  \bibitem[{\citenamefont{Gaulin \emph{et~al.}}(1992)}]
{Gaulin:1992}
\bibinfo{author}{\bibnamefont{Gaulin}, \bibfnamefont{B. D.}},
  \bibinfo{author}{\bibfnamefont{J. N.}~\bibnamefont{Reimers}},
  \bibinfo{author}{\bibfnamefont{T. E.}~\bibnamefont{Mason}},
  \bibinfo{author}{\bibfnamefont{J. E.}~\bibnamefont{Greedan}}, and
  \bibinfo{author}{\bibfnamefont{Z.}~\bibnamefont{Tun}},
  \bibinfo{year}{1992},
  \bibinfo{journal}{Phys. Rev. Lett.} \textbf{\bibinfo{volume}{69}},
  \bibinfo{pages}{3244}.



\bibitem[{\citenamefont{Gaulin}(1994)}]
{Gaulin:1994}
\bibinfo{author}{\bibnamefont{Gaulin},~\bibfnamefont{B. D.}},
 \bibinfo{year}{1994},
 \bibinfo{journal}{Hyperfine Interactions} \textbf{\bibinfo{volume}{85}},
 \bibinfo{pages}{159}.



\bibitem[{\citenamefont{Ghosh} \emph{et~al.}(2002)}]
{Ghosh:2002}
  \bibinfo{author}{\bibfnamefont{Ghosh}, \bibnamefont{S.}},
  \bibinfo{author}{\bibfnamefont{R.}, \bibnamefont{Parthasarathy}},
  \bibinfo{author}{\bibnamefont{T. F.}~\bibfnamefont{Rosenbaum}}, and
  \bibinfo{author}{\bibnamefont{G.}~\bibfnamefont{Aeppli}},
  \bibinfo{year}{2002},
  \bibinfo{journal}{Science } \textbf{\bibinfo{volume}{296}},
  \bibinfo{pages}{2195}.



 \bibitem[{\citenamefont{Gingras \emph{et~al.}}(1996)}]
{Gingras:1996}
\bibinfo{author}{\bibnamefont{Gingras}, \bibfnamefont{M. J. P.}},
  \bibinfo{author}{\bibfnamefont{C. V.}~\bibnamefont{Stager}},
  \bibinfo{author}{\bibfnamefont{B. D.}~\bibnamefont{Gaulin}},
  \bibinfo{author}{\bibfnamefont{N. P.}~\bibnamefont{Raju}}, and
  \bibinfo{author}{\bibfnamefont{J. E.} ~\bibnamefont{Greedan}},
  \bibinfo{year}{1996},
  \bibinfo{journal}{J. Appl. Phys.} \textbf{\bibinfo{volume}{79}},
  \bibinfo{pages}{6170}.


  \bibitem[{\citenamefont{Gingras \emph{et~al.}}(1997)}]
{Gingras:1997}
\bibinfo{author}{\bibnamefont{Gingras}, \bibfnamefont{M. J. P.}},
  \bibinfo{author}{\bibfnamefont{C. V.}~\bibnamefont{Stager}},
  \bibinfo{author}{\bibfnamefont{N. P.}~\bibnamefont{Raju}},
 \bibinfo{author}{\bibfnamefont{B. D.}~\bibnamefont{Gaulin}}, and
  \bibinfo{author}{\bibfnamefont{J. E.}~\bibnamefont{Greedan}},
  \bibinfo{year}{1997},
  \bibinfo{journal}{Phys. Rev. Lett.} \textbf{\bibinfo{volume}{78}},
  \bibinfo{pages}{947}.



   \bibitem[{\citenamefont{Gingras \emph{et~al.}}(2000)}]
{Gingras:2000}
\bibinfo{author}{\bibnamefont{Gingras}, \bibfnamefont{M. J. P.}},
  \bibinfo{author}{\bibfnamefont{B. C. den}~\bibnamefont{Hertog}},
  \bibinfo{author}{\bibfnamefont{M.}~\bibnamefont{Faucher}},
  \bibinfo{author}{\bibfnamefont{J. S.}~\bibnamefont{Gardner}},
  \bibinfo{author}{\bibfnamefont{S. R.}~\bibnamefont{Dunsiger}},
  \bibinfo{author}{\bibfnamefont{L. J.}~\bibnamefont{Chang}},
 \bibinfo{author}{\bibfnamefont{B. D.}~\bibnamefont{Gaulin}},
  \bibinfo{author}{\bibfnamefont{N. P.}~\bibnamefont{Raju}}, and
  \bibinfo{author}{\bibfnamefont{J. E.}~\bibnamefont{Greedan}},
  \bibinfo{year}{2000},
  \bibinfo{journal}{Phys. Rev. B.} \textbf{\bibinfo{volume}{68}},
  \bibinfo{pages}{6496}.




\bibitem[{\citenamefont{Gingras and den Hertog}(2001)}]
{Gingras:2001}
  \bibinfo{author}{\bibnamefont{Gingras}~\bibfnamefont{M. J. P.}},   and
  \bibinfo{author}{\bibnamefont{B. C. }~\bibfnamefont{den Hertog}},
   \bibinfo{year}{2001},
  \bibinfo{journal}{Can. J. Phys.} \textbf{\bibinfo{volume}{79}},
  \bibinfo{pages}{1339}.



   \bibitem[{\citenamefont{Greedan \emph{et~al.}}(1986)}]
{Greedan:1986}
\bibinfo{author}{\bibnamefont{Greedan}, \bibfnamefont{J. E.}},
  \bibinfo{author}{\bibfnamefont{M.}~\bibnamefont{Sato}},
  \bibinfo{author}{\bibfnamefont{X.}~\bibnamefont{Yan}}, and
  \bibinfo{author}{\bibfnamefont{F. S.}~\bibnamefont{Razavi}},
  \bibinfo{year}{1986},
  \bibinfo{journal}{Sol. State Comm.} \textbf{\bibinfo{volume}{59}},
  \bibinfo{pages}{895}.



  \bibitem[{\citenamefont{Greedan \emph{et~al.}}(1987)}]
{Greedan:1987}
\bibinfo{author}{\bibnamefont{Greedan}, \bibfnamefont{J. E.}},
  \bibinfo{author}{\bibfnamefont{M.}~\bibnamefont{Sato}},
  \bibinfo{author}{\bibfnamefont{N.}~\bibnamefont{Ali}}, and
  \bibinfo{author}{\bibfnamefont{W. R.}~\bibnamefont{Datars}},
  \bibinfo{year}{1987},
  \bibinfo{journal}{J. Solid State Chem.} \textbf{\bibinfo{volume}{68}},
  \bibinfo{pages}{300}.



 %TbMo

 \bibitem[{\citenamefont{Greedan \emph{et~al.}}(1990)}]
{Greedan:1990}
\bibinfo{author}{\bibnamefont{Greedan}, \bibfnamefont{J. E.}},
 \bibinfo{author}{\bibfnamefont{J. N.}~\bibnamefont{Reimers}},
  \bibinfo{author}{\bibfnamefont{S. L.}~\bibnamefont{Penny}}, and
  \bibinfo{author}{\bibfnamefont{C. V.}~\bibnamefont{Stager}},
  \bibinfo{year}{1990},
  \bibinfo{journal}{J. Appl. Phys.} \textbf{\bibinfo{volume}{67}},
  \bibinfo{pages}{5967}.



 %moly Nd, Tb and Y

  \bibitem[{\citenamefont{Greedan \emph{et~al.}}(1991)}]
{Greedan:1991}
\bibinfo{author}{\bibnamefont{Greedan}, \bibfnamefont{J. E.}},
  \bibinfo{author}{\bibfnamefont{J. N.}~\bibnamefont{Reimers}},
  \bibinfo{author}{\bibfnamefont{C. V.}~\bibnamefont{Stager}}, and
  \bibinfo{author}{\bibfnamefont{S. L.}~\bibnamefont{Penny}},
  \bibinfo{year}{1991},
  \bibinfo{journal}{Phys. Rev. B} \textbf{\bibinfo{volume}{43}},
  \bibinfo{pages}{5682}.



\bibitem[{\citenamefont{Greedan}(1992a)}]
{Greedan:1992a}
\bibinfo{author}{\bibnamefont{Greedan}, \bibfnamefont{J. E.}},
\bibinfo{year}{1992a}, in
\emph{\bibinfo{booktitle}{Landolt-Bornstein New Series}},
ed. \bibinfo{editor}{\bibfnamefont{H. P. J.}~\bibnamefont{Wijn}}
(\bibinfo{publisher}{Springer-Verlag Berlin Heidelberg.}),
\textbf{\bibinfo{volume}{27}},
\bibinfo{pages}{100}.



 %Y2Mn2O7

 \bibitem[{\citenamefont{Greedan \emph{et~al.}}(1992b)}]
{Greedan:1992b}
\bibinfo{author}{\bibnamefont{Greedan}, \bibfnamefont{J. E.}},
  \bibinfo{author}{\bibfnamefont{J.}~\bibnamefont{Avelar}} and
  \bibinfo{author}{\bibfnamefont{M. A.}~\bibnamefont{Subramanian}},
  \bibinfo{year}{1992b},
  \bibinfo{journal}{Sol. State Comm.} \textbf{\bibinfo{volume}{82}},
  \bibinfo{pages}{797}.




%Y2Mn2O7, Ho2Mn2O7, and Yb2Mn2O7: Bulk magnetism and magnetic microstructure

\bibitem[{\citenamefont{Greedan \emph{et~al.}}(1996)}]
{Greedan:1996}
\bibinfo{author}{\bibnamefont{Greedan}, \bibfnamefont{J. E.}},
  \bibinfo{author}{\bibfnamefont{N. P.}~\bibnamefont{Raju}},
  \bibinfo{author}{\bibfnamefont{A.}~\bibnamefont{Maignan}},
  \bibinfo{author}{\bibfnamefont{C.}~\bibnamefont{Simon}},
  \bibinfo{author}{\bibfnamefont{J. S.}~\bibnamefont{Pedersen}},
  \bibinfo{author}{\bibfnamefont{A. M.}~\bibnamefont{Niraimathi}},
  \bibinfo{author}{\bibfnamefont{E.}~\bibnamefont{Gmelin}} and
  \bibinfo{author}{\bibfnamefont{M. A.}~\bibnamefont{Subramanian}},
  \bibinfo{year}{1996},
  \bibinfo{journal}{Phys. Rev. B.} \textbf{\bibinfo{volume}{54}},
  \bibinfo{pages}{7189}.


 %only Sc2Mn2O7

 \bibitem[{\citenamefont{Greedan \emph{et~al.}}(1996a)}]
{Greedan:1996a}
\bibinfo{author}{\bibnamefont{Greedan}, \bibfnamefont{J. E.}},
  \bibinfo{author}{\bibfnamefont{N. P.}~\bibnamefont{Raju}}, and
  \bibinfo{author}{\bibfnamefont{M. A.}~\bibnamefont{Subramanian}},
  \bibinfo{year}{1996a},
  \bibinfo{journal}{Solid State Comm.} \textbf{\bibinfo{volume}{99}},
  \bibinfo{pages}{399}.


\bibitem[{\citenamefont{Greedan}(2001)}]
{Greedan:2001}
\bibinfo{author}{\bibnamefont{Greedan},~\bibfnamefont{J. E.}},
 \bibinfo{year}{2001},
 \bibinfo{journal}{J. Materials Chem.} \textbf{\bibinfo{volume}{11}},
 \bibinfo{pages}{37}.


\bibitem[{\citenamefont{Greedan}(2006)}]
{Greedan:2006}
\bibinfo{author}{\bibnamefont{Greedan},~\bibfnamefont{J. E.}},
 \bibinfo{year}{2006},
 \bibinfo{journal}{J.  Alloys and Compounds.} \textbf{\bibinfo{volume}{408}},
 \bibinfo{pages}{444}.



%  \bibitem[{\citenamefont{Grohol \emph{et~al.}}(2003)}]
%{Grohol:2003}
%\bibinfo{author}{\bibnamefont{Grohol}, \bibfnamefont{D.}},
%  \bibinfo{author}{\bibfnamefont{D. G.}~\bibnamefont{Nocera}}, and
%  \bibinfo{author}{\bibfnamefont{D.}~\bibnamefont{Papoutsakis}},
%  \bibinfo{year}{2003},
%  \bibinfo{journal}{Phys. Rev. B.} \textbf{\bibinfo{volume}{67}},
%  \bibinfo{pages}{064401}.



   \bibitem[{\citenamefont{Gurgul \emph{et~al.}}(2007)}]
{Gurgul:2007}
\bibinfo{author}{\bibnamefont{Gurgul}, \bibfnamefont{J.}},
 \bibinfo{author}{\bibfnamefont{M.}~\bibnamefont{Rams}},
 \bibinfo{author}{\bibfnamefont{\.{Z}.}~\bibnamefont{\'{S}wi\c{a}tkowska}},
 \bibinfo{author}{\bibfnamefont{R.}~\bibnamefont{Kmie\'{c}}}, and
 \bibinfo{author}{\bibfnamefont{K.}~\bibnamefont{Tomala}},
 \bibinfo{year}{2007},
 \bibinfo{journal}{Phys. Rev. B.} \textbf{\bibinfo{volume}{75}},
 \bibinfo{pages}{064426}.




%
%\bibitem[{\citenamefont{Hagemann} \emph{et~al.}(2001)}]
%{Hagemann:2001}
%  \bibinfo{author}{\bibfnamefont{Hagemann}, \bibnamefont{I. S.}},
%  \bibinfo{author}{\bibnamefont{Q.}~\bibfnamefont{Huang}},
%  \bibinfo{author}{\bibnamefont{X. P. A.}~\bibfnamefont{Gao}},
%  \bibinfo{author}{\bibnamefont{A. P.}~\bibfnamefont{Ramirez}},  and
%  \bibinfo{author}{\bibnamefont{R. J.}~\bibfnamefont{Cava}},
% \bibinfo{year}{2001},
%  \bibinfo{journal}{Phys. Rev. Lett.} \textbf{\bibinfo{volume}{86}},
%  \bibinfo{pages}{894}.


 \bibitem[{\citenamefont{Han \emph{et~al.}}(2004)}]
{Han:2004}
\bibinfo{author}{\bibnamefont{Han}, \bibfnamefont{S.-W.}},
  \bibinfo{author}{\bibfnamefont{J. S.}~\bibnamefont{Gardner}}, and
  \bibinfo{author}{\bibfnamefont{C. H.}~\bibnamefont{Booth}},
  \bibinfo{year}{2004},
  \bibinfo{journal}{Phys. Rev. B.} \textbf{\bibinfo{volume}{69}},
  \bibinfo{pages}{024416}.



 \bibitem[{\citenamefont{Haldane}(1983)}]
{Haldane:1983}
\bibinfo{author}{\bibnamefont{Haldane},~\bibfnamefont{F. D. M.}},
 \bibinfo{year}{1983},
 \bibinfo{journal}{Phys. Rev. Lett.} \textbf{\bibinfo{volume}{50}},
 \bibinfo{pages}{1153}.


 \bibitem[{\citenamefont{Hanasaki \emph{et~al.}}(2006)}]
{Hanasaki:2006}
\bibinfo{author}{\bibnamefont{Hanasaki}, \bibfnamefont{N.}},
  \bibinfo{author}{\bibfnamefont{M.}~\bibnamefont{Kinuhara}},
  \bibinfo{author}{\bibfnamefont{I.}~\bibnamefont{K\'ezm\'arki}},
  \bibinfo{author}{\bibfnamefont{S.}~\bibnamefont{Iguchi}},
  \bibinfo{author}{\bibfnamefont{S.}~\bibnamefont{Miyasaka}},
  \bibinfo{author}{\bibfnamefont{N.}~\bibnamefont{Takeshita}},
  \bibinfo{author}{\bibfnamefont{C.}~\bibnamefont{Terakura}},
  \bibinfo{author}{\bibfnamefont{H.}~\bibnamefont{Takagi}}, and
  \bibinfo{author}{\bibfnamefont{Y.}~\bibnamefont{Tokura}},
  \bibinfo{year}{2006},
  \bibinfo{journal}{Phys. Rev. Lett.} \textbf{\bibinfo{volume}{96}},
  \bibinfo{pages}{116403}.



 \bibitem[{\citenamefont{Hanasaki \emph{et~al.}}(2007)}]
{Hanasaki:2007}
\bibinfo{author}{\bibnamefont{Hanasaki}, \bibfnamefont{N.}},
  \bibinfo{author}{\bibfnamefont{K.}~\bibnamefont{Watanabe}},
  \bibinfo{author}{\bibfnamefont{T.}~\bibnamefont{Ohtsuka}},
  \bibinfo{author}{\bibfnamefont{I.}~\bibnamefont{Kezmarki}},
  \bibinfo{author}{\bibfnamefont{S.}~\bibnamefont{Iguchi}},
  \bibinfo{author}{\bibfnamefont{S.}~\bibnamefont{Miyasaka}}, and
  \bibinfo{author}{\bibfnamefont{Y.}~\bibnamefont{Tokura}},
  \bibinfo{year}{2007},
  \bibinfo{journal}{Phys. Rev. Lett.} \textbf{\bibinfo{volume}{99}},
  \bibinfo{pages}{086401}.


\bibitem[{\citenamefont{Hanawa \emph{et~al.}}(2001)}]
{Hanawa:2001}
\bibinfo{author}{\bibnamefont{Hanawa}, \bibfnamefont{M.}},
  \bibinfo{author}{\bibfnamefont{Y.}~\bibnamefont{Muraoka}},
  \bibinfo{author}{\bibfnamefont{T.}~\bibnamefont{Sakakibara}},
  \bibinfo{author}{\bibfnamefont{T.}~\bibnamefont{Yamaura}}, and
  \bibinfo{author}{\bibfnamefont{Z.}~\bibnamefont{Hiroi}},
  \bibinfo{year}{2001},
  \bibinfo{journal}{Phys. Rev. Lett.} \textbf{\bibinfo{volume}{87}},
  \bibinfo{pages}{187001}.



\bibitem[{\citenamefont{Harris} \emph{et~al.}(1991)}]
{Harris:1991}
  \bibinfo{author}{\bibfnamefont{Harris}, \bibnamefont{A. B.}},
  \bibinfo{author}{\bibnamefont{A. J.}~\bibfnamefont{Berlinsky}}, and
 \bibinfo{author}{\bibnamefont{C.}~\bibfnamefont{Bruder}},
 \bibinfo{year}{1991},
  \bibinfo{journal}{J. Appl. Phys.} \textbf{\bibinfo{volume}{69}},
  \bibinfo{pages}{5200}.


\bibitem[{\citenamefont{Harris \emph{et~al.}}(1994)}]
{Harris:1994}
\bibinfo{author}{\bibnamefont{Harris}, \bibfnamefont{M. J.}},
  \bibinfo{author}{\bibfnamefont{M. P.}~\bibnamefont{Zinkin}},
  \bibinfo{author}{\bibfnamefont{Z.}~\bibnamefont{Tun}},
  \bibinfo{author}{\bibfnamefont{B. M.}~\bibnamefont{Wanklyn}}, and
  \bibinfo{author}{\bibfnamefont{I. P.}~\bibnamefont{Swainson}},
  \bibinfo{year}{1994},
  \bibinfo{journal}{Phys. Rev. Lett.} \textbf{\bibinfo{volume}{73}},
  \bibinfo{pages}{189}.


  \bibitem[{\citenamefont{Harris} \emph{et~al.}(1997)}]
{Harris:1997}
  \bibinfo{author}{\bibfnamefont{Harris}, \bibnamefont{M. J.}},
  \bibinfo{author}{\bibnamefont{S. T.}~\bibfnamefont{Bramwell}},
  \bibinfo{author}{\bibnamefont{D. F.}~\bibfnamefont{McMorrow}},
  \bibinfo{author}{\bibnamefont{T.}~\bibfnamefont{Zeiske}},	and
  \bibinfo{author}{\bibnamefont{K. W.}~\bibfnamefont{Godfrey}},
 \bibinfo{year}{1997},
  \bibinfo{journal}{Phys. Rev. Lett.} \textbf{\bibinfo{volume}{79}},
  \bibinfo{pages}{2554}.



\bibitem[{\citenamefont{Harris} \emph{et~al.}(1998)}]
{Harris:1998}
  \bibinfo{author}{\bibfnamefont{Harris}, \bibnamefont{M. J.}},
  \bibinfo{author}{\bibnamefont{S. T.}~\bibfnamefont{Bramwell}},
  \bibinfo{author}{\bibnamefont{P. C. W.}~\bibfnamefont{Holdsworth}},  and
  \bibinfo{author}{\bibnamefont{J. D.}~\bibfnamefont{Champion}},
 \bibinfo{year}{1998},
  \bibinfo{journal}{Phys. Rev. Lett.} \textbf{\bibinfo{volume}{81}},
  \bibinfo{pages}{4496}.


 \bibitem[{\citenamefont{Harris} \emph{et~al.}(1998a)}]
{Harris:1998a}
  \bibinfo{author}{\bibfnamefont{Harris}, \bibnamefont{M. J.}},
  \bibinfo{author}{\bibnamefont{S. T.}~\bibfnamefont{Bramwell}},
  \bibinfo{author}{\bibnamefont{T.}~\bibfnamefont{Zeiske}},
  \bibinfo{author}{\bibnamefont{D. F.}~\bibfnamefont{McMorrow}}, and
  \bibinfo{author}{\bibnamefont{P. J. C.}~\bibfnamefont{King}},
 \bibinfo{year}{1998a},
  \bibinfo{journal}{J. Magn. Magn. Mater.} \textbf{\bibinfo{volume}{177}},
  \bibinfo{pages}{757}.



\bibitem[{\citenamefont{Harrison}(2004)}]
{Harrison:2004}
  \bibinfo{author}{\bibfnamefont{Harrison}, \bibnamefont{A.}},
  \bibinfo{year}{2004},
  \bibinfo{journal}{J. Phys.: Condens. Matter} \textbf{\bibinfo{volume}{16}},
  \bibinfo{pages}{S553}.


\bibitem[{\citenamefont{Hassan \emph{et~al.}}(2003)}]
{Hassan:2003}
\bibinfo{author}{\bibnamefont{Hassan}, \bibfnamefont{A. K.}},
 \bibinfo{author}{\bibfnamefont{L. P.}~\bibnamefont{Levy}},
 \bibinfo{author}{\bibfnamefont{C.}~\bibnamefont{Darie}}, and
 \bibinfo{author}{\bibfnamefont{P.}~\bibnamefont{Strobel}},
 \bibinfo{year}{2003},
 \bibinfo{journal}{Phys. Rev. B.} \textbf{\bibinfo{volume}{67}},
 \bibinfo{pages}{214432}.


\bibitem[{\citenamefont{He \emph{et~al.}}(2006)}]
{He:2006}
\bibinfo{author}{\bibnamefont{He}, \bibfnamefont{J.}},
 \bibinfo{author}{\bibfnamefont{R.}~\bibnamefont{Jin}},
 \bibinfo{author}{\bibfnamefont{B. C.}~\bibnamefont{Chakoumakos}},
 \bibinfo{author}{\bibfnamefont{J. S.}~\bibnamefont{Gardner}},
 \bibinfo{author}{\bibfnamefont{D.}~\bibnamefont{Mandrus}}, and
 \bibinfo{author}{\bibfnamefont{T. M.}~\bibnamefont{Tritt}},
 \bibinfo{year}{2007},
 \bibinfo{journal}{J. Electronic Materials} \textbf{\bibinfo{volume}{36}},
 \bibinfo{pages}{740}.
%

%\bibitem[{\citenamefont{Helton} \emph{et~al.}(2007)}]
%{Helton:2007}
%  \bibinfo{author}{\bibfnamefont{Helton}, \bibnamefont{J. S.}},
%  \bibinfo{author}{\bibnamefont{K.}~\bibfnamefont{Matan}},
%  \bibinfo{author}{\bibnamefont{M. P.}~\bibfnamefont{Shores}},
%  \bibinfo{author}{\bibnamefont{E. A.}~\bibfnamefont{Nytko}},
%  \bibinfo{author}{\bibnamefont{B. M.}~\bibfnamefont{Bartlett}},
%  \bibinfo{author}{\bibnamefont{Y.}~\bibfnamefont{Yoshida}},
%  \bibinfo{author}{\bibnamefont{Y.}~\bibfnamefont{Takano}},
%  \bibinfo{author}{\bibnamefont{A.}~\bibfnamefont{Suslov}},
%   \bibinfo{author}{\bibnamefont{Y.}~\bibfnamefont{Qiu}},
%  \bibinfo{author}{\bibnamefont{J.-H.}~\bibfnamefont{Chung}},
%  \bibinfo{author}{\bibnamefont{D. G.}~\bibfnamefont{Nocera}},  and
%  \bibinfo{author}{\bibnamefont{Y. S.}~\bibfnamefont{Lee}},
% \bibinfo{year}{2007},
%  \bibinfo{journal}{Phys. Rev. Lett.} \textbf{\bibinfo{volume}{98}},
%  \bibinfo{pages}{107204}.




%Specific heat of Dy2Ti2O7 in magnetic fields: comparison between single-crystalline and polycrystalline

\bibitem[{\citenamefont{Higashinaka} \emph{et~al.}(2002)}]
{Higashinaka:2002}
  \bibinfo{author}{\bibfnamefont{Higashinaka}, \bibnamefont{R.}},
  \bibinfo{author}{\bibnamefont{H.}~\bibfnamefont{Fukazawa}},
  \bibinfo{author}{\bibnamefont{D.}~\bibfnamefont{Yanagishima}},  and
  \bibinfo{author}{\bibnamefont{Y.}~\bibfnamefont{Maeno}},
 \bibinfo{year}{2002},
  \bibinfo{journal}{J. Phys. Chem. Solids} \textbf{\bibinfo{volume}{63}},
  \bibinfo{pages}{1043}.



%Anisotropic release of the residual zero-point entropy in the spin
% ice compound Dy2Ti2O7: Kagome ice behavior
\bibitem[{\citenamefont{Higashinaka} \emph{et~al.}(2003)}]
{Higashinaka:2003}
  \bibinfo{author}{\bibfnamefont{Higashinaka}, \bibnamefont{R.}},
  \bibinfo{author}{\bibnamefont{H.}~\bibfnamefont{Fukazawa}}, and
  \bibinfo{author}{\bibnamefont{Y.}~\bibfnamefont{Maeno}},
 \bibinfo{year}{2003},
  \bibinfo{journal}{Phys. Rev. B	} \textbf{\bibinfo{volume}{68}},
  \bibinfo{pages}{014415}.



%Specific heat of single crystal of spin ice compound Dy2Ti2O7
\bibitem[{\citenamefont{Higashinaka} \emph{et~al.}(2003a)}]
{Higashinaka:2003a}
  \bibinfo{author}{\bibfnamefont{Higashinaka}, \bibnamefont{R.}},
  \bibinfo{author}{\bibnamefont{H.}~\bibfnamefont{Fukazawa}}, and
  \bibinfo{author}{\bibnamefont{Y.}~\bibfnamefont{Maeno}},
 \bibinfo{year}{2003a},
  \bibinfo{journal}{Physica B} \textbf{\bibinfo{volume}{329}},
  \bibinfo{pages}{1040}.




 \bibitem[{\citenamefont{Higashinaka \emph{et~al.}}(2004)}]
{Higashinaka:2004}
\bibinfo{author}{\bibnamefont{Higashinaka}, \bibfnamefont{R.}},
 \bibinfo{author}{\bibfnamefont{H.}~\bibnamefont{Fukazawa}},
 \bibinfo{author}{\bibfnamefont{K.}~\bibnamefont{Deguchi}}, and
 \bibinfo{author}{\bibfnamefont{Y.}~\bibnamefont{Maeno}},
 \bibinfo{year}{2004},
 \bibinfo{journal}{J. Phys.: Condens. Matter} \textbf{\bibinfo{volume}{16}},
 \bibinfo{pages}{S679}.




%

%\bibitem[{\citenamefont{Higashinaka and Maeno}(2005)}]
%{Higashinaka:2005}
%  \bibinfo{author}{\bibfnamefont{Higashinaka}, \bibnamefont{R.}}, and
%  \bibinfo{author}{\bibnamefont{Y.}~\bibfnamefont{Maeno}},  and
%  \bibinfo{year}{2005},
%  \bibinfo{journal}{Phys. Rev. Lett.  } \textbf{\bibinfo{volume}{95}},
%%  \bibinfo{pages}{237208}.

%

%
%\bibitem[{\citenamefont{Hiroi} \emph{et~al.}(2001)}]
%{Hiroi:2001}
%  \bibinfo{author}{\bibfnamefont{Hiroi}, \bibnamefont{Z.}},
%  \bibinfo{author}{\bibnamefont{M.}~\bibfnamefont{Hanawa}},
%  \bibinfo{author}{\bibnamefont{N.}~\bibfnamefont{Kobayashi}},
%  \bibinfo{author}{\bibnamefont{M.}~\bibfnamefont{Nohara}},
%  \bibinfo{author}{\bibnamefont{H.}~\bibfnamefont{Takagi}},
%  \bibinfo{author}{\bibnamefont{Y.}~\bibfnamefont{Kato}},  and
%  \bibinfo{author}{\bibnamefont{M.}~\bibfnamefont{Takigawa}},
% \bibinfo{year}{2001},
%  \bibinfo{journal}{J. Phys. Soc. Jpn.} \textbf{\bibinfo{volume}{70}},
%  \bibinfo{pages}{3377}.


%
% \bibitem[{\citenamefont{Hiroi \emph{et~al.}}(2002)}]
%{Hiroi:2002}
%\bibinfo{author}{\bibnamefont{Hiroi }, \bibfnamefont{Z.}},
%  \bibinfo{author}{\bibfnamefont{J.-I.}~\bibnamefont{Yamaura}},
%  \bibinfo{author}{\bibfnamefont{Y.}~\bibnamefont{Muraoka}}, and
%  \bibinfo{author}{\bibfnamefont{M.}~\bibnamefont{Hanawa}},
%  \bibinfo{year}{2002},
%  \bibinfo{journal}{J. Phys. Soc. Japan} \textbf{\bibinfo{volume}{71}},
%  \bibinfo{pages}{1634}.



%Specific heat of kagome ice in the pyrochlore oxide Dy2Ti2O7

\bibitem[{\citenamefont{Hiroi} \emph{et~al.}(2003)}]
{Hiroi:2003}
  \bibinfo{author}{\bibfnamefont{Hiroi}, \bibnamefont{Z.}},
  \bibinfo{author}{\bibnamefont{K.}~\bibfnamefont{Matsuhira}},
  \bibinfo{author}{\bibnamefont{S.}~\bibfnamefont{Takagi}},
  \bibinfo{author}{\bibnamefont{T.}~\bibfnamefont{Tayama}}, and
  \bibinfo{author}{\bibnamefont{T.}~\bibfnamefont{Sakakibara}},
 \bibinfo{year}{2003},
  \bibinfo{journal}{J. Phys. Soc. Jpn. } \textbf{\bibinfo{volume}{72}},
  \bibinfo{pages}{411}.





\bibitem[{\citenamefont{Hiroi} \emph{et~al.}(2003a)}]
{Hiroi:2003a}
  \bibinfo{author}{\bibfnamefont{Hiroi}, \bibnamefont{Z.}},
  \bibinfo{author}{\bibnamefont{K.}~\bibfnamefont{Matsuhira}}, and
  \bibinfo{author}{\bibnamefont{M.}~\bibfnamefont{Ogata}},
 \bibinfo{year}{2003a},
  \bibinfo{journal}{J. Phys. Soc. Jpn.	} \textbf{\bibinfo{volume}{72}},
  \bibinfo{pages}{3045}.



\bibitem[{\citenamefont{Hizi and Henley}(2007)}]
{Hizi:2007}
  \bibinfo{author}{\bibfnamefont{Hizi}, \bibnamefont{U.}}, and
\bibinfo{author}{\bibnamefont{C. L.}~\bibfnamefont{Henley}},
 \bibinfo{year}{2007},
  \bibinfo{journal}{J. Phys.: Condens. Matter} \textbf{\bibinfo{volume}{19}},
  \bibinfo{pages}{145268}.



 \bibitem[{\citenamefont{Hodges \emph{et~al.}}(2001)}]
{Hodges:2001}
\bibinfo{author}{\bibnamefont{Hodges}, \bibfnamefont{J. A.}},
 \bibinfo{author}{\bibfnamefont{P.}~\bibnamefont{Bonville}},
 \bibinfo{author}{\bibfnamefont{A.}~\bibnamefont{Forget}},
 \bibinfo{author}{\bibfnamefont{M.}~\bibnamefont{Rams}},
  \bibinfo{author}{\bibfnamefont{K.}~\bibnamefont{Kr\'{o}las}}, and
 \bibinfo{author}{\bibfnamefont{G.}~\bibnamefont{Dhalenne}},
 \bibinfo{year}{2001},
 \bibinfo{journal}{J. Phys.: Condens. Matter} \textbf{\bibinfo{volume}{13}},
 \bibinfo{pages}{9301}.



 \bibitem[{\citenamefont{Hodges \emph{et~al.}}(2002)}]
{Hodges:2002}
\bibinfo{author}{\bibnamefont{Hodges}, \bibfnamefont{J. A.}},
 \bibinfo{author}{\bibfnamefont{P.}~\bibnamefont{Bonville}},
 \bibinfo{author}{\bibfnamefont{A.}~\bibnamefont{Forget}},
 \bibinfo{author}{\bibfnamefont{A.}~\bibnamefont{Yaouanc}},
 \bibinfo{author}{\bibfnamefont{P.}~\bibnamefont{Dalmas de R\'{e}otier}},
 \bibinfo{author}{\bibfnamefont{G.}~\bibnamefont{Andr\'{e}}},
 \bibinfo{author}{\bibfnamefont{M.}~\bibnamefont{Rams}},
 \bibinfo{author}{\bibfnamefont{K.}~\bibnamefont{Kr\'{o}las}},
 \bibinfo{author}{\bibfnamefont{C.}~\bibnamefont{Ritter}},
 \bibinfo{author}{\bibfnamefont{P. C. M.}~\bibnamefont{Gubbens}},
  \bibinfo{author}{\bibfnamefont{C. T.}~\bibnamefont{Kaiser}},
  \bibinfo{author}{\bibfnamefont{P. C.}~\bibnamefont{King}}, and
 \bibinfo{author}{\bibfnamefont{C.}~\bibnamefont{Baines}},
 \bibinfo{year}{2002},
 \bibinfo{journal}{Phys. Rev. Lett.} \textbf{\bibinfo{volume}{88}},
 \bibinfo{pages}{077204}.



\bibitem[{\citenamefont{Hodges \emph{et~al.}}(2003)}]
{Hodges:2003}
\bibinfo{author}{\bibnamefont{Hodges}, \bibfnamefont{J. A.}},
  \bibinfo{author}{\bibfnamefont{P.}~\bibnamefont{Bonville}},
  \bibinfo{author}{\bibfnamefont{A.}~\bibnamefont{Forget}},
    \bibinfo{author}{\bibfnamefont{J. P.}~\bibnamefont{Sanchez}},
  \bibinfo{author}{\bibfnamefont{P.}~\bibnamefont{Vulliet}},
  \bibinfo{author}{\bibfnamefont{M.}~\bibnamefont{Rams}}, and
  \bibinfo{author}{\bibfnamefont{K.}~\bibnamefont{Kr\'olas}},
  \bibinfo{year}{2003},
  \bibinfo{journal}{Eur. Phys. J. B} \textbf{\bibinfo{volume}{33}},
  \bibinfo{pages}{173}.


\bibitem[{\citenamefont{Hoekstra and Gallagher}(1968)}]
{Hoekstra:1968}
\bibinfo{author}{\bibnamefont{Hoekstra},~\bibfnamefont{H. R.}}, and
\bibinfo{author}{\bibfnamefont{F.}~\bibnamefont{Gallagher}},
 \bibinfo{year}{1968},
 \bibinfo{journal}{Inorg. Chem.} \textbf{\bibinfo{volume}{7}},
 \bibinfo{pages}{2553}.


%
%\bibitem[{\citenamefont{Hopkinson} \emph{et~al.}(2007)}]
%{Hopkinson:2007}
%  \bibinfo{author}{\bibfnamefont{Hopkinson}, \bibnamefont{J. M.}},
%  \bibinfo{author}{\bibnamefont{S. V.}~\bibfnamefont{Isakov}},
%  \bibinfo{author}{\bibnamefont{H.-Y.}~\bibfnamefont{Kee}},  and
%  \bibinfo{author}{\bibnamefont{Y.-B.}~\bibfnamefont{Kim}},
% \bibinfo{year}{2007},
%  \bibinfo{journal}{Phys. Rev. Lett.} \textbf{\bibinfo{volume}{99}},
%  \bibinfo{pages}{037201}.



\bibitem[{\citenamefont{Horowitz \emph{et~al.}}(1983)}]
{Horowitz:1983}
\bibinfo{author}{\bibnamefont{Horowitz}, \bibfnamefont{H. S.}},
\bibinfo{author}{\bibfnamefont{J. M.}~\bibnamefont{Longo}}, and
\bibinfo{author}{\bibfnamefont{H. H.}~\bibnamefont{Horowitz}},
\bibinfo{year}{1983},
\bibinfo{journal}{J. Electochem. Soc.} \textbf{\bibinfo{volume}{130}},
\bibinfo{pages}{1851}.



\bibitem[{\citenamefont{Houtappel}(1950)}]
{Houtappel:1950}
  \bibinfo{author}{\bibfnamefont{Houtappel}, \bibnamefont{R. M. F.}},
  \bibinfo{year}{1950},
  \bibinfo{journal}{Physica B} \textbf{\bibinfo{volume}{16}},
  \bibinfo{pages}{425}.



 \bibitem[{\citenamefont{Hubert}(1974)}]
{Hubert:1974}
\bibinfo{author}{\bibnamefont{Hubert},~\bibfnamefont{P. H.}},
 \bibinfo{year}{1974},
 \bibinfo{journal}{Bull. Chim. Soc. Fr.}
\textbf{\bibinfo{volume}{11}},
 \bibinfo{pages}{2385}.



\bibitem[{\citenamefont{Hubert}(1975)}]
{Hubert:1975}
\bibinfo{author}{\bibnamefont{Hubert},~\bibfnamefont{P. H.}},
 \bibinfo{year}{1975},
 \bibinfo{journal}{Bull. Chim. Soc. Fr.}
\textbf{\bibinfo{volume}{11-1}},
 \bibinfo{pages}{2463}.



\bibitem[{\citenamefont{H\"{u}fner}(1978)}]
{Hufner:1978}
\bibinfo{author}{\bibnamefont{H\"{u}fner}, \bibfnamefont{S.}},
  \bibinfo{year}{1978},
  \emph{\bibinfo{booktitle}{Optical Spectra of Transparent Rare Earth Compounds}},
  (\bibinfo{publisher}{Academic Press}),
 \bibinfo{pages}{38}.




\bibitem[{\citenamefont{Iikubo \emph{et~al.}}(2001)}]
{Iikubo:2001}
\bibinfo{author}{\bibnamefont{Iikubo}, \bibfnamefont{S.}},
 \bibinfo{author}{\bibfnamefont{S.}~\bibnamefont{Yoshii}},
 \bibinfo{author}{\bibfnamefont{T.}~\bibnamefont{Kageyama}},
 \bibinfo{author}{\bibfnamefont{K.}~\bibnamefont{Oda}},
 \bibinfo{author}{\bibfnamefont{Y.}~\bibnamefont{Kondo}},
 \bibinfo{author}{\bibfnamefont{K.}~\bibnamefont{Murata}}, and
 \bibinfo{author}{\bibfnamefont{M.}~\bibnamefont{Sato}},
 \bibinfo{year}{2001},
 \bibinfo{journal}{J. Phys. Soc. Jpn.} \textbf{\bibinfo{volume}{70}},
 \bibinfo{pages}{212}.



%\bibitem[{\citenamefont{Isakov} \emph{et~al.}(2004)}]
%{Isakov:2004}
% \bibinfo{author}{\bibfnamefont{Isakov}, \bibnamefont{S. V.}},
% \bibinfo{author}{\bibnamefont{K.}~\bibfnamefont{Gregor}},
%  \bibinfo{author}{\bibnamefont{R.}~\bibfnamefont{Moessner}},  and
%  \bibinfo{author}{\bibnamefont{S. L.}~\bibfnamefont{Sondhi}},
% \bibinfo{year}{2004},
%  \bibinfo{journal}{Phys. Rev. Lett.} \textbf{\bibinfo{volume}{93}},
%  \bibinfo{pages}{167204}.


\bibitem[{\citenamefont{Isakov} \emph{et~al.}(2005)}]
{Isakov:2005}
  \bibinfo{author}{\bibfnamefont{Isakov}, \bibnamefont{S. V.}},
  \bibinfo{author}{\bibnamefont{R.}~\bibfnamefont{Moessner}},  and
  \bibinfo{author}{\bibnamefont{S. L.}~\bibfnamefont{Sondhi}},
 \bibinfo{year}{2005},
  \bibinfo{journal}{Phys. Rev. Lett.} \textbf{\bibinfo{volume}{95}},
  \bibinfo{pages}{217201}.



 \bibitem[{\citenamefont{Ito \emph{et~al.}}(2001)}]
{Ito:2001}
\bibinfo{author}{\bibnamefont{Ito}, \bibfnamefont{M.}},
 \bibinfo{author}{\bibfnamefont{Y.}~\bibnamefont{Yasui}},
 \bibinfo{author}{\bibfnamefont{M.}~\bibnamefont{Kanada}},
 \bibinfo{author}{\bibfnamefont{H.}~\bibnamefont{Harashina}},
 \bibinfo{author}{\bibfnamefont{S.}~\bibnamefont{Yoshii}},
 \bibinfo{author}{\bibfnamefont{K.}~\bibnamefont{Murata}},
 \bibinfo{author}{\bibfnamefont{M.}~\bibnamefont{Sato}},
 \bibinfo{author}{\bibfnamefont{H.}~\bibnamefont{Okumura}}, and
 \bibinfo{author}{\bibfnamefont{K.}~\bibnamefont{Kakurai}},
 \bibinfo{year}{2001},
 \bibinfo{journal}{J. Phys. Chem. Solids} \textbf{\bibinfo{volume}{62}},
 \bibinfo{pages}{337}.



%Estimation of single ion anisotropy in pyrochlore Dy2Ti2O7, a
%geometrically frustrated system, using crystal field theory

\bibitem[{\citenamefont{Jana} \emph{et~al.}(2002)}]
{Jana:2002}
  \bibinfo{author}{\bibfnamefont{Jana}, \bibnamefont{Y. M.}},
  \bibinfo{author}{\bibnamefont{A.}~\bibfnamefont{Sengupta}}, and
  \bibinfo{author}{\bibnamefont{D.}~\bibfnamefont{Ghosh}},
 \bibinfo{year}{2002},
  \bibinfo{journal}{J. Magn. Magn. Materials} \textbf{\bibinfo{volume}{248}},
  \bibinfo{pages}{7}.




\bibitem[{\citenamefont{Jaubert}  \emph{et~al.}(2007)}]
{Jaubert:2007}
  \bibinfo{author}{\bibfnamefont{Jaubert}, \bibnamefont{L.}},
  \bibinfo{author}{\bibfnamefont{J. T.}, \bibnamefont{Chalker}},
  \bibinfo{author}{\bibfnamefont{P. C. W.}, \bibnamefont{Holdsworth}}, and
  \bibinfo{author}{\bibfnamefont{R.}, \bibnamefont{Moessner}},
\bibinfo{year}{2008},
 \bibinfo{journal}{Phys. Rev. Lett.} \textbf{\bibinfo{volume}{100}},
  \bibinfo{pages}{067207}.

 

 \bibitem[{\citenamefont{Jin \emph{et~al.}}(2001)}]
{Jin:2001}
\bibinfo{author}{\bibnamefont{Jin}, \bibfnamefont{R.}},
  \bibinfo{author}{\bibfnamefont{J.}~\bibnamefont{He}},
  \bibinfo{author}{\bibfnamefont{S.}~\bibnamefont{McCall}},
  \bibinfo{author}{\bibfnamefont{C. S.}~\bibnamefont{Alexander}},
  \bibinfo{author}{\bibfnamefont{F.}~\bibnamefont{Drymiotis}}, and
  \bibinfo{author}{\bibfnamefont{D.}~\bibnamefont{Mandrus}},
  \bibinfo{year}{2001},
  \bibinfo{journal}{Phys. Rev. B} \textbf{\bibinfo{volume}{64}},
  \bibinfo{pages}{180503}.





\bibitem[{\citenamefont{J\"onsson} \emph{et~al.}(2007)}]
{Jonsson:2007}
  \bibinfo{author}{\bibfnamefont{J\"onsson}, \bibnamefont{P. E.}},
  \bibinfo{author}{\bibnamefont{R.}~\bibfnamefont{Mathieu}},
  \bibinfo{author}{\bibnamefont{W.}~\bibfnamefont{Wernsdorfer}},
  \bibinfo{author}{\bibnamefont{A. M.}~\bibfnamefont{Tkachuk}}, and
  \bibinfo{author}{\bibnamefont{B.}~\bibfnamefont{Barbara}},
 \bibinfo{year}{2007},
  \bibinfo{journal}{Phys. Rev. Lett.} \textbf{\bibinfo{volume}{98}},
  \bibinfo{pages}{256403}.



\bibitem[{\citenamefont{Kadowaki \emph{et~al.}}(2002)}]
{Kadowaki:2002}
\bibinfo{author}{\bibnamefont{Kadowaki}, \bibfnamefont{H.}},
  \bibinfo{author}{\bibfnamefont{Y.}~\bibnamefont{Ishii}},
  \bibinfo{author}{\bibfnamefont{K.}~\bibnamefont{Matsuhira}}, and
  \bibinfo{author}{\bibfnamefont{Y.}~\bibnamefont{Hinatsu}},
  \bibinfo{year}{2002},
  \bibinfo{journal}{Phys. Rev. B} \textbf{\bibinfo{volume}{65}},
  \bibinfo{pages}{144421}.



\bibitem[{\citenamefont{Kageyama \emph{et~al.}}(2001)}]
{Kageyama:2001}
\bibinfo{author}{\bibnamefont{Kageyama}, \bibfnamefont{T.}},
  \bibinfo{author}{\bibfnamefont{S.}~\bibnamefont{Iikubo}},
  \bibinfo{author}{\bibfnamefont{S.}~\bibnamefont{Yoshii}},
  \bibinfo{author}{\bibfnamefont{Y.}~\bibnamefont{Kondo}},
  \bibinfo{author}{\bibfnamefont{M.}~\bibnamefont{Sato}}, and
  \bibinfo{author}{\bibfnamefont{Y.}~\bibnamefont{Iye}},
  \bibinfo{year}{2001},
  \bibinfo{journal}{J. Phys. Soc. Jpn.} \textbf{\bibinfo{volume}{70}},
  \bibinfo{pages}{3006}.




\bibitem[{\citenamefont{Kao} \emph{et~al.}(2003)}]
{Kao:2003}
  \bibinfo{author}{\bibfnamefont{Kao}, \bibnamefont{Y.-J.}},
  \bibinfo{author}{\bibfnamefont{M}, \bibnamefont{Enjalran}},
  \bibinfo{author}{\bibnamefont{A. G.}~\bibfnamefont{Del Maestro}},
  \bibinfo{author}{\bibnamefont{H. R.}~\bibfnamefont{Molavian}},  and
  \bibinfo{author}{\bibnamefont{M. J. P.}~\bibfnamefont{Gingras}},
  \bibinfo{year}{2003},
  \bibinfo{journal}{Phys. Rev. B} \textbf{\bibinfo{volume}{68}},
  \bibinfo{pages}{172407}.




\bibitem[{\citenamefont{Katsufuji \emph{et~al.}}(2000)}]
{Katsufuji:2000}
\bibinfo{author}{\bibnamefont{Katsufuji}, \bibfnamefont{T.}},
  \bibinfo{author}{\bibfnamefont{H. Y.}~\bibnamefont{Hwang}}, and
  \bibinfo{author}{\bibfnamefont{S.-W.}~\bibnamefont{Cheong}},
  \bibinfo{year}{2000},
  \bibinfo{journal}{Phys. Rev. Lett.} \textbf{\bibinfo{volume}{84}},
  \bibinfo{pages}{1998}.



\bibitem[{\citenamefont{Kawamura}(1988)}]
{Kawamura:1988}
\bibinfo{author}{\bibnamefont{Kawamura},~\bibfnamefont{H.}},
 \bibinfo{year}{1988},
 \bibinfo{journal}{Phys. Rev. B} \textbf{\bibinfo{volume}{38}},
 \bibinfo{pages}{4916}.




  %Nonmonotonic Zero-Point Entropy in Diluted Spin Ice

\bibitem[{\citenamefont{Ke} \emph{et~al.}(2007)}]
{Ke:2007}
  \bibinfo{author}{\bibfnamefont{Ke}, \bibnamefont{X.}},
  \bibinfo{author}{\bibfnamefont{R. S.}, \bibnamefont{Freitas}},
  \bibinfo{author}{\bibnamefont{B. G.}~\bibfnamefont{Ueland}},
  \bibinfo{author}{\bibnamefont{G. C.}~\bibfnamefont{Lau}},
  \bibinfo{author}{\bibnamefont{M. L.}~\bibfnamefont{Dahlberg}},
  \bibinfo{author}{\bibnamefont{R. J.}~\bibfnamefont{Cava}},
  \bibinfo{author}{\bibnamefont{R.}~\bibfnamefont{Moessner}}, and
  \bibinfo{author}{\bibnamefont{P.}~\bibfnamefont{Schiffer}},
  \bibinfo{year}{2007},
  \bibinfo{journal}{Phys. Rev. Lett.} \textbf{\bibinfo{volume}{99}},
  \bibinfo{pages}{137203}.




\bibitem[{\citenamefont{Keren and Gardner}(2001)}]
{Keren:2001}
\bibinfo{author}{\bibnamefont{Keren},~\bibfnamefont{A.}}, and
\bibinfo{author}{\bibfnamefont{J. S.}~\bibnamefont{Gardner}},
 \bibinfo{year}{2001},
 \bibinfo{journal}{Phys. Rev. Lett.} \textbf{\bibinfo{volume}{87}},
 \bibinfo{pages}{177201}.


\bibitem[{\citenamefont{Keren \emph{et~al.}}(2004)}]
{Keren:2004}
\bibinfo{author}{\bibnamefont{Keren}, \bibfnamefont{A.}},
  \bibinfo{author}{\bibfnamefont{J. S.}~\bibnamefont{Gardner}},
  \bibinfo{author}{\bibfnamefont{G.}~\bibnamefont{Ehlers}},
  \bibinfo{author}{\bibfnamefont{A.}~\bibnamefont{Fukaya}}
  \bibinfo{author}{\bibfnamefont{E.}~\bibnamefont{Segal}},  and
  \bibinfo{author}{\bibfnamefont{Y. J.}~\bibnamefont{Uemura}},
  \bibinfo{year}{2004},
  \bibinfo{journal}{Phys. Rev. Lett.} \textbf{\bibinfo{volume}{92}},
  \bibinfo{pages}{107204}.




\bibitem[{\citenamefont{K\'{e}zsm\'{a}rki \emph{et~al.}}(2004)}]
{Kezsmarki:2004}
\bibinfo{author}{\bibnamefont{K\'{e}zsm\'{a}rki}, \bibfnamefont{I.}},
  \bibinfo{author}{\bibnamefont{N.}, \bibfnamefont{Hanasaki}},
  \bibinfo{author}{\bibnamefont{D.}, \bibfnamefont{Hashimoto}},
  \bibinfo{author}{\bibnamefont{S.}, \bibfnamefont{Iguchi}},
  \bibinfo{author}{\bibnamefont{Y.}, \bibfnamefont{Taguchi}},
  \bibinfo{author}{\bibfnamefont{S.}~\bibnamefont{Miyasaka}}, and
  \bibinfo{author}{\bibfnamefont{Y.}~\bibnamefont{Tokura}},
 \bibinfo{year}{2004},
 \bibinfo{journal}{Phys. Rev. Lett.} \textbf{\bibinfo{volume}{93}},
 \bibinfo{pages}{266401}.


   \bibitem[{\citenamefont{K\'{e}zsm\'{a}rki \emph{et~al.}}(2005)}]
{Kezsmarki:2005}
\bibinfo{author}{\bibnamefont{K\'{e}zsm\'{a}rki}, \bibfnamefont{I.}},
  \bibinfo{author}{\bibnamefont{S.}, \bibfnamefont{Onoda}},
   \bibinfo{author}{\bibnamefont{Y.}, \bibfnamefont{Taguchi}},
  \bibinfo{author}{\bibnamefont{T.}, \bibfnamefont{Ogasawara}},
   \bibinfo{author}{\bibnamefont{M.}, \bibfnamefont{Matsubara}},
  \bibinfo{author}{\bibnamefont{S.}, \bibfnamefont{Iguchi}},
  \bibinfo{author}{\bibnamefont{N.}, \bibfnamefont{Hanasaki}},
  \bibinfo{author}{\bibfnamefont{N.}~\bibnamefont{Nagaosa}}, and
  \bibinfo{author}{\bibfnamefont{Y.}~\bibnamefont{Tokura}},
 \bibinfo{year}{2005},
 \bibinfo{journal}{Phys. Rev. B} \textbf{\bibinfo{volume}{72}},
 \bibinfo{pages}{094427}.



  \bibitem[{\citenamefont{K\'{e}zsm\'{a}rki \emph{et~al.}}(2006)}]
{Kezsmarki:2006}
\bibinfo{author}{\bibnamefont{K\'{e}zsm\'{a}rki}, \bibfnamefont{I.}},
  \bibinfo{author}{\bibfnamefont{N.}~\bibnamefont{Hanasaki}},
  \bibinfo{author}{\bibfnamefont{K.}~\bibnamefont{Watanabe}},
  \bibinfo{author}{\bibfnamefont{S.}~\bibnamefont{Iguchi}},
  \bibinfo{author}{\bibfnamefont{Y.}~\bibnamefont{Taguchi}},
  \bibinfo{author}{\bibfnamefont{S.}~\bibnamefont{Miyasaka}},and
  \bibinfo{author}{\bibfnamefont{Y.}~\bibnamefont{Tokura}},
  \bibinfo{year}{2006},
  \bibinfo{journal}{Phys. Rev. B} \textbf{\bibinfo{volume}{73}},
  \bibinfo{pages}{125122}.



\bibitem[{\citenamefont{Kido \emph{et~al.}}(1991)}]
{Kido:1991}
\bibinfo{author}{\bibnamefont{Kido}, \bibfnamefont{H.}},
  \bibinfo{author}{\bibfnamefont{S.}~\bibnamefont{Komarneni}},  and
  \bibinfo{author}{\bibfnamefont{R.}~\bibnamefont{Roy}},
  \bibinfo{year}{1991},
  \bibinfo{journal}{J. Am. Cer. Soc.} \textbf{\bibinfo{volume}{74}},
  \bibinfo{pages}{422}.



\bibitem[{\citenamefont{Kim \emph{et~al.}}(2005)}]
{Kim:2005}
\bibinfo{author}{\bibnamefont{Kim}, \bibfnamefont{H. C.}},
  \bibinfo{author}{\bibfnamefont{Y.}~\bibnamefont{Jo}},
  \bibinfo{author}{\bibfnamefont{J. G.}~\bibnamefont{Park}},
  \bibinfo{author}{\bibfnamefont{S. W.}~\bibnamefont{Cheong}},
  \bibinfo{author}{\bibfnamefont{M.}~\bibnamefont{Ulharz}},
   \bibinfo{author}{\bibfnamefont{C.}~\bibnamefont{Pfleiderer}},  and
  \bibinfo{author}{\bibfnamefont{H. V.}~\bibnamefont{Lohneysen}},
  \bibinfo{year}{2005},
  \bibinfo{journal}{Physica B} \textbf{\bibinfo{volume}{359}},
  \bibinfo{pages}{1246}.




\bibitem[{\citenamefont{Kmiec \emph{et~al.}}(2006)}]
{Kmiec:2006}
\bibinfo{author}{\bibnamefont{Kmie\'{c}}, \bibfnamefont{R.}},
  \bibinfo{author}{\bibfnamefont{\.{Z}.}~\bibnamefont{\'{S}wi\c{a}tkowska}},
  \bibinfo{author}{\bibfnamefont{J.}~\bibnamefont{Gurgul}},
  \bibinfo{author}{\bibfnamefont{M.}~\bibnamefont{Rams}},
  \bibinfo{author}{\bibfnamefont{A.}~\bibnamefont{Zarzycki}},  and
  \bibinfo{author}{\bibfnamefont{K.}~\bibnamefont{Tomala}},
  \bibinfo{year}{2006},
  \bibinfo{journal}{Phys. Rev. B} ~\textbf{\bibinfo{volume}{74}},
  \bibinfo{pages}{104425}.


%
%\bibitem[{\citenamefont{Kunes and Pickett}(2006)}]
%{Kunes:2006}
%\bibinfo{author}{\bibnamefont{Kunes},~\bibfnamefont{J.}}, and
%\bibinfo{author}{\bibfnamefont{W. E.}~\bibnamefont{Pickett}},
% \bibinfo{year}{2006},
% \bibinfo{journal}{Phys. Rev. B} ~\textbf{\bibinfo{volume}{74}},
% \bibinfo{pages}{094302}.




  \bibitem[{\citenamefont{Kwei \emph{et~al.}}(1997)}]
{Kwei:1997}
\bibinfo{author}{\bibnamefont{Kwei}, \bibfnamefont{G. H.}},
  \bibinfo{author}{\bibfnamefont{C. H.}~\bibnamefont{Booth}},
  \bibinfo{author}{\bibfnamefont{F.}~\bibnamefont{Bridges}}, and
  \bibinfo{author}{\bibfnamefont{M. A.}~\bibnamefont{Subramanian}},
  \bibinfo{year}{1997}  \bibinfo{journal}{Phys. Rev.B} \textbf{\bibinfo{volume}{55}},
  \bibinfo{pages}{R688}.




 \bibitem[{\citenamefont{Ladieu \emph{et~al.}}(2004)}]
{Ladieu:2004}
\bibinfo{author}{\bibnamefont{Ladieu}, \bibfnamefont{F.}},
  \bibinfo{author}{\bibfnamefont{F.}~\bibnamefont{Bert}},
  \bibinfo{author}{\bibfnamefont{V.}~\bibnamefont{Dupuis}},
  \bibinfo{author}{\bibfnamefont{E.}~\bibnamefont{Vincent}}, and
  \bibinfo{author}{\bibfnamefont{J.}~\bibnamefont{Hammann}},
  \bibinfo{year}{2004},
  \bibinfo{journal}{J. Phys. Condens. Matter} \textbf{\bibinfo{volume}{16}},
  \bibinfo{pages}{S735}.



\bibitem[{\citenamefont{Lago \emph{et~al.}}(2005)}]
{Lago:2005}
\bibinfo{author}{\bibnamefont{Lago}, \bibfnamefont{J.}},
 \bibinfo{author}{\bibfnamefont{T.}~\bibnamefont{Lancaster}},
 \bibinfo{author}{\bibfnamefont{S. J.}~\bibnamefont{Blundell}},
 \bibinfo{author}{\bibfnamefont{S. T.}~\bibnamefont{Bramwell}},
 \bibinfo{author}{\bibfnamefont{F. L.}~\bibnamefont{Pratt}},
 \bibinfo{author}{\bibfnamefont{M.}~\bibnamefont{Shirai}}, and
 \bibinfo{author}{\bibfnamefont{C.}~\bibnamefont{Baines}},
 \bibinfo{year}{2005},
 \bibinfo{journal}{J. Phys.: Condens. Matter} \textbf{\bibinfo{volume}{17}},
 \bibinfo{pages}{979}.




\bibitem[{\citenamefont{Lago} \emph{et~al.}(2007)}]
{Lago:2007}
  \bibinfo{author}{\bibfnamefont{Lago}, \bibnamefont{J.}},
  \bibinfo{author}{\bibnamefont{S. J.}~\bibfnamefont{Blundell}}, and
  \bibinfo{author}{\bibnamefont{C.}~\bibfnamefont{Baines}},
  \bibinfo{year}{2007},
  \bibinfo{journal}{J. Phys.: Condens. Matter  } \textbf{\bibinfo{volume}{19}},  
  \bibinfo{pages}{326210}.





\bibitem[{\citenamefont{L\"{a}uchli} \emph{et~al.}(2007)}]
{Lauchli:2007}
  \bibinfo{author}{\bibfnamefont{L\"{a}uchli}, \bibnamefont{A.}},
  \bibinfo{author}{\bibnamefont{S.}~\bibfnamefont{Dommanger}},
    \bibinfo{author}{\bibnamefont{B.}~\bibfnamefont{Normand}},  and
  \bibinfo{author}{\bibnamefont{F.}~\bibfnamefont{Mila}},
 \bibinfo{year}{2007},
  \bibinfo{journal}{Phys. Rev. B} \textbf{\bibinfo{volume}{76}},
  \bibinfo{pages}{144413}.


\bibitem[{\citenamefont{Lau \emph{et~al.}}(2006)}]
{Lau:2006}
\bibinfo{author}{\bibnamefont{Lau}, \bibfnamefont{G. C.}},
 \bibinfo{author}{\bibfnamefont{R. S.}~\bibnamefont{Freitas}},
 \bibinfo{author}{\bibfnamefont{B. G.}~\bibnamefont{Ueland}},
 \bibinfo{author}{\bibfnamefont{B. D.}~\bibnamefont{Muegge}},
 \bibinfo{author}{\bibfnamefont{E. L.}~\bibnamefont{Duncan}},
 \bibinfo{author}{\bibfnamefont{P.}~\bibnamefont{Schiffer}}, and
 \bibinfo{author}{\bibfnamefont{R. J.}~\bibnamefont{Cava}},
 \bibinfo{year}{2006},
 \bibinfo{journal}{Nature Physics} \textbf{\bibinfo{volume}{2}},
 \bibinfo{pages}{249}.
 
 
\bibitem[{\citenamefont{Lau \emph{et~al.}}(2006a)}]
{Lau:2006a}
\bibinfo{author}{\bibnamefont{Lau}, \bibfnamefont{G. C.}},
 \bibinfo{author}{\bibfnamefont{B. D.}~\bibnamefont{Muegge}},
 \bibinfo{author}{\bibfnamefont{T. M.}~\bibnamefont{McQueen}},
 \bibinfo{author}{\bibfnamefont{E. L.}~\bibnamefont{Duncan}}, and
 \bibinfo{author}{\bibfnamefont{R. J.}~\bibnamefont{Cava}},
 \bibinfo{year}{2006a},
 \bibinfo{journal}{J. Solid State Chem.} \textbf{\bibinfo{volume}{179}},
 \bibinfo{pages}{3126}.


\bibitem[{\citenamefont{Lau \emph{et~al.}}(2007)}]
{Lau:2007}
\bibinfo{author}{\bibnamefont{Lau}, \bibfnamefont{G. C.}},
 \bibinfo{author}{\bibfnamefont{R. S.}~\bibnamefont{Freitas}},
 \bibinfo{author}{\bibfnamefont{B. G.}~\bibnamefont{Ueland}},
 \bibinfo{author}{\bibfnamefont{M. L.}~\bibnamefont{Dahlberg}},
 \bibinfo{author}{\bibfnamefont{Q.}~\bibnamefont{Huang}},
  \bibinfo{author}{\bibfnamefont{H. W.}~\bibnamefont{Zandbergen}},
 \bibinfo{author}{\bibfnamefont{P.}~\bibnamefont{Schiffer}}, and
 \bibinfo{author}{\bibfnamefont{R. J.}~\bibnamefont{Cava}},
 \bibinfo{year}{2007},
 \bibinfo{journal}{Phys. Rev. B.} \textbf{\bibinfo{volume}{76}},
 \bibinfo{pages}{054430}.


  \bibitem[{\citenamefont{Lazarev and Shaplygin}(1978)}]
{Lazarev:1978}
\bibinfo{author}{\bibnamefont{Lazarev},~\bibfnamefont{V. B.}}, and
\bibinfo{author}{\bibfnamefont{V. B.}~\bibnamefont{Shaplygin}},
 \bibinfo{year}{1978},
 \bibinfo{journal}{Mat. Res. Bull.} \textbf{\bibinfo{volume}{13}},
 \bibinfo{pages}{229}.



\bibitem[{\citenamefont{Lecheminant} \emph{et~al.}(1997)}]
{Lecheminant:1997}
  \bibinfo{author}{\bibfnamefont{Lecheminant}, \bibnamefont{P.}},
\bibinfo{author}{\bibnamefont{P.}~\bibfnamefont{Bernu}},
 \bibinfo{author}{\bibnamefont{C.}~\bibfnamefont{Lhuillier}},
  \bibinfo{author}{\bibnamefont{L.}~\bibfnamefont{Pierre}}, and
  \bibinfo{author}{\bibnamefont{P.}~\bibfnamefont{Sindzingre}},
 \bibinfo{year}{1997},
  \bibinfo{journal}{Phys. Rev. B} \textbf{\bibinfo{volume}{56}},
  \bibinfo{pages}{2521}.


 \bibitem[{\citenamefont{Lee \emph{et~al.}}(2000)}]
{Lee:2000}
\bibinfo{author}{\bibnamefont{Lee}, \bibfnamefont{S. H.}},
  \bibinfo{author}{\bibfnamefont{C.}~\bibnamefont{Broholm}},
  \bibinfo{author}{\bibfnamefont{T. H.}~\bibnamefont{Kim}},
  \bibinfo{author}{\bibfnamefont{W.}~\bibnamefont{Ratcliff}}, and
  \bibinfo{author}{\bibfnamefont{S.-W.}~\bibnamefont{Cheong}},
  \bibinfo{year}{2000},
  \bibinfo{journal}{Phys. Rev. Lett.} \textbf{\bibinfo{volume}{84}},
  \bibinfo{pages}{3718}.




 \bibitem[{\citenamefont{Lee \emph{et~al.}}(2002)}]
{Lee:2002}
\bibinfo{author}{\bibnamefont{Lee}, \bibfnamefont{S. H.}},
  \bibinfo{author}{\bibfnamefont{C.}~\bibnamefont{Broholm}},
  \bibinfo{author}{\bibfnamefont{W.}~\bibnamefont{Ratcliff}},
  \bibinfo{author}{\bibfnamefont{G.}~\bibnamefont{Gasparovic}},
  \bibinfo{author}{\bibfnamefont{Q.}~\bibnamefont{Huang}},
   \bibinfo{author}{\bibfnamefont{T. H.}~\bibnamefont{Kim}}, and
  \bibinfo{author}{\bibfnamefont{S.-W.}~\bibnamefont{Cheong}},
  \bibinfo{year}{2002},
  \bibinfo{journal}{Nature} \textbf{\bibinfo{volume}{418}},
  \bibinfo{pages}{856}.



\bibitem[{\citenamefont{Lee and Young}(2003)}]
{Lee:SG2003}
  \bibinfo{author}{\bibfnamefont{Lee}, \bibnamefont{L. W.}} and
\bibinfo{author}{\bibnamefont{A. P.}~\bibfnamefont{Young}},
 \bibinfo{year}{2003},
  \bibinfo{journal}{Phys. Rev. Lett.} \textbf{\bibinfo{volume}{90}},
  \bibinfo{pages}{227203}.


\bibitem[{\citenamefont{Lee \emph{et~al.}}(2006)}]
{Lee:2006}
\bibinfo{author}{\bibnamefont{Lee}, \bibfnamefont{Seongsu}},
  \bibinfo{author}{\bibfnamefont{J.-G..}~\bibnamefont{Park}},
  \bibinfo{author}{\bibfnamefont{D. T.}~\bibnamefont{Adroja}},
  \bibinfo{author}{\bibfnamefont{D.}~\bibnamefont{Khomskii}},
  \bibinfo{author}{\bibfnamefont{S.}~\bibnamefont{Streltsov}},
  \bibinfo{author}{\bibfnamefont{K. A.}~\bibnamefont{McEwen}},
  \bibinfo{author}{\bibfnamefont{H.}~\bibnamefont{Sakai}},
  \bibinfo{author}{\bibfnamefont{K.}~\bibnamefont{Yoshimura}},
  \bibinfo{author}{\bibfnamefont{V. I.}~\bibnamefont{Anisimov}},
  \bibinfo{author}{\bibfnamefont{D.}~\bibnamefont{Mori}},
  \bibinfo{author}{\bibfnamefont{R.}~\bibnamefont{Kanno}}, and
  \bibinfo{author}{\bibfnamefont{R. I.}~\bibnamefont{Ibberson}},
  \bibinfo{year}{2006},
  \bibinfo{journal}{Nature Materials} \textbf{\bibinfo{volume}{5}},
  \bibinfo{pages}{471}.


\bibitem[{\citenamefont{Lee and Young}(2007)}]
{Lee:SG2007}
  \bibinfo{author}{\bibfnamefont{Lee}, \bibnamefont{L. W.}}, and
\bibinfo{author}{\bibnamefont{A. P.}~\bibfnamefont{Young}},
 \bibinfo{year}{2007},
  \bibinfo{journal}{Phys. Rev. B} \textbf{\bibinfo{volume}{76}},
  \bibinfo{pages}{024405}.





\bibitem[{\citenamefont{Lozano \emph{et~al.}}(2007)}]
{Lozano:2007}
\bibinfo{author}{\bibnamefont{Lozano}, \bibfnamefont{A. D.}},
  \bibinfo{author}{\bibfnamefont{J. E.}~\bibnamefont{Greedan}}, and
  \bibinfo{author}{\bibfnamefont{T.}~\bibnamefont{Proffen}},
  \bibinfo{year}{2007},
  \bibinfo{journal}{unpublished}.


%  \bibitem[{\citenamefont{Lu \emph{et~al.}}(2004)}]
%{Lu:2004}
%\bibinfo{author}{\bibnamefont{Lu}, \bibfnamefont{C.}},
%  \bibinfo{author}{\bibfnamefont{J.}~\bibnamefont{Zhang}},
%  \bibinfo{author}{\bibfnamefont{R.}~\bibnamefont{Jin}},
%  \bibinfo{author}{\bibfnamefont{H.}~\bibnamefont{Qu}},
%  \bibinfo{author}{\bibfnamefont{J.}~\bibnamefont{He}},
%  \bibinfo{author}{\bibfnamefont{D.}~\bibnamefont{Mandrus}},
%  \bibinfo{author}{\bibfnamefont{K.-D.}~\bibnamefont{Tsuei}},
%  \bibinfo{author}{\bibfnamefont{C.-T.}~\bibnamefont{Tzeng}},
%  \bibinfo{author}{\bibfnamefont{L.-C.}~\bibnamefont{Lin}}, and
%  \bibinfo{author}{\bibfnamefont{E. W.}~\bibnamefont{Plummer}},
%  \bibinfo{year}{2004},
%  \bibinfo{journal}{Phys. Rev. B} \textbf{\bibinfo{volume}{70}},
%  \bibinfo{pages}{092506}.


% \bibitem[{\citenamefont{Lumsden \emph{et~al.}}(2002)}]
%{Lumsden:2002}
%\bibinfo{author}{\bibnamefont{Lumsden}, \bibfnamefont{M. D.}},
%  \bibinfo{author}{\bibfnamefont{S. R.}~\bibnamefont{Dunsiger}},
%  \bibinfo{author}{\bibfnamefont{J. E.}~\bibnamefont{Sonier}},
%  \bibinfo{author}{\bibfnamefont{R. I.}~\bibnamefont{Miller}},
%  \bibinfo{author}{\bibfnamefont{H.}~\bibnamefont{Kiefl}},
%  \bibinfo{author}{\bibfnamefont{R.}~\bibnamefont{Jin}},
%  \bibinfo{author}{\bibfnamefont{J.}~\bibnamefont{He}},
%  \bibinfo{author}{\bibfnamefont{D.}~\bibnamefont{Mandrus}},
%  \bibinfo{author}{\bibfnamefont{S. T.}~\bibnamefont{Bramwell}},and
%  \bibinfo{author}{\bibfnamefont{J. S.}~\bibnamefont{Gardner}},
%  \bibinfo{year}{2002},
%  \bibinfo{journal}{Phys. Rev. Lett. } \textbf{\bibinfo{volume}{89}},
%  \bibinfo{pages}{147002}.





 \bibitem[{\citenamefont{Luo \emph{et~al.}}(2001)}]
{Luo:2001}
\bibinfo{author}{\bibnamefont{Luo}, \bibfnamefont{G.}},
  \bibinfo{author}{\bibfnamefont{S. T.}~\bibnamefont{Hess}}, and
  \bibinfo{author}{\bibfnamefont{L. R.}~\bibnamefont{Corruccini}},
  \bibinfo{year}{2001},
  \bibinfo{journal}{Phys. Lett. A} \textbf{\bibinfo{volume}{291}},
  \bibinfo{pages}{306}.



\bibitem[{\citenamefont{Lynn \emph{et~al.}}(1998)}]
{Lynn:1998}
\bibinfo{author}{\bibnamefont{Lynn}, \bibfnamefont{J. W.}},
  \bibinfo{author}{\bibfnamefont{L.}~\bibnamefont{Vasiliu-Doloc}}, and
  \bibinfo{author}{\bibfnamefont{M. A.}~\bibnamefont{Subramanian}},
  \bibinfo{year}{1998},
  \bibinfo{journal}{Phys. Rev. Lett.} \textbf{\bibinfo{volume}{80}},
  \bibinfo{pages}{4582}.


\bibitem[{\citenamefont{Machida \emph{et~al.}}(2005)}]
{Machida:2005}
\bibinfo{author}{\bibnamefont{Machida}, \bibfnamefont{Y.}},
  \bibinfo{author}{\bibfnamefont{S.}~\bibnamefont{Nakatsuji}},
  \bibinfo{author}{\bibfnamefont{H.}~\bibnamefont{Tonomura}},
  \bibinfo{author}{\bibfnamefont{T.}~\bibnamefont{Tayama}},
  \bibinfo{author}{\bibfnamefont{T.}~\bibnamefont{Sakakibara}},
  \bibinfo{author}{\bibfnamefont{J.}~\bibnamefont{van Duijn}},
  \bibinfo{author}{\bibfnamefont{C.}~\bibnamefont{Broholm}}, and
  \bibinfo{author}{\bibfnamefont{Y.}~\bibnamefont{Maeno}}, 
  \bibinfo{year}{2005},
  \bibinfo{journal}{J. Phys. Chem. Solids} \textbf{\bibinfo{volume}{66}},
  \bibinfo{pages}{1435}.


\bibitem[{\citenamefont{Machida \emph{et~al.}}(2007)}]
{Machida:2007}
\bibinfo{author}{\bibnamefont{Machida}, \bibfnamefont{Y.}},
  \bibinfo{author}{\bibfnamefont{S.}~\bibnamefont{Nakatsuji}},
  \bibinfo{author}{\bibfnamefont{Y.}~\bibnamefont{Maeno}},
  \bibinfo{author}{\bibfnamefont{T.}~\bibnamefont{Yamada}},
  \bibinfo{author}{\bibfnamefont{T.}~\bibnamefont{Tayama}}, and
  \bibinfo{author}{\bibfnamefont{T.}~\bibnamefont{Sakakibara}},
  \bibinfo{year}{2007},
  \bibinfo{journal}{J. Magn. Magn. Mater.} \textbf{\bibinfo{volume}{310}},
  \bibinfo{pages}{1079}.


%
%\bibitem[{\citenamefont{Magishi \emph{et~al.}}(2005)}]
%{Magishi:2005}
%\bibinfo{author}{\bibnamefont{Magishi}, \bibfnamefont{K.}},
% \bibinfo{author}{\bibfnamefont{J. L.}~\bibnamefont{Gavilano}},
%  \bibinfo{author}{\bibfnamefont{B.}~\bibnamefont{Pedrini}},
%  \bibinfo{author}{\bibfnamefont{J.}~\bibnamefont{Hindered}},
%  \bibinfo{author}{\bibfnamefont{M.}~\bibnamefont{Weller}},
%  \bibinfo{author}{\bibfnamefont{H. R.}~\bibnamefont{Ott}},
%  \bibinfo{author}{\bibfnamefont{H.}~\bibnamefont{Ohno}},
%  \bibinfo{author}{\bibfnamefont{S. M.}~\bibnamefont{Kazakov}},and
%  \bibinfo{author}{\bibfnamefont{J.}~\bibnamefont{Karpinski}},
%  \bibinfo{year}{2005},
%  \bibinfo{journal}{Phys. Rev. B } \textbf{\bibinfo{volume}{71}},
%  \bibinfo{pages}{024524}.



\bibitem[{\citenamefont{Mailhot and Plumer}(1993)}]
{Mailhot:1993}
  \bibinfo{author}{\bibfnamefont{Mailhot}, \bibnamefont{A.}} and
 \bibinfo{author}{\bibnamefont{M. L.}~\bibfnamefont{Plumer}},
 \bibinfo{year}{1993},
  \bibinfo{journal}{Phys. Rev. B} \textbf{\bibinfo{volume}{48}},
  \bibinfo{pages}{9881}.


 \bibitem[{\citenamefont{Mandiram and Gopalakrishnan}(1980)}]
{Mandiram:1980}
\bibinfo{author}{\bibnamefont{Mandiram},~\bibfnamefont{A.}} and
\bibinfo{author}{\bibfnamefont{A.}~\bibnamefont{Gopalakrishnan}},
 \bibinfo{year}{1980},
 \bibinfo{journal}{Indian J. Chem.} \textbf{\bibinfo{volume}{19A}},
 \bibinfo{pages}{1042}.


\bibitem[{\citenamefont{Mandrus \emph{et~al.}}(2001)}]
{Mandrus:2001}
\bibinfo{author}{\bibnamefont{Mandrus}, \bibfnamefont{D.}},
  \bibinfo{author}{\bibfnamefont{J. R..}~\bibnamefont{Thompson}},
  \bibinfo{author}{\bibfnamefont{R.}~\bibnamefont{Gaal}},
  \bibinfo{author}{\bibfnamefont{L.}~\bibnamefont{Forro}},
  \bibinfo{author}{\bibfnamefont{J. C.}~\bibnamefont{Bryan}},
  \bibinfo{author}{\bibfnamefont{B. C.}~\bibnamefont{Chakoumakos}},
 \bibinfo{author}{\bibfnamefont{L. M.}~\bibnamefont{Woods}},
  \bibinfo{author}{\bibfnamefont{B.C.}~\bibnamefont{Sales}},
  \bibinfo{author}{\bibfnamefont{R. S.}~\bibnamefont{Fishman}}, and
  \bibinfo{author}{\bibfnamefont{V.}~\bibnamefont{Keppens}},
  \bibinfo{year}{2001},
  \bibinfo{journal}{Phys. Rev. B} \textbf{\bibinfo{volume}{63}},
  \bibinfo{pages}{195104}.


\bibitem[{\citenamefont{Matsuda \emph{et~al.}}(2007)}]
{Matsuda:2007}
\bibinfo{author}{\bibnamefont{Matsuda}, \bibfnamefont{M.}},
  \bibinfo{author}{\bibfnamefont{H.}~\bibnamefont{Ueda}},
  \bibinfo{author}{\bibfnamefont{A.}~\bibnamefont{Kikkawa}},
  \bibinfo{author}{\bibfnamefont{Y.}~\bibnamefont{Tanaka}},
  \bibinfo{author}{\bibfnamefont{K.}~\bibnamefont{Katsumata}},
  \bibinfo{author}{\bibfnamefont{Y.}~\bibnamefont{Narumi}},
 \bibinfo{author}{\bibfnamefont{T.}~\bibnamefont{Inami}},
  \bibinfo{author}{\bibfnamefont{Y.}~\bibnamefont{Ueda}}, and
  \bibinfo{author}{\bibfnamefont{S.-H.}~\bibnamefont{Lee}},
  \bibinfo{year}{2007},
  \bibinfo{journal}{Nature Physics} \textbf{\bibinfo{volume}{3}},
  \bibinfo{pages}{397}.



 \bibitem[{\citenamefont{Matsuhira \emph{et~al.}}(2000)}]
{Matsuhira:2000}
\bibinfo{author}{\bibnamefont{Matsuhira}, \bibfnamefont{K.}},
  \bibinfo{author}{\bibfnamefont{Y.}~\bibnamefont{Hinatsu}},
  \bibinfo{author}{\bibfnamefont{K.}~\bibnamefont{Tenya}}, and
  \bibinfo{author}{\bibfnamefont{T.}~\bibnamefont{Sakakibara}},
  \bibinfo{year}{2000},
  \bibinfo{journal}{J. Phys.: Condens. Matter} \textbf{\bibinfo{volume}{12}},
  \bibinfo{pages}{L649}.


\bibitem[{\citenamefont{Matsuhira} \emph{et~al.}(2001)}]
{Matsuhira:2001}
  \bibinfo{author}{\bibfnamefont{Matsuhira}, \bibnamefont{K.}},
  \bibinfo{author}{\bibnamefont{Y.}~\bibfnamefont{Hinatsu}}, and
  \bibinfo{author}{\bibnamefont{T.}~\bibfnamefont{Sakakibara}},
  \bibinfo{year}{2001},
  \bibinfo{journal}{J. Phys.: Condens. Matter} \textbf{\bibinfo{volume}{13}},
  \bibinfo{pages}{L737}.



%stannates
 \bibitem[{\citenamefont{Matsuhira \emph{et~al.}}(2002)}]
{Matsuhira:2002}
\bibinfo{author}{\bibnamefont{Matsuhira}, \bibfnamefont{K.}},
  \bibinfo{author}{\bibfnamefont{Y.}~\bibnamefont{Hinatsu}},
  \bibinfo{author}{\bibfnamefont{K.}~\bibnamefont{Tenya}},
  \bibinfo{author}{\bibfnamefont{H.}~\bibnamefont{Amitsuka}}, and
  \bibinfo{author}{\bibfnamefont{T.}~\bibnamefont{Sakakibara}},
  \bibinfo{year}{2002},
  \bibinfo{journal}{J. Phys. Soc. Jpn.} \textbf{\bibinfo{volume}{71}},
  \bibinfo{pages}{1576}.



  %kagome ice
  
\bibitem[{\citenamefont{Matsuhira \emph{et~al.}}(2002a)}]
{Matsuhira:2002a}
\bibinfo{author}{\bibnamefont{Matsuhira}, \bibfnamefont{K.}},
  \bibinfo{author}{\bibfnamefont{Z.}~\bibnamefont{Hiroi}},
  \bibinfo{author}{\bibfnamefont{T.}~\bibnamefont{Tayama}},
  \bibinfo{author}{\bibfnamefont{S.}~\bibnamefont{Takagi}}, and
  \bibinfo{author}{\bibfnamefont{T.}~\bibnamefont{Sakakibara}},
  \bibinfo{year}{2002a},
  \bibinfo{journal}{J. Phys.: Condens. Matter} \textbf{\bibinfo{volume}{14}},
  \bibinfo{pages}{L559}.



  %Pr2Sn2O7
\bibitem[{\citenamefont{Matsuhira \emph{et~al.}}(2004)}]
{Matsuhira:2004}
\bibinfo{author}{\bibnamefont{Matsuhira}, \bibfnamefont{K.}},
 \bibinfo{author}{\bibfnamefont{C.}~\bibnamefont{Sekine}},
  \bibinfo{author}{\bibfnamefont{C.}~\bibnamefont{Paulsen}}, and
  \bibinfo{author}{\bibfnamefont{Y.}~\bibnamefont{Hinatsu}},
  \bibinfo{year}{2004},
  \bibinfo{journal}{J. Magn. Magn. Matter.} \textbf{\bibinfo{volume}{272}},
  \bibinfo{pages}{E981}.



\bibitem[{\citenamefont{Matsuhira \emph{et~al.}}(2007)}]
{Matsuhira:2007}
\bibinfo{author}{\bibnamefont{Matsuhira}, \bibfnamefont{K.}},
  \bibinfo{author}{\bibfnamefont{M.}~\bibnamefont{Wakeshima}},
  \bibinfo{author}{\bibfnamefont{R.}~\bibnamefont{Nakanishi}},
  \bibinfo{author}{\bibfnamefont{T.}~\bibnamefont{Yamada}},
  \bibinfo{author}{\bibfnamefont{W.}~\bibnamefont{Kawano}},
  \bibinfo{author}{\bibfnamefont{S.}~\bibnamefont{Tagaki}},and
  \bibinfo{author}{\bibfnamefont{Y.}~\bibnamefont{Hinatsu}},
  \bibinfo{year}{2007},
  \bibinfo{journal}{J. Phys. Soc. Jpn.} \textbf{\bibinfo{volume}{76}},
  \bibinfo{pages}{043706}.


\bibitem[{\citenamefont{McCarthy}(1971)}]
{McCarthy:1971}
\bibinfo{author}{\bibnamefont{McCarthy},~\bibfnamefont{G. J.}},
 \bibinfo{year}{1971},
 \bibinfo{journal}{Mat. Res. Bull.} \textbf{\bibinfo{volume}{6}},
 \bibinfo{pages}{31}.



\bibitem[{\citenamefont{Melko} \emph{et~al.}(2001)}]
{Melko:2001}
 \bibinfo{author}{\bibfnamefont{Melko}, \bibnamefont{R. G.}},
 \bibinfo{author}{\bibnamefont{B. C.}~\bibfnamefont{den Hertog}}, and
   \bibinfo{author}{\bibnamefont{M. J. P.}~\bibfnamefont{Gingras}},
\bibinfo{year}{2001},
  \bibinfo{journal}{Phys. Rev. Lett. } \textbf{\bibinfo{volume}{87}},
  \bibinfo{pages}{067203}.




\bibitem[{\citenamefont{Melko} \emph{et~al.}(2004)}]
{Melko:2004}
  \bibinfo{author}{\bibfnamefont{Melko}, \bibnamefont{R. G.}}  and
  \bibinfo{author}{\bibnamefont{M. J. P.}~\bibfnamefont{Gingras}},
 \bibinfo{year}{2004},
  \bibinfo{journal}{J. Phys.: Condens. Matter  } \textbf{\bibinfo{volume}{16}},
  \bibinfo{pages}{R1277}.


\bibitem[{\citenamefont{Mila}(2008)}]
{Mila:2008}
\bibinfo{author}{\bibnamefont{ Mila}, \bibfnamefont{F.}, \bibfnamefont {Ed.}}, 
  \bibinfo{year}{2008},
  \emph{\bibinfo{booktitle}{Highly Frustrated Magnetism}},
  (\bibinfo{publisher}{Springer, Berlin}).


\bibitem[{\citenamefont{Millican \emph{et~al.}}(2007)}]
{Millican:2007}
\bibinfo{author}{\bibnamefont{Millican}, \bibfnamefont{J. N.}},
  \bibinfo{author}{\bibfnamefont{R. T.}~\bibnamefont{Macaluso}},
  \bibinfo{author}{\bibfnamefont{S.}~\bibnamefont{Nakatsuji}},
  \bibinfo{author}{\bibfnamefont{Y.}~\bibnamefont{Machida}},
  \bibinfo{author}{\bibfnamefont{Y.}~\bibnamefont{Maeno}},  and
  \bibinfo{author}{\bibfnamefont{J. Y.}~\bibnamefont{Chan}},
  \bibinfo{year}{2007},
  \bibinfo{journal}{Mat. Res. Bull.} \textbf{\bibinfo{volume}{42}},
  \bibinfo{pages}{928}.



\bibitem[{\citenamefont{Minervini \emph{et~al.}}(2002)}]
{Minervini:2002}
\bibinfo{author}{\bibnamefont{Minervini}, \bibfnamefont{L.}},
  \bibinfo{author}{\bibfnamefont{R. W.}~\bibnamefont{Grimes}},
  \bibinfo{author}{\bibfnamefont{Y.}~\bibnamefont{Tabira}},
  \bibinfo{author}{\bibfnamefont{R. L.}~\bibnamefont{Withers}}, and
  \bibinfo{author}{\bibfnamefont{K. E.}~\bibnamefont{Sickafus}},
  \bibinfo{year}{2002},
  \bibinfo{journal}{Philos. Magazine A} \textbf{\bibinfo{volume}{82}},
  \bibinfo{pages}{123}.



\bibitem[{\citenamefont{Mirebeau \emph{et~al.}}(2002)}]
{Mirebeau:2002}
\bibinfo{author}{\bibnamefont{Mirebeau}, \bibfnamefont{I.}},
  \bibinfo{author}{\bibfnamefont{I. N.}~\bibnamefont{Goncharenko}},
  \bibinfo{author}{\bibfnamefont{P.}~\bibnamefont{Cadavez-Peres}},
  \bibinfo{author}{\bibfnamefont{S. T.}~\bibnamefont{Bramwell}},
  \bibinfo{author}{\bibfnamefont{M. J. P.}~\bibnamefont{Gingras}}, and
  \bibinfo{author}{\bibfnamefont{J. S.}~\bibnamefont{Gardner}},
  \bibinfo{year}{2002},
  \bibinfo{journal}{Nature} \textbf{\bibinfo{volume}{420}},
  \bibinfo{pages}{54}.



\bibitem[{\citenamefont{Mirebeau and Goncharenko}(2004)}]
{Mirebeau:2004}
\bibinfo{author}{\bibnamefont{Mirebeau}, \bibfnamefont{I.}} and
 \bibinfo{author}{\bibfnamefont{I. N.}~\bibnamefont{Goncharenko}},
 \bibinfo{year}{2004},
 \bibinfo{journal}{Physica B} \textbf{\bibinfo{volume}{350}},
 \bibinfo{pages}{250}.



\bibitem[{\citenamefont{Mirebeau \emph{et~al.}}(2004a)}]
{Mirebeau:2004a}
\bibinfo{author}{\bibnamefont{Mirebeau}, \bibfnamefont{I.}},
  \bibinfo{author}{\bibfnamefont{I. N.}~\bibnamefont{Goncharenko}},
  \bibinfo{author}{\bibfnamefont{D.}~\bibnamefont{Dhalenne}}, and
  \bibinfo{author}{\bibfnamefont{A.}~\bibnamefont{Revcolevschi}},
  \bibinfo{year}{2004a},
  \bibinfo{journal}{Phys. Rev. Lett.} \textbf{\bibinfo{volume}{93}},
  \bibinfo{pages}{187204}.



\bibitem[{\citenamefont{Mirebeau \emph{et~al.}}(2005)}]
{Mirebeau:2005}
\bibinfo{author}{\bibnamefont{Mirebeau}, \bibfnamefont{I.}},
  \bibinfo{author}{\bibfnamefont{A.}~\bibnamefont{Apetrei}},
  \bibinfo{author}{\bibfnamefont{J.}~\bibnamefont{Rodr\'igues-Carvajal}},
  \bibinfo{author}{\bibfnamefont{P.}~\bibnamefont{Bonville}},
  \bibinfo{author}{\bibfnamefont{A.}~\bibnamefont{Forget}},
  \bibinfo{author}{\bibfnamefont{D.}~\bibnamefont{Colson}},
  \bibinfo{author}{\bibfnamefont{V.}~\bibnamefont{Glazkov}},
  \bibinfo{author}{\bibfnamefont{J. P.}~\bibnamefont{Sanchez}},
  \bibinfo{author}{\bibfnamefont{O.}~\bibnamefont{Isnard}}, and
  \bibinfo{author}{\bibfnamefont{E.}~\bibnamefont{Suard}},
  \bibinfo{year}{2005},
  \bibinfo{journal}{Phys. Rev. Lett.} \textbf{\bibinfo{volume}{94}},
  \bibinfo{pages}{246402}.





\bibitem[{\citenamefont{Mirebeau \emph{et~al.}}(2006)}]
{Mirebeau:2006}
\bibinfo{author}{\bibnamefont{Mirebeau}, \bibfnamefont{I.}},
  \bibinfo{author}{\bibfnamefont{A.}~\bibnamefont{Apetrei}},
  \bibinfo{author}{\bibfnamefont{I. N.}~\bibnamefont{Goncharenko}}, and
  \bibinfo{author}{\bibfnamefont{R.}~\bibnamefont{Moessner}},
  \bibinfo{year}{2006},
  \bibinfo{journal}{Physica B} \textbf{\bibinfo{volume}{385}},
  \bibinfo{pages}{307}.



\bibitem[{\citenamefont{Mirebeau \emph{et~al.}}(2006a)}]
{Mirebeau:2006a}
\bibinfo{author}{\bibnamefont{Mirebeau}, \bibfnamefont{I.}},
  \bibinfo{author}{\bibfnamefont{A. }~\bibnamefont{Apetrei}},
  \bibinfo{author}{\bibfnamefont{I. }~\bibnamefont{Goncharenko}},
  \bibinfo{author}{\bibfnamefont{D.}~\bibnamefont{Andreica}},
  \bibinfo{author}{\bibfnamefont{P.}~\bibnamefont{Bonville}},
  \bibinfo{author}{\bibfnamefont{J. P.}~\bibnamefont{Sanchez}},
  \bibinfo{author}{\bibfnamefont{A.}~\bibnamefont{Amato}},
  \bibinfo{author}{\bibfnamefont{E.}~\bibnamefont{Suard}},
  \bibinfo{author}{\bibfnamefont{W. A.}~\bibnamefont{Crichton}},
  \bibinfo{author}{\bibfnamefont{A.}~\bibnamefont{Forget}},and
  \bibinfo{author}{\bibfnamefont{D.}~\bibnamefont{Colson}},
 \bibinfo{year}{2006a},
  \bibinfo{journal}{Phys. Rev. B} \textbf{\bibinfo{volume}{74}},
  \bibinfo{pages}{174414}.





\bibitem[{\citenamefont{Mirebeau \emph{et~al.}}(2007)}]
{Mirebeau:2007}
\bibinfo{author}{\bibnamefont{Mirebeau}, \bibfnamefont{I.}},
  \bibinfo{author}{\bibfnamefont{P.}~\bibnamefont{Bonville}}, and
  \bibinfo{author}{\bibfnamefont{M.}~\bibnamefont{Hennion}},
 \bibinfo{year}{2007},
  \bibinfo{journal}{Phys. Rev. B} \textbf{\bibinfo{volume}{76}},
  \bibinfo{pages}{184436}.



\bibitem[{\citenamefont{Mito \emph{et~al.}}(2007)}]
{Mito:2007}
\bibinfo{author}{\bibnamefont{Mito}, \bibfnamefont{M.}},
  \bibinfo{author}{\bibfnamefont{S.}~\bibnamefont{Kuwabara}},
  \bibinfo{author}{\bibfnamefont{K.}~\bibnamefont{Matsuhira}},
   \bibinfo{author}{\bibfnamefont{H.}~\bibnamefont{Deguchi}},
  \bibinfo{author}{\bibfnamefont{S.}~\bibnamefont{Takagi}}, and
  \bibinfo{author}{\bibfnamefont{Z.}~\bibnamefont{Hiroi}},
  \bibinfo{year}{2007},
  \bibinfo{journal}{J. Magn. Mag. Mater.} \textbf{\bibinfo{volume}{310}},
  \bibinfo{pages}{e432}.



\bibitem[{\citenamefont{Miyoshi \emph{et~al.}}(2000)}]
{Miyoshi:2000}
\bibinfo{author}{\bibnamefont{Miyoshi}, \bibfnamefont{K.}},
  \bibinfo{author}{\bibfnamefont{Y.}~\bibnamefont{Nishimura}},
  \bibinfo{author}{\bibfnamefont{K.}~\bibnamefont{Honda}},
  \bibinfo{author}{\bibfnamefont{K.}~\bibnamefont{Fujiwara}}, and
  \bibinfo{author}{\bibfnamefont{J.}~\bibnamefont{Takeuchi}},
  \bibinfo{year}{2000},
  \bibinfo{journal}{J. Phys. Soc. Jpn.} \textbf{\bibinfo{volume}{69}},
  \bibinfo{pages}{3517}.



 \bibitem[{\citenamefont{Miyoshi \emph{et~al.}}(2002)}]
{Miyoshi:2001}
\bibinfo{author}{\bibnamefont{Miyoshi}, \bibfnamefont{K.}},
  \bibinfo{author}{\bibfnamefont{T.}~\bibnamefont{Yamashita}},
  \bibinfo{author}{\bibfnamefont{K.}~\bibnamefont{Fujiwara}}, and
  \bibinfo{author}{\bibfnamefont{J.}~\bibnamefont{Takeuchi}},
  \bibinfo{year}{2002},
  \bibinfo{journal}{Physica B} \textbf{\bibinfo{volume}{312}},
  \bibinfo{pages}{706}.



\bibitem[{\citenamefont{Miyoshi \emph{et~al.}}(2006)}]
{Miyoshi:2006}
\bibinfo{author}{\bibnamefont{Miyoshi}, \bibfnamefont{K.}},
  \bibinfo{author}{\bibfnamefont{Y.}~\bibnamefont{Takamatsu}}, and
  \bibinfo{author}{\bibfnamefont{J.}~\bibnamefont{Takeuchi}},
  \bibinfo{year}{2006},
  \bibinfo{journal}{J. Phys. Soc. Jpn.} \textbf{\bibinfo{volume}{75}},
  \bibinfo{pages}{065001}.



%also called {Moessner:PRB1998}
\bibitem[{\citenamefont{Moessner and Chalker}(1998)}]
{Moessner:1998}
\bibinfo{author}{\bibnamefont{Moessner},~\bibfnamefont{R.}} and
\bibinfo{author}{\bibfnamefont{J. T.}~\bibnamefont{Chalker}},
 \bibinfo{year}{1998},
 \bibinfo{journal}{Phys. Rev. B.} \textbf{\bibinfo{volume}{58}},
 \bibinfo{pages}{12049}.



\bibitem[{\citenamefont{Moessner and Chalker}(1998a)}]
{Moessner:1998a}
  \bibinfo{author}{\bibfnamefont{Moessner}, \bibnamefont{R}} and
  \bibinfo{author}{\bibnamefont{J. T.} ~\bibfnamefont{Chalker}},
 \bibinfo{year}{1998a},
  \bibinfo{journal}{Phys. Rev. Lett.} \textbf{\bibinfo{volume}{80}},
  \bibinfo{pages}{2929}.



\bibitem[{\citenamefont{Moessner}(2001)}]
{Moessner:2001}
\bibinfo{author}{\bibnamefont{Moessner},~\bibfnamefont{R.}},
 \bibinfo{year}{2001},
 \bibinfo{journal}{ Can. J. Phys.} \textbf{\bibinfo{volume}{79}},
 \bibinfo{pages}{1283}.


\bibitem[{\citenamefont{Moessner and Sondhi}(2003)}]
{Moessner:2003}
\bibinfo{author}{\bibnamefont{Moessner},~\bibfnamefont{R.}} and
\bibinfo{author}{\bibfnamefont{S. L.}~\bibnamefont{Sondhi}},
 \bibinfo{year}{2003},
 \bibinfo{journal}{Phys. Rev. B.} \textbf{\bibinfo{volume}{68}},
 \bibinfo{pages}{064411}.



\bibitem[{\citenamefont{Moessner} \emph{et~al.}(2006)}]
{Moessner:2006}
  \bibinfo{author}{\bibfnamefont{Moessner}, \bibnamefont{R.}},
\bibinfo{author}{\bibnamefont{S. L.}~\bibfnamefont{Sondhi}}, and
 \bibinfo{author}{\bibnamefont{M. O.}~\bibfnamefont{Goerbig}},
 \bibinfo{year}{2006},
  \bibinfo{journal}{Phys. Rev. B} \textbf{\bibinfo{volume}{73}},
  \bibinfo{pages}{094430}.





\bibitem[{\citenamefont{Molavian} \emph{et~al.}(2007)}]
{Molavian:2007}
  \bibinfo{author}{\bibnamefont{Molavian}~\bibfnamefont{H. R.}},
  \bibinfo{author}{\bibnamefont{M. J. P.}~\bibfnamefont{Gingras}}, and
  \bibinfo{author}{\bibnamefont{B. }~\bibfnamefont{Canals}},
   \bibinfo{year}{2007},
  \bibinfo{journal}{Phys. Rev. Lett.} \textbf{\bibinfo{volume}{98}},
  \bibinfo{pages}{157204}.





 \bibitem[{\citenamefont{Moritomo \emph{et~al.}}(2001)}]
{Moritomo:2001}
\bibinfo{author}{\bibnamefont{Moritomo}, \bibfnamefont{Y.}},
  \bibinfo{author}{\bibfnamefont{Sh.}~\bibnamefont{Xu}},
  \bibinfo{author}{\bibfnamefont{A.}~\bibnamefont{Machida}},
  \bibinfo{author}{\bibfnamefont{T.}~\bibnamefont{Katsufuji}},
  \bibinfo{author}{\bibfnamefont{E.}~\bibnamefont{Nishibori}},
  \bibinfo{author}{\bibfnamefont{M.}~\bibnamefont{Takata}},
  \bibinfo{author}{\bibfnamefont{S.}~\bibnamefont{Sakata}}, and
  \bibinfo{author}{\bibfnamefont{S.-W.}~\bibnamefont{Cheong}},
  \bibinfo{year}{2001},
  \bibinfo{journal}{Phys. Rev. B.} \textbf{\bibinfo{volume}{63}},
  \bibinfo{pages}{144425}.



\bibitem[{\citenamefont{Munenaka and Sato}(2006)}]
{Munenaka:2006}
\bibinfo{author}{\bibnamefont{Munenaka},~\bibfnamefont{T,}} and
\bibinfo{author}{\bibfnamefont{Hirohiko}~\bibnamefont{S.}},
 \bibinfo{year}{2006},
 \bibinfo{journal}{J. Phys. Soc. Jpn.} \textbf{\bibinfo{volume}{75}},
 \bibinfo{pages}{103801}.


%
%\bibitem[{\citenamefont{Nakatsuji} \emph{et~al.}(2005)}]
%{Nakatsuji:2005}
%  \bibinfo{author}{\bibfnamefont{Nakatsuji}, \bibnamefont{S.}},
%  \bibinfo{author}{\bibnamefont{Y.}~\bibfnamefont{Nambu}},
%  \bibinfo{author}{\bibnamefont{H.}~\bibfnamefont{Tonomura}},
%  \bibinfo{author}{\bibnamefont{O.}~\bibfnamefont{Sakai}},
%  \bibinfo{author}{\bibnamefont{S.}~\bibfnamefont{Jonas}},
%  \bibinfo{author}{\bibnamefont{C.}~\bibfnamefont{Broholm}},
%  \bibinfo{author}{\bibnamefont{H.}~\bibfnamefont{Tsunetsugu}},
%  \bibinfo{author}{\bibnamefont{Y.}~\bibfnamefont{Qiu}},  and
%  \bibinfo{author}{\bibnamefont{Y.}~\bibfnamefont{Maeno}},
%  \bibinfo{year}{2005},
%  \bibinfo{journal}{Science} \textbf{\bibinfo{volume}{309}},
%  \bibinfo{pages}{1697}.



\bibitem[{\citenamefont{Nakatsuji \emph{et~al.}}(2006)}]
{Nakatsuji:2006}
\bibinfo{author}{\bibnamefont{Nakatsuji}, \bibfnamefont{S.}},
  \bibinfo{author}{\bibfnamefont{Y. }~\bibnamefont{Machida}},
  \bibinfo{author}{\bibfnamefont{Y. }~\bibnamefont{Maeno}},
  \bibinfo{author}{\bibfnamefont{T.}~\bibnamefont{Tayama}},
  \bibinfo{author}{\bibfnamefont{T.}~\bibnamefont{Sakakibara}},
  \bibinfo{author}{\bibfnamefont{J.}~\bibnamefont{van Duijn}},
  \bibinfo{author}{\bibfnamefont{L.}~\bibnamefont{Balicas}},
  \bibinfo{author}{\bibfnamefont{J. N.}~\bibnamefont{Millican}},
  \bibinfo{author}{\bibfnamefont{R. T.}~\bibnamefont{Macaluso}}, and
  \bibinfo{author}{\bibfnamefont{J. Y.}~\bibnamefont{Chan}},
  \bibinfo{year}{2006},
  \bibinfo{journal}{Phys. Rev. Lett.} \textbf{\bibinfo{volume}{96}},
  \bibinfo{pages}{087204}.




\bibitem[{\citenamefont{N\'eel}(1948)}]
{Neel:1948}
  \bibinfo{author}{\bibfnamefont{N\'eel}, \bibnamefont{L.}},
  \bibinfo{year}{1948},
  \bibinfo{journal}{Ann. Phys.} \textbf{\bibinfo{volume}{3}},
  \bibinfo{pages}{137}.


\bibitem[{\citenamefont{Nussinov} \emph{et~al.}(2007)}]
{Nussinov:2007}
\bibinfo{author}{\bibfnamefont{Nussinov}, \bibnamefont{Z.}},
\bibinfo{author}{\bibnamefont{C. D.}~\bibfnamefont{Batista}},
 \bibinfo{author}{\bibnamefont{B.}~\bibfnamefont{Normand}},  and
\bibinfo{author}{\bibnamefont{S. A.}~\bibfnamefont{Trugman}},
 \bibinfo{year}{2007},
  \bibinfo{journal}{Phys. Rev. B} \textbf{\bibinfo{volume}{75}},
  \bibinfo{pages}{094411}.

%

%\bibitem[{\citenamefont{Oba} \emph{et~al.}(2006)}]
%{Oba:2006}
%  \bibinfo{author}{\bibfnamefont{Oba}, \bibnamefont{N.}},
%  \bibinfo{author}{\bibnamefont{H.}~\bibfnamefont{Kageyama}},
%  \bibinfo{author}{\bibnamefont{T.}~\bibfnamefont{Kitano}},
%  \bibinfo{author}{\bibnamefont{J.}~\bibfnamefont{Yasuda}},
%  \bibinfo{author}{\bibnamefont{Y.}~\bibfnamefont{Baba}},
%  \bibinfo{author}{\bibnamefont{M.}~\bibfnamefont{Nishi}},
%  \bibinfo{author}{\bibnamefont{K.}~\bibfnamefont{Hirota}},
%  \bibinfo{author}{\bibnamefont{Y.}~\bibfnamefont{Narumi}},
%  \bibinfo{author}{\bibnamefont{M.}~\bibfnamefont{Hagiwara}},
%  \bibinfo{author}{\bibnamefont{K.}~\bibfnamefont{Kindo}},
%  \bibinfo{author}{\bibnamefont{T.}~\bibfnamefont{Saito}},
%  \bibinfo{author}{\bibnamefont{Y.}~\bibfnamefont{Ajiro}},  and
%  \bibinfo{author}{\bibnamefont{K.}~\bibfnamefont{Yoshimura}},
%  \bibinfo{year}{2006},
%  \bibinfo{journal}{J. Phys. Soc. Jpn.} \textbf{\bibinfo{volume}{75}},
%  \bibinfo{pages}{113601}.





\bibitem[{\citenamefont{Obradors} \emph{et~al.}(1988)}]
{Obradors:1988}
  \bibinfo{author}{\bibfnamefont{Obradors}, \bibnamefont{X.}},
  \bibinfo{author}{\bibnamefont{A.}~\bibfnamefont{Labarta}},
  \bibinfo{author}{\bibnamefont{A.}~\bibfnamefont{Isalgue}},
  \bibinfo{author}{\bibnamefont{J.}~\bibfnamefont{Tejada}},
  \bibinfo{author}{\bibnamefont{J.}~\bibfnamefont{Rodriguez}},  and
  \bibinfo{author}{\bibnamefont{M.}~\bibfnamefont{Pernet}},
    \bibinfo{year}{1988},
  \bibinfo{journal}{Solid State Commun.} \textbf{\bibinfo{volume}{65}},
  \bibinfo{pages}{189}.



\bibitem[{\citenamefont{Okamoto} \emph{et~al.}(2007)}]
{Okamoto:2007}
  \bibinfo{author}{\bibfnamefont{Okamoto}, \bibnamefont{Y.}},
  \bibinfo{author}{\bibnamefont{M.}~\bibfnamefont{Nohara}},
  \bibinfo{author}{\bibnamefont{H.}~\bibfnamefont{Aruga-Katori}},  and
    \bibinfo{author}{\bibnamefont{H.}~\bibfnamefont{Takagi}},
 \bibinfo{year}{2007},
  \bibinfo{journal}{Phys. Rev. Lett.} \textbf{\bibinfo{volume}{99}},
  \bibinfo{pages}{137207}.



\bibitem[{\citenamefont{Orbach}(1961)}]
{Orbach:1961}
   \bibinfo{author}{\bibnamefont{Orbach}~\bibfnamefont{R.}},
 \bibinfo{year}{1961},
  \bibinfo{journal}{Proc. Phys. Soc. } \textbf{\bibinfo{volume}{77}},
  \bibinfo{pages}{821}.


\bibitem[{\citenamefont{Orend\'ac} \emph{et~al.}(2007)}]
{Orendac:2007}
  \bibinfo{author}{\bibfnamefont{Orend\'ac}, \bibnamefont{M.}},
  \bibinfo{author}{\bibnamefont{J.}~\bibfnamefont{Hanko}},
  \bibinfo{author}{\bibfnamefont{E.}~\bibnamefont{{\v C}i$\check{\mbox{z}}$m{\'a}r}},
  \bibinfo{author}{\bibfnamefont{A.}~\bibnamefont{Orend{\'a}$\check{\mbox{c}}$ov{\'a}}},
  \bibinfo{author}{\bibnamefont{J.}~\bibfnamefont{Shirai}},  and
  \bibinfo{author}{\bibnamefont{S. T.}~\bibfnamefont{Bramwell}},
  \bibinfo{year}{2007},
  \bibinfo{journal}{Phys. Rev. B.} \textbf{\bibinfo{volume}{75}},
  \bibinfo{pages}{104425}.



\bibitem[{\citenamefont{Palmer and Chalker}(2000)}]
{Palmer:2000}
\bibinfo{author}{\bibnamefont{Palmer},~\bibfnamefont{S. E.}} and
\bibinfo{author}{\bibfnamefont{J. T.}~\bibnamefont{Chalker}},
 \bibinfo{year}{2000},
 \bibinfo{journal}{Phys. Rev. B} \textbf{\bibinfo{volume}{62}},
 \bibinfo{pages}{488}.



\bibitem[{\citenamefont{Park \emph{et~al.}}(2003)}]
{Park:2003}
\bibinfo{author}{\bibnamefont{Park}, \bibfnamefont{J. G.}},
  \bibinfo{author}{\bibfnamefont{Y.}~\bibnamefont{Jo}},
  \bibinfo{author}{\bibfnamefont{J.}~\bibnamefont{Park}},
  \bibinfo{author}{\bibfnamefont{H. C.}~\bibnamefont{Kim}},
  \bibinfo{author}{\bibfnamefont{H. C.}~\bibnamefont{Ri}},
  \bibinfo{author}{\bibfnamefont{S.}~\bibnamefont{Xu}},
  \bibinfo{author}{\bibfnamefont{Y.}~\bibnamefont{Moritomo}}, and
  \bibinfo{author}{\bibfnamefont{S. W.}~\bibnamefont{Cheong}},
  \bibinfo{year}{2003},
  \bibinfo{journal}{Physica B} \textbf{\bibinfo{volume}{328}},
  \bibinfo{pages}{90}.




\bibitem[{\citenamefont{Pauling}(1935)}]
{Pauling:1935}
\bibinfo{author}{\bibnamefont{Pauling},~\bibfnamefont{L.}},
 \bibinfo{year}{1935},
 \bibinfo{journal}{ J. Am. Chem. Soc.} \textbf{\bibinfo{volume}{57}},
 \bibinfo{pages}{2680}.



\bibitem[{\citenamefont{Penc} \emph{et~al.}(2004)}]
{Penc:2004}
  \bibinfo{author}{\bibfnamefont{Penc}, \bibnamefont{K.}},
  \bibinfo{author}{\bibnamefont{N.}~\bibfnamefont{Shannon}}, and
 \bibinfo{author}{\bibnamefont{H.}~\bibfnamefont{Shiba}},
 \bibinfo{year}{2004},
  \bibinfo{journal}{Phys. Rev. Lett.} \textbf{\bibinfo{volume}{93}},
  \bibinfo{pages}{197203}.


\bibitem[{\citenamefont{Petrenko} \emph{et~al.}(1998)}]
{Petrenko:1998}
  \bibinfo{author}{\bibnamefont{Petrenko}~\bibfnamefont{O. A.}},
\bibinfo{author}{\bibnamefont{C. }~\bibfnamefont{Ritter}},
\bibinfo{author}{\bibnamefont{M. }~\bibfnamefont{Yethiraj}},  and
\bibinfo{author}{\bibnamefont{D. M. }~\bibfnamefont{Paul}},
 \bibinfo{year}{1998},
 \bibinfo{journal}{Phys. Rev. Lett.} \textbf{\bibinfo{volume}{80}},
 \bibinfo{pages}{4570}.



\bibitem[{\citenamefont{Petrenko \emph{et~al.}}(2004)}]
{Petrenko:2004}
\bibinfo{author}{\bibnamefont{Petrenko}, \bibfnamefont{O. A.}},
\bibinfo{author}{\bibfnamefont{M. R.}~\bibnamefont{Lees}},
 \bibinfo{author}{\bibfnamefont{G.}~\bibnamefont{Balakrishnan}}, and
\bibinfo{author}{\bibfnamefont{D. M$^c$K}~\bibnamefont{Paul}},
  \bibinfo{year}{2004},
 \bibinfo{journal}{Phys. Rev. B} \textbf{\bibinfo{volume}{70}},
   \bibinfo{pages}{012402}.



\bibitem[{\citenamefont{Pike and Seager}(1977)}]
{Pike:1977}
\bibinfo{author}{\bibnamefont{Pike},~\bibfnamefont{G. E.}} and
\bibinfo{author}{\bibfnamefont{C. H.}~\bibnamefont{Seager}},
 \bibinfo{year}{1977},
 \bibinfo{journal}{J. Appl. Phys.} \textbf{\bibinfo{volume}{53}},
 \bibinfo{pages}{5152}.


 \bibitem[{\citenamefont{Pinettes \emph{et~al.}}(2002)}]
{Pinettes:2002}
\bibinfo{author}{\bibnamefont{Pinettes}, \bibfnamefont{C.}},
  \bibinfo{author}{\bibfnamefont{B.}~\bibnamefont{Canals}}, and
  \bibinfo{author}{\bibfnamefont{C.}~\bibnamefont{Lacroix}},
  \bibinfo{year}{2002},
  \bibinfo{journal}{Phys. Rev. B} \textbf{\bibinfo{volume}{66}},
  \bibinfo{pages}{024422}.




\bibitem[{\citenamefont{Poole} \emph{et~al.}(2007)}]
{Poole:2007}
 \bibinfo{author}{\bibfnamefont{Poole}, \bibnamefont{A.}},
  \bibinfo{author}{\bibnamefont{A. S.}~\bibfnamefont{Wills}}, and
  \bibinfo{author}{\bibnamefont{E.}~\bibfnamefont{Leli\`evre-Berna}},
  \bibinfo{year}{2007},
  \bibinfo{journal}{J. Phys.: Condens. Matter } \textbf{\bibinfo{volume}{19}},
  \bibinfo{pages}{452201}.



%\bibitem[{\citenamefont{Powell and McKenzie}(2006)}]
%{Powell:2006}
% \bibinfo{author}{\bibfnamefont{Powell}, \bibnamefont{B. J.}} and
%  \bibinfo{author}{\bibnamefont{R. H.}~\bibfnamefont{McKenzie}},   \bibinfo{year}{2006},
%  \bibinfo{journal}{J. Phys.: Condens. Matter} \textbf{\bibinfo{volume}{18}},
%  \bibinfo{pages}{R827}.



\bibitem[{\citenamefont{Proffen \emph{et~al.}}(2003)}]
{Proffen:2003}
\bibinfo{author}{\bibnamefont{Proffen},~\bibfnamefont{Th.}},
 \bibinfo{author}{\bibfnamefont{S. J. L.}~\bibnamefont{Billinge}},
  \bibinfo{author}{\bibfnamefont{T.}~\bibnamefont{Egami}},and
  \bibinfo{author}{\bibfnamefont{D.} ~\bibnamefont{Louca}},
  \bibinfo{year}{2003},
  \bibinfo{journal}{Z. Kristallogr.} \textbf{\bibinfo{volume}{218}},
  \bibinfo{pages}{132}.



\bibitem[{\citenamefont{Quilliam \emph{et~al.}}(2007)}]
{Quilliam:2007}
\bibinfo{author}{\bibnamefont{Quilliam}, \bibfnamefont{J. A.}},
  \bibinfo{author}{\bibfnamefont{K. A.}~\bibnamefont{Ross}},
  \bibinfo{author}{\bibfnamefont{A. G.}~\bibnamefont{Del Maestro}},
  \bibinfo{author}{\bibfnamefont{M. J. P.}~\bibnamefont{Gingras}},
  \bibinfo{author}{\bibfnamefont{L. R.}~\bibnamefont{Corruccini}}, and
  \bibinfo{author}{\bibfnamefont{J. B.}~\bibnamefont{Kycia}},
  \bibinfo{year}{2007},
  \bibinfo{journal}{Phys. Rev. Lett.} \textbf{\bibinfo{volume}{99}},
  \bibinfo{pages}{097201}.




%\bibitem[{\citenamefont{Ramirez} \emph{et~al.}(1990)}]
%{Ramirez:1990}
%  \bibinfo{author}{\bibfnamefont{Ramirez}, \bibnamefont{A. P.}},
%  \bibinfo{author}{\bibnamefont{G. P.}~\bibfnamefont{Espinosa}},  and
%  \bibinfo{author}{\bibnamefont{A. S.}~\bibfnamefont{Cooper}},
% \bibinfo{year}{1990},
%  \bibinfo{journal}{Phys. Rev. Lett.} \textbf{\bibinfo{volume}{64}},
%  \bibinfo{pages}{2070}.




\bibitem[{\citenamefont{Ramirez}(1994)}]
{Ramirez:1994}
\bibinfo{author}{\bibnamefont{Ramirez},~\bibfnamefont{A.P.}},
 \bibinfo{year}{1994},
 \bibinfo{journal}{Ann. Rev. Mater. Sci.} \textbf{\bibinfo{volume}{24}},
 \bibinfo{pages}{453}.


 \bibitem[{\citenamefont{Ramirez and Subramanian}(1997)}]
{Ramirez:1997}
\bibinfo{author}{\bibnamefont{Ramirez},~\bibfnamefont{A. P.}} and
\bibinfo{author}{\bibfnamefont{M. A.}~\bibnamefont{Subramanian}},
 \bibinfo{year}{1997},
 \bibinfo{journal}{Science} \textbf{\bibinfo{volume}{277}},
 \bibinfo{pages}{546}.


\bibitem[{\citenamefont{Ramirez} \emph{et~al.}(1999)}]
{Ramirez:1999}
\bibinfo{author}{\bibfnamefont{Ramirez}, \bibnamefont{A. P.}},
\bibinfo{author}{\bibnamefont{A.}~\bibfnamefont{Hayashi}},
\bibinfo{author}{\bibnamefont{R. J.}~\bibfnamefont{Cava}},
\bibinfo{author}{\bibnamefont{R.}~\bibfnamefont{Siddharthan}},  and
\bibinfo{author}{\bibnamefont{B. S.}~\bibfnamefont{Shastry}},
\bibinfo{year}{1999},
 \bibinfo{journal}{Nature} \textbf{\bibinfo{volume}{399}},
\bibinfo{pages}{333}.



\bibitem[{\citenamefont{Ramirez \emph{et~al.}}(2002)}]
{Ramirez:2002}
\bibinfo{author}{\bibnamefont{Ramirez}, \bibfnamefont{A. P.}},
  \bibinfo{author}{\bibfnamefont{B. S.}~\bibnamefont{Shastry}},
  \bibinfo{author}{\bibfnamefont{A.}~\bibnamefont{Hayashi}},
    \bibinfo{author}{\bibfnamefont{J. J.}~\bibnamefont{Krajewski}},
  \bibinfo{author}{\bibfnamefont{D. A.}~\bibnamefont{Huse}}, and
  \bibinfo{author}{\bibfnamefont{R. J.}~\bibnamefont{Cava}},
  \bibinfo{year}{2002},
  \bibinfo{journal}{Phys. Rev. Lett.} \textbf{\bibinfo{volume}{89}},
  \bibinfo{pages}{067202}.


\bibitem[{\citenamefont{Ranganathan \emph{et~al.}}(1983)}]
{Ranganathan:1983}
\bibinfo{author}{\bibnamefont{Ranganathan}, \bibfnamefont{R.}},
  \bibinfo{author}{\bibfnamefont{G.}~\bibnamefont{Rangarajan}},
  \bibinfo{author}{\bibfnamefont{R.}~\bibnamefont{Srinivasan}},
  \bibinfo{author}{\bibfnamefont{M. A.}~\bibnamefont{Subramanian}}, and
  \bibinfo{author}{\bibfnamefont{G. V.}~\bibnamefont{Subba Rao}},
  \bibinfo{year}{1983},
  \bibinfo{journal}{J. Low Temp. Phys.} \textbf{\bibinfo{volume}{52}},
  \bibinfo{pages}{481}.



\bibitem[{\citenamefont{Raju \emph{et~al.}}(1992)}]
{Raju:1992}
\bibinfo{author}{\bibnamefont{Raju}, \bibfnamefont{N. P.}},
  \bibinfo{author}{\bibfnamefont{E.}~\bibnamefont{Gmelin}}, and
  \bibinfo{author}{\bibfnamefont{R.}~\bibnamefont{Kremer}},
  \bibinfo{year}{1992},
  \bibinfo{journal}{Phys. Rev. B} \textbf{\bibinfo{volume}{46}},
  \bibinfo{pages}{5405}.



 %SANS Tl and In Mn2O7

\bibitem[{\citenamefont{Raju \emph{et~al.}}(1994)}]
{Raju:1994}
\bibinfo{author}{\bibnamefont{Raju}, \bibfnamefont{N. P.}},
 \bibinfo{author}{\bibfnamefont{J. E.}~\bibnamefont{Greedan}}, and
   \bibinfo{author}{\bibfnamefont{M. A.}~\bibnamefont{Subramanian}},
  \bibinfo{year}{1994},
  \bibinfo{journal}{Phys. Rev. B} \textbf{\bibinfo{volume}{49}},
  \bibinfo{pages}{1086}.


\bibitem[{\citenamefont{Raju and Gougeon}(1995)}]
{Raju:1995}
\bibinfo{author}{\bibnamefont{Raju}, ~\bibfnamefont{N. P.}} and
\bibinfo{author}{\bibfnamefont{P.}~\bibnamefont{Gougeon}},
 \bibinfo{year}{1995},
 \bibinfo{journal}{unpublished}


\bibitem[{\citenamefont{Raju \emph{et~al.}}(1999)}]
{Raju:1999}
\bibinfo{author}{\bibnamefont{Raju}, \bibfnamefont{N. P.}},
  \bibinfo{author}{\bibfnamefont{M.}~\bibnamefont{Dion}},
  \bibinfo{author}{\bibfnamefont{M. J. P.}~\bibnamefont{Gingras}},
  \bibinfo{author}{\bibfnamefont{T. E.}~\bibnamefont{Mason}}, and
  \bibinfo{author}{\bibfnamefont{J. E.}~\bibnamefont{Greedan}},
  \bibinfo{year}{1999},
  \bibinfo{journal}{Phys. Rev. B} \textbf{\bibinfo{volume}{59}},
  \bibinfo{pages}{14489}.



\bibitem[{\citenamefont{Reich} \emph{et~al.}(1987)}]
{Reich:1987}
 \bibinfo{author}{\bibfnamefont{Reich}, \bibnamefont{D. H.}},
 \bibinfo{author}{\bibnamefont{T. F.}~\bibfnamefont{Rosenbaum}}, and
  \bibinfo{author}{\bibnamefont{G.}~\bibfnamefont{Aeppli}},
  \bibinfo{year}{1987},
  \bibinfo{journal}{Phys. Rev. Lett.  } \textbf{\bibinfo{volume}{59}},
  \bibinfo{pages}{1969}.



\bibitem[{\citenamefont{Reich} \emph{et~al.}(1990)}]
{Reich:1990}
  \bibinfo{author}{\bibfnamefont{Reich}, \bibnamefont{D. H.}},
  \bibinfo{author}{\bibnamefont{B.}~\bibfnamefont{Ellman}},
  \bibinfo{author}{\bibnamefont{Y.}~\bibfnamefont{Yang}},
  \bibinfo{author}{\bibnamefont{T. F.}~\bibfnamefont{Rosenbaum}},
  \bibinfo{author}{\bibnamefont{G.}~\bibfnamefont{Aeppli}},  and
  \bibinfo{author}{\bibnamefont{D. P.}~\bibfnamefont{Belanger}},
 \bibinfo{year}{1990},
  \bibinfo{journal}{Phys. Rev. B } \textbf{\bibinfo{volume}{42}},
  \bibinfo{pages}{4631}.




\bibitem[{\citenamefont{Reimers \emph{et~al.}}(1988)}]
{Reimers:1988}
\bibinfo{author}{\bibnamefont{Reimers}, \bibfnamefont{J. N.}},
  \bibinfo{author}{\bibfnamefont{J. E.}~\bibnamefont{Greedan}}, and
  \bibinfo{author}{\bibfnamefont{M.}~\bibnamefont{Sato}},
  \bibinfo{year}{1988},
  \bibinfo{journal}{J. Solid State Chem.} \textbf{\bibinfo{volume}{72}},
  \bibinfo{pages}{390}.




\bibitem[{\citenamefont{Reimers \emph{et~al.}}(1991)}]
{Reimers:1991}
\bibinfo{author}{\bibnamefont{Reimers}, \bibfnamefont{J. N.}},
  \bibinfo{author}{\bibfnamefont{J. E.}~\bibnamefont{Greedan}},
  \bibinfo{author}{\bibfnamefont{R. K.}~\bibnamefont{Kremer}},
  \bibinfo{author}{\bibfnamefont{E.}~\bibnamefont{Gmelin}}, and
  \bibinfo{author}{\bibfnamefont{M. A.}~\bibnamefont{Subramanian}},
  \bibinfo{year}{1991},
  \bibinfo{journal}{Phys. Rev. B} \textbf{\bibinfo{volume}{43}},
  \bibinfo{pages}{3387}.




\bibitem[{\citenamefont{Reimers} \emph{et~al.}(1991a)}]
{Reimers:1991a}
  \bibinfo{author}{\bibfnamefont{Reimers}, \bibnamefont{J. N.}},
  \bibinfo{author}{\bibnamefont{A. J.}~\bibfnamefont{Berlinsky}},  and
  \bibinfo{author}{\bibnamefont{A.-C.}~\bibfnamefont{Shi}},
 \bibinfo{year}{1991a},
  \bibinfo{journal}{Phys. Rev. B} \textbf{\bibinfo{volume}{43}},
  \bibinfo{pages}{865}.



\bibitem[{\citenamefont{Reimers}(1992)}]
{Reimers:1992}
  \bibinfo{author}{\bibfnamefont{Reimers}, \bibnamefont{J. N.}},
 \bibinfo{year}{1992},
  \bibinfo{journal}{Phys. Rev. B} \textbf{\bibinfo{volume}{45}},
  \bibinfo{pages}{7287}.



\bibitem[{\citenamefont{Reimers} \emph{et~al.}(1992)}]
{Reimers:crit}
  \bibinfo{author}{\bibfnamefont{Reimers}, \bibnamefont{J. N.}},
\bibinfo{author}{\bibnamefont{J. E.}~\bibfnamefont{Greedan}}, and
 \bibinfo{author}{\bibnamefont{M.}~\bibfnamefont{Bj\" orgvinsson}},
 \bibinfo{year}{1992},
  \bibinfo{journal}{Phys. Rev. B} \textbf{\bibinfo{volume}{45}},
  \bibinfo{pages}{7295}.



\bibitem[{\citenamefont{Reimers and Berlinsky}(1993)}]
{Reimers:kagome}
  \bibinfo{author}{\bibfnamefont{Reimers}, \bibnamefont{J. N.}}, and
  \bibinfo{author}{\bibnamefont{A. J.}~\bibfnamefont{Berlinsky}},
  \bibinfo{year}{1993},
  \bibinfo{journal}{Phys. Rev. B} \textbf{\bibinfo{volume}{48}},
  \bibinfo{pages}{9539}.



\bibitem[{\citenamefont{Rhyne}(1985)}]
{Rhyne:1985}
\bibinfo{author}{\bibnamefont{Rhyne},~\bibfnamefont{J. J.}},
 \bibinfo{year}{1985},
 \bibinfo{journal}{IEEE Trans. Magn.} \textbf{\bibinfo{volume}{21}},
 \bibinfo{pages}{1990}.



\bibitem[{\citenamefont{Rhyne and Fish}(1985)}]
{Rhyne:1985a}
\bibinfo{author}{\bibnamefont{Rhyne},~\bibfnamefont{J. J.}} and
\bibinfo{author}{\bibfnamefont{G. E.}~\bibnamefont{Fish}},
 \bibinfo{year}{1985},
 \bibinfo{journal}{J. Appl. Phys.} \textbf{\bibinfo{volume}{57}},
 \bibinfo{pages}{3407}.




\bibitem[{\citenamefont{Rosenfeld and Subramanian}(1996)}]
{Rosenfeld:1996}
\bibinfo{author}{\bibnamefont{Rosenfeld},~\bibfnamefont{H. D.}} and
\bibinfo{author}{\bibfnamefont{M. A.}~\bibnamefont{Subramanian}},
 \bibinfo{year}{1996},
 \bibinfo{journal}{J. Solid State Chem.} \textbf{\bibinfo{volume}{125}},
 \bibinfo{pages}{278}.




\bibitem[{\citenamefont{Rosenkranz \emph{et~al.}}(2000)}]
{Rosenkranz:2000}
\bibinfo{author}{\bibnamefont{Rosenkranz}, \bibfnamefont{S.}},
 \bibinfo{author}{\bibfnamefont{A. P.}~\bibnamefont{Ramirez}},
  \bibinfo{author}{\bibfnamefont{A.}~\bibnamefont{Hayashi}},
 \bibinfo{author}{\bibfnamefont{R. J.}~\bibnamefont{Cava}},
 \bibinfo{author}{\bibfnamefont{R.}~\bibnamefont{Siddharthan}}, and
\bibinfo{author}{\bibfnamefont{B. S.}~\bibnamefont{Shastry}},
  \bibinfo{year}{2000},
  \bibinfo{journal}{J. Appl. Phys.} \textbf{\bibinfo{volume}{87}},
  \bibinfo{pages}{5914}.




\bibitem[{\citenamefont{Roth}(1956)}]
{Roth:1956}
\bibinfo{author}{\bibnamefont{Roth},~\bibfnamefont{R. S.}},
 \bibinfo{year}{1956},
 \bibinfo{journal}{ J. Res. Natl. Bur. Stds..} \textbf{\bibinfo{volume}{56}},
 \bibinfo{pages}{17}.



\bibitem[{\citenamefont{Ruff} \emph{et~al.}(2005)}]
{Ruff:2005}
  \bibinfo{author}{\bibfnamefont{Ruff}, \bibnamefont{J. P. C.}},
  \bibinfo{author}{\bibnamefont{R. G.}~\bibfnamefont{Melko}},  and
  \bibinfo{author}{\bibnamefont{M. J. P.}~\bibfnamefont{Gingras}},
  \bibinfo{year}{2005},
  \bibinfo{journal}{Phys. Rev. Lett.  } \textbf{\bibinfo{volume}{95}},
  \bibinfo{pages}{097202}.



\bibitem[{\citenamefont{Ruff \emph{et~al.}}(2007)}]
{Ruff:2007}
\bibinfo{author}{\bibnamefont{Ruff}, \bibfnamefont{J. P. C.}},
  \bibinfo{author}{\bibfnamefont{B. D.}~\bibnamefont{Gaulin}},
  \bibinfo{author}{\bibfnamefont{J. P.}~\bibnamefont{Castellan}},
  \bibinfo{author}{\bibfnamefont{K. C.}~\bibnamefont{Rule}},
  \bibinfo{author}{\bibfnamefont{J. P.}~\bibnamefont{Clancy}},
  \bibinfo{author}{\bibfnamefont{J.}~\bibnamefont{Rodrigues}},  and
  \bibinfo{author}{\bibfnamefont{H. A.}~\bibnamefont{Dabkowska}},
  \bibinfo{year}{2007},
  \bibinfo{journal}{Phys. Rev. Lett.} \textbf{\bibinfo{volume}{99}},
  \bibinfo{pages}{237202}.



\bibitem[{\citenamefont{Rule \emph{et~al.}}(2006)}]
{Rule:2006}
\bibinfo{author}{\bibnamefont{Rule}, \bibfnamefont{K. C.}},
  \bibinfo{author}{\bibfnamefont{J. P. C.}~\bibnamefont{Ruff}},
  \bibinfo{author}{\bibfnamefont{B. D.}~\bibnamefont{Gaulin}},
  \bibinfo{author}{\bibfnamefont{S. R.}~\bibnamefont{Dunsiger}},
   \bibinfo{author}{\bibfnamefont{J. S.}~\bibnamefont{Gardner}},
  \bibinfo{author}{\bibfnamefont{J. P.}~\bibnamefont{Clancy}},
  \bibinfo{author}{\bibfnamefont{M. J.}~\bibnamefont{Lewis}},
  \bibinfo{author}{\bibfnamefont{H. A.}~\bibnamefont{Dabkowska}},
  \bibinfo{author}{\bibfnamefont{I.}~\bibnamefont{Mirebeau}},
  \bibinfo{author}{\bibfnamefont{P.}~\bibnamefont{Manuel}},
  \bibinfo{author}{\bibfnamefont{Y.}~\bibnamefont{Qiu}}, and
  \bibinfo{author}{\bibfnamefont{J. R. D.}~\bibnamefont{Copley}},  
  \bibinfo{year}{2006},
  \bibinfo{journal}{Phys. Rev. Lett.} \textbf{\bibinfo{volume}{96}},
  \bibinfo{pages}{177201}.




\bibitem[{\citenamefont{Rushbrooke and Wood}(1958)}]
{Rushbrooke:1958}
\bibinfo{author}{\bibnamefont{Rushbrooke},~\bibfnamefont{G. S.}}, and
\bibinfo{author}{\bibfnamefont{P. J.}~\bibnamefont{Wood}},
 \bibinfo{year}{1958},
 \bibinfo{journal}{Mol. Phys.} \textbf{\bibinfo{volume}{1}},
 \bibinfo{pages}{257}.



\bibitem[{\citenamefont{Sagi \emph{et~al.}}(2005)}]
{Sagi:2005}
\bibinfo{author}{\bibnamefont{Sagi},~\bibfnamefont{E.}},
 \bibinfo{author}{\bibfnamefont{O.}~\bibnamefont{Ofer}},
  \bibinfo{author}{\bibfnamefont{A.}~\bibnamefont{Keren}}, and
  \bibinfo{author}{\bibfnamefont{J. S.} ~\bibnamefont{Gardner}},
  \bibinfo{year}{2005},
  \bibinfo{journal}{Phys. Rev. Lett.} \textbf{\bibinfo{volume}{94}},
  \bibinfo{pages}{237202}.




\bibitem[{\citenamefont{Saha \emph{et~al.}}(2006)}]
{Saha:2006}
\bibinfo{author}{\bibnamefont{Saha}, \bibfnamefont{S.}},
  \bibinfo{author}{\bibfnamefont{D. V. S.}~\bibnamefont{Muthu}},
  \bibinfo{author}{\bibfnamefont{C.}~\bibnamefont{Pascanut}},
  \bibinfo{author}{\bibfnamefont{N.}~\bibnamefont{Dragoe}},
  \bibinfo{author}{\bibfnamefont{R.}~\bibnamefont{Suryanarayanan}},
  \bibinfo{author}{\bibfnamefont{G.}~\bibnamefont{Dhalenne}},
  \bibinfo{author}{\bibfnamefont{A.}~\bibnamefont{Revcolevschi}},
  \bibinfo{author}{\bibfnamefont{S. M.}~\bibnamefont{Karmakar}},
  \bibinfo{author}{\bibfnamefont{S. }~\bibnamefont{Sharma}}, and
  \bibinfo{author}{\bibfnamefont{A. K.}~\bibnamefont{Sood}},
  \bibinfo{year}{2006},
  \bibinfo{journal}{Phys. Rev. B.} \textbf{\bibinfo{volume}{74}},
  \bibinfo{pages}{064109}.




\bibitem[{\citenamefont{Sakai \emph{et~al.}}(2001)}]
{Sakai:2001}
\bibinfo{author}{\bibnamefont{Sakai}, \bibfnamefont{H.}},
  \bibinfo{author}{\bibfnamefont{K.}~\bibnamefont{Yoshimura}},
  \bibinfo{author}{\bibfnamefont{H.}~\bibnamefont{Ohno}},
  \bibinfo{author}{\bibfnamefont{H.}~\bibnamefont{Kato}},
  \bibinfo{author}{\bibfnamefont{S.}~\bibnamefont{Kambe}},
  \bibinfo{author}{\bibfnamefont{R. E.}~\bibnamefont{Walstedt}},
  \bibinfo{author}{\bibfnamefont{T. D.}~\bibnamefont{Matsuda}}, and
  \bibinfo{author}{\bibfnamefont{Y.}~\bibnamefont{Haga}},
  \bibinfo{year}{2001},
  \bibinfo{journal}{J. Phys.: Condens. Matter} \textbf{\bibinfo{volume}{13}},
  \bibinfo{pages}{L785}.



%\bibitem[{\citenamefont{Sakai \emph{et~al.}}(2004)}]
%{Sakai:2004}
%\bibinfo{author}{\bibnamefont{Sakai}, \bibfnamefont{H.}},
% \bibinfo{author}{\bibfnamefont{Y.}~\bibnamefont{Tokunaga}},
%  \bibinfo{author}{\bibfnamefont{S.}~\bibnamefont{Kambe}},
%  \bibinfo{author}{\bibfnamefont{K.}~\bibnamefont{Kitagawa}},
%  \bibinfo{author}{\bibfnamefont{H.}~\bibnamefont{Murakawa}},
%  \bibinfo{author}{\bibfnamefont{K.}~\bibnamefont{Ishida}},
%  \bibinfo{author}{\bibfnamefont{H.}~\bibnamefont{Ohno}},
%  \bibinfo{author}{\bibfnamefont{M.}~\bibnamefont{Kato}},
%  \bibinfo{author}{\bibfnamefont{K.}~\bibnamefont{Yoshimura}}, and
%  \bibinfo{author}{\bibfnamefont{R. E.}~\bibnamefont{Walstedt}},
%  \bibinfo{year}{2004},
%  \bibinfo{journal}{J. Phys. Soc. Jpn. } \textbf{\bibinfo{volume}{73}},
%  \bibinfo{pages}{2940}.

%



\bibitem[{\citenamefont{Sakakibara \emph{et~al.}}(2004)}]
{Sakakibara:2004}
\bibinfo{author}{\bibnamefont{Sakakibara}, \bibfnamefont{T.}},
  \bibinfo{author}{\bibfnamefont{T.}~\bibnamefont{Tayama}},
  \bibinfo{author}{\bibfnamefont{K.}~\bibnamefont{Matsuhira}},
  \bibinfo{author}{\bibfnamefont{S.}~\bibnamefont{Takagi}}, and
  \bibinfo{author}{\bibfnamefont{Z.}~\bibnamefont{Hiroi}},
  \bibinfo{year}{2004},
  \bibinfo{journal}{J. Magn. Mag. Matter.} \textbf{\bibinfo{volume}{272}},
  \bibinfo{pages}{1312}.



\bibitem[{\citenamefont{Samara \emph{et~al.}}(2006)}]
{Samara:2006}
\bibinfo{author}{\bibnamefont{Samara}, \bibfnamefont{G. A.}},
  \bibinfo{author}{\bibfnamefont{E. L.}~\bibnamefont{Venturini}}, and
  \bibinfo{author}{\bibfnamefont{L. A.}~\bibnamefont{Boatner}},
  \bibinfo{year}{2006},
  \bibinfo{journal}{J. Appl. Phys.} \textbf{\bibinfo{volume}{100}},
  \bibinfo{pages}{074112}.




\bibitem[{\citenamefont{Sato \emph{et~al.}}(1986)}]
{Sato:1986}
\bibinfo{author}{\bibnamefont{Sato}, \bibfnamefont{M.}},
 \bibinfo{author}{\bibfnamefont{X.}~\bibnamefont{Yan}}, and
 \bibinfo{author}{\bibfnamefont{J. E.}~\bibnamefont{Greedan}},
\bibinfo{year}{1986},
\bibinfo{journal}{Z. Anorg. Allg. Chem.} \textbf{\bibinfo{volume}{540}},
\bibinfo{pages}{177}.




\bibitem[{\citenamefont{Sato and Greedan}(1987)}]
{Sato:1987}
\bibinfo{author}{\bibnamefont{Sato},~\bibfnamefont{M.}} and
\bibinfo{author}{\bibfnamefont{J. E.}~\bibnamefont{Greedan}},
 \bibinfo{year}{1987},
 \bibinfo{journal}{J. Sol. State Chem.} \textbf{\bibinfo{volume}{67}},
 \bibinfo{pages}{248}.




\bibitem[{\citenamefont{Saunders and Chalker}(2007)}]
{Saunders:2007}
 \bibinfo{author}{\bibfnamefont{Saunders}, \bibnamefont{T. E.}} and
  \bibinfo{author}{\bibnamefont{J. T.}~\bibfnamefont{Chalker}},
 \bibinfo{year}{2007},
  \bibinfo{journal}{Phys. Rev. Lett.} \textbf{\bibinfo{volume}{98}},
  \bibinfo{pages}{157201}.




\bibitem[{\citenamefont{Schnelle and Kremer}(2004)}]
{Schnelle:2004}
\bibinfo{author}{\bibnamefont{Schnelle},~\bibfnamefont{W.}} and
\bibinfo{author}{\bibfnamefont{R. K.}~\bibnamefont{Kremer}},
 \bibinfo{year}{2004},
 \bibinfo{journal}{J. Phys. Condens. Matter} \textbf{\bibinfo{volume}{16}},
 \bibinfo{pages}{S685}.



\bibitem[{\citenamefont{Schuck \emph{et~al.}}(2006)}]
{Schuck:2006}
\bibinfo{author}{\bibnamefont{Schuck}, \bibfnamefont{G.}},
  \bibinfo{author}{\bibfnamefont{S. M.}~\bibnamefont{Kazakov}},
  \bibinfo{author}{\bibfnamefont{K.}~\bibnamefont{Rogacki}},
  \bibinfo{author}{\bibfnamefont{N. D.}~\bibnamefont{Zhigadlo}}, and
  \bibinfo{author}{\bibfnamefont{J. }~\bibnamefont{Karpinski}},
  \bibinfo{year}{2006},
  \bibinfo{journal}{Phys. Rev. B} \textbf{\bibinfo{volume}{73}},
  \bibinfo{pages}{144506}.




\bibitem[{\citenamefont{Schiffer} \emph{et~al.}(1994)}]
{Schiffer:1994}
  \bibinfo{author}{\bibfnamefont{Schiffer}, \bibnamefont{P.}},
  \bibinfo{author}{\bibnamefont{A. P.}~\bibfnamefont{Ramirez}},
  \bibinfo{author}{\bibnamefont{D. A.}~\bibfnamefont{Huse}},  and
    \bibinfo{author}{\bibnamefont{A. J.}~\bibfnamefont{Valentino}},
 \bibinfo{year}{1994},
  \bibinfo{journal}{Phys. Rev. Lett.} \textbf{\bibinfo{volume}{73}},
  \bibinfo{pages}{2500}.



\bibitem[{\citenamefont{Schiffer \emph{et~al.}}(1995)}]
{Schiffer:1995}
\bibinfo{author}{\bibnamefont{Schiffer},~\bibfnamefont{P.}},
\bibinfo{author}{\bibfnamefont{A. P.}~\bibnamefont{Ramirez}},
\bibinfo{author}{\bibnamefont{W.},~\bibfnamefont{Bao}}, and
\bibinfo{author}{\bibfnamefont{S.-W.}~\bibnamefont{Cheong}},
 \bibinfo{year}{1995},
 \bibinfo{journal}{Phys. Rev. Lett.} \textbf{\bibinfo{volume}{75}},
 \bibinfo{pages}{3336}.

\bibitem[{\citenamefont{Schiffer \emph{et~al.}}(1995a)}]
{Schiffer:1995a}
\bibinfo{author}{\bibnamefont{Schiffer},~\bibfnamefont{P.}},
\bibinfo{author}{\bibfnamefont{A. P.}~\bibnamefont{Ramirez}},
\bibinfo{author}{\bibnamefont{D. H.},~\bibfnamefont{Huse}}, 
\bibinfo{author}{\bibfnamefont{P. L.}~\bibnamefont{Gammel}},
\bibinfo{author}{\bibnamefont{U.},~\bibfnamefont{Yaron}}, 
\bibinfo{author}{\bibfnamefont{D. J.}~\bibnamefont{Bishop}}, and
\bibinfo{author}{\bibfnamefont{A. J.}~\bibnamefont{Valentino}},
 \bibinfo{year}{1995a},
 \bibinfo{journal}{Phys. Rev. Lett.} \textbf{\bibinfo{volume}{74}},
 \bibinfo{pages}{2379}.


\bibitem[{\citenamefont{Schiffer and Ramirez}(1996)}]
{Schiffer:1996}
\bibinfo{author}{\bibnamefont{Schiffer},~\bibfnamefont{P.}}, and
\bibinfo{author}{\bibfnamefont{A. P.}~\bibnamefont{Ramirez}},
 \bibinfo{year}{1996},
 \bibinfo{journal}{Comments Condens. Matter Phys.} \textbf{\bibinfo{volume}{18}},
 \bibinfo{pages}{21}.

\bibitem[{\citenamefont{Schnelle and Kremer}(2004)}]
{Schnell:2004}
\bibinfo{author}{\bibnamefont{Schnelle},~\bibfnamefont{W.}}, and
\bibinfo{author}{\bibfnamefont{R. K.}~\bibnamefont{Kremer}},
 \bibinfo{year}{2004},
 \bibinfo{journal}{J. Phys.: Condens. Matter} \textbf{\bibinfo{volume}{16}},
 \bibinfo{pages}{S685}.




%\bibitem[{\citenamefont{Sergienko \emph{et~al.}}(2004)}]
%{Sergienko:2004}
%\bibinfo{author}{\bibnamefont{Sergienko}, \bibfnamefont{I. A.}},
%  \bibinfo{author}{\bibfnamefont{V.}~\bibnamefont{Keppens}},
%  \bibinfo{author}{\bibfnamefont{M.}~\bibnamefont{McGuire}},
%  \bibinfo{author}{\bibfnamefont{R.}~\bibnamefont{Jin}},
%  \bibinfo{author}{\bibfnamefont{S. H.}~\bibnamefont{Curnoe}},
%  \bibinfo{author}{\bibfnamefont{B. C.}~\bibnamefont{Sales}},
%  \bibinfo{author}{\bibfnamefont{P.}~\bibnamefont{Blaha}},
%  \bibinfo{author}{\bibfnamefont{D.}~\bibnamefont{Singh}},
%  \bibinfo{author}{\bibfnamefont{K.}~\bibnamefont{Schwarz}}, and
%  \bibinfo{author}{\bibfnamefont{D.}~\bibnamefont{Mandrus}},
%  \bibinfo{year}{2004},
%  \bibinfo{journal}{Phys. Rev. Lett.} \textbf{\bibinfo{volume}{92}},
%  \bibinfo{pages}{065501}.



\bibitem[{\citenamefont{Shender} \emph{et~al.}(1993)}]
{Shender:1993}
  \bibinfo{author}{\bibfnamefont{Shender}, \bibnamefont{E. F.}},
  \bibinfo{author}{\bibnamefont{V. P.}~\bibfnamefont{Cherepanov}},
 \bibinfo{author}{\bibnamefont{P. C. W.}~\bibfnamefont{Holdsworth}}, and
\bibinfo{author}{\bibnamefont{A. J.}~\bibfnamefont{Berlinsky}},
 \bibinfo{year}{1993},
  \bibinfo{journal}{Phys. Rev. Lett.} \textbf{\bibinfo{volume}{70}},
  \bibinfo{pages}{3812}.



\bibitem[{\citenamefont{Sherrington and Southern}(1975)}]
{Sherrington:1975}
\bibinfo{author}{\bibnamefont{Sherrington},~\bibfnamefont{D.}} and
\bibinfo{author}{\bibfnamefont{B. W.}~\bibnamefont{Southern}},
 \bibinfo{year}{1975},
 \bibinfo{journal}{J. Phys. F: Met. Phys.} \textbf{\bibinfo{volume}{5}},
 \bibinfo{pages}{L49}.




\bibitem[{\citenamefont{Shi} \emph{et~al.}(2007)}]
{Shi:2007}
  \bibinfo{author}{\bibfnamefont{Shi}, \bibnamefont{J.}},
  \bibinfo{author}{\bibnamefont{Z.}~\bibfnamefont{Tang}},
  \bibinfo{author}{\bibnamefont{B. P.}~\bibfnamefont{Zhu}},
  \bibinfo{author}{\bibnamefont{P.}~\bibfnamefont{Huang}},
  \bibinfo{author}{\bibnamefont{D.}~\bibfnamefont{Yin}},
\bibinfo{author}{\bibnamefont{C. Z.}~\bibfnamefont{Li}},
  \bibinfo{author}{\bibnamefont{Y.}~\bibfnamefont{Wang}}, and
\bibinfo{author}{\bibnamefont{H.}~\bibfnamefont{Wen}},
 \bibinfo{year}{2007},
  \bibinfo{journal}{J. Magn. Mag. Mate.} \textbf{\bibinfo{volume}{310}},
  \bibinfo{pages}{1322}.


\bibitem[{\citenamefont{Shiga \emph{et~al.}}(1993)}]
{Shiga:1993}
\bibinfo{author}{\bibnamefont{Shiga},~\bibfnamefont{M.}},
 \bibinfo{author}{\bibfnamefont{K.}~\bibnamefont{Fujisawa}}, and
  \bibinfo{author}{\bibfnamefont{H.}~\bibnamefont{Wada}},
\bibinfo{year}{1993},
\bibinfo{journal}{J. Phys. Soc. Jpn.} \textbf{\bibinfo{volume}{62}},
\bibinfo{pages}{1329}.




\bibitem[{\citenamefont{Shimakawa \emph{et~al.}}(1996)}]
{Shimakawa:1996}
\bibinfo{author}{\bibnamefont{Shimakawa},~\bibfnamefont{Y.}},
 \bibinfo{author}{\bibfnamefont{Y.}~\bibnamefont{Kubo}}, and
  \bibinfo{author}{\bibfnamefont{T.}~\bibnamefont{Manako}},
  \bibinfo{year}{1996},
\bibinfo{journal}{Nature} \textbf{\bibinfo{volume}{379}},
\bibinfo{pages}{53}.




\bibitem[{\citenamefont{Shimakawa \emph{et~al.}}(1999)}]
{Shimakawa:1999}
\bibinfo{author}{\bibnamefont{Shimakawa},~\bibfnamefont{Y.}},
 \bibinfo{author}{\bibfnamefont{Y.}~\bibnamefont{Kubo}},
 \bibinfo{author}{\bibfnamefont{N.}~\bibnamefont{Hamada}},
 \bibinfo{author}{\bibfnamefont{J. D.}~\bibnamefont{Jorgensen}},
  \bibinfo{author}{\bibfnamefont{Z.}~\bibnamefont{Hu}},
  \bibinfo{author}{\bibfnamefont{S.}~\bibnamefont{Short}},
  \bibinfo{author}{\bibfnamefont{M.}~\bibnamefont{Nohara}}, and
  \bibinfo{author}{\bibfnamefont{H.}~\bibnamefont{Takagi}},
\bibinfo{year}{1999},
\bibinfo{journal}{Phys. Rev. B} \textbf{\bibinfo{volume}{59}},
\bibinfo{pages}{1249}.



\bibitem[{\citenamefont{Shimizu} \emph{et~al.}(2006)}]
{Shimizu:2006}
  \bibinfo{author}{\bibfnamefont{Shimizu}, \bibnamefont{Y.}},
  \bibinfo{author}{\bibnamefont{K.}~\bibfnamefont{Miyagawa}},
  \bibinfo{author}{\bibnamefont{K.}~\bibfnamefont{Kanoda}},
  \bibinfo{author}{\bibnamefont{M.}~\bibfnamefont{Mitsuhiko}},  and
  \bibinfo{author}{\bibnamefont{M.}~\bibfnamefont{Saito}},
  \bibinfo{year}{2006},
  \bibinfo{journal}{Phys. Rev. B} \textbf{\bibinfo{volume}{73}},
  \bibinfo{pages}{140407(R)}.




\bibitem[{\citenamefont{Shin-ike \emph{et~al.}}(1977)}]
{Shinike:1977}
\bibinfo{author}{\bibnamefont{Shin-ike},~\bibfnamefont{T.}},
 \bibinfo{author}{\bibfnamefont{G.}~\bibnamefont{Adachi}}, and
  \bibinfo{author}{\bibfnamefont{J.}~\bibnamefont{Shiokawa}},
\bibinfo{year}{1977},
\bibinfo{journal}{Mat. Res. Bull.} \textbf{\bibinfo{volume}{12}},
\bibinfo{pages}{1149}.



 \bibitem[{\citenamefont{Shirai and Bramwell}(2007)}]
{Shirai:2007}
\bibinfo{author}{\bibnamefont{Shirai},~\bibfnamefont{M.}} and
\bibinfo{author}{\bibfnamefont{S. T.}~\bibnamefont{Bramwell}},
 \bibinfo{year}{2007}, Thesis Chapter, UCL and private communication.




\bibitem[{\citenamefont{Shlyakhtina \emph{et~al.}}(2004)}]
{Shlyakhtina:2004}
\bibinfo{author}{\bibnamefont{Shlyakhtina},~\bibfnamefont{A. V.}},
\bibinfo{author}{\bibnamefont{L. G.} ~\bibfnamefont{Shcherbakova}},
 \bibinfo{author}{\bibfnamefont{A. V.}~\bibnamefont{Knotko}}, and
  \bibinfo{author}{\bibfnamefont{A. V.}~\bibnamefont{Steblevskii}},
\bibinfo{year}{2004},
\bibinfo{journal}{J. Solid State Electrochem.} \textbf{\bibinfo{volume}{8}},
\bibinfo{pages}{661}.



\bibitem[{\citenamefont{Sickafus \emph{et~al.}}(2000)}]
{Sickafus:2000}
\bibinfo{author}{\bibnamefont{Sickafus}, \bibfnamefont{K. E.}},
  \bibinfo{author}{\bibfnamefont{L.}~\bibnamefont{Minervini}},
  \bibinfo{author}{\bibfnamefont{R. W.}~\bibnamefont{Grimes}},
  \bibinfo{author}{\bibfnamefont{J. A.}~\bibnamefont{Valdez}},
  \bibinfo{author}{\bibfnamefont{M.}~\bibnamefont{Ishimaru}},
  \bibinfo{author}{\bibfnamefont{F.}~\bibnamefont{Li}},
 \bibinfo{author}{\bibfnamefont{K. J.}~\bibnamefont{McClellan}}, and
  \bibinfo{author}{\bibfnamefont{T.}~\bibnamefont{Hartmann}},
  \bibinfo{year}{2000},
  \bibinfo{journal}{Science} \textbf{\bibinfo{volume}{289}},
  \bibinfo{pages}{748}.




\bibitem[{\citenamefont{Siddharthan} \emph{et~al.}(1999)}]
{Siddharthan:1999}
  \bibinfo{author}{\bibfnamefont{Siddharthan}, \bibnamefont{R.}},
  \bibinfo{author}{\bibnamefont{B. S.}~\bibfnamefont{Shastry}},
  \bibinfo{author}{\bibnamefont{A. P.}~\bibfnamefont{Ramirez}},
  \bibinfo{author}{\bibnamefont{A. }~\bibfnamefont{Hayashi.}},
  \bibinfo{author}{\bibnamefont{R. J.}~\bibfnamefont{Cava}}, and
  \bibinfo{author}{\bibnamefont{S.}~\bibfnamefont{Rosenkranz}},
 \bibinfo{year}{1999},
  \bibinfo{journal}{Phys. Rev. Lett.} \textbf{\bibinfo{volume}{83}},
  \bibinfo{pages}{1854}.


%
%\bibitem[{\citenamefont{Singh and Huse}(1992)}]
%{Singh:1992}
%  \bibinfo{author}{\bibfnamefont{Singh}, \bibnamefont{R. R. P.}},  and
%  \bibinfo{author}{\bibnamefont{D. A.}~\bibfnamefont{Huse}},
% \bibinfo{year}{1992},
%  \bibinfo{journal}{Phys. Rev. Lett.},\textbf{\bibinfo{volume}{68}},
%  \bibinfo{pages}{1766}.



%\bibitem[{\citenamefont{Singh and Huse}(2007)}]
%{Singh:2007}
%  \bibinfo{author}{\bibfnamefont{Singh}, \bibnamefont{R. R. P.}}  and
%  \bibinfo{author}{\bibnamefont{D. A.}~\bibfnamefont{Huse}},
% \bibinfo{year}{2007},
%\bibinfo{journal}{Phys. Rev. B},\textbf{\bibinfo{volume}{76}},
% \bibinfo{pages}{180407}.




 \bibitem[{\citenamefont{Sleight and Bouchard}(1972)}]
{Sleight:1972}
\bibinfo{author}{\bibnamefont{Sleight},~\bibfnamefont{A. W.}}, and
\bibinfo{author}{\bibfnamefont{P. J.}~\bibnamefont{Bouchard}},
 \bibinfo{year}{1972},
 \bibinfo{journal}{Proc. 5th Mater. Res. Symp.} \textbf{\bibinfo{volume}{364}},
 \bibinfo{pages}{227}.


\bibitem[{\citenamefont{Sleight} \emph{et~al.}(1974)}]
{Sleight:1974}
\bibinfo{author}{\bibnamefont{Sleight},~\bibfnamefont{A. W.}},
\bibinfo{author}{\bibfnamefont{J. L.}~\bibnamefont{Gillson}},
\bibinfo{author}{\bibfnamefont{J. F.}~\bibnamefont{Weiher}}, and
\bibinfo{author}{\bibfnamefont{W.}~\bibnamefont{Bindloss}},
 \bibinfo{year}{1974},
 \bibinfo{journal}{Solid State Commun.} \textbf{\bibinfo{volume}{14}},
 \bibinfo{pages}{357}.


\bibitem[{\citenamefont{Snyder} \emph{et~al.}(2001)}]
{Snyder:2001}
  \bibinfo{author}{\bibfnamefont{Snyder}, \bibnamefont{J.}},
  \bibinfo{author}{\bibnamefont{J. S.}~\bibfnamefont{Slusky}},
  \bibinfo{author}{\bibnamefont{R. J.}~\bibfnamefont{Cava}},  and
  \bibinfo{author}{\bibnamefont{P.}~\bibfnamefont{Schiffer}},
 \bibinfo{year}{2001},
  \bibinfo{journal}{Nature } \textbf{\bibinfo{volume}{413}},
  \bibinfo{pages}{48}.


%Dirty spin ice: The effect of dilution on spin freezing in Dy2Ti2O7

\bibitem[{\citenamefont{Snyder} \emph{et~al.}(2002)}]
{Snyder:2002}
  \bibinfo{author}{\bibfnamefont{Snyder}, \bibnamefont{J.}},
  \bibinfo{author}{\bibnamefont{J. S.}~\bibfnamefont{Slusky}},
  \bibinfo{author}{\bibnamefont{R. J.}~\bibfnamefont{Cava}},  and
  \bibinfo{author}{\bibnamefont{P.}~\bibfnamefont{Schiffer}},
  \bibinfo{year}{2002},
  \bibinfo{journal}{Phys. Rev. B} \textbf{\bibinfo{volume}{66}},
  \bibinfo{pages}{064432}.




%Quantum-Classical Reentrant Relaxation Crossover in Dy2Ti2O7 Spin Ice

\bibitem[{\citenamefont{Snyder} \emph{et~al.}(2003)}]
{Snyder:2003}
  \bibinfo{author}{\bibfnamefont{Snyder}, \bibnamefont{J.}},
  \bibinfo{author}{\bibnamefont{B. G.}~\bibfnamefont{Ueland}},
  \bibinfo{author}{\bibnamefont{J. S.}~\bibfnamefont{Slusky}},
  \bibinfo{author}{\bibnamefont{H.}~\bibfnamefont{Karunadasa}},
  \bibinfo{author}{\bibnamefont{R. J.}~\bibfnamefont{Cava}},
  \bibinfo{author}{\bibnamefont{A.}~\bibfnamefont{Mizel}},  and
  \bibinfo{author}{\bibnamefont{P.}~\bibfnamefont{Schiffer}},
  \bibinfo{year}{2003},
  \bibinfo{journal}{Phys. Rev. Lett.} \textbf{\bibinfo{volume}{91}},
  \bibinfo{pages}{107201}.



%Low-temperature spin freezing in the Dy2Ti2O7 spin ice

\bibitem[{\citenamefont{Snyder} \emph{et~al.}(2004)}]
{Snyder:2004}
  \bibinfo{author}{\bibfnamefont{Snyder}, \bibnamefont{J.}},
  \bibinfo{author}{\bibnamefont{B. G.}~\bibfnamefont{Ueland}},
  \bibinfo{author}{\bibnamefont{J. S.}~\bibfnamefont{Slusky}},
  \bibinfo{author}{\bibnamefont{H.}~\bibfnamefont{Karunadasa}},
  \bibinfo{author}{\bibnamefont{R. J.}~\bibfnamefont{Cava}},  and
  \bibinfo{author}{\bibnamefont{P.}~\bibfnamefont{Schiffer}},
  \bibinfo{year}{2004},
  \bibinfo{journal}{Phys. Rev. B.} \textbf{\bibinfo{volume}{69}},
  \bibinfo{pages}{064414}.



%Quantum and thermal spin relaxation in the diluted spin ice Dy2?xMxTi2O7 (M=Lu,Y)

\bibitem[{\citenamefont{Snyder} \emph{et~al.}(2004a)}]
{Snyder:2004a}
  \bibinfo{author}{\bibfnamefont{Snyder}, \bibnamefont{J.}},
  \bibinfo{author}{\bibnamefont{B. G.}~\bibfnamefont{Ueland}},
  \bibinfo{author}{\bibnamefont{A.}~\bibfnamefont{Mizel}},
  \bibinfo{author}{\bibnamefont{J. S.}~\bibfnamefont{Slusky}},
  \bibinfo{author}{\bibnamefont{H.}~\bibfnamefont{Karunadasa}},
  \bibinfo{author}{\bibnamefont{R. J.}~\bibfnamefont{Cava}},  and
  \bibinfo{author}{\bibnamefont{P.}~\bibfnamefont{Schiffer}},
  \bibinfo{year}{2004a},
  \bibinfo{journal}{Phys. Rev. B} \textbf{\bibinfo{volume}{70}},
  \bibinfo{pages}{184431}.



\bibitem[{\citenamefont{Soderholm \emph{et~al.}}(1980)}]
{Soderholm:1980}
\bibinfo{author}{\bibnamefont{Soderholm},~\bibfnamefont{L.}},
\bibinfo{author}{\bibnamefont{J. E.} ~\bibfnamefont{Greedan}}, and
  \bibinfo{author}{\bibfnamefont{M. F.}~\bibnamefont{Collins}},
\bibinfo{year}{1980},
\bibinfo{journal}{J. Solid State Chem.} \textbf{\bibinfo{volume}{35}},
\bibinfo{pages}{385}.




\bibitem[{\citenamefont{Solovyev}(2003)}]
{Solovyev:2003}
\bibinfo{author}{\bibnamefont{Solovyev},~\bibfnamefont{I.}},
 \bibinfo{year}{2003},
 \bibinfo{journal}{Phys. Rev. B} \textbf{\bibinfo{volume}{67}},
 \bibinfo{pages}{174406}



\bibitem[{\citenamefont{Sosin \emph{et~al.}}(2006)}]
{Sosin:2006}
\bibinfo{author}{\bibnamefont{Sosin}, \bibfnamefont{S. S.}},
 \bibinfo{author}{\bibfnamefont{A. I.}~\bibnamefont{Smirnov}},
 \bibinfo{author}{\bibfnamefont{L. A.}~\bibnamefont{Prozorova}},
   \bibinfo{author}{\bibfnamefont{G.}~\bibnamefont{Balakrishnan}},  and
 \bibinfo{author}{\bibfnamefont{M. E.}~\bibnamefont{Zhitomirsky}},
  \bibinfo{year}{2006},
  \bibinfo{journal}{Phys. Rev. B} \textbf{\bibinfo{volume}{73}},
  \bibinfo{pages}{212402}.



\bibitem[{\citenamefont{Sosin \emph{et~al.}}(2006a)}]
{Sosin:2006a}
\bibinfo{author}{\bibnamefont{Sosin}, \bibfnamefont{S. S.}},
  \bibinfo{author}{\bibfnamefont{A. I.}~\bibnamefont{Smirnov}},
  \bibinfo{author}{\bibfnamefont{L. A.}~\bibnamefont{Prozorova}},
  \bibinfo{author}{\bibfnamefont{O. A.}~\bibnamefont{Petrenko}},
  \bibinfo{author}{\bibfnamefont{M. E.}~\bibnamefont{Zhitomirsky}},  and
  \bibinfo{author}{\bibfnamefont{J. -P.}~\bibnamefont{Sanchez}},
  \bibinfo{year}{2006a},
  \bibinfo{journal}{J. Magn. Magn. Mater.} \textbf{\bibinfo{volume}{310}},
  \bibinfo{pages}{1590}.



\bibitem[{\citenamefont{Sosin \emph{et~al.}}(2007)}]
{Sosin:2007}
\bibinfo{author}{\bibnamefont{Sosin}, \bibfnamefont{S. S.}},
  \bibinfo{author}{\bibfnamefont{L. A.}~\bibnamefont{Prozorova}},
  \bibinfo{author}{\bibfnamefont{A. I.}~\bibnamefont{Smirnov}},
  \bibinfo{author}{\bibfnamefont{P.}~\bibnamefont{Bonville}},
  \bibinfo{author}{\bibfnamefont{G.}~\bibnamefont{Jasmin -Le Bras}}, and
  \bibinfo{author}{\bibfnamefont{O. A.}~\bibnamefont{Petrenko}},
  \bibinfo{year}{2007},
  \bibinfo{journal}{Cond-Mat. 0709.4379}



\bibitem[{\citenamefont{Stevens}(1952)}]
 {Stevens:1952}
 \bibinfo{author}{\bibnamefont{Stevens}, \bibfnamefont{K. W. H.}},
   \bibinfo{year}{1952},
   \bibinfo{journal}{Proc. Phys. Soc. (London)} \textbf{\bibinfo{volume}{A65}},
   \bibinfo{pages}{209}.




\bibitem[{\citenamefont{Stewart \emph{et~al.}}(2004)}]
{Stewart:2004}
\bibinfo{author}{\bibnamefont{Stewart}, \bibfnamefont{J. R.}},
  \bibinfo{author}{\bibfnamefont{G.}~\bibnamefont{Ehlers}},
  \bibinfo{author}{\bibfnamefont{A. S.}~\bibnamefont{Wills}},
  \bibinfo{author}{\bibfnamefont{S. T.}~\bibnamefont{Bramwell}},  and
  \bibinfo{author}{\bibfnamefont{J. S.}~\bibnamefont{Gardner}},
  \bibinfo{year}{2004},
  \bibinfo{journal}{J. Phys.: Condens. Matter} \textbf{\bibinfo{volume}{16}},
  \bibinfo{pages}{L321}.




\bibitem[{\citenamefont{Stewart \emph{et~al.}}(2004a)}]
{Stewart:2004a}
\bibinfo{author}{\bibnamefont{Stewart}, \bibfnamefont{J. R.}},
  \bibinfo{author}{\bibfnamefont{G.}~\bibnamefont{Ehlers}},  and
  \bibinfo{author}{\bibfnamefont{J. S.}~\bibnamefont{Gardner}},
  \bibinfo{year}{2004a}, private communication.



\bibitem[{\citenamefont{Subramanian \emph{et~al.}}(1983)}]
{Subramanian:1983}
\bibinfo{author}{\bibnamefont{Subramanian},~\bibfnamefont{M. A.}},
 \bibinfo{author}{\bibfnamefont{G.}~\bibnamefont{Aravamudan}}, and
  \bibinfo{author}{\bibfnamefont{G. V.}~\bibnamefont{Subba Rao}},
\bibinfo{year}{1983},
\bibinfo{journal}{Prog. Solid State Chem.} \textbf{\bibinfo{volume}{15}},
\bibinfo{pages}{55}.



% FERROMAGNETIC DY-LU2MN2O7, Y2MN2O7 PYROCHLORES

\bibitem[{\citenamefont{Subramanian \emph{et~al.}}(1988)}]
{Subramanian:1988}
\bibinfo{author}{\bibnamefont{Subramanian},~\bibfnamefont{M. A.}},
 \bibinfo{author}{\bibfnamefont{C. C.}~\bibnamefont{Torardi}},
 \bibinfo{author}{\bibfnamefont{D. C.}~\bibnamefont{Johnson}},
 \bibinfo{author}{\bibfnamefont{J.}~\bibnamefont{Pannetier}}, and
 \bibinfo{author}{\bibfnamefont{A. W.}~\bibnamefont{Sleight}},
\bibinfo{year}{1988},
\bibinfo{journal}{J. Solid State Chem.} \textbf{\bibinfo{volume}{72}},
\bibinfo{pages}{24}.



\bibitem[{\citenamefont{Subramanian and Sleight}(1993)}]
{Subramanian:1993}
\bibinfo{author}{\bibnamefont{Subramanian}, \bibfnamefont{M. A.}}, and
\bibinfo{author}{\bibnamefont{A. W.}, \bibfnamefont{Sleight}},
  \bibinfo{year}{1993}, in
\emph{\bibinfo{booktitle}{Handbook on the Physics and Chemistry of Rare Earths}},
  ed. \bibinfo{editor}{\bibfnamefont{K. A.}~\bibnamefont{Gschneidner}} and
  \bibinfo{editor}{\bibfnamefont{L.}~\bibnamefont{Eyring}}
  (\bibinfo{publisher}{Elsevier Science Publishers B.V.}), p. \bibinfo{pages}{225}.




\bibitem[{\citenamefont{Subramanian \emph{et~al.}}(1996)}]
{Subramanian:1996}
\bibinfo{author}{\bibnamefont{Subramanian},~\bibfnamefont{M. A.}},
 \bibinfo{author}{\bibfnamefont{B. H.}~\bibnamefont{Toby}},
 \bibinfo{author}{\bibfnamefont{A. P.}~\bibnamefont{Ramirez}},
 \bibinfo{author}{\bibfnamefont{W. J.}~\bibnamefont{Marshall}},
 \bibinfo{author}{\bibfnamefont{A. W.}~\bibnamefont{Sleight}}, and
 \bibinfo{author}{\bibfnamefont{G. H.}~\bibnamefont{Kwei}},
\bibinfo{year}{1996},
\bibinfo{journal}{J. Solid State Chem.} \textbf{\bibinfo{volume}{72}},
\bibinfo{pages}{24}.




\bibitem[{\citenamefont{Sushko \emph{et~al.}}(1996)}]
{Sushko:1996}
\bibinfo{author}{\bibnamefont{Sushko},~\bibfnamefont{Yu. V.}},
 \bibinfo{author}{\bibfnamefont{Y.}~\bibnamefont{Kubo}},
 \bibinfo{author}{\bibfnamefont{Y.}~\bibnamefont{Shimakawa}}, and
 \bibinfo{author}{\bibfnamefont{T.}~\bibnamefont{Manako}},
\bibinfo{year}{1996},
\bibinfo{journal}{Czechoslovak J. Phys.} \textbf{\bibinfo{volume}{46}},
\bibinfo{pages}{2003}.



\bibitem[{\citenamefont{Sushkov \emph{et~al.}}(2006)}]
{Sushkov:2005}
\bibinfo{author}{\bibnamefont{Sushkov},~\bibfnamefont{A. B.}},
 \bibinfo{author}{\bibfnamefont{O.}~\bibnamefont{Tchernyshyov}},
  \bibinfo{author}{\bibfnamefont{W.}~\bibnamefont{Ratcliff II}},
 \bibinfo{author}{\bibfnamefont{S. W.}~\bibnamefont{Cheong}}, and
 \bibinfo{author}{\bibfnamefont{H. D.}~\bibnamefont{Drew}},
\bibinfo{year}{2005},
\bibinfo{journal}{Phys. Rev Lett.} \textbf{\bibinfo{volume}{94}},
\bibinfo{pages}{137202}.


%
%\bibitem[{\citenamefont{Syozi}(1951)}]
%{Syozi:1951}
%  \bibinfo{author}{\bibfnamefont{Syozi}, \bibnamefont{I.}},
%  \bibinfo{year}{1951},
%  \bibinfo{journal}{Prog. Theor. Phys. } \textbf{\bibinfo{volume}{6}},
%  \bibinfo{pages}{306}.




\bibitem[{\citenamefont{Tabata \emph{et~al.}}(2006)}]
{Tabata:2006}
\bibinfo{author}{\bibnamefont{Tabata}, \bibfnamefont{Y.}},
  \bibinfo{author}{\bibfnamefont{H.}~\bibnamefont{Kadowaki}},
  \bibinfo{author}{\bibfnamefont{K.}~\bibnamefont{Matsuhira}},
  \bibinfo{author}{\bibfnamefont{Z.}~\bibnamefont{Hiroi}},
  \bibinfo{author}{\bibfnamefont{N.}~\bibnamefont{Aso}},
  \bibinfo{author}{\bibfnamefont{E.}~\bibnamefont{Ressouche}}, and
  \bibinfo{author}{\bibfnamefont{B.}~\bibnamefont{F\aa k}},
  \bibinfo{year}{2006},
  \bibinfo{journal}{Phys. Rev. Lett.} \textbf{\bibinfo{volume}{97}},
  \bibinfo{pages}{257205}.



\bibitem[{\citenamefont{Taguchi and Tokura}(1999)}]
{Taguchi:1999}
\bibinfo{author}{\bibnamefont{Taguchi},~\bibfnamefont{Y.}} and
\bibinfo{author}{\bibfnamefont{Y.}~\bibnamefont{Tokura}},
 \bibinfo{year}{1999},
 \bibinfo{journal}{Phys. Rev. B} \textbf{\bibinfo{volume}{60}},
 \bibinfo{pages}{10280}.



 \bibitem[{\citenamefont{Taguchi and Tokura}(2000)}]
{Taguchi:2000}
\bibinfo{author}{\bibnamefont{Taguchi},~\bibfnamefont{Y.}} and
\bibinfo{author}{\bibfnamefont{Y.}~\bibnamefont{Tokura}},
 \bibinfo{year}{2000},
 \bibinfo{journal}{Physica B} \textbf{\bibinfo{volume}{284}},
 \bibinfo{pages}{1448}.



\bibitem[{\citenamefont{Taguchi \emph{et~al.}}(2001)}]
{Taguchi:2001}
\bibinfo{author}{\bibnamefont{Taguchi}, \bibfnamefont{Y.}},
  \bibinfo{author}{\bibfnamefont{Y.}~\bibnamefont{Oohara}},
  \bibinfo{author}{\bibfnamefont{H.}~\bibnamefont{Yoshizawa}},
  \bibinfo{author}{\bibfnamefont{N.}~\bibnamefont{Nagaosa}}, and
  \bibinfo{author}{\bibfnamefont{Y.}~\bibnamefont{Tokura}},
  \bibinfo{year}{2001},
  \bibinfo{journal}{Science} \textbf{\bibinfo{volume}{291}},
  \bibinfo{pages}{2573}.



  \bibitem[{\citenamefont{Taguchi \emph{et~al.}}(2002)}]
{Taguchi:2002}
\bibinfo{author}{\bibnamefont{Taguchi}, \bibfnamefont{Y.}},
  \bibinfo{author}{\bibfnamefont{K.}~\bibnamefont{Ohgushi}}, and
  \bibinfo{author}{\bibfnamefont{Y.}~\bibnamefont{Tokura}},
  \bibinfo{year}{2002},
  \bibinfo{journal}{Phys. Rev. B} \textbf{\bibinfo{volume}{65}},
  \bibinfo{pages}{115102}.



\bibitem[{\citenamefont{Taguchi \emph{et~al.}}(2004)}]
{Taguchi:2004}
\bibinfo{author}{\bibnamefont{Taguchi}, \bibfnamefont{Y.}},
  \bibinfo{author}{\bibfnamefont{Y. }~\bibnamefont{Oohara}},
  \bibinfo{author}{\bibfnamefont{H. }~\bibnamefont{Yoshizawa}},
  \bibinfo{author}{\bibfnamefont{N. }~\bibnamefont{Nagaosa}},
  \bibinfo{author}{\bibfnamefont{T.}~\bibnamefont{Sasaki}},
  \bibinfo{author}{\bibfnamefont{S.}~\bibnamefont{Awaji}},
  \bibinfo{author}{\bibfnamefont{Y.}~\bibnamefont{Iwasa}},
  \bibinfo{author}{\bibfnamefont{T.}~\bibnamefont{Tayama}},
  \bibinfo{author}{\bibfnamefont{T.}~\bibnamefont{Sakakibara}},
  \bibinfo{author}{\bibfnamefont{S.}~\bibnamefont{Iguchi}},
  \bibinfo{author}{\bibfnamefont{K.}~\bibnamefont{Ohgushi}},
  \bibinfo{author}{\bibfnamefont{T.}~\bibnamefont{Ito}}, and
  \bibinfo{author}{\bibfnamefont{Y.}~\bibnamefont{Tokura}},
  \bibinfo{year}{2004},
  \bibinfo{journal}{J. Phys.: Condens. Matter} \textbf{\bibinfo{volume}{16}},
  \bibinfo{pages}{S599}.



\bibitem[{\citenamefont{Taira \emph{et~al.}}(1999)}]
{Taira:1999}
\bibinfo{author}{\bibnamefont{Taira}, \bibfnamefont{N.}},
  \bibinfo{author}{\bibfnamefont{M.}~\bibnamefont{Wakeshima}}, and
  \bibinfo{author}{\bibfnamefont{Y.}~\bibnamefont{Hinatsu}},
  \bibinfo{year}{1999},
  \bibinfo{journal}{J. Solid State Chem.} \textbf{\bibinfo{volume}{144}},
  \bibinfo{pages}{216}.



\bibitem[{\citenamefont{Taira \emph{et~al.}}(2001)}]
{Taira:2001}
\bibinfo{author}{\bibnamefont{Taira}, \bibfnamefont{N.}},
  \bibinfo{author}{\bibfnamefont{M.}~\bibnamefont{Wakeshima}}, and
  \bibinfo{author}{\bibfnamefont{Y.}~\bibnamefont{Hinatsu}},
  \bibinfo{year}{2001},
  \bibinfo{journal}{J. Phys.: Condens. Matter} \textbf{\bibinfo{volume}{13}},
  \bibinfo{pages}{5527}.



\bibitem[{\citenamefont{Taira \emph{et~al.}}(2003)}]
{Taira:2003}
\bibinfo{author}{\bibnamefont{Taira}, \bibfnamefont{N.}},
  \bibinfo{author}{\bibfnamefont{Makoto}~\bibnamefont{W.}},
  \bibinfo{author}{\bibfnamefont{Yukio.}~\bibnamefont{H.}},
  \bibinfo{author}{\bibfnamefont{Aya}~\bibnamefont{T.}}, and
  \bibinfo{author}{\bibfnamefont{Kenji}~\bibnamefont{O.}},
  \bibinfo{year}{2003},
  \bibinfo{journal}{J. Solid State Chem.} \textbf{\bibinfo{volume}{176}},
  \bibinfo{pages}{165}.




\bibitem[{\citenamefont{Tang \emph{et~al.}}(2006)}]
{Tang:2006}
\bibinfo{author}{\bibnamefont{Tang}, \bibfnamefont{Z.}},
  \bibinfo{author}{\bibfnamefont{C. Z.}~\bibnamefont{Li}},
  \bibinfo{author}{\bibfnamefont{D.}~\bibnamefont{Yin}},
  \bibinfo{author}{\bibfnamefont{B. P.}~\bibnamefont{Zhu}},
  \bibinfo{author}{\bibfnamefont{L. L.}~\bibnamefont{Wang}},
  \bibinfo{author}{\bibfnamefont{J. F.}~\bibnamefont{Wang}},
  \bibinfo{author}{\bibfnamefont{R.}~\bibnamefont{Xiong}},
  \bibinfo{author}{\bibfnamefont{Q. Q.}~\bibnamefont{Wang}}, and
  \bibinfo{author}{\bibfnamefont{J.}~\bibnamefont{Shi}},
  \bibinfo{year}{2006},
  \bibinfo{journal}{Acta Physica Sinica} \textbf{\bibinfo{volume}{55}},
  \bibinfo{pages}{6532}.




\bibitem[{\citenamefont{Tchernyshyov} \emph{et~al.}(2002)}]
{Tchernyshyov:2002}
  \bibinfo{author}{\bibfnamefont{Tchernyshyov}, \bibnamefont{O.}},
  \bibinfo{author}{\bibfnamefont{R.} \bibnamefont{Moessner}}, and
  \bibinfo{author}{\bibnamefont{S. L.}~\bibfnamefont{Sondhi}},
  \bibinfo{year}{2002},
  \bibinfo{journal}{Phys. Rev. B.} \textbf{\bibinfo{volume}{66}},
  \bibinfo{pages}{064403}.




\bibitem[{\citenamefont{Tchernyshyov} \emph{et~al.}(2006)}]
{Tchernyshyov:2006}
  \bibinfo{author}{\bibfnamefont{Tchernyshyov}, \bibnamefont{O.}},
\bibinfo{author}{\bibnamefont{R.}~\bibfnamefont{Moessner}}, and
 \bibinfo{author}{\bibnamefont{S. L.}~\bibfnamefont{Sondhi}},
 \bibinfo{year}{2006},
  \bibinfo{journal}{Europhys. Lett.} \textbf{\bibinfo{volume}{73}},
  \bibinfo{pages}{278}.



\bibitem[{\citenamefont{Toulouse}(1977)}]
{Toulouse:1977}
  \bibinfo{author}{\bibfnamefont{Toulouse}, \bibnamefont{G.}},
  \bibinfo{year}{1977},
  \bibinfo{journal}{Commun. Phys.} \textbf{\bibinfo{volume}{2}},
  \bibinfo{pages}{115}.




\bibitem[{\citenamefont{Troyanchuk  \emph{et~al.}}(1988)}]
{Troyanchuk:1988}
\bibinfo{author}{\bibnamefont{Troyanchuk},~\bibfnamefont{I. O.}}, and
\bibinfo{author}{\bibfnamefont{V. N.}~\bibnamefont{Derkachenko}},
 \bibinfo{year}{1988},
 \bibinfo{journal}{Sov. Phys. Solid State} \textbf{\bibinfo{volume}{30}},
 \bibinfo{pages}{2003}.


\bibitem[{\citenamefont{Troyanchuk}(1990)}]
{Troyanchuk:1990}
\bibinfo{author}{\bibnamefont{Troyanchuk},~\bibfnamefont{I. O.}},
\bibinfo{year}{1990},
 \bibinfo{journal}{Inorganic Materials} \textbf{\bibinfo{volume}{26}},
 \bibinfo{pages}{182}.



\bibitem[{\citenamefont{Troyanchuk \emph{et~al.}}(1998)}]
{Troyanchuk:1998}
\bibinfo{author}{\bibnamefont{Troyanchuk}, \bibfnamefont{I. O.}},
  \bibinfo{author}{\bibfnamefont{N. V.}~\bibnamefont{Kasper}},
  \bibinfo{author}{\bibfnamefont{D. D.}~\bibnamefont{Khalyavin}},
  \bibinfo{author}{\bibfnamefont{H.}~\bibnamefont{Szymczak}}, and
  \bibinfo{author}{\bibfnamefont{A.}~\bibnamefont{Nabialek}},
  \bibinfo{year}{1998},
  \bibinfo{journal}{Phys. Stat. Sol. A } \textbf{\bibinfo{volume}{167}},
  \bibinfo{pages}{151}.




\bibitem[{\citenamefont{Tsuneishi} \emph{et~al.}(2007)}]
{Tsuneishi:2007}
  \bibinfo{author}{\bibfnamefont{Tsuneishi}, \bibnamefont{D.}},
  \bibinfo{author}{\bibnamefont{M.}~\bibfnamefont{Ioki}},  and
  \bibinfo{author}{\bibnamefont{H.}~\bibfnamefont{Kawamura}},
  \bibinfo{year}{2007},
  \bibinfo{journal}{J. Phys.: Condens. Matter} \textbf{\bibinfo{volume}{19}},
  \bibinfo{pages}{145273}.



\bibitem[{\citenamefont{Ueda \emph{et~al.}}(2005)}]
{Ueda:2005}
\bibinfo{author}{\bibnamefont{Ueda}, \bibfnamefont{H.}},
  \bibinfo{author}{\bibfnamefont{H. A.}~\bibnamefont{Katori}},
  \bibinfo{author}{\bibfnamefont{H.}~\bibnamefont{Mitamura}},
  \bibinfo{author}{\bibfnamefont{T.}~\bibnamefont{Goto}}, and
  \bibinfo{author}{\bibfnamefont{H.}~\bibnamefont{Takagi}},
  \bibinfo{year}{2005},
  \bibinfo{journal}{Phys. Rev. Lett.} \textbf{\bibinfo{volume}{94}},
  \bibinfo{pages}{047202}.




\bibitem[{\citenamefont{Ueland \emph{et~al.}}(2006)}]
{Ueland:2006}
\bibinfo{author}{\bibnamefont{Ueland}, \bibfnamefont{B. G.}},
  \bibinfo{author}{\bibfnamefont{G. C.}~\bibnamefont{Lau}},
  \bibinfo{author}{\bibfnamefont{R. J.}~\bibnamefont{Cava}},
  \bibinfo{author}{\bibfnamefont{J. R.}~\bibnamefont{O'Brien}}, and
  \bibinfo{author}{\bibfnamefont{P.}~\bibnamefont{Schiffer}},
  \bibinfo{year}{2006},
  \bibinfo{journal}{Phys. Rev. Lett.} \textbf{\bibinfo{volume}{96}},
  \bibinfo{pages}{027216}.


\bibitem[{\citenamefont{Ueland}(2008)}]
{Schiffer:2007}
\bibinfo{author}{\bibnamefont{Ueland},~\bibfnamefont{B. G.}},
\bibinfo{author}{\bibnamefont{G. C.}~\bibfnamefont{Lau}},
\bibinfo{author}{\bibnamefont{R. S.}~\bibfnamefont{Freitas}},
\bibinfo{author}{\bibnamefont{J.}~\bibfnamefont{Snyder}},
\bibinfo{author}{\bibnamefont{M. L.}~\bibfnamefont{Dahlberg}},
\bibinfo{author}{\bibnamefont{B. D.}~\bibfnamefont{Muegge}},
\bibinfo{author}{\bibnamefont{E. L.}~\bibfnamefont{Duncan}},
\bibinfo{author}{\bibnamefont{R. J.}~\bibfnamefont{Cava}} and
\bibinfo{author}{\bibnamefont{P.}~\bibfnamefont{Schiffer}},
 \bibinfo{year}{2008},
 \bibinfo{journal}{Phys. Rev. B.} \textbf{\bibinfo{volume}{77}},
 \bibinfo{pages}{144412}.



\bibitem[{\citenamefont{van Duijn \emph{et~al.}}(2007)}]
{vanDuijn:2007}
\bibinfo{author}{\bibnamefont{van Duijn}, \bibfnamefont{J.}},
  \bibinfo{author}{\bibfnamefont{N.}~\bibnamefont{Hur}},
  \bibinfo{author}{\bibfnamefont{J. W.}~\bibnamefont{Taylor}},
  \bibinfo{author}{\bibfnamefont{Y.}~\bibnamefont{Qiu}},
  \bibinfo{author}{\bibfnamefont{Q.}~\bibnamefont{Huang}},
  \bibinfo{author}{\bibfnamefont{S. W.}~\bibnamefont{Cheong}},
  \bibinfo{author}{\bibfnamefont{C.}~\bibnamefont{Broholm}}, and
  \bibinfo{author}{\bibfnamefont{T.}~\bibnamefont{Perring}},
  \bibinfo{year}{2008},
  \bibinfo{journal}{Phys. Rev. B.} \textbf{\bibinfo{volume}{77}},
  \bibinfo{pages}{020405}.




 \bibitem[{\citenamefont{Velasco \emph{et~al.}}(2002)}]
{Velasco:2002}
\bibinfo{author}{\bibnamefont{Velasco}, \bibfnamefont{P.}},
  \bibinfo{author}{\bibfnamefont{J. A.}~\bibnamefont{Alonso}},
  \bibinfo{author}{\bibfnamefont{M. T.}~\bibnamefont{Casais}},
  \bibinfo{author}{\bibfnamefont{M. J.}~\bibnamefont{Mart\'inez-Lope}},
  \bibinfo{author}{\bibfnamefont{J. L.}~\bibnamefont{Mart\'inez}}, and
  \bibinfo{author}{\bibfnamefont{M. T.}~\bibnamefont{Fern\'andez-D\'iaz}},
  \bibinfo{year}{2002},
  \bibinfo{journal}{Phys. Rev. B} \textbf{\bibinfo{volume}{66}},
  \bibinfo{pages}{174408}.


%
%\bibitem[{\citenamefont{Vernay} \emph{et~al.}(2004)}]
%{Vernay:2004}
%  \bibinfo{author}{\bibfnamefont{Vernay}, \bibnamefont{F.}},
%  \bibinfo{author}{\bibnamefont{K.}~\bibfnamefont{Penc}},
%  \bibinfo{author}{\bibnamefont{P.}~\bibfnamefont{Fazekas}},  and
%  \bibinfo{author}{\bibnamefont{F.}~\bibfnamefont{Mila}},
%  \bibinfo{year}{2004},
%  \bibinfo{journal}{Phys. Rev. B} \textbf{\bibinfo{volume}{70}},
%  \bibinfo{pages}{014428}.



%ODD Rule
\bibitem[{\citenamefont{Villain}(1977)}]
{Villain:1977}
  \bibinfo{author}{\bibfnamefont{Villain}, \bibnamefont{J.}},
  \bibinfo{year}{1977},
  \bibinfo{journal}{J. Phys. C: Solid State Phys} \textbf{\bibinfo{volume}{10}},
  \bibinfo{pages}{1717}.




\bibitem[{\citenamefont{Villain}(1979)}]
{Villain:1979}
  \bibinfo{author}{\bibfnamefont{Villain}, \bibnamefont{J.}},
  \bibinfo{year}{1979},
  \bibinfo{journal}{Z. Phys. B} \textbf{\bibinfo{volume}{33}},
  \bibinfo{pages}{31}.


\bibitem[{\citenamefont{Wang and Sleight}(1998)}]
{Wang:1998}
\bibinfo{author}{\bibnamefont{Wang},~\bibfnamefont{R.}} and
\bibinfo{author}{\bibfnamefont{A. W.}~\bibnamefont{Sleight}},
 \bibinfo{year}{1998},
 \bibinfo{journal}{Mater. Res. Bull.} \textbf{\bibinfo{volume}{33}},
 \bibinfo{pages}{1005}.



\bibitem[{\citenamefont{Wanklyn}(1968)}]
{Wanklyn:1968}
\bibinfo{author}{\bibnamefont{Wanklyn},~\bibfnamefont{B.}},
 \bibinfo{year}{1968},
 \bibinfo{journal}{ J. Mater. Sci.} \textbf{\bibinfo{volume}{3}},
 \bibinfo{pages}{395}.



\bibitem[{\citenamefont{Wannier}(1950)}]
{Wannier:1950}
  \bibinfo{author}{\bibfnamefont{Wannier}, \bibnamefont{G. H.}},
  \bibinfo{year}{1950},
  \bibinfo{journal}{Phys. Rev.} \textbf{\bibinfo{volume}{79}},
  \bibinfo{pages}{357}.



\bibitem[{\citenamefont{Wannier}(1973)}]
{Wannier:1973}
  \bibinfo{author}{\bibfnamefont{Wannier}, \bibnamefont{G. H.}},
  \bibinfo{year}{1973},
  \bibinfo{journal}{Phys. Rev. B} \textbf{\bibinfo{volume}{7}},
  \bibinfo{pages}{5017}.



\bibitem[{\citenamefont{White} \emph{et~al.}(1993)}]
 {White:1993}
   \bibinfo{author}{\bibfnamefont{White}, \bibnamefont{S. J.}},
   \bibinfo{author}{\bibnamefont{M. R.}~\bibfnamefont{Roser}},
   \bibinfo{author}{\bibnamefont{J.}~\bibfnamefont{Xu}},
     \bibinfo{author}{\bibnamefont{J. T.}~\bibfnamefont{van der Norrdaa}}, and
   \bibinfo{author}{\bibnamefont{L. R.}~\bibfnamefont{Corruccini}},
  \bibinfo{year}{1993},
   \bibinfo{journal}{Phys. Rev. Lett.} \textbf{\bibinfo{volume}{71}},
   \bibinfo{pages}{3553}.


%
%\bibitem[{\citenamefont{Wiebe} \emph{et~al.}(2002)}]
%{Wiebe:SCaReO2002}
%  \bibinfo{author}{\bibfnamefont{Wiebe}, \bibnamefont{C. R.}},
%  \bibinfo{author}{\bibnamefont{J. E.}~\bibfnamefont{Greedan}},
%  \bibinfo{author}{\bibnamefont{G. M.}~\bibfnamefont{Luke}},  and
%  \bibinfo{author}{\bibnamefont{J. S.}~\bibfnamefont{Gardner}},
% \bibinfo{year}{2002},
%  \bibinfo{journal}{Phys. Rev. B.} \textbf{\bibinfo{volume}{65}},
%  \bibinfo{pages}{144413}.

%

%\bibitem[{\citenamefont{Wiebe} \emph{et~al.}(2003)}]
%{Wiebe:SMgReO2003}
%  \bibinfo{author}{\bibfnamefont{Wiebe}, \bibnamefont{C. R.}},
%  \bibinfo{author}{\bibnamefont{J. E.}~\bibfnamefont{Greedan}},
%  \bibinfo{author}{\bibnamefont{P. P.}~\bibfnamefont{Kyriakou}},
%  \bibinfo{author}{\bibnamefont{G. M.}~\bibfnamefont{Luke}},
%  \bibinfo{author}{\bibnamefont{J. S.}~\bibfnamefont{Gardner}},
%  \bibinfo{author}{\bibnamefont{A.}~\bibfnamefont{Fuyaka}},
%  \bibinfo{author}{\bibnamefont{I. M.}~\bibfnamefont{Gat-Malureanu}},
%  \bibinfo{author}{\bibnamefont{P. L.}~\bibfnamefont{Russo}},
%  \bibinfo{author}{\bibnamefont{A. T.}~\bibfnamefont{Savici}}, and
%  \bibinfo{author}{\bibnamefont{Y. J.}~\bibfnamefont{Uemura}},
% \bibinfo{year}{2003},
%  \bibinfo{journal}{Phys. Rev. B.} \textbf{\bibinfo{volume}{68}},
%  \bibinfo{pages}{134410}.





\bibitem[{\citenamefont{Wiebe \emph{et~al.}}(2004)}]
{Wiebe:2004}
\bibinfo{author}{\bibnamefont{Wiebe}, \bibfnamefont{C. R.}},
  \bibinfo{author}{\bibfnamefont{J. S.}~\bibnamefont{Gardner}},
  \bibinfo{author}{\bibfnamefont{S.-J.}~\bibnamefont{Kim}},
  \bibinfo{author}{\bibfnamefont{G. M.}~\bibnamefont{Luke}},
  \bibinfo{author}{\bibfnamefont{A. S.}~\bibnamefont{Wills}},
  \bibinfo{author}{\bibfnamefont{B. D.}~\bibnamefont{Gaulin}},
  \bibinfo{author}{\bibfnamefont{J. E.}~\bibnamefont{Greedan}},
  \bibinfo{author}{\bibfnamefont{I.}~\bibnamefont{Swainson}},
  \bibinfo{author}{\bibfnamefont{Y.}~\bibnamefont{Qiu}}, and
  \bibinfo{author}{\bibfnamefont{C. Y.}~\bibnamefont{Jones}},
  \bibinfo{year}{2004},
  \bibinfo{journal}{Phys. Rev. Lett.} \textbf{\bibinfo{volume}{93}},
  \bibinfo{pages}{076403}.



%
%\bibitem[{\citenamefont{Wills} \emph{et~al.}(2000)}]
%{Wills:2000}
%  \bibinfo{author}{\bibfnamefont{Wills}, \bibnamefont{A. S.}},
%  \bibinfo{author}{\bibnamefont{V.}~\bibfnamefont{Dupuis}},
%  \bibinfo{author}{\bibnamefont{E.}~\bibfnamefont{Vincent}},
%  \bibinfo{author}{\bibnamefont{J.}~\bibfnamefont{Hammann}},  and
%  \bibinfo{author}{\bibnamefont{R.}~\bibfnamefont{Calemczuk}},
% \bibinfo{year}{2000},
%  \bibinfo{journal}{Phys. Rev. B.} \textbf{\bibinfo{volume}{62}},
%  \bibinfo{pages}{R9264}.



\bibitem[{\citenamefont{Wills \emph{et~al.}}(2002)}]
{Wills:2002}
\bibinfo{author}{\bibnamefont{Wills}, \bibfnamefont{A. S.}},
  \bibinfo{author}{\bibfnamefont{R.}~\bibnamefont{Ballou}}, and
  \bibinfo{author}{\bibfnamefont{C.}~\bibnamefont{Lacroix}},
  \bibinfo{year}{2002},
  \bibinfo{journal}{Phys. Rev. B.} \textbf{\bibinfo{volume}{66}},
  \bibinfo{pages}{144407}.



\bibitem[{\citenamefont{Wills \emph{et~al.}}(2006)}]
{Wills:2006}
\bibinfo{author}{\bibnamefont{Wills}, \bibfnamefont{A. S.}},
  \bibinfo{author}{\bibfnamefont{M. E.}~\bibnamefont{Zhitomirsky}},
  \bibinfo{author}{\bibfnamefont{B.}~\bibnamefont{Canals}},
  \bibinfo{author}{\bibfnamefont{J. P.}~\bibnamefont{Sanchez}},
  \bibinfo{author}{\bibfnamefont{P.}~\bibnamefont{Bonville}},
  \bibinfo{author}{\bibfnamefont{P.}~\bibnamefont{ Dalmas de R\'eotier}} and
  \bibinfo{author}{\bibfnamefont{A.}~\bibnamefont{Yaouanc}},
  \bibinfo{year}{2006},
  \bibinfo{journal}{J. Phys.: Condens. Matter} \textbf{\bibinfo{volume}{18}},
    \bibinfo{pages}{L37}.


\bibitem[{\citenamefont{Wu} \emph{et~al.}(1993)}]
{Wu:1993}
  \bibinfo{author}{\bibfnamefont{Wu}, \bibnamefont{W.}},
  \bibinfo{author}{\bibnamefont{D.}~\bibfnamefont{Bitko}}, 
    \bibinfo{author}{\bibnamefont{T. F.}~\bibfnamefont{Rosenbaum}}, and
\bibinfo{author}{\bibnamefont{G.}~\bibfnamefont{Aeppli}},
 \bibinfo{year}{1993},
  \bibinfo{journal}{Phys. Rev. Lett. } \textbf{\bibinfo{volume}{71}},
  \bibinfo{pages}{1919}.



\bibitem[{\citenamefont{Yafet and Kittel}(1952)}]
{Yafet:1952}
\bibinfo{author}{\bibnamefont{Yafet}, \bibfnamefont{Y.}} and
\bibinfo{author}{\bibfnamefont{C.}~\bibnamefont{Kittel}},
\bibinfo{year}{1952},
\bibinfo{journal}{Phys. Rev.} \textbf{\bibinfo{volume}{87}},
\bibinfo{pages}{290}.



\bibitem[{\citenamefont{Yamamoto \emph{et~al.}}(2007)}]
{Yamamoto:2007}
\bibinfo{author}{\bibnamefont{Yamamoto}, \bibfnamefont{A.}},
  \bibinfo{author}{\bibfnamefont{P. A.}~\bibnamefont{Sharma}},
  \bibinfo{author}{\bibfnamefont{Y.}~\bibnamefont{Okamoto}},
  \bibinfo{author}{\bibfnamefont{A.}~\bibnamefont{Nakao}},
  \bibinfo{author}{\bibfnamefont{H. A.}~\bibnamefont{Katori}},
  \bibinfo{author}{\bibfnamefont{S.}~\bibnamefont{Niitaka}},
  \bibinfo{author}{\bibfnamefont{D. }~\bibnamefont{Hashizume}}, and
  \bibinfo{author}{\bibfnamefont{H. }~\bibnamefont{Takagi}},
  \bibinfo{year}{2007},
  \bibinfo{journal}{J. Phys. Soc. Jpn.} \textbf{\bibinfo{volume}{76}},
  \bibinfo{pages}{043703}.




%\bibitem[{\citenamefont{Yamamura} \emph{et~al.}(2006)}]
%{Yamamura:2006}
%  \bibinfo{author}{\bibfnamefont{Yamamura}, \bibnamefont{K.}},
%  \bibinfo{author}{\bibnamefont{M.}~\bibfnamefont{Wakeshima}}, and
%  \bibinfo{author}{\bibnamefont{Y.}~\bibfnamefont{Hinatsu}},
% \bibinfo{year}{2006},
%  \bibinfo{journal}{J. Solid State Chem.} \textbf{\bibinfo{volume}{179}},
%  \bibinfo{pages}{605}.




\bibitem[{\citenamefont{Yamashita and Ueda}(2000)}]
{Yamashita:2000}
  \bibinfo{author}{\bibfnamefont{Yamashita}, \bibnamefont{Y.}}, and
  \bibinfo{author}{\bibnamefont{K.}~\bibfnamefont{Ueda}}, 
 \bibinfo{year}{2000},
  \bibinfo{journal}{Phys. Rev. Lett. } \textbf{\bibinfo{volume}{85}},
  \bibinfo{pages}{4960}.



%
%\bibitem[{\citenamefont{Yamaura and Hiroi}(2002)}]
%{Yamaura:2002}
%\bibinfo{author}{\bibnamefont{Yamaura},~\bibfnamefont{J.-I.}} and
%\bibinfo{author}{\bibfnamefont{Z.}~\bibnamefont{Hiroi}},
% \bibinfo{year}{2002},
% \bibinfo{journal}{J. Phys. Soc. Jpn.} \textbf{\bibinfo{volume}{71}},
%% \bibinfo{pages}{2598}.

%

%
%\bibitem[{\citenamefont{Yamaura \emph{et~al.}}(2006)}]
%{Yamaura:2006}
%\bibinfo{author}{\bibnamefont{Yamaura }, \bibfnamefont{J.-I.}},
%  \bibinfo{author}{\bibfnamefont{S.}~\bibnamefont{Yonezawa}},
%  \bibinfo{author}{\bibfnamefont{Y.}~\bibnamefont{Muraoka}}, and
%  \bibinfo{author}{\bibfnamefont{Z.}~\bibnamefont{Hiroi}},
%  \bibinfo{year}{2006},
%  \bibinfo{journal}{J. Solid State Chem.} \textbf{\bibinfo{volume}{179}},
%  \bibinfo{pages}{336}.





\bibitem[{\citenamefont{Yanagishima and Maeno}(2001)}]
{Yanagishima:2001}
\bibinfo{author}{\bibnamefont{Yanagishima},~\bibfnamefont{D.}} and
\bibinfo{author}{\bibfnamefont{Y.}~\bibnamefont{Maeno}},
 \bibinfo{year}{2001},
 \bibinfo{journal}{J. Phys. Soc. Jpn.} \textbf{\bibinfo{volume}{70}},
 \bibinfo{pages}{2880}.




\bibitem[{\citenamefont{Yaouanc, \emph{et~al.}}(2005)}]
{Yaouanc:2005}
\bibinfo{author}{\bibnamefont{Yaouanc}, \bibfnamefont{A.}},
  \bibinfo{author}{\bibfnamefont{P.}~\bibnamefont{Dalmas de R\'{e}otier}},
  \bibinfo{author}{\bibfnamefont{V.}~\bibnamefont{Glazkov}},
  \bibinfo{author}{\bibfnamefont{C.}~\bibnamefont{Marin}},
  \bibinfo{author}{\bibfnamefont{P.}~\bibnamefont{Bonville}},
  \bibinfo{author}{\bibfnamefont{J. A.}~\bibnamefont{Hodges}},
  \bibinfo{author}{\bibfnamefont{P. C. M.}~\bibnamefont{Gubbens}},
  \bibinfo{author}{\bibfnamefont{S.}~\bibnamefont{Sakarya}}, and
  \bibinfo{author}{\bibfnamefont{C.}~\bibnamefont{Baines}},
  \bibinfo{year}{2005},
  \bibinfo{journal}{Phys. Rev. Lett.} \textbf{\bibinfo{volume}{95}},
  \bibinfo{pages}{047203}.




\bibitem[{\citenamefont{Yasui \emph{et~al.}}(2001)}]
{Yasui:2001}
\bibinfo{author}{\bibnamefont{Yasui}, \bibfnamefont{Y.}},
  \bibinfo{author}{\bibfnamefont{Y.}~\bibnamefont{Kondo}},
  \bibinfo{author}{\bibfnamefont{M.}~\bibnamefont{Kanada}},
  \bibinfo{author}{\bibfnamefont{M.}~\bibnamefont{Ito}},
  \bibinfo{author}{\bibfnamefont{H.}~\bibnamefont{Harashina}},
  \bibinfo{author}{\bibfnamefont{M.}~\bibnamefont{Sato}}, and
  \bibinfo{author}{\bibfnamefont{K.}~\bibnamefont{Kakurai}},
  \bibinfo{year}{2001},
  \bibinfo{journal}{J. Phys. Soc. Japan} \textbf{\bibinfo{volume}{70}},
  \bibinfo{pages}{284}.




  \bibitem[{\citenamefont{Yasui \emph{et~al.}}(2002)}]
{Yasui:2002}
\bibinfo{author}{\bibnamefont{Yasui}, \bibfnamefont{Y.}},
  \bibinfo{author}{\bibfnamefont{M.}~\bibnamefont{Kanada}},
  \bibinfo{author}{\bibfnamefont{M.}~\bibnamefont{Ito}},
  \bibinfo{author}{\bibfnamefont{H.}~\bibnamefont{Harashina}},
  \bibinfo{author}{\bibfnamefont{M.}~\bibnamefont{Sato}},
  \bibinfo{author}{\bibfnamefont{H.}~\bibnamefont{Okumura}},
  \bibinfo{author}{\bibfnamefont{K.}~\bibnamefont{Kakurai}}, and
  \bibinfo{author}{\bibfnamefont{H.}~\bibnamefont{Kadowaki}},
  \bibinfo{year}{2002},
  \bibinfo{journal}{J. Phys. Soc. Jpn.} \textbf{\bibinfo{volume}{71}},
  \bibinfo{pages}{599}.



%Yb2Ti2O7
\bibitem[{\citenamefont{Yasui \emph{et~al.}}(2003)}]
{Yasui:2003}
\bibinfo{author}{\bibnamefont{Yasui}, \bibfnamefont{Y.}},
  \bibinfo{author}{\bibfnamefont{M.}~\bibnamefont{Soda}},
  \bibinfo{author}{\bibfnamefont{S.}~\bibnamefont{Iikubo}},
  \bibinfo{author}{\bibfnamefont{M.}~\bibnamefont{Ito}},
  \bibinfo{author}{\bibfnamefont{M.}~\bibnamefont{Sato}},
  \bibinfo{author}{\bibfnamefont{N.}~\bibnamefont{Hamaguchi}},
  \bibinfo{author}{\bibfnamefont{T.}~\bibnamefont{Matsushita}},
  \bibinfo{author}{\bibfnamefont{N.}~\bibnamefont{Wada}},
  \bibinfo{author}{\bibfnamefont{T.}~\bibnamefont{Takeuchi}},
  \bibinfo{author}{\bibfnamefont{N.}~\bibnamefont{Aso}}, and
  \bibinfo{author}{\bibfnamefont{K.}~\bibnamefont{Kakurai}},
  \bibinfo{year}{2003},
  \bibinfo{journal}{J. Phys. Soc. Jpn.} \textbf{\bibinfo{volume}{72}},
  \bibinfo{pages}{3014}.



%Nd2Mo2O7

\bibitem[{\citenamefont{Yasui \emph{et~al.}}(2003a)}]
{Yasui:2003a}
\bibinfo{author}{\bibnamefont{Yasui}, \bibfnamefont{Y.}},
  \bibinfo{author}{\bibfnamefont{S.}~\bibnamefont{Iikubo}},
  \bibinfo{author}{\bibfnamefont{H.}~\bibnamefont{Harashina}},
  \bibinfo{author}{\bibfnamefont{T.}~\bibnamefont{Kageyama}},
  \bibinfo{author}{\bibfnamefont{M.}~\bibnamefont{Ito}},
  \bibinfo{author}{\bibfnamefont{M.}~\bibnamefont{Sato}}, and
  \bibinfo{author}{\bibfnamefont{K.}~\bibnamefont{Kakurai}},
  \bibinfo{year}{2003a},
  \bibinfo{journal}{J. Phys. Soc. Jpn.} \textbf{\bibinfo{volume}{72}},
  \bibinfo{pages}{865}.



\bibitem[{\citenamefont{Yasui \emph{et~al.}}(2006)}]
{Yasui:2006}
\bibinfo{author}{\bibnamefont{Yasui}, \bibfnamefont{Y.}},
  \bibinfo{author}{\bibfnamefont{T.}~\bibnamefont{Kageyama}},
  \bibinfo{author}{\bibfnamefont{T.}~\bibnamefont{Moyoshi}},
  \bibinfo{author}{\bibfnamefont{H.}~\bibnamefont{Harashina}},
  \bibinfo{author}{\bibfnamefont{M.}~\bibnamefont{Soda}},
  \bibinfo{author}{\bibfnamefont{M.}~\bibnamefont{Sato}}, and
  \bibinfo{author}{\bibfnamefont{K.}~\bibnamefont{Kakurai}},
  \bibinfo{year}{2006},
  \bibinfo{journal}{J. Phys. Soc. Jpn.} \textbf{\bibinfo{volume}{75}},
  \bibinfo{pages}{084711}.



\bibitem[{\citenamefont{Yavors'kii} \emph{et~al.}(2006)}]
{Yavorskii:2006}
  \bibinfo{author}{\bibfnamefont{Yavors'kii}, \bibnamefont{T.}},
  \bibinfo{author}{\bibnamefont{M.}~\bibfnamefont{Enjalran}},  and
  \bibinfo{author}{\bibnamefont{M. J. P.}~\bibfnamefont{Gingras}},
 \bibinfo{year}{2006},
  \bibinfo{journal}{Phys. Rev. Lett.} \textbf{\bibinfo{volume}{97}},
  \bibinfo{pages}{267203}.



\bibitem[{\citenamefont{Yavors'kii} \emph{et~al.}(2007)}]
{Yavorskii:2007}
  \bibinfo{author}{\bibfnamefont{Yavors'kii}, \bibnamefont{T.}},
  \bibinfo{author}{\bibnamefont{W.}~\bibfnamefont{Apel}},  and
  \bibinfo{author}{\bibnamefont{H.-U.}~\bibfnamefont{Everts}},
 \bibinfo{year}{2007},
  \bibinfo{journal}{Phys. Rev. B.} \textbf{\bibinfo{volume}{76}},
  \bibinfo{pages}{064430}.



%reCITE

\bibitem[{\citenamefont{Yavors'kii}  \emph{et~al.}(2008)}]
{Yavorskii-hexagons}
  \bibinfo{author}{\bibfnamefont{Yavors'kii}, \bibnamefont{T.}}, 
  \bibinfo{author}{\bibnamefont{T.}~\bibfnamefont{Fennell}},
  \bibinfo{author}{\bibnamefont{M. J. P.}~\bibfnamefont{Gingras}}, and
  \bibinfo{author}{\bibnamefont{S. T.}~\bibfnamefont{Bramwell}},
\bibinfo{year}{2008},
  \bibinfo{journal}{Phys. Rev. Lett.} \textbf{\bibinfo{volume}{101}},
  \bibinfo{pages}{037204}.


\bibitem[{\citenamefont{Yonezawa \emph{et~al.}}(2004)}]
{Yonezawa:2004}
\bibinfo{author}{\bibnamefont{Yonezawa}, \bibfnamefont{S.}},
  \bibinfo{author}{\bibfnamefont{Y.}~\bibnamefont{Muraoka}},
  \bibinfo{author}{\bibfnamefont{Y.}~\bibnamefont{Matsushita}}, and
  \bibinfo{author}{\bibfnamefont{Z.}~\bibnamefont{Hiroi}},
  \bibinfo{year}{2004},
  \bibinfo{journal}{J. Phys.: Condens. Matter} \textbf{\bibinfo{volume}{16}},
  \bibinfo{pages}{L9}.




\bibitem[{\citenamefont{Yonezawa \emph{et~al.}}(2004a)}]
{Yonezawa:2004a}
\bibinfo{author}{\bibnamefont{Yonezawa}, \bibfnamefont{S.}},
  \bibinfo{author}{\bibfnamefont{Y.}~\bibnamefont{Muraoka}},
  \bibinfo{author}{\bibfnamefont{Y.}~\bibnamefont{Matsushita}}, and
  \bibinfo{author}{\bibfnamefont{Z.}~\bibnamefont{Hiroi}},
  \bibinfo{year}{2004a},
  \bibinfo{journal}{J. Phys. Soc. Jpn.} \textbf{\bibinfo{volume}{73}},
  \bibinfo{pages}{819}.



\bibitem[{\citenamefont{Yonezawa \emph{et~al.}}(2004b)}]
{Yonezawa:2004b}
\bibinfo{author}{\bibnamefont{Yonezawa}, \bibfnamefont{S.}},
  \bibinfo{author}{\bibfnamefont{Y.}~\bibnamefont{Muraoka}}, and
  \bibinfo{author}{\bibfnamefont{Z.}~\bibnamefont{Hiroi}}  \bibinfo{year}{2004b},
  \bibinfo{journal}{J. Phys. Soc. Jpn.} \textbf{\bibinfo{volume}{73}},
  \bibinfo{pages}{1655}.



\bibitem[{\citenamefont{Yoshida} \emph{et~al.}(2004)}]
{Yoshida:2004}
  \bibinfo{author}{\bibfnamefont{Yoshida}, \bibnamefont{S.-i.}},
  \bibinfo{author}{\bibnamefont{K.}~\bibfnamefont{Nemoto}}, and
  \bibinfo{author}{\bibnamefont{K.}~\bibfnamefont{Wada}},
 \bibinfo{year}{2004},
  \bibinfo{journal}{J. Phys. Soc. Jpn.} \textbf{\bibinfo{volume}{73}},
  \bibinfo{pages}{1619}.




%
%\bibitem[{\citenamefont{Yoshida \emph{et~al.}}(2007)}]
%{Yoshida:2007}
%\bibinfo{author}{\bibnamefont{Yoshida}, \bibfnamefont{M.}},
%  \bibinfo{author}{\bibfnamefont{K.}~\bibnamefont{Arai}},
%  \bibinfo{author}{\bibfnamefont{R.}~\bibnamefont{Kaido}},
%  \bibinfo{author}{\bibfnamefont{M.}~\bibnamefont{Takigawa}},
%  \bibinfo{author}{\bibfnamefont{S.}~\bibnamefont{Yonezawa}},
%  \bibinfo{author}{\bibfnamefont{Y.}~\bibnamefont{Muraoka}}, and
%  \bibinfo{author}{\bibfnamefont{Z.}~\bibnamefont{Hiroi}},
%  \bibinfo{year}{2007},
%  \bibinfo{journal}{Phys. Rev. Lett.} \textbf{\bibinfo{volume}{98}},
%  \bibinfo{pages}{197002}.



\bibitem[{\citenamefont{Yoshii and Sato}(1999)}]
{Yoshii:1999}
\bibinfo{author}{\bibnamefont{Yoshii}, \bibfnamefont{S.}} and
\bibinfo{author}{\bibfnamefont{M.}~\bibnamefont{Sato}},
\bibinfo{year}{1999},
\bibinfo{journal}{J. Phys. Soc. Jpn.} \textbf{\bibinfo{volume}{68}},
\bibinfo{pages}{3034}.



\bibitem[{\citenamefont{Yoshii \emph{et~al.}}(2000)}]
{Yoshii:2000}
\bibinfo{author}{\bibnamefont{Yoshii}, \bibfnamefont{S.}},
 \bibinfo{author}{\bibfnamefont{S.}~\bibnamefont{Iikubo}},
 \bibinfo{author}{\bibfnamefont{T.}~\bibnamefont{Kageyama}},
  \bibinfo{author}{\bibfnamefont{K.}~\bibnamefont{Oda}},
    \bibinfo{author}{\bibfnamefont{Y.}~\bibnamefont{Kondo}},
  \bibinfo{author}{\bibfnamefont{K.}~\bibnamefont{Murata}}, and
  \bibinfo{author}{\bibfnamefont{M.}~\bibnamefont{Sato}},
  \bibinfo{year}{2000},
  \bibinfo{journal}{J. Phys. Soc. Jpn.} \textbf{\bibinfo{volume}{69}},
  \bibinfo{pages}{3777}.



  \bibitem[{\citenamefont{Zhang \emph{et~al.}}(2007)}]
{Zhang:2007}
\bibinfo{author}{\bibnamefont{Zhang}, \bibfnamefont{F. X.}},
  \bibinfo{author}{\bibfnamefont{J.}~\bibnamefont{Lian}},
  \bibinfo{author}{\bibfnamefont{U.}~\bibnamefont{Becker}},
  \bibinfo{author}{\bibfnamefont{L. M.}~\bibnamefont{Wang}},
  \bibinfo{author}{\bibfnamefont{J. Z.}~\bibnamefont{Hu}},
  \bibinfo{author}{\bibfnamefont{S.}~\bibnamefont{Saxena}}, and
  \bibinfo{author}{\bibfnamefont{R. C.}~\bibnamefont{Ewing}},
  \bibinfo{year}{2007},
  \bibinfo{journal}{Chem. Phys. Lett.} \textbf{\bibinfo{volume}{441}},
  \bibinfo{pages}{216}.



\bibitem[{\citenamefont{Zhang \emph{et~al.}}(2006)}]
{Zhang:2006}
\bibinfo{author}{\bibnamefont{Zhang}, \bibfnamefont{F. X.}},
  \bibinfo{author}{\bibfnamefont{J.}~\bibnamefont{Lian}},
  \bibinfo{author}{\bibfnamefont{U.}~\bibnamefont{Becker}},
  \bibinfo{author}{\bibfnamefont{R. C.}~\bibnamefont{Ewing}},
  \bibinfo{author}{\bibfnamefont{L. M.}~\bibnamefont{Wang}},
   \bibinfo{author}{\bibfnamefont{L. A.}~\bibnamefont{Boatner}},
  \bibinfo{author}{\bibfnamefont{J. Z.}~\bibnamefont{Hu}},  and
  \bibinfo{author}{\bibfnamefont{S.}~\bibnamefont{Saxena}},
  \bibinfo{year}{2006},
  \bibinfo{journal}{Phys. Rev. B} \textbf{\bibinfo{volume}{74}},
  \bibinfo{pages}{174116}.




\bibitem[{\citenamefont{Zhou \emph{et~al.}}(2007)}]
{Zhou:2007}
\bibinfo{author}{\bibnamefont{Zhou}, \bibfnamefont{H. D.}},
  \bibinfo{author}{\bibfnamefont{C. R.}~\bibnamefont{Wiebe}},
  \bibinfo{author}{\bibfnamefont{Y. J.}~\bibnamefont{Jo}},
  \bibinfo{author}{\bibfnamefont{L.}~\bibnamefont{Balicas}},
  \bibinfo{author}{\bibfnamefont{Y.}~\bibnamefont{Qiu}},
  \bibinfo{author}{\bibfnamefont{J. R. D.}~\bibnamefont{Copley}},
  \bibinfo{author}{\bibfnamefont{G.}~\bibnamefont{Ehlers}},
  \bibinfo{author}{\bibfnamefont{P.}~\bibnamefont{Fouquet}},and
  \bibinfo{author}{\bibfnamefont{J. S.}~\bibnamefont{Gardner}},
  \bibinfo{year}{2007},
  \bibinfo{journal}{J. Phys.: Condens. Matter} \textbf{\bibinfo{volume}{10}},
  \bibinfo{pages}{342201}.




%
%\bibitem[{\citenamefont{Zhou \emph{et~al.}}(2007a)}]
%{Zhou:2007a}
%\bibinfo{author}{\bibnamefont{Zhou}, \bibfnamefont{H. D.}},
%  \bibinfo{author}{\bibfnamefont{B. W.}~\bibnamefont{Vogt}},
%  \bibinfo{author}{\bibfnamefont{J. A.}~\bibnamefont{Janik}},
%  \bibinfo{author}{\bibfnamefont{Y. J.}~\bibnamefont{Jo}},
%  \bibinfo{author}{\bibfnamefont{L.}~\bibnamefont{Balicas}},
%  \bibinfo{author}{\bibfnamefont{Y.}~\bibnamefont{Qiu}},
%  \bibinfo{author}{\bibfnamefont{J. R. D.}~\bibnamefont{Copley}},
%   \bibinfo{author}{\bibfnamefont{J. S.}~\bibnamefont{Gardner}}, and
% \bibinfo{author}{\bibfnamefont{C. R.}~\bibnamefont{Wiebe}},
%  \bibinfo{year}{2007a},
%  \bibinfo{journal}{Phys. Rev. Lett.} \textbf{\bibinfo{volume}{99}},
%  \bibinfo{pages}{236401}.




\end{thebibliography}

%\end{document}

\end{document}